\documentclass[12pt,a4paper,twoside,final]{book}

\usepackage[BCOR=10mm,
  DIV=18,
  headinclude=true,
  footinclude=false]{typearea}
\usepackage[electronic]{optional}
\usepackage[T1]{fontenc}
\usepackage[latin1]{inputenc}
\usepackage[ngerman,english]{babel}
\usepackage{titlesec}
\usepackage{fancyhdr}
\usepackage{graphicx}
\usepackage[bookmarks,hidelinks]{hyperref}
\usepackage{amsmath,amsthm,amsfonts,amssymb}
\allowdisplaybreaks
\usepackage{cite}
\usepackage{empheq}
\usepackage{mathdots}
\usepackage{colortbl}
\usepackage[nottoc,notlot,notlof]{tocbibind}
\DeclareMathOperator{\Tr}{Tr}
\DeclareMathOperator{\sign}{sign}
\DeclareMathOperator{\diag}{diag}
\DeclareMathOperator{\adj}{ad}
\DeclareMathOperator{\rank}{rank}

\newcommand{\compliI}{{\scriptscriptstyle\mathrm I}}
\newcommand{\compliR}{{\scriptscriptstyle\mathrm R}}
\usepackage{dsfont}
\usepackage{tikz}
\usetikzlibrary{calc,shapes,arrows,patterns,positioning,decorations.markings,trees}
\tikzset{>=stealth'}
\tikzstyle point=[minimum size=1mm,inner sep=0pt,outer sep=0pt,shape=circle,fill=black]
\newcommand{\HIDDEN}[1]{}
\usepackage[textsize=small,shadow]{todonotes}
\usepackage{ulem}

\definecolor{fillcolor}{HTML}{C4C4C4}
\definecolor{plotgray}{HTML}{626262}

\makeatletter
\renewcommand\chapter{\if@openright\cleardoublepage\else\clearpage\fi
                    \thispagestyle{empty}
                    \global\@topnum\z@
                    \@afterindentfalse
                    \secdef\@chapter\@schapter}
\makeatother
\titleformat{\chapter}[display]{\normalfont\huge\bfseries}{\chaptertitlename\ \thechapter}{20pt}{\Huge}

\newcommand{\leadingzero}[1]{\ifnum #1<10 0\the#1\else\the#1\fi}
\newcommand{\LMUTitle}[9]{%
\opt{electronic}{%
  \vspace*{\stretch{1}}
  {\parindent0cm
   \rule{\linewidth}{.5ex}}
   \begin{center}

    \vspace*{\stretch{0.4}}
    \sffamily\bfseries\Huge
    #1\\
    \vspace*{\stretch{0.75}}
    \sffamily\bfseries\Large
    #2
    \vspace*{\stretch{0.4}}
  \end{center}
  \rule{\linewidth}{.5ex}
  \vspace*{\stretch{5}}
  \begin{center}
    \includegraphics[width=2in]{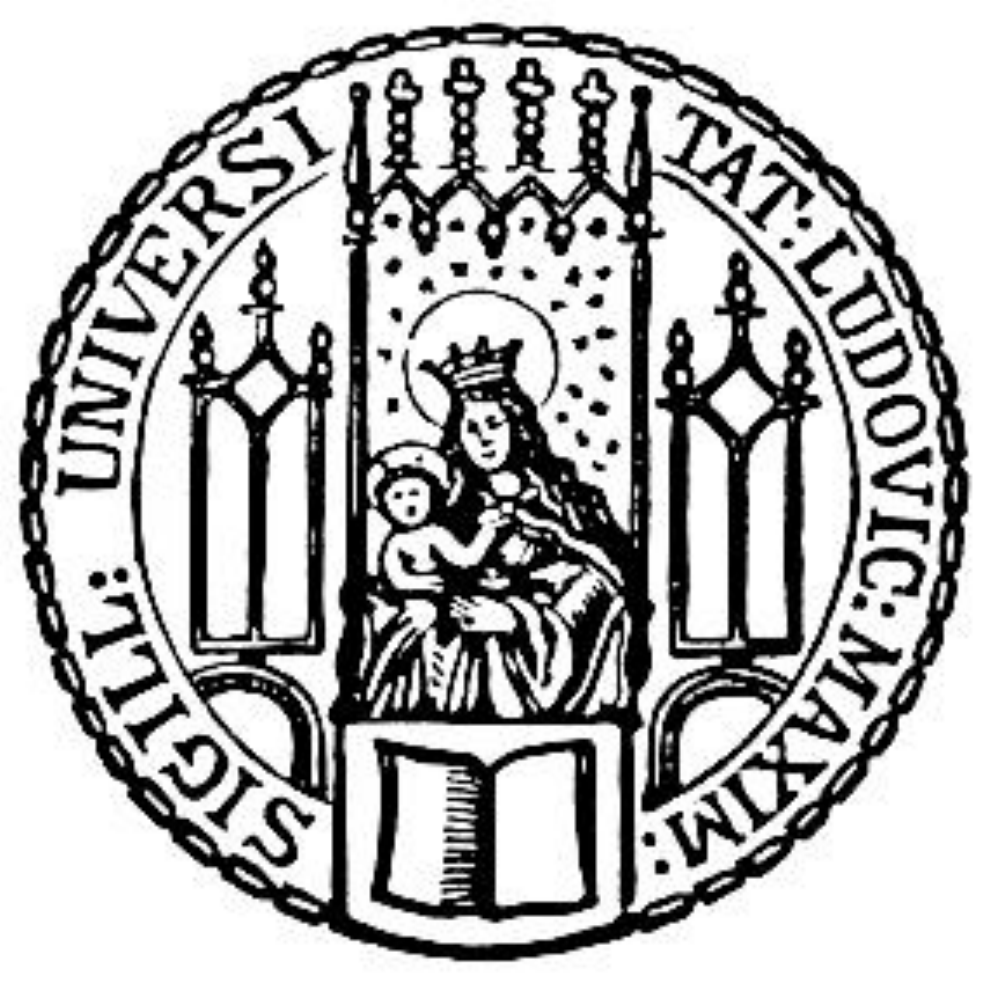}
  \end{center}
  \vspace*{\stretch{1}}
  \begin{center}\sffamily\Large{#5}\end{center}
  \cleardoublepage}%
  \vspace*{\stretch{1}}
  {\parindent0cm
  \rule{\linewidth}{.5ex}}
  \begin{center}
    \vspace*{\stretch{0.4}}
    \sffamily\bfseries\Huge
    #1\\
    \vspace*{\stretch{0.75}}
    \sffamily\bfseries\Large
    #2
    \vspace*{\stretch{0.4}}
  \end{center}
  \rule{\linewidth}{.5ex}

  \vspace*{\stretch{3}}
  \begin{center}
    \Large Dissertation\\
    \Large an der #4\\
    \Large der Ludwig--Maximilians--Universität\\
    \Large M\"unchen\\
    \vspace*{\stretch{1}}
    \Large vorgelegt von\\
    \Large #2\\
    \Large aus #3\\
    \vspace*{\stretch{2}}
    \Large München, den #6
  \end{center}

  \newpage

  \vspace*{\stretch{1}}

  \begin{flushleft}
    \Large Erstgutachter:  #7 \\[1mm]
    \Large Zweitgutachter: #8 \\[1mm]
    \Large Tag der mündlichen Prüfung: #9\\
  \end{flushleft}

  \cleardoublepage
}

\hypersetup {
  pdftitle = {Double Field Theory on Group Manifolds},
  pdfauthor = {Falk Haßler}
} 

\begin{document}

  \frontmatter

  \LMUTitle
      {Double Field Theory\\on Group Manifolds}            
      {Falk Haßler}                                        
      {Eberswalde-Finow}                                   
      {Fakultät für Physik}                                
      {München 2015}                                       
      {01.06.2015}                                         
      {Prof. Dr. Dieter Lüst}                              
      {Priv.-Doz. Dr. Ralph Blumenhagen}                   
      {15.07.2015}                                         

  \pagestyle{empty}
  \titlespacing*{\chapter}{0pt}{-20pt}{20pt}

\chapter*{Zusammenfassung}
\begin{otherlanguage}{ngerman}
Die vorliegende Arbeit befasst sich mit Double Field Theory (DFT), einer effektiven Feldtheorie, die die Dynamik von geschlossenen Strings mit niedriger Energie auf einem Torus beschreibt. Alle Observablen, die aus dieser Dynamik entstehen, stimmen in bestimmten Familien von Hintergrundraumzeiten überein. Solche verschiedenen Hintergründe sind durch T-Dualität miteinander verbunden. DFT macht T-Dualität auf einem Torus explizit sichtbar, indem sie zusätzlich zu den $D$ Koordinaten der Raumzeit weitere $D$ Windungskoordinaten einführt. Der Strong Constraint, eine wichtige Konsistenzbedingung der Theorie, schränkt allerdings die Abhängigkeit aller physikalischen Felder auf die Hälfte der Koordinaten des entstehenden gedoppelten Raumes ein.

Eine wichtige Anwendung findet DFT bei verallgemeinerten Scherk-Schwarz Kompaktifizierungen. Sie führen zu halbmaximalen, elektrisch geeichten Supergravitationen, welche durch den Embedding Tensor Formalismus klassifiziert werden. Er beschreibt die Einbettung ihrer Eichsymmetrie in die Gruppe O($n,n$). Da der Strong Constraint nicht mit allen Lösungen des Embedding Tensors kompatibel ist, wird er in der DFT Flussformulierung durch den schwächeren Closure Constraint ersetzt. Dadurch werden Kompaktifizierungen auf Hintergründen möglich, die nicht T-dual zu geometrisch wohldefinierten Hintergründen sind. Sie können nur mit Hilfe von nicht-geometrischen Flüssen beschrieben werden und sind auf Grund ihrer speziellen Eigenschaften von großem phänomenologischem Interesse. Allerdings verschleiert die Verletzung des Strong Constraints ihren Uplift zur vollen String Theorie. Des Weiteren gibt es technische Unklarheiten bei der Verallgemeinerung von herkömmlichen Scherk-Schwarz Kompaktifizierungen auf den gedoppelten Raum der DFT. So ist zum Beispiel nicht bekannt, wie der Twist der Kompaktifizierung im Allgemeinen zu konstruieren ist.

Nachdem die grundlegenden Konzepte von DFT und verallgemeinerten Scherk-Schwarz Kompaktifizierungen wiederholt wurden, wird DFT${}_\mathrm{WZW}$, eine Verallgemeinerung der bisherigen Formulierung, vorgestellt. Sie beschreibt die Niederenergiedynamik von geschlossenen, bosonischen Strings auf einer kompakten Gruppenmanigfaltigkeit und erlaubt die genannten Probleme zu lösen. Ihre klassische Wirkung und die dazugehörenden Eichtransformationen ergeben sich aus Closed String Field Theory bis zur kubischen Ordnung in den masselosen Feldern. Die so gewonnen Ergebnisse werden durch eine verallgemeinerte Metrik ausgedrückt und auf alle Ordnungen erweitert. Es zeigt sich eine explizite Trennung zwischen Hintergrund und Fluktuationen. Damit die Eichalgebra schließt, müssen die Fluktuationen einem modifiziertem Strong Constraint genügen, während für den Hintergrund der Closure Constraint ausreicht. Zusätzlich zu den aus der bisherigen Formulierung bekannten verallgemeinerten Diffeomorphismen ist DFT${}_\mathrm{WZW}$ auch unter gewöhnlichen Diffeomorphismen des gedoppelten Raumes invariant. Werden diese durch den extended Strong Constraint, eine optionale Zusatzbedingung, gebrochen, ergibt sich die traditionelle DFT. Für die neue Theorie wird eine Flussformulierung hergeleitet und der Zusammenhang zu verallgemeinerten Scherk-Schwarz Kompaktifizierungen beleuchtet. Ein möglicher tree-level Uplift von einem wahrhaft nicht-geometrischen Hintergrund (nicht T-dual zu geometrischen) wird präsentiert. Weiterhin können die Unklarheiten bei der Konstruktion von geeigneten Twists beseitigt werden. Auf diese Weise ensteht ein allgemeineres Bild von DFT und der ihr zugrunde liegenden Strukturen.
\end{otherlanguage}

\chapter{Abstract}
This thesis deals with Double Field Theory (DFT), an effective field theory capturing the low energy dynamics of closed strings on a torus. All observables arising from those dynamics match on certain families of background space times. These different backgrounds are connected by T-duality. DFT renders T-duality on a torus manifest by adding  $D$ winding coordinates in addition to the $D$ space time coordinates. An essential consistency constraint of the theory, the strong constraint, only allows for fields which depend on half of the coordinates of the arising doubled space.

An important application of DFT are generalized Scherk-Schwarz compactifications. They give rise to half-maximal, electrically gauged supergravities which are classified by the embedding tensor formalism, specifying the embedding of their gauge group into O($n,n$). Because it is not compatible with all solutions of the embedding tensor, the strong constraint is replaced by the closure constraint of DFT's flux formulation. This allows for compactifications on backgrounds which are not T-dual to well-defined geometric ones. Their description requires non-geometric fluxes. Due to their special properties, they are also of particular phenomenological interest. However, the violation of the strong constraint obscures their uplift to full string theory. Moreover, there is an ambiguity in generalizing traditional Scherk-Schwarz compactifications to the doubled space of DFT: There is a lack of a general procedure to construct the twist of the compactification.

After reviewing DFT and generalized Scherk-Schwarz compactifications, DFT${}_\mathrm{WZW}$, a generalization of the current formalism is presented. It captures the low energy dynamics of a closed bosonic string propagating on a compact group manifold and it allows to solve the problems mentioned above. Its classical action and the corresponding gauge transformations arise from Closed String Field Theory up to cubic order in the massless fields. These results are rewritten in terms of a generalized metric and extended to all orders in the fields. There is an explicit distinction between background and fluctuations. For the gauge algebra to close, the latter have to fulfill a modified strong constraint, while for the former the closure constraint is sufficient. Besides the generalized diffeomorphism invariance known from the traditional formulation, DFT${}_\mathrm{WZW}$ is invariant under standard diffeomorphisms of the doubled space. They are broken by imposing the totally optional extended strong constraint. In doing so, the traditional formulation is restored. A flux formulation for the new theory is derived and its connection to generalized Scherk-Schwarz compactifications is discussed. Further, a possible tree-level uplift of a genuinely non-geometric background (not T-dual to any geometric configuration) is presented. Finally, the ambiguity in constructing the compactification's twist is eliminated. Altogether, a more general picture of DFT and the structures it is based on emerges.

\chapter{Acknowledgments}
First, I would like to express my special gratitude to my supervisor Dieter L\"ust for his irreplaceable support, encouragement and trust. He offered me, a total string theory newcomer, the great opportunity to make my first contact with this wonderfully diverse field of research. Working with him, I constantly benefited from his vivid ideas, physical intuition, striking knowledge of the literature and last but not least his widespread connections to the scientific community.

Further, I am deeply indebted to Ralph Blumenhagen who answered my numerous questions with impressive patience. His ability to spot an ambiguity immediately -- even in the most complicated setup -- was essential for our fruitful collaborations. I also had the pleasure to collaborate with Pascal du Bosque whom I would like to thank especially. His unbreakable endurance while fighting with me through many very technical calculations was indispensable.

I also want to thank (in alphabetical order)
Lara Anderson,
Andr\'{e} Betz,
Ilka Brunner,
Livia Ferro,
Michael Fuchs,
James Gray,
Michael Haack,
Daniela Herschmann,
Olaf Hohm,
Daniel Junghans,
Florian Kurz,
Stefano Massai,
Christoph Mayrhofer,
Stefan Groot Nibbelink,
Peter Patalong,
Erik Plauschinn,
Daniel Plencner,
Felix Rennecke,
Cornelius Schmidt-Colinet,
Maximilian Schmidt-Sommerfeld and
Rui Sun
for many great discussions, patience while answering my questions and removing my concerns. All members of the String Theory Group in Munich and of the Max-Planck-Institute for Physics made it fun to work here. The three years of my doctoral studies went by much too fast.

Finally, I want to mention that all the work I have done here would have been hardly possible without the support and understanding my wife and my parents gave me. Living with a physicist who sometimes forgets the most basic things in everyday life while thinking about some abstract question is a challenge. Thank you for mastering it everyday.
  \cleardoublepage
  
  \pagestyle{fancyplain}
  \titlespacing*{\chapter}{0pt}{50pt}{40pt}
  \tableofcontents

  \mainmatter\setcounter{page}{1}
\chapter{Introduction}
\section{Unification}\label{sec:unification}
The quest for unification is nearly as old as physics itself. An illustrative example of this concept is the formulation of classical electrodynamics by Maxwell. In 1865, he gave a theoretical, unified description of magnetism and electricity \cite{Maxwell:1865zz} based on two observations made by \O{}rsted and Faraday. A great success of this new theory was the prediction of electromagnetic waves traveling at the finite speed of light $c$. More than twenty years later, Hertz proved this conjecture by several experiments. The success story of unification went on for the following decades. Inspired by the symmetries of Maxwell's theory, Lorentz, Poincar\'{e} and Einstein started to unify the concepts of space and time. First, this idea gave rise to the theory of special relativity \cite{Einstein:1905ve}. After Einstein was able to implement gravity in this new framework, it finally culminated in general relativity \cite{Einstein:1916vd}.

General relativity and electrodynamics allowed to study gravity and electromagnetism, two of the four different interactions known in nature, properly at the classical level. The remaining two interactions, the weak and the strong nuclear force, became accessible after the advent of quantum mechanics. Again electrodynamics paved the way for this development. Quantum electrodynamics (QED) was the role model for a quantum formulation of a field theory, a Quantum Field Theory (QFT). In general, quantum field theories describe subatomic particles, like e.g. electrons, and their interaction. A well established experimental method to study them are particle colliders. In a particle collision with an appropriate high center of mass energy, new particles are created. Their properties, like charge and momentum, are analyzed in several detectors which are arranged around the collision point. From these data one is able to reconstruct the fundamental interaction between the colliding particles which is governed by a quantum field theory. In this procedure, one distinguishes two different kind of particles: fermions which form the matter content of our universe and bosons with mediate interactions between fermions. Besides the massless photon, the boson of QED, collider experiments discovered additional bosons. There are the three weak bosons $\mathrm{W}^+$, $\mathrm{W}^-$, $\mathrm{Z}$ and eight massless gluons. While the former mediate the weak nuclear force, the latter are responsible for the strong nuclear force. All these bosons arise naturally in the framework of gauge theories. There, they are associated to the different generators of the gauge group, a semisimple Lie group describing local symmetries of the theory.

Again, the concept of unification is applicable to the electromagnetic and the weak nuclear force. Both arise from a gauge theory with gauge group SU($2$)$\times$U($1$)${}_\mathrm{Y}$, which is spontaneously broken by the Higgs mechanism\cite{Higgs:1964pj,Englert:1964et,Guralnik:1964eu} to U($1$), the gauge group of QED. As a consequence the weak bosons acquire a mass which is in perfect alignment with the experimental findings. Further, the Higgs mechanism predicts an additional scalar field denoted as Higgs field. Its excitations are massive and called Higgs bosons. Recently, enormous experimental efforts where made to detect these bosons. In Juli 2012, the Atlas and CMS collaborations at CERN announced the discovery of a new elementary particle \cite{Aad:2012tfa,Chatrchyan:2012ufa} whose properties match the ones expected for the Higgs boson. The non-vanishing vacuum expectation value of the Higgs field fixes the energy scale $m_\mathrm{ew}=246$\,GeV at which the unification of electromagnetic and weak force occurs. The resulting theory, which contains besides the electroweak the strong interaction, but not gravity, is called the standard model of particle physics.
\begin{figure}[t]
  \centering
  \begin{tikzpicture}[grow=left,growth parent anchor=west,
   level distance=5em,edge from parent path={(\tikzparentnode.west) .. controls +(-1,0) and +(1,0) .. (\tikzchildnode.east)}]
    \node[anchor=west] (st) {}
      child{ child{ child {node[anchor=east] {electromagnetism} }
                    child {node[anchor=east] {weak nuclear force} } }
      child{ child {node[anchor=east] {strong nuclear force} } } }
      child{ child{ child {node[anchor=east] {gravity} } } } ;
    \draw (st.west) -- +(1,0) node[anchor=west] {theory of everything};
    \draw[->] ($(st-2-1-1.east)+(0,-0.75)$) node[name=eaxis] {} -- +(8,0) node[anchor=west] {energy};
    \draw[dashed] (st-1-1) -- (eaxis -| st-1-1);
    \draw (eaxis -| st-1-1) -- +(0,-0.5ex) node[anchor=north] {$m_\mathrm{ew}$};
    \draw[dashed] (st-1) -- (eaxis -| st-1);
    \draw (eaxis -| st-1) -- +(0,-0.5ex) node[anchor=north] {$m_\mathrm{GUT}$};
    \draw[dashed] (st) -- (eaxis -| st);
    \draw (eaxis -| st) -- +(0,-0.5ex) node[anchor=north] {$m_\mathrm{Pl}$};
  \end{tikzpicture}
  \caption{Energy scales for the unification of the four fundamental forces in nature. Everything above $m_\mathrm{ew}$ is conjectured.}\label{fig:unification}
\end{figure}
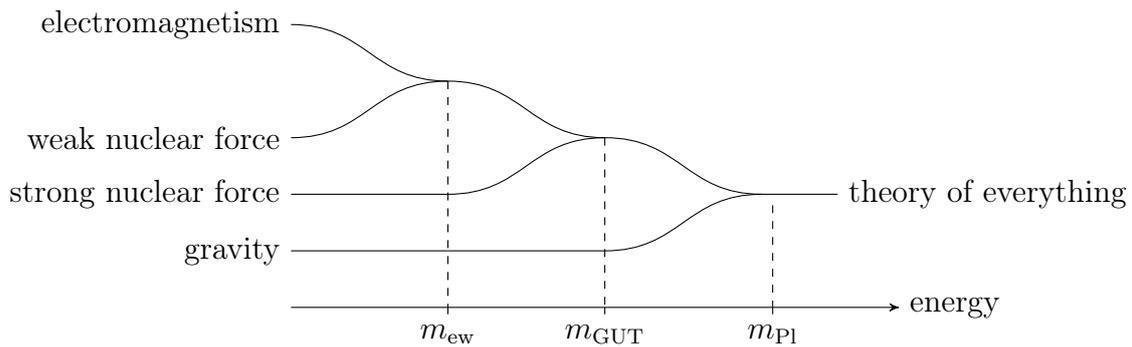

Inspired by the recent successes of unification, there is a justified hope that all four fundamental forces in nature can be unified. Taking into account a minimal supersymmetric extension of the standard model called MSSM, one assumes that the strong and electroweak interaction are unified at an energy scale $m_\mathrm{GUT}=10^{16}$\,GeV \cite{Dimopoulos:1981zb,Dimopoulos:1981yj}. Compared to the energy scale of $10^4$\,GeV\,, which is accessible with contemporary particle colliders like the LHC at CERN, this scale seems completely out of reach. However, it is essential for processes that took place shortly after the Big Bang. At even higher energies, namely the so called Planck scale $m_\mathrm{Pl}=1,22\cdot10^{19}$\,GeV, gravity is assumed to finally join in so that all four interactions are unified. Figure~\ref{fig:unification} visualizes the energy scales at which the different fundamental interactions merge.

Because of the lack of any experimental inputs, it is very hard to say how a so called theory of everything which unifies all four interactions may look like. Still, there are several candidates which give an idea about the dynamics at the Planck scale. String theory, loop quantum gravity\footnote{Normally, loop quantum gravity includes gravity only. Thus, it should not be called theory of everything. Still, there are efforts \cite{BilsonThompson:2006yc} to implement gauge interactions in this framework, too.}, for an introduction see e.g. \cite{Nicolai:2006id,Rovelli:2011eq}, and non-commutative geometry \cite{Connes:1994yd,Szabo:2001kg} are the most prominent among them. In general these theories aim at
\begin{itemize}
  \item Reproducing the physics of the standard model at low energies.
  \item Predicting new, experimentally accessible observations beyond the standard model.
  \item Requiring as few as possible additional assumptions.
\end{itemize}
At the moment string theory seems to be the most mature candidate for a theory of everything.

\section{String theory}
Instead of point like elementary particles as they appear in quantum field theory, string theory considers one-dimensional extended objects called strings. Originally, it was developed in the late 1960s and early 1970s as a theory of the strong interaction. Around 1974, some of the spin two excitations, which arise naturally in closed string theory, were related to gravitons. They reproduce the field equations of general relativity at low energies. While quantum chromodynamics (QCD) was recognized as the appropriate theory for the strong interaction, string theory was reinterpreted as a possible candidate for a theory of quantum gravity. Besides the gravitons, there are many other excitations of the string. Some of them are massless and can be regarded as gauge bosons. In this way, closed string theory unifies gravity and gauge interactions which justifies its role as a theory of everything. There are two different perspectives one can take on string theory, the worldsheet and the target space perspective. Both are inevitable to address string theory related questions. Note that this section only discusses bosonic string theory. Its supersymmetric extension, superstring theory, is treated in the next section.

\subsection{Worldsheet}
String theory can be stated in terms of a two-dimensional conformal field theory on an euclidean Riemann surface $\Sigma$ called the worldsheet\footnote{This worldsheet arises after Wick rotation of a worldsheet with Minkowski signature. Precisely, there are also string theories whose worldsheets are not orientable and thus no Riemann surfaces. They arise after orientifolding. However, they are not of special interest for this thesis.}. Depending on whether $\Sigma$ with all punctures removed is compact or not, one distinguishes between open and closed string theory. A puncture is a point which is cut out from the worldsheet. In this thesis, we are mainly interested in closed string theory with $D$ bosonic fields $x^i$ on the worldsheet. Their dynamics is governed by the Polyakov action
\begin{equation}\label{eqn:SP}
  S_\mathrm{P} = -\frac{1}{4 \pi \alpha'} \int_\Sigma d^2 \sigma \sqrt{h}  h^{\alpha\beta} \partial_\alpha x^i \partial_\beta x^j (g_{ij} + B_{ij}) + S_\chi 
    \quad \text{with} \quad
  S_\chi = \frac{1}{4 \pi} \int_\Sigma d^2 \sigma \sqrt{h} \phi R\,,
\end{equation}
where $h_{\alpha\beta}$ is the euclidean metric on the worldsheet and its determinant is denoted by $h$. This metric gives rise to the scalar curvature $R$ on $\Sigma$. Further, there are three different, in general $x^i$-dependent, coupling constants. They are the symmetric target space metric $g_{ij}$, the antisymmetric $B$-field $B_{ij}$ and the dilaton $\phi$. The only scale entering the action is encoded in the slope parameter $\alpha'$ which specifies the length
\begin{equation}
  l_\mathrm{s} =2\pi \sqrt{\alpha'}
\end{equation}
of the string. Assuming that the dilaton only changes on much larger length scales, $S_\chi$ can be expressed in terms of the Euler characteristic $\chi(\Sigma)$ as
\begin{equation}
  S_\chi = \frac{\phi}{4\pi} \int_\Sigma d^2 \sigma \sqrt{h} R = \phi \chi(\Sigma)\,.
\end{equation}
It only depends on the topology of the worldsheet. A straightforward way to calculate the Euler characteristic is
\begin{equation}
  \chi(\Sigma) = 2 - 2 g - b\,,
\end{equation}%
\begin{figure}
  \centering
  \begin{tikzpicture}[x=-1cm,y=-1cm]
    \begin{scope}[shift={(0,0)}]
      \node[at={(0,-1.3)},anchor=south] {$g=1$\,, $\chi=-1$};
      \draw[fill=fillcolor] (0,0) ellipse (1.5 and 1);
      \node[at={(1.2, 0)},point] {};
      \clip (1, -0.2) arc (0:360:1 and 0.5);
      \draw[fill=white] (0,  0.2) ellipse (1 and 0.5);
      \draw[thick] (1, -0.2) arc (0:180:1 and 0.5);
    \end{scope}
    \begin{scope}[shift={(4,0)}]
      \node[at={(0,-1.3)},anchor=south] {$g=1$\,, $\chi=0$};
      \draw[fill=fillcolor] (0,0) ellipse (1.5 and 1);
      \clip (1, -0.2) arc (0:360:1 and 0.5);
      \draw[fill=white] (0,  0.2) ellipse (1 and 0.5);
      \draw[thick] (1, -0.2) arc (0:180:1 and 0.5);
    \end{scope}
    \begin{scope}[shift={(8,0)}]
      \draw[fill=fillcolor] (0,0) circle (1.2);
      \draw[dashed] (-1.2,0) arc (180:360:1.2 and 0.5);
      \draw (1.2,0) arc (0:180:1.2 and 0.5);
      \node[at={(0,-1.3)},anchor=south] {$g=0$\,, $\chi=-1$};
      \node[at={(0,-0.8)},point] {};
      \node[at={(0, 0.8)},point] {};
      \node[at={(-0.6, 0)},point] {};
    \end{scope}
    \begin{scope}[shift={(12,0)}]
      \draw[fill=fillcolor] (0,0) circle (1.2);
      \draw[dashed] (-1.2,0) arc (180:360:1.2 and 0.5);
      \draw (1.2,0) arc (0:180:1.2 and 0.5);
      \node[at={(0,-1.3)},anchor=south] {$g=0$\,, $\chi=0$};
      \node[at={(0,-0.8)},point] {};
      \node[at={(0, 0.8)},point] {};
    \end{scope}
  \end{tikzpicture}
  \caption{Relevant closed string worldsheets for Euler characteristic $\chi(\Sigma) \ge -1$. The small black dots denote punctures.}
  \label{fig:worldsheets}
\end{figure}
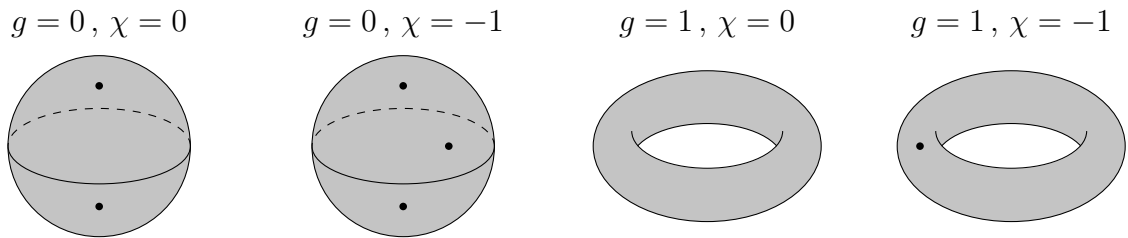%
where $g$ denotes the genus of $\Sigma$ and $b$ counts its boundaries. These two quantities allow to classify all worldsheets relevant for string theory. Figure~\ref{fig:worldsheets} depicts them for closed string theory and $\chi(\Sigma)\ge -1$. The black dots mark punctures which count as boundaries. We excluded the one-punctured sphere and the unpunctured sphere, because they do not give a contribution to any observables of the theory.

A challenge in quantizing the Polyakov action \eqref{eqn:SP} are its local symmetries. One approach to treat them in a proper way is BRST quantization: First, the Faddeev-Popov procedure is applied to obtain a well-defined path integral formulation without divergences. To this end, local symmetries are fixed and auxiliary fields, the ghosts, are introduced. These ghosts preserve the information about local symmetries which would otherwise be lost due to the fixing necessary to evaluate the path integral. On the quantum level, the BRST charge $Q$ reimplements the local symmetries by acting on the ghosts. In an anomaly free quantization, $Q$ has to be nilpotent and fixes $D=26$ for the bosonic string.

\subsection{Target space}\label{sec:targetspace}
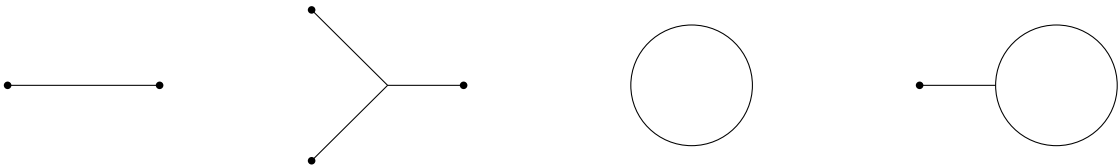
\begin{figure}
  \centering
  \begin{tikzpicture}
    \begin{scope}[shift={(0,0)}]
      \draw (-1,0) node[point] {} -- (1,0) node[point] {};
    \end{scope}
    \begin{scope}[shift={(4,0)}]
      \draw (-1,1)  node[point] {} -- (0,0) -- (1,0) node[point] {};
      \draw (-1,-1) node[point] {} -- (0,0);
    \end{scope}
    \begin{scope}[shift={(8,0)}]
      \draw (0,0) circle (0.8);
    \end{scope}
    \begin{scope}[shift={(12,0)}]
      \draw (-1,0) node[point] {} -- (0,0);
      \draw (0.8,0) circle (0.8);
    \end{scope}
  \end{tikzpicture}
  \caption{Target space Feynman diagrams corresponding to the four different worldsheet topologies depicted in figure~\ref{fig:worldsheets}.}\label{fig:feynmandiag}
\end{figure}

One can interpret the bosonic field $x^i$ as tracing out the string in a $D$-dimensional target space. To underpin this intuitive picture, we consider the two punctured sphere for which $S_\chi$ vanishes and neglect the $B$-field in \eqref{eqn:SP}. Solving the equations of motion for the worldsheet metric $h_{\alpha\beta}$ and plugging the solution back into to the Polyakov action, we obtain the Nambu-Goto action
\begin{equation}
  S_\mathrm{NG} = -\frac{1}{2\pi \alpha'} \int_\Sigma d^2 \sigma \sqrt{\gamma}\,,
\end{equation}
where $\gamma$ denotes the determinant of the induced metric
\begin{equation}
  \gamma_{\alpha\beta} = \frac{\partial x^i}{\partial \sigma^\alpha} 
    \frac{\partial x^j}{\partial \sigma^\beta} g_{ij}
\end{equation}
arising as the pull back of the target space metric $g_{ij}$ on the string worldsheet $\Sigma$. Hence, the two punctured sphere gives rise to a string propagating in a $D$-dimensional target space specified by the worldsheet coupling constants $g_{ij}$ and $B_{ij}$. In a quantum field theory, this situation is captured by a propagator. Following this analogy, the worldsheet topologies in figure~\ref{fig:worldsheets} can be interpreted as the target space Feynman diagrams depicted in figure~\ref{fig:feynmandiag}.

From the target space perspective, also the difference between open and closed strings becomes obvious. Take, e.g. the explicit parameterization
\begin{equation}\label{eqn:paramSigma}
  \sigma^0 = \tau \in \mathds{R} \quad \text{and} \quad \sigma^1 = \sigma \in [0, 2\pi)
\end{equation}
for the worldsheet, where $\sigma$ labels all points of the string at a fixed time $\tau$. For a closed string the boundary condition
\begin{equation}
  x^i(\tau, 0) = x^i(\tau, 2 \pi)
\end{equation}
has to hold. Hence, the worldsheet is a cylinder. Up to conformal transformations, which leave the theory invariant, this cylinder is equivalent to the two punctured sphere. For an open string, the parameterization \eqref{eqn:paramSigma} gives rise to an infinite, extended band. It is conformally mapped to a unit disk with two punctures on its boundary.

\subsection{T-duality}\label{sec:t-duality}
While probed by a closed string, there is a family of different target spaces which are indistinguishable. They are connected by so called T(arget space)-duality transformations. Because these transformations originate from the extended nature of the string, no analogous concept exists for point particles. Thus, it is not surprising that T-duality is an important tool to study string theory. Its implementation on general backgrounds is still a topic of active research addressing a wide range of different questions stretching from physical to purely mathematical ones.

Here, we focus on the most basic example, a $D$-dimensional Minkowski space with $n$ dimensions coiled up to a flat torus
\begin{equation}
  \mathrm{T}^n = \mathds{R}^n / 2\pi \Gamma_n \quad \text{with} \quad
  \Gamma_n = \{ n^i \vec e_i\, |\, \forall \,n_i \in \mathds{Z} \}\,.
\end{equation}
Following \cite{Blumenhagen:2013fgp}, we construct the torus by identifying points of $\mathds{R}^n$ which are shifted by elements of the discrete lattice $2 \pi \Gamma_n$. Calculating the closed string spectrum, one obtains
\begin{equation}
  \alpha' m^2 = 2(N_\mathrm{L} + N_\mathrm{R} - 2) + \alpha' {\vec p}^{\;2} + \frac{1}{\alpha'} {\vec L}^2 \,.
\end{equation}
This equation is governed by several quantum numbers: $N_\mathrm{L}$/$N_\mathrm{R}$ counts the left/right handed creation operators acting on the string's vacuum states which carry the additional quantum numbers $\vec p$ and $\vec L\in \Gamma_n$. The former quantifies the target space momentum of the string and the latter describes how it is wound around the torus. The left/right splitting one encounters originates from the most general solution of the string's equations of motion. Further, there is the so called level matching condition
\begin{equation}\label{eqn:levelmatchintro}
  N_\mathrm{R} - N_\mathrm{L} = \vec p \cdot \vec L \in \mathds{Z}\,,
\end{equation}
which holds for all physical states.

To show how T-duality affects the torus, it is convenient to combine $\vec p$ and $\vec L$ into a doubled, $2n$-dimensional vector
\begin{equation}
  \vec P = \begin{pmatrix} \vec p_\mathrm{L} & \vec p_\mathrm{R} \end{pmatrix}^T\,,
  \quad \text{where} \quad
    \vec p_\mathrm{L} = \frac{1}{\sqrt{2\alpha'}}(\alpha'\vec p + \vec L) 
  \quad \text{and} \quad 
    \vec p_\mathrm{R} = \frac{1}{\sqrt{2\alpha'}}(\alpha' \vec p - \vec L)
\end{equation}
denote the momenta of the string's left and right moving parts. These momenta are conjugate to the coordinates $x_\mathrm{L}$ and $x_\mathrm{R}$. Whereas their linear combinations
\begin{equation}
  \vec x = \sqrt{\frac{\alpha'}{2}} (\vec x_\mathrm{L} + \vec x_\mathrm{R})
    \quad \text{and} \quad
  \vec{\tilde x} = \frac{1}{\sqrt{2 \alpha'}} (\vec x_\mathrm{L} - \vec x_\mathrm{R})
\end{equation}
are conjugate to the momentum $\vec p$ and the winding $\vec L$ along the torus again. Further, we introduce the metric
\begin{equation}
  \eta = \begin{pmatrix}
    \mathbf{1}_{n} & 0 \\
    0 & -\mathbf{1}_{n}
  \end{pmatrix}
    \quad \text{to define the scalar product} \quad
  \vec P \cdot \vec P' = \vec P^T \eta \vec P'
\end{equation}
between doubled vectors. A quick calculation, which takes the level matching \eqref{eqn:levelmatching} into account, proves that $\vec P$ is an element of an even, self-dual, Lorentzian lattice $\Gamma_{n,n}$. T-duality acts as automorphism of this lattice and is mediated by a discrete O($n,n$,$\mathds{Z}$) transformation which leaves $\eta$ invariant, e.g.
\begin{equation}
  \vec P \rightarrow O \vec P  \quad \text{with} \quad O^T \eta O = \eta\,.
\end{equation}
Finally, let us consider $O=\eta$ as an explicit example of a T-duality transformation. By flipping the sign of the right moving sector
\begin{equation}
  \vec p_\mathrm{R} \to - \vec p_\mathrm{R}
    \quad \text{and} \quad
  \vec x_\mathrm{R} \to - \vec x_\mathrm{R}
\end{equation}
while keeping $\vec p_\mathrm{L}$ and $\vec x_\mathrm{L}$ untouched, it swaps
\begin{equation}\label{eqn:tdualityeta}
  \vec L \leftrightarrow \alpha' \vec p
    \quad \text{and} \quad
  \alpha' \vec{\tilde x} \leftrightarrow \vec x\,.
\end{equation}
For a circle with the radius $R$, we obtain
\begin{equation}
  L = R n \quad \text{and} \quad \alpha' p = \frac{\alpha'}{R} m \quad \text{with} \quad
  m\,,n \in \mathds{Z}
\end{equation}
and the transformation \eqref{eqn:tdualityeta} reads
\begin{equation}
  m \leftrightarrow n \quad \text{and} \quad R \rightarrow \frac{\alpha'}{R}\,,
\end{equation}
telling us that T-duality identifies a circle with a large radius (compared to the string length) with another one with a small radius by exchanging momentum and winding excitations, $m$ and $n$, along the circle. This identification is not limited to the spectrum. It holds for all observables. Hence for an experiment, it is impossible to distinguish between these two circles identified by T-duality.

\section{Superstring theory}
The bosonic string theory we discussed so far has two major issues. First, its spectrum contains a state with a negative mass squared. Excitations of this state are called tachyons and render the full, interacting theory instable. Second, it only gives rise to bosonic excitations in target-space. These are well suited to study the fundamental interactions discussed in section~\ref{sec:unification}, but fermions which describe matter are missing. Superstring theory solves both these problems. It assigns to each of the $D$ worldsheet bosons $x^i$ a Majorana worldsheet fermion. In two dimensions, these fermions represent $2D$ real degrees of freedom. Again one splits the bosons $x^i$ into a left- and a right-moving part as discussed in section~\ref{sec:t-duality}. Then, $D$ of the $2D$ fermionic degrees of freedom are the superpartners of the left-movers and the remaining ones correspond to the right-movers. We denote them as $\psi^i_\mathrm{L/R}$. The resulting theory exhibits $N=(1,1)$ supersymmetry on the worldsheet.

The boundary conditions for the bosonic degrees of freedom we studied in section~\ref{sec:targetspace} are completely unchanged. In addition, there are the boundary conditions
\begin{equation}
  \psi^i_\mathrm{L}(\tau,0) = \pm \psi^i_\mathrm{L}(\tau,2\pi) \quad \text{and} \quad
  \psi^i_\mathrm{R}(\tau,0) = \pm \psi^i_\mathrm{R}(\tau,2\pi)
\end{equation}
for the closed string worldsheet fermions. They give rise to four different sectors, NS/NS, NS/R, R/NS, R/R, where NS is an abbreviation for Neveu-Schwarz and R stands for Ramond.

\begin{figure}[b]
  \centering
  \begin{tikzpicture}
    \node[star,star points=5,star point ratio=0.6,draw,fill=fillcolor,
      minimum size=2.5cm,anchor=center,at={(0,0)}] {M-theory};
    \node[at={(60:2)},anchor=south west,name=SO32] {$SO(32)$  heterotic};
    \node[at={(120:2)},anchor=south east,name=I] {Type I};
    \node[at={(200:2)},anchor=east,name=IIA] {Type IIA};
    \node[at={(270:2)},anchor=north,name=IIB] {Type IIB};
    \node[at={(340:2)},anchor=west,name=E8] {$\mathrm{E}_8\times\mathrm{E}_8$ heterotic};
    \draw[<->] (205:2.5) to[bend right] node[midway,below,xshift=-0.5em] {T} (245:2.5);
    \draw[<->] (355:2.5) to[bend right] node[midway,right] {T} (40:2.5);
    \draw[<->] (70 :2.5) to[bend right] node[midway,above] {S} (110:2.5);
    \draw[<->] (198:2.1) to node[midway,above] {S} (198:0.8);
    \draw[<->] (342:2.1) to node[midway,above] {S} (342:0.8);
    \draw[<->] ($(270:2)+(0.5,0)$) arc (150:-155:0.7);
    \node[at={($(270:2)+(1.8,-0.3)$)},anchor=west] {S};
    \node[at={(5,-2)},anchor=west] {$\begin{aligned} \text{T} &= \text{T-duality} \\ 
                                                    \text{S} &= \text{S-duality} \end{aligned}$};
  \end{tikzpicture}
  \caption{Web of dualities which connects the five superstring theories with M-theory.}\label{fig:mtheory}
\end{figure}
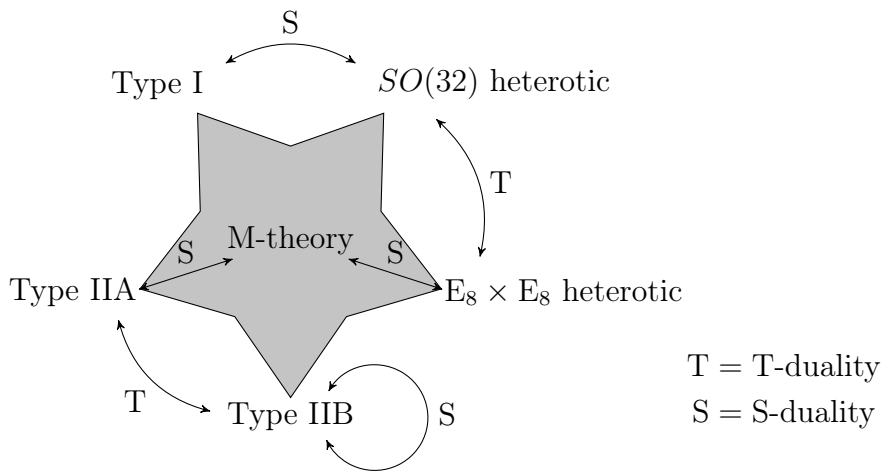
A GSO projection eliminates tachyonic states from the spectrum and gives rise to a modular invariant torus partition function. For the setup we discussed so far, there are two different GSO projections leading to type IIA and IIB superstring theory with $\mathcal{N}=2$ target space supersymmetry. Limiting the supersymmetric extension to the left-moving part of the string while keeping the right-movers bosonic, one obtains heterotic string theories with $\mathcal{N}=1$ target space supersymmetry. One of them exhibits the gauge group $\mathrm{E}_8\times\mathrm{E}_8$ and the other one $SO(32)$. Finally, there is type I. It is obtained from type IIB by identifying the two different orientations of the closed string. All these five superstring theories share the critical dimension $D=10$.

As figure~\ref{fig:mtheory} shows, they are connected by duality transformations. Besides T-duality, which we discussed in section~\ref{sec:t-duality}, superstring theories exhibit S-duality. This duality identifies theories with strong and weak string coupling constant and thus allows to go beyond the perturbative regime. Further, it connects type IIA and $\mathrm{E}_8\times\mathrm{E}_8$ heterotic theories with the eleven dimensional M-theory. M-theory still lacks a complete quantum mechanical description, but its low energy effective action is known to be 11-dimensional supergravity (SUGRA). It allows to unify all five string theories into a theory with membranes as physical degrees of freedom. The combination of S-duality and T-duality is called U(nified)-duality. It is an important tool to obtain a better understanding of M-theory beyond its low energy approximation.

\section{Low energy effective theory}\label{sec:loweneffaction}
String theory has to reproduce the standard model physics at low energies to be a theory of everything in the sense of figure~\ref{fig:unification}. As outlined in section~\ref{sec:unification}, the standard model is formulated in terms of a quantum field theory with a finite number of interacting particles. To connect this paradigm to string theory, one identifies each string excitation with a different particle propagating in the target space. In general this procedure would lead to an infinite number of particles. However, considering only the lightest of them, which can be excited at low energies, allows to formulate a low energy effective field theory capturing the low energy dynamics of the string. A natural choice for an energy cut-off is the string mass
\begin{equation}
  m_s = \frac{1}{\sqrt{\alpha'}}\,.
\end{equation}
There are only very limited experimental restrictions on this mass. It definitely is larger than the energy scales probed by collider experiments like the LHC, because no signatures of string resonances \cite{Anchordoqui:2014wha} where detected yet. A possible assumption is to choose the string mass comparable to the Planck mass $m_\mathrm{Pl}$.

There are two different approaches to calculate an effective action starting from the string's worldsheet theory:
\begin{itemize}
  \item First, one calculates string amplitudes on worldsheets with different topologies. An appropriate effective field theory reproduces these amplitudes if the corresponding target space Feynman diagrams are evaluated. Looking for a field theory which reproduces the classic behavior of the string at weak string coupling, it is sufficient to consider only the two and three punctured sphere depicted in figure~\ref{fig:worldsheets}. A general ansatz for the corresponding effective action contains terms quadratic and cubic in the fields with arbitrary coupling constants. Finally, these constants are fixed to match the amplitudes arising from the target space tree-level Feynman diagrams in figure~\ref{fig:feynmandiag} and the amplitudes calculated on the worldsheet.
  \item One can also calculate the one-loop $\beta$-function for the coupling constants on the worldsheet. Fluctuations of these coupling constants, e.g. around a flat background, correspond to massless string excitations. Seeking for a worldsheet theory whose conformal symmetry is not broken at the quantum level, the $\beta$-function has to vanish. This condition gives rise to the equations of motion of the effective theory. The corresponding action can be derived from them.
\end{itemize}
Both approaches give rise to the same results. However, in general they reproduce the classical behavior of the string only. String Field Theory provides a framework allowing to capture quantum effects in the effective action, too. We use covariant Closed String Field Theory (CSFT) \cite{Siegel:1988yz,Saadi:1989tb,Sonoda:1989wa,Zwiebach:1990ni,Zwiebach:1992ie} to derive an effective action in chapter~\ref{chap:groupdft}.

Thus, we leave the details for later and focus on a concrete example. There is one part in the effective action all five superstring theories share. It reads
\begin{equation}\label{eqn:nsnsaction}
  S_\mathrm{NS} = \int\mathrm{d}^{D}x\,
    \sqrt{-g} e^{-2\phi} \big(\mathcal R + 4 \partial_i \phi \partial^i \phi 
    - \frac{1}{12} H_{ijk} H^{ijk} \big)
\end{equation}
and is denoted as NS/NS action, because it e.g. arises from the NS/NS sector in type IIA and type IIB. The curvature scalar $\mathcal{R}$ of the metric $g_{ij}$ gives rise to an Einstein-Hilbert term implementing gravity in target space. Further there is the abelian field strength
\begin{equation}
  H_{ijk} = 3 \partial_{[i} B_{jk]}
\end{equation}
for the $B$-field. Intriguingly, \eqref{eqn:nsnsaction} matches the effective action of the closed bosonic string after substituting the critical dimension $D=10$ by $D=26$. Adding the remaining sectors, a full supergravity in ten dimensions arises as effective theory for each of the five superstring theories. All effective actions receive higher derivative corrections which are suppressed by $\alpha'^{n/2-1}$ where $n$ counts the derivatives \cite{Metsaev:1987zx,Meissner:1996sa}.

\subsection{T-duality}\label{sec:eftdualities}
We already discussed the important role dualities play in string theory. Hence, it is natural to ask whether they leave any imprints on the low energy effective theory. Especially T-duality is interesting in this context, because it depends on the closed string's extended nature. Finding properties of quantum field theories which originate from T-duality and observe them experimentally would be an important evidence in favor of string theory.

By gauging a U($1$) isometry in the worldsheet action, Buscher \cite{Buscher:1987sk} obtained a prescription on how T-duality acts on target space fields. Assume that the isometry stretches along the distinguished target space direction $a$. In this case, the so-called Buscher rules for the metric, the dilaton and the $B$-field read
\begin{align}
  g_{aa} &\rightarrow \frac{1}{g_{aa}} & 
  g_{aj} &\rightarrow \frac{B_{aj}}{g_{aa}} &
  g_{ij} &\rightarrow g_{ij} - \frac{g_{ia} g_{aj} + B_{ia} B_{aj}}{g_{aa}} \nonumber \\
  \phi &\rightarrow \phi - \frac{1}{2} \ln g_{aa} &
  B_{aj} &\rightarrow \frac{g_{aj}}{g_{aa}} &
  B_{ij} &\rightarrow B_{ij} - \frac{g_{ia} B_{aj} + B_{ia} g_{aj}}{g_{aa}}\,.
  \label{eqn:buscherrules}
\end{align}
They map solutions of the equations of motion of the action \eqref{eqn:nsnsaction} to other solutions. In doing so, they implement a discrete symmetry of the effective theory. However, due to the complicated, non-linear form of the transformations \eqref{eqn:buscherrules}, this symmetry is not manifest.

Motivated by this observation, Double Field Theory (DFT) was created \cite{Hull:2009mi,Hull:2009zb,Hohm:2010jy,Hohm:2010pp}. It is reviewed in chapter~\ref{chap:DFTreview}. Here, we only emphasize its salient features. It combines the metric and the $B$-field into the generalized metric
\begin{equation}
  \mathcal{H} = \begin{pmatrix}
    g - B g^{-1} B & - B g^{-1} \\
    g^{-1} B & g^{-1}
  \end{pmatrix}\,.
\end{equation}
Further, the dilaton is replaced by the generalized dilaton
\begin{equation}
  d = \phi - \frac{1}{2} \log \sqrt{-g}\,.
\end{equation}
To reproduce the Buscher rules, one applies the O($D,D,\mathds{Z}$) transformation
\begin{equation}
  \mathcal{H} \rightarrow O^T \mathcal{H} O \quad \text{and} \quad
  d \rightarrow d
  \quad \text{with} \quad
  O = \begin{pmatrix}
    m & 1-m \\
    1-m & m
  \end{pmatrix} \quad\text{and}\quad
  m_{ij} = \begin{cases} 1 & i=j=a \\
    0 & \text{otherwise.}
  \end{cases}
\end{equation}
In terms of these objects, an effective action and the corresponding gauge transformations were derived \cite{Hull:2009mi,Hohm:2010pp}. To this end, $D$ additional target space coordinates are required. Further, the so called strong constraint has to be imposed. It avoids that the additional coordinates give rise to unphysical degrees of freedom. Imposing a special realization of the strong constraint, DFT reduces to the NS/NS action \eqref{eqn:nsnsaction}. Besides the original CSFT derivation of DFT \cite{Hull:2009mi}, there exists a geometric approach in which the action takes an Einstein-Hilbert like form \cite{Hohm:2011si}. Its scalar curvature can be derived from a doubled geometry based on a generalization of the Lie derivative of ordinary geometry. However, this approach comes with some issues. E.g., it is not possible to fully fix the connection on the doubled space. Recently, Extended Field Theory (EFT) was presented \cite{Hohm:2013vpa,Hohm:2013uia,Hohm:2014fxa,Godazgar:2014nqa}. It extends the formalism of DFT from T-duality to full U-duality.

\subsection{Compactification}\label{sec:introcomp}
A low energy effective theory is not sufficient to connect string theory with experimental observation. Until now, only four dimensions were observed. What happens to the remaining six which are needed to define a consistent string theory in $D=10$? Perhaps, they are compact and occupy such a small volume that they evade detection at the energy scale accessible by contemporary particle colliders. This idea goes back much before the advent of string theory. It was first proposed by Klein to explain an unobserved fifth dimension proposed by Kaluza to unify electrodynamics with general relativity in 1921.

Going from a higher dimensional theory to a lower dimensional one by assuming small compact dimensions is called compactification. Even when the compact, internal manifold can not be probed directly at low energies, its shape governs the properties of the effective theory arising in four dimensions. E.g., one is seeking for a four dimensional theory with Minkowski vacuum and minimal supersymmetry to implement the MSSM mentioned in section~\ref{sec:unification}. In this case the internal manifold can be chosen to be a Calabi-Yau threefold. Still, there is an infinite number of such manifolds. They are distinguished by parameters called moduli. The moduli are counted by the hodge numbers $h^{1,1}$ and $h^{2,1}$ in Dolbeault cohomology. In the four-dimensional theory, each of the moduli gives rise to a massless scalar field. Even if these fields would decouple from the three fundamental forces governed by the standard model of particle physics, they at least couple to gravity and affect the cosmology of our universe. It is very difficult to allow them while preserving the predictions of standard cosmology. Thus, severe efforts are made to give masses to the moduli and stabilize them at fixed vacuum expectation values. This procedure is called moduli stabilization.

At tree-level, a well established tool to stabilize moduli is flux compactification \cite{Giddings:2001yu,Grana:2005jc,Douglas:2006es,Denef:2008wq}. By assigning non-vanishing vacuum expectation values to fluxes, like e.g. the $H$-flux, a scalar potential for the moduli arises. In the ideal case, this potential would have at least one minimum which stabilizes all moduli. Normally, not all moduli can be stabilized by tree-level fluxes, the scalar potential has at least one flat or runaway direction. Non-perturbative effects can be applied to the remaining moduli, but in general there is no prescription how to stabilize all moduli.

To make contact with some basic ingredients of a flux compactification, let us consider a type IIB, orientifold compactification on $\mathrm{T}^6/\mathds{Z}_2\times\mathds{Z}_2$. It gives rise to a four-dimensional theory with $\mathcal{N}=1$ supersymmetry and Minkowski vacuum. This background is a toy model for more general Calabi-Yau compactifications. Next, we switch on the 3-form flux
\begin{equation}
  G_3 = F_3 - i e^{-\phi} H
\end{equation}
which combines the NS/NS $H$-flux with the R/R $F_3$-flux. Most vacuum expectation values for $G_3$ spoil the consistency of the theory. But still there are some configurations giving rise to a valid theory at tree-level. These are non-trivial elements of the third cohomology
\begin{equation}
  H^3 = H^{(3,0)} \oplus H^{(2,1)} \oplus H^{(1,2)} \oplus H^{(0,3)}
\end{equation}
of the complex, internal threefold. Its dimension is given by the sum of the hodge numbers
\begin{equation}
  h^{(3,0)} = h^{(0,3)} = 1 \quad \text{and} \quad
  h^{(2,1)} = h^{(1,2)} = 3
\end{equation}
for the untwisted sector \cite{Lust:2005dy}. Each Calabi-Yau threefold has a unique covariantly constant $(3,0)$-form $\Omega$ which spans $H^{(3,0)}$. It can be used to calculate the superpotential \cite{Gukov:1999ya}
\begin{equation}
  W = \frac{1}{\kappa^2} \int G_3 \wedge \Omega
\end{equation}
arising due to the non-vanishing $G_3$ flux. In our example, there are $h^{(2,1)}=3$ complex structure and $h^{(1,1)}=3$ Kähler moduli. They arise naturally, if we decompose $\mathrm{T}^6$ into three different $\mathrm{T}^2$ and parameterize each of the resulting two-tori in terms of a complex structure $\tau_i$ and a Kähler parameter $\rho_i$. For simplicity, we further assume that $\mathrm{T}^6$ is symmetric, meaning $\tau_i=\tau$ and $\rho_i=\rho$. In this case, the superpotential has the simple form \cite{Shelton:2005cf}
\begin{equation}\label{eqn:superpotIIB}
  W = P_1(\tau) + e^{-\phi} P_2(\tau)
\end{equation}
where $P_{1,2}$ denote cubic polynomials in $\tau$. All coefficients in these polynomials are NS/NS or R/R fluxes. From the superpotential $W$, the scalar potential
\begin{equation}
  V = e^K ( K^{ij} D_i W \overline{D_j W} - 3 | W |^2 )
\end{equation}
can be derived using the Kähler potential
\begin{equation}
  K = - 3 \ln( -i(\tau - \bar \tau) ) - 3 \ln( -(\rho - \bar \rho) ) - \phi\,,
\end{equation}
the Kähler metric $K_{ij} = \partial_i \bar\partial_j K$ and its covariant derivative $D_i W = \partial_i W + \partial_i K W$. Note that the tadpole cancellation condition for orientifold compactifications requires also to introduce localized sources in addition to the topological fluxes arising from the cohomology.

\subsection{Non-geometric backgrounds}\label{sec:nongeombackgr}
In the last subsection, we have seen that fluxes create a scalar potential for complex structure moduli in type IIB compactifications. All Kähler moduli, which include the volume of the compact space, are flat directions of the scalar potential and thus cannot be stabilized. At this point T-duality is useful again. It allows to exchange Kähler and complex structure moduli, while going from a Calabi-Yau manifold to its mirror. Adding additional terms to the superpotential \eqref{eqn:superpotIIB} to make it invariant under an even number of T-dualities gives rise to \cite{Shelton:2005cf}
\begin{equation}
  W = P_1(\tau) + e^{-2 \phi} P_2(\tau) + \rho P_3(\tau)\,,
\end{equation}
where the coefficients of the new cubic polynomial $P_3$ are interpreted as the non-geometric $Q$-flux. As shown in figure~\ref{fig:mtheory}, an odd number of T-dualities transforms type IIB into type IIA. By making a type IIA superpotential T-duality invariant, one obtains the so called $R$-flux \cite{Shelton:2005cf} in the NS/NS sector.

A nice toy model where all these fluxes appear is the T-duality chain
\begin{equation}\label{eqn:t-dualitychain}
  \tikz[baseline=0pt]{ 
    \node[name=H, at={(0,0)}] {$H_{ijk}$};
    \node[name=f, at={($(H.east)+(1,0)$)}, anchor=west] {$f_{ij}^k$};
    \node[name=Q, at={($(f.east)+(1,0)$)}, anchor=west] {$Q_i^{jk}$};
    \node[name=R, at={($(Q.east)+(1,0)$)}, anchor=west] {$R_{ijk}$};
    \draw[->] (H.east) -- (f.west) node[midway, above] {$T_1$};
    \draw[->] (f.east) -- (Q.west) node[midway, above] {$T_2$};
    \draw[->] (Q.east) -- (R.west) node[midway, above] {$T_3$};
  }
\end{equation}
for a flat, three-torus with $H$-flux. $T_i$ denotes a T-duality transformation along the $i$th direction of the torus. After the first transformation $T_1$, a twisted torus with vanishing $H$-flux arises. Its twist is captured by the geometric $f$-flux. A further T-duality gives rise to a $Q$-flux background. It represents the canonical example of a T-fold \cite{Dabholkar:2005ve,Hull:2006va,Hull:2006qs}, a space which is globally patched by a T-duality transformation. In the effective theory \eqref{eqn:nsnsaction}, it lacks a globally well-defined metric and is thus called non-geometric. The $R$-flux even evades a local geometric description \cite{Shelton:2006fd}. Further, note that the third T-duality $T_3$ goes beyond the Buscher procedure because the $Q$-flux background lacks the required U($1$) isometry. As a theory with manifest T-duality on the torus, DFT is an optimal tool to describe all these backgrounds in a consistent way. It e.g. gives rise to field redefinitions which allow to formulate well defined ten dimensional actions with $Q$- and $R$-flux \cite{Andriot:2012an,Andriot:2011uh,Andriot:2012wx}.

\begin{figure}
  \centering
  \begin{tikzpicture}[>=stealth',node distance=4em]
    \node[draw,ellipse,at={($(current page.north west)+(8.5em,-4.5em)$)}] (st) {string theory};
    \node[draw,rectangle] (eft) [below of=st,yshift=-2.5em] 
      {SUGRA};
    \path[->] (st.south) edge node[anchor=east]
        {\parbox{6em}{\raggedleft matching amplitudes}}
      (eft.north);
    \node (dft) [above right of=eft,xshift=8em,yshift=0.75em,minimum width=8em]
      {DFT};
    \node[at={(dft.south)},anchor=north,minimum width=8em] (fluxform) {flux formulation};
    \draw (dft.north west) rectangle (fluxform.south east);
    \draw[dashed] (fluxform.north west) -- (fluxform.north east);
    \path[->] (st.south) edge 
      node[anchor=south,rotate=-16,xshift=0.4em] 
      {\hspace{1em}CSFT}(dft.west);
    \path[->] (dft.west) edge node[anchor=west,yshift=-1.5em,xshift=-1.5em] 
      {$\tilde\partial_i \cdot= 0$} ($(eft.north)+(1em,0)$);
    \node[draw,rectangle] (sugra) [below of=eft,yshift=-6em] {$D-n$ SUGRA};
    \begin{scope}
      \clip (dft|-sugra) ellipse (2 and 0.8);
      \draw[fill=fillcolor,dashed] ($(dft|-sugra)+(-0.75,0.5)$) circle (0.5);
    \end{scope}
    \path[->] (eft.south) edge 
      node[anchor=south,rotate=90] {without fluxes} (sugra.north);
    \node[draw,ellipse,at=(dft|-sugra),minimum width=4cm,minimum height=1.6cm] (gsugra)
      {\parbox{6em}{\centering gauged SUGRA}};
    \path[->] ($(eft.south)+(1em,0)$) edge node[anchor=north, 
      rotate=-45] {with fluxes} ($(dft|-sugra)+(-0.75,0.5)$);
    \path[->] (sugra.east) edge 
      node[anchor=north,yshift=-0.5em,xshift=-2em] {\parbox{12em}{\centering embedding tensor}} 
    (gsugra.west);
    \path[->] (fluxform.south) edge node[anchor=west] 
     {\parbox{15em}{generalized\\Scherk-Schwarz\\compactification}} (gsugra.north);
    \node[draw,rectangle,right of=gsugra,xshift=7em] (vacuum) {vacuum};
    \path[->] (gsugra.east) edge node[anchor=north] {eom} (vacuum.west);
    \draw[->] (vacuum.north) --
      node[anchor=north,rotate=90] {uplift} (vacuum.north|-fluxform.east)
      -- (fluxform.east);
    \draw[->,dashed] (vacuum.north|-fluxform.east) -- (vacuum.north|-st.east)
    -- (st.east) node[midway, above] {only few hints};
    \draw[->] ($(st.west)+(-2em,0)$) -- +(0,-18em)
        node[midway,rotate=90,anchor=south] {simplification (truncation)}; 
  \end{tikzpicture}
  \caption{Mutual dependence of string theory, DFT flux formulation, supergravity and half-maximal, electrically gauged supergravities. The small, shaded region denotes gaugings embedded in the subgroup $G_\mathrm{geom}$ of the full O($n,n$).}\label{fig:bigpicture}
\end{figure}
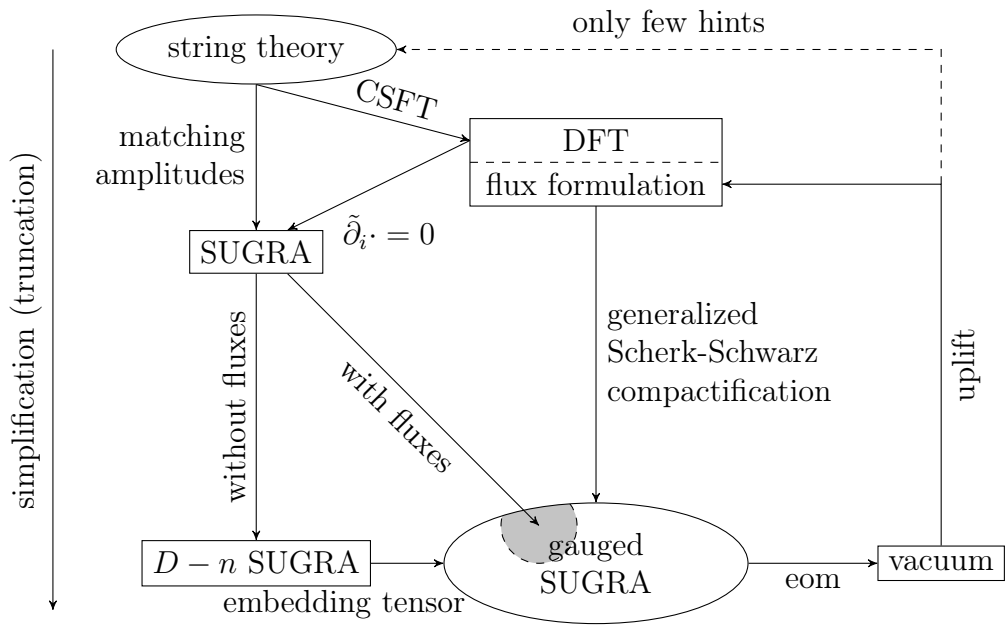

Unfortunately, T-duality chains do not lead to any improvements in moduli stabilization, because T-duality by definition does not change the physical properties of a system. Even when it looks completely different, the $R$-flux background gives rise to the same physics as the torus with $H$-flux. Thus, backgrounds with $Q$- and $R$-flux which are not T-dual to geometric onces are even more interesting. They are called genuinely non-geometric and do not admit a full ten dimensional supergravity description \cite{Dibitetto:2012rk,Hassler:2014sba}. In DFT, they violate the strong constraint, which is essential for the consistency of the theory. However, there are still indications giving hope that these backgrounds can be treated in a consistent way:
\begin{itemize}
  \item The gauge group of $D-n < 10$ dimensional, half-maximal, electrically gauged supergravities \cite{Schon:2006kz,Samtleben:2008pe} can be embedded in O($n,n$). Geometric and non-geometric fluxes form the structure coefficients of the gauge algebra \cite{Geissbuhler:2011mx,Aldazabal:2011nj,Condeescu:2013yma}. Choosing the fluxes T-dual to geometric $H$- and $f$-fluxes only limits the gauge group to embeddings in subgroup
    \begin{equation}
      \mathrm{GL}(n) \ltimes \Lambda_2 = G_\mathrm{geom} \subset \mathrm{O}(n,n)\,,
    \end{equation}
    where $\Lambda_2$ denotes a $n(n-1)/2$ dimensional, abelian subgroup of O($n,n$). But the embedding tensor formalism, which starts from a $n$-dimensional, ungauged supergravity and classifies all gaugings consistent with supersymmetry, also gives rise to gaugings outside this subset. They do not admit an uplift to ten-dimensional supergravity and are assumed to arise from compactifications on genuinely non-geometric backgrounds  \cite{Dibitetto:2012rk}.
  \item There are asymmetric orbifold constructions with a modular invariant partition function who reproduce the gauge algebra expected from a genuinely non-geometric compactification \cite{Condeescu:2012sp,Condeescu:2013yma}.
\end{itemize}
These observations have triggered the development of the DFT flux formulation \cite{Geissbuhler:2011mx,Geissbuhler:2013uka,Aldazabal:2013sca}. It implements a weaker version of the strong constraint, the closure constraint, to obtain a gauge invariant action. Performing a generalized Scherk-Schwarz compactification \cite{Geissbuhler:2011mx,Aldazabal:2011nj,Grana:2012rr} in the flux formulation, all consistent gaugings in O($n,n$) are accessible. These gauging are encoded in the embedding tensor formalism which allows to obtain all maximal and half-maximal gauged supergravities from the ungauged theory. Figure~\ref{fig:bigpicture} illustrates the relations between the different theories discussed yet.

Let us point out two major problems which arise in this picture:
\begin{itemize}\label{page:questions}
  \item While the original generalized metric formulation of DFT was derived top down using CSFT on a torus \cite{Hull:2009mi,Hull:2009zb,Hohm:2010jy}, the flux formulation follows a bottom up approach. Thus, it is not clear whether its solutions can be uplifted to full string theory.
  \item Even generalized Scherk-Schwarz compactifications are problematic. In supergravity, Scherk-Schwarz compactifications rely on the Riemannian geometry of group manifolds. But such a geometric description does not exist for the doubled space of DFT. E.g., there are issues with undetermined components of the Levi-Civita connection and the Riemann tensor \cite{Jeon:2010rw,Jeon:2011cn,Hohm:2011si}. A direct confrontation with these problems can be avoided by expressing the compactification ansatz in terms of a formal twist tensor. However, in contrast to the twist in ordinary Scherk-Schwarz compactifications, there is no direct prescription how to construct this twist on a doubled space.
\end{itemize}
In the course of this thesis, we will address these problems. A generalization of DFT, called double field theory on group manifolds or for short DFT${}_\mathrm{WZW}$ \cite{Blumenhagen:2014gva,Blumenhagen:2015zma}, is constructed in order to cope with them. An uplift to string theory is discussed at tree-level only. In general, it breaks down at loop-level due to the failure of the torus partition function to be modular invariant. Still there are some examples, like asymmetric toroidal orbifolds, where a full uplift is possible.

\subsection{Applications}
There is a very wide range of different applications for the techniques we presented so far in this section. E.g., the scalar potential which arises in flux compactifications can be used to build models of cosmic inflation. A nice review on this topic is given by \cite{McAllister:2007bg}. A recent announcement from the BICEP collaboration \cite{Ade:2014xna} has triggered especially interest in slow-roll inflation with trans-Planckian field range. While such setups are in general difficult to implement in string theory, a promising approach is axion monodromy inflation \cite{Silverstein:2008sg,McAllister:2008hb}. Like the simple toy model in \cite{Hassler:2014mla} shows, it benefits from non-geometric fluxes on genuinely non-geometric backgrounds. More sophisticated scenarios, which include non-geometric fluxes as an essential ingredient, are discussed in \cite{Blumenhagen:2015qda,Blumenhagen:2015kja}. Effective field theories are also an excellent tool to study the structure of the target space. In this context non-geometric fluxes can be connected with non-commutative and even non-associate deformations of a target space \cite{Lust:2010iy,Blumenhagen:2010hj,Blumenhagen:2011ph,Bakas:2013jwa,Bakas:2015gia}. Corresponding effects where studied directly on the worldsheet as well as from the DFT perspective \cite{Blumenhagen:2013zpa}.

All these results suggest that in the long term it is inevitable to go beyond the landscape of geometric string backgrounds to a more general one, taking into account the extended nature of a closed string. To address the problems stated in the last subsection is a small step into this direction.

\section{Outline and summary}
This thesis is organized as follows: We begin with a review of DFT, its flux formulation and its connection to non-geometric fluxes in chapter~\ref{chap:DFTreview}. Subsequently, chapter~\ref{chap:genScherkSchwarz} introduces generalized Scherk-Schwarz compactifications and discusses their properties. It presents an explicit example of a genuinely non-geometric background stabilizing complex structure and Kähler moduli at the same time, while leaving still some moduli unstabilized. During this chapter, we also face the ambiguities discussed above. Section~\ref{sec:questiongenSS} suggests to consider a string propagating on a semisimple, compact group manifold to address them.

Following this idea, we derive DFT${}_\mathrm{WZW}$ in the chapters~\ref{chap:groupdft} and \ref{chap:genmetric}. In the former, we start from a Wess-Zumino-Witten model on the worldsheet and derive a CSFT tree-level action up to cubic order in the fields. We further compute the corresponding gauge transformations. Afterwards, the action is rewritten in terms of a generalized metric and extrapolated to all orders in the fields, giving rise to
\begin{align}
  S = & \int d^{2D} X e^{-2d} \Big(  \frac{1}{8} \mathcal{H}^{KL} \nabla_K \mathcal{H}_{IJ} \nabla_L \mathcal{H}^{IJ} -\frac{1}{2} \mathcal{H}^{IJ} \nabla_{J} \mathcal{H}^{KL} \nabla_L \mathcal{H}_{IK} \nonumber \\
  & - 2 \nabla_I d \nabla_J \mathcal{H}^{IJ} + 4 \mathcal{H}^{IJ} \nabla_I d \nabla_J d + \frac{1}{6} F_{IKL} F_J{}^{KL} \mathcal{H}^{IJ} \Big)\,.
\end{align}
It differs in two important points from the DFT action in generalized metric formulation:
\begin{itemize}
  \item It possesses covariant derivatives $\nabla_I$ instead of partial derivatives $\partial_I$.
  \item The last term on the second line is missing in the traditional formulation.
\end{itemize}
We check explicitly that this action is invariant under the generalized diffeomorphisms
\begin{align}
  \delta_\xi \mathcal{H}^{IJ} = \mathcal{L}_\xi \mathcal{H}^{IJ} &= \lambda^K \nabla_K \mathcal{H}^{IJ} + (\nabla^I \lambda_K - \nabla_K \lambda^I) \mathcal{H}^{KJ} + (\nabla^J \lambda_K - \nabla_K \lambda^J) \mathcal{H}^{IK} \nonumber \\
  \delta_\xi d = \mathcal{L}_\xi d & = \xi^K \nabla_K d - \frac{1}{2} \nabla_K \xi^K
\end{align}
if the strong constraint 
\begin{equation}
  \nabla^I \partial_I \cdot = 0
\end{equation}
holds. Here, $\cdot$ is a placeholder for fields, parameters of gauge transformations and arbitrary products of them. They have to be treated like scalars in this expression. Further, the background fluxes $F_{IJK}$ are assumed to be covariantly constraint and to fulfill the Jacobi identity
\begin{equation}
  F_{IJ}{}^M F_{MK}{}^L + F_{KI}{}^M F_{MJ}{}^L + F_{JK}{}^M F_{MI}{}^L = 0\,.
\end{equation}
In this respect, they only have to satisfy the flux formulation's closure constraint. The covariant derivatives, appearing everywhere, transform covariantly with respect to standard $2D$-diffeomorphisms. Thus in contrast to traditional DFT, the framework presented here is manifestly invariant under ordinary diffeomorphisms, too. By imposing an additional constraint linking the background fields with the fluctuations, this additional symmetry is broken and the traditional generalized metric formulation is recovered from DFT${}_\mathrm{WZW}$. Hence, after chapter~\ref{chap:genmetric}, we conclude:
\begin{itemize}
  \item DFT${}_\mathrm{WZW}$ is a generalization of DFT. The latter arises as a special case from the former after imposing an additional constraint, the extended strong constraint, which links background and fluctuations.
  \item The background fluxes $F_{IJK}$ capture all gauge groups properly embedded into O($n,n$). They are not restricted to the geometric subgroup $G_\mathrm{geom}$.
  \item The doubled space our theory is formulated on is completely governed by pseudo-Riemannian geometry. It possesses a metric $\eta_{IJ}$ with split signature and a metric compatible, torsion-full connection.
\end{itemize}

In order to apply these results to the generalized Scherk-Schwarz compactifications discussed in chapter~\ref{chap:genScherkSchwarz}, we derive the flux formulation \cite{Bosque:2015jda}
\begin{equation}
  S = \int d^{2 D}X \, e^{-2d} \big( S^{AB} \mathcal{F}_A \mathcal{F}_B + \frac{1}{4} \mathcal{F}_{ACD} \, \mathcal{F}_B{}^{CD} \, S^{AB} - \frac{1}{12} \mathcal{F}_{ACE} \, \mathcal{F}_{BDF} \, S^{AB} S^{CD} S^{EF} \big)
\end{equation}
of DFT${}_\mathrm{WZW}$ in chapter~\ref{chap:fluxform}. To this end, we introduce covariant fluxes
\begin{equation}
  \mathcal{F}_{ABC} = \tilde F_{ABC} + F_{ABC}
\end{equation}
combining a background part $F_{ABC}$, the flat version of $F_{IJK}$, with a fluctuation part $\tilde F_{ABC}$. The latter is restricted by the strong constraint. Note that we do not reproduce the strong constraint violation term
\begin{equation}
  \frac{1}{6} F_{ABC} F^{ABC}\,,
\end{equation}
which was added to the action of the traditional flux formulation by hand. In section~\ref{sec:missingFABCFABCterm}, we explain why it is absent. Furthermore, we show that a generalized Scherk-Schwarz reduction of DFT${}_\mathrm{WZW}$ gives rise to the expected bosonic sector of a half-maximal, electrically gauged supergravity. Due to the well-defined doubled background geometry of our theory the twist, which gives rise to the background fluxes $F_{ABC}$, is constructed like in ordinary Scherk-Schwarz compactifications. Thus, the second ambiguity pointed out on page~\pageref{page:questions} is solved. Further, the complete derivation of DFT${}_\mathrm{WZW}$ is top down. Hence, it suggests an uplift of genuinely non-geometric backgrounds to full string theory as it is discussed in section~\ref{sec:uplift}. Remember, this uplift is restricted to tree-level results. We close with a conclusion and present several ideas how to develop DFT${}_\mathrm{WZW}$ further in chapter~\ref{chap:conclusion}.

\chapter{Double Field Theory}\label{chap:DFTreview}
In this chapter, we review some important aspects of DFT, which will be relevant for the following chapters. We start with introducing the action in the generalized metric formulation and show its various symmetries. Afterwards we present the equations of motion which arise from the variation of this action. Finally, we discuss the flux formulation. It allows to substitute the strong constraint of the generalized metric formulation with the weaker closure constraint. Further, we comment on how geometric and non-geometric fluxes arise naturally in DFT.

\section{Action and its symmetries}\label{sec:dftsym}
DFT is a low energy effective description of closed string theory that takes into account both momentum and winding modes in a compact space time. Hence in addition to the $D$ space time coordinates $x^i$ (conjugate to the momentum modes), it introduces $D$ new coordinates $\tilde x_i$ (conjugate to the winding modes of the string). In total there are now $2D$ coordinates which are combined into the $2D$-dimensional vector $X^M=\begin{pmatrix}\tilde x_i & x^i\end{pmatrix}$. To lower and raise the index $M$ of this vector, the O($D,D$) invariant metric
\begin{equation}\label{eqn:etaMN}
    \eta_{MN}=\begin{pmatrix}
      0 & \delta^i_j \\
      \delta_i^j & 0
    \end{pmatrix} \quad \text{and its inverse} \quad
    \eta^{MN}=\begin{pmatrix}
      0 & \delta_i^j \\
      \delta^i_j & 0
    \end{pmatrix}
\end{equation}
are used. Further, one defines the partial derivative $\partial^M=\begin{pmatrix}\partial_i & \tilde{\partial}^i\end{pmatrix}$.

\subsection{Action}
The DFT action in the generalized metric formulation \cite{Hohm:2010pp} reads
\begin{equation}\label{eqn:dftaction}
  S_\mathrm{DFT} = \int \mathrm{d}^{2D} X\,e^{-2 d} \mathcal{R}
\end{equation}
where
\begin{align}\label{eqn:genricciscalar}
  \mathcal{R} = 4 \mathcal{H}^{MN} \partial_M d \partial_N d
    - \partial_M \partial_N \mathcal{H}^{MN} &- 4\mathcal{H}^{MN}
      \partial_M d \partial_N d + 4 \partial_M \mathcal{H}^{MN}
      \partial_N d \nonumber \\
  + \frac{1}{8} \mathcal{H}^{MN} \partial_M \mathcal{H}^{KL}  
    \partial_N \mathcal{H}_{KL} &- \frac{1}{2} \mathcal{H}^{MN}\partial_N
    \mathcal{H}^{KL}\partial_L\mathcal{H}_{MK}
\end{align}
is called the generalized Ricci or curvature scalar and
\begin{equation}\label{eqn:genmetricBg}
  \mathcal{H}^{MN}=\begin{pmatrix}
    g_{ij} - B_{ik}g^{kl}B_{lj} & -B_{ik}g^{kj} \\
    g^{ik} B_{kj} & g^{ij}
  \end{pmatrix}
\end{equation}
is the generalized metric, we mentioned already in section~\ref{sec:eftdualities}. It combines the metric $g_{ij}$ and the $B$-field $B_{ij}$ into an O($D,D$) valued, symmetric tensor fulfilling
\begin{equation}\label{eqn:genmetricO(D,D)}
  \mathcal{H}^{MN} \eta_{ML} \mathcal{H}^{LK} = \eta^{NK}\,.
\end{equation}
The NS/NS sector dilaton $\phi$ is encoded in the O($D,D$) singlet
\begin{equation}
  d = \phi - \frac{1}{2} \log \sqrt{-g}
\end{equation}
which is called generalized dilaton.

\subsection{Generalized diffeomorphisms}
The action \eqref{eqn:dftaction} possesses a manifest, global $O(D,D)$ symmetry
\begin{equation}
  \mathcal{H}^{MN} \rightarrow \mathcal{H}^{LK} M_L{}^M M_K{}^N
    \quad \text{and} \quad
  X^M \rightarrow X^N M_N{}^M 
\end{equation}
where $M_L{}^K$ is a constant tensor which leaves $\eta^{MN}$ invariant, namely
\begin{equation}
  \eta^{LK} M_L{}^M M_K{}^N = \eta^{MN}\,.
\end{equation}
If broken to the discrete O($D,D,\mathds{Z}$), it can be interpreted as a T-duality transformation acting on the background torus DFT is defined on. This global symmetry extends to a local O($D,D$) symmetry called generalized diffeomorphisms. Its infinitesimal version
\begin{equation}
  V^M \rightarrow V^M + \delta_\xi V^M
    \quad \text{and} \quad
  X^M \rightarrow X^M - \xi^M
\end{equation}
is mediated by the generalized Lie derivative
\begin{equation}\label{eqn:genliederiv}
  \delta_\xi V^M = \mathcal{L}_\xi V^M = \xi^N \partial_N V^M + (\partial^M \xi_N - \partial_N \xi^M) V^N
\end{equation}
and its generalization for higher rank tensors. The generalized metric $\mathcal{H}^{MN}$ and the generalized dilaton $d$ transform as
\begin{align}\label{eqn:genLieH}
  \delta_\xi \mathcal{H}^{MN} = \mathcal{L}_\xi \mathcal{H}^{MN} &=
    \xi^L \partial_L \mathcal{H}^{MN} + (\partial^M \xi_L - \partial_L \xi^M) \mathcal{H}^{LN}
    + (\partial^N \xi_L - \partial_L \xi^N) \mathcal{H}^{ML} \\
  \delta_\xi d = \mathcal{L}_\xi d &= \xi^M \partial_M d - \frac{1}{2} \partial_M \xi^M \,.
\end{align}
These transformations give rise to the algebra
\begin{equation}
  [\mathcal{L}_{\xi_1}, \mathcal{L}_{\xi_2}] = \mathcal{L}_{\xi_1} \mathcal{L}_{\xi_2} -
  \mathcal{L}_{\xi_2} \mathcal{L}_{\xi_1} = \mathcal{L}_{[\xi_1, \xi_2]_\mathrm{C}}
\end{equation}
which is governed by the C-bracket
\begin{equation}\label{eqn:Cbracket}
  \left[ \xi_1, \xi_2 \right]_\mathrm{C}^M = \xi_1^N \partial_N \xi_2^M -
    \frac{1}{2} \xi_{1 N} \partial^M \xi_2^N - \left( \xi_1 
    \leftrightarrow \xi_2 \right)\,,
\end{equation}
provided we impose the strong constraint
\begin{equation}\label{eqn:strongconstraint}
  \partial_N \partial^N \cdot = 0
\end{equation}
where $\cdot$ is a place holder for fields, gauge parameters and arbitrary products of them. In general this algebra does not satisfy the Jacobi identity and so the generalized diffeomorphisms do not form a Lie algebroid. However, the failure to satisfy the Jacobi identity is a trivial gauge transformation which leave all fields fulfilling the strong constraint invariant. A trivial way to solve \eqref{eqn:strongconstraint} is to set $\tilde \partial^i = 0$. In this case, the DFT action \eqref{eqn:dftaction} transforms into the low energy effective action
\begin{equation}
  S_\mathrm{NS} = \left. S_\mathrm{DFT} \right|_{\tilde \partial^i = 0} = \int\mathrm{d}^{D}x\,
    \sqrt{-g} e^{-2\phi} \big(\mathcal R + 4 \partial_\mu \phi \partial^\mu \phi 
    - \frac{1}{12} H_{\mu\nu\rho} H^{\mu\nu\rho} \big)
\end{equation}
discussed in section~\ref{sec:loweneffaction}.

Because it contains partial instead of covariant derivatives, the DFT action \eqref{eqn:dftaction} is not manifestly invariant under generalized diffeomorphisms. One is still able to show after a lengthy calculation \cite{Hohm:2010jy} that
\begin{equation}\label{eqn:deltaxieR}
  \delta_\xi ( e^{-2d} \mathcal{R} ) = \partial_I ( \xi^I e^{-2d} \mathcal{R} )
\end{equation}
holds. Being a total derivative, this variation vanishes under the action integral and the action remains invariant. Like it is common in field theory, the variation $\delta_\xi$ commutes with partial derivative. It further is linear and fulfills the Leibnitz rule.

\subsection{Double Lorentz symmetry}\label{sec:doublelorentz}
Further, there is a local double Lorentz symmetry. In order to study it, we employ a vielbein formalism, as originally introduced by Siegel in \cite{Siegel:1993th} and applied to DFT in \cite{Hohm:2010pp,Hohm:2010xe}. To this end, we express the generalized metric in terms of the generalized vielbein $E_A{}^M$ via
\begin{equation}\label{eqn:genmetricformvielbein}
  \mathcal{H}^{MN} = E_A{}^M S^{AB} E_B{}^N \,. 
\end{equation}
Using a frame formalism, we distinguish between flat and curved indices. The former are labeled by $A, B, \dots$ and the latter by $I, J, \dots\,$. In the literature, there are two different ways of defining the generalized vielbein: In the flux formulation of DFT \cite{Geissbuhler:2011mx,Aldazabal:2013sca}, $E_A{}^M$ is an O($D,D$) valued matrix, whereas the original frame formalism of \cite{Siegel:1993th,Hohm:2010xe} uses GL($D$)$\times$ GL($D$) vielbeins. Both conventions will be relevant later on. We begin with the one used in the flux formulation. Here, the flat generalized metric is given by
\begin{equation}
  S^{AB} = \begin{pmatrix} \eta_{ab} & 0 \\ 
      0 & \eta^{ab}
    \end{pmatrix}\,,
\end{equation}
where $\eta_{ab}$ and its inverse $\eta^{ab}$ denote the $D$-dimensional Minkowski metric. Due to the O($D,D$) valued generalized vielbein, the flat version
\begin{equation}\label{eqn:vielbeinodd}
  \eta^{AB} = E^A{}_M \eta^{MN} E^B{}_N
  \quad \text{with} \quad
  \eta^{AB}=\begin{pmatrix}
    0 & \delta_a^b \\
    \delta^a_b & 0
  \end{pmatrix}\,.
\end{equation}
of $\eta_{MN}$ does not differ from the curved one.

Now, consider the local double Lorentz transformation of the generalized vielbein
\begin{equation}\label{eqn:localodxodsym}
  E^A{}_M \rightarrow T^A{}_B E^B{}_M\,.
\end{equation}
We require that it leaves the generalized metric invariant. Hence, the transformation has to fulfill
\begin{equation}\label{eqn:trafoo2d}
  T^A{}_C S^{CD} T^B{}_D = S^{AB}\,.
\end{equation}
In addition, the transformed generalized vielbein has still to satisfy \eqref{eqn:vielbeinodd}, which gives rise to the further constraint
\begin{equation}\label{eqn:trafoodd}
  T^A{}_C \eta^{CD} T^B{}_D = \eta^{AB}\,.
\end{equation}
Transformations that simultaneously solve \eqref{eqn:trafoo2d} and \eqref{eqn:trafoodd} are of the form
\begin{equation}\label{eqn:trafoonxon}
  T^A{}_B = 
  \begin{pmatrix}
    u_a{}^b + v_a{}^b & u_{ab} - v_{ab} \\
    u^{ab} - v^{ab} & u^a{}_b + v^a{}_b
  \end{pmatrix}
\end{equation}
where $u_a{}^b$ and $v_a{}^b$ denote two independent O($1,D-1$) transformations with the defining properties
\begin{equation}
  u_a{}^c \eta_{cd} u_b{}^d = \eta_{ab}
  \quad \text{and} \quad
  v_a{}^c \eta_{cd} v_b{}^d = \eta_{ab}\,.
\end{equation}
By leaving the generalized and the $\eta$ metric invariant, double Lorentz transformations are a manifest symmetry of the action \eqref{eqn:dftaction}.

Except for the dilaton, the generalized vielbein combines all fields of the theory. As an element of O($D,D$) it has $D(2D-1)$ independent degrees of freedom. By gauge fixing the local double Lorentz symmetry only $D^2$ of them remain. A possible parameterization of the generalized vielbein is given by
\begin{equation}\label{eqn:EAMfixed}
  E^A{}_M = \begin{pmatrix}
    e_a{}^i & e_a{}^l B_{li} \\
    0 & e^a{}_i
  \end{pmatrix}
\end{equation}
in terms of the metric's vielbein $e^a{}_i$ with $e^a{}_i \eta_{ab} e^b{}_j = g_{ij}$ and the antisymmetric $B$-field $B_{ij}$. If $e^a{}_i$ is restricted to be an upper triangular matrix, this parameterization fixes the double Lorentz symmetry completely. An O($D,D$) vielbein without any gauge fixing is
\begin{equation}\label{eqn:EAMgeneral}
  E^A{}_M = \begin{pmatrix}
    e_a{}^i & e_a{}^l B_{li} \\
    e^a{}_l \beta^{li} & e^a{}_i + e^a{}_l \beta^{lk} B_{ki}
  \end{pmatrix}
\end{equation}
where $e^a{}_i$ is an unrestricted vielbein of $g_{ij}$ and $\beta^{ij}$ is an antisymmetric bi-vector.

In the frame formalism of \cite{Siegel:1993th, Hohm:2010pp, Hohm:2010xe}, the flat generalized metric reads
\begin{equation}
  S_{\bar A\bar B} = \begin{pmatrix}
    \eta_{a b} & 0 \\
    0 & \eta_{\bar a\bar b} \\
  \end{pmatrix}\,.
\end{equation}
We have introduced two new set of indices: unbared and bared indices. They are in one-to-one correspondence with the closed string's left- and right-moving parts. Two identical, $D$-dimensional Minkowski metrics $\eta_{ab}$, $\eta_{\bar a\bar b}$ and their inverse $\eta^{ab}$, $\eta^{\bar a\bar b}$ lower and raise them. The $2D$-dimensional $\eta$ metric, which lowers and raises doubled indices, is defined as
\begin{equation}
  \eta_{\bar A\bar B} = \begin{pmatrix}
    \eta_{ab} & 0 \\
    0 & - \eta_{\bar a\bar b}
  \end{pmatrix}
  \quad \text{and} \quad
  \eta^{\bar A\bar B} = \begin{pmatrix}
    \eta^{ab} & 0 \\
    0 & - \eta^{\bar a\bar b}
  \end{pmatrix}\,.
\end{equation}
In order to link quantities in this frame formalism with the corresponding ones in the flux formulation, we apply the transformation
\begin{equation}\label{eqn:bartounbar}
  T^{\bar A}{}_B = \frac{1}{\sqrt{2}}
  \begin{pmatrix}
    u^{ba} & u_b{}^a \\
    -v^{b\bar a} & v_b{}^{\bar a} 
  \end{pmatrix} \quad \text{and} \quad
  T_{\bar A}{}^B = \frac{1}{\sqrt{2}}
  \begin{pmatrix}
    u_{ba} & u^b{}_a \\
    -v_{b\bar{a}} & v^b{}_{\bar a} 
  \end{pmatrix}\,,
\end{equation}
respectively. Again, $u_a{}^b$ and $v_a{}^{\bar b}$ are two independent Lorentz transformations fulfilling
\begin{equation}
  u_a{}^c \eta_{cd} u_b{}^d = \eta_{ab}
  \quad \text{and} \quad
  v_a{}^{\bar c} \eta_{\bar c\bar d} v_b{}^{\bar d} = \eta_{ab}\,.
\end{equation}
Applying \eqref{eqn:bartounbar} to the generalized and the $\eta$ metric, we obtain
\begin{equation}
  T^{\bar A}{}_C S_{\bar A\bar B} T^{\bar B}{}_D = S_{CD} 
    \quad \text{and} \quad
  T^{\bar A}{}_C \eta_{\bar A\bar B} T^{\bar B}{}_D = \eta_{CD} \,.
\end{equation}
Further, we calculate the generalized vielbein
\begin{equation}
  E_{\bar A}{}^M = T_{\bar A}{}^B E_B{}^M = \frac{1}{\sqrt{2}} \begin{pmatrix}
    e_{ai} + e_a{}^l B_{li} & e_a{}^i \\
    - e_{\bar ai} + e_a{}^l B_{li} & e_{\bar a}{}^i \\
  \end{pmatrix}
\end{equation}
starting from \eqref{eqn:EAMfixed} and fixing $u_a{}^b = \delta_a^b$, $v_a{}^{\bar b}=\delta_a^{\bar b}$.

\section{Equations of motion}\label{sec:dfteom}
The field equations of DFT are obtained by the variation of the DFT action with respect to the generalized metric and the generalized dilaton. The variation with respect to the generalized metric yields
\begin{equation}  
  \delta S_\mathrm{DFT} = \int d^{2D} X e^{-2 d}  \mathcal{K}_{MN} \delta \mathcal{H}^{MN}
\end{equation}
with\footnote{We use the abbreviations
\begin{equation*}
  T_{[a_1 \dots a_n]} = \frac{1}{n!}  \sum\limits_{\sigma\in P} \sign(\sigma) T_{\sigma_1 \dots \sigma_n}
  \quad \text{and} \quad 
  T_{(a_1 \dots a_n)} = \frac{1}{n!}  \sum\limits_{\sigma\in P} 
    T_{\sigma_1 \dots \sigma_n}\,,
\end{equation*}
where $P$ is the set of all permutations of the indices $a_1,\dots,a_n$,
for the (anti)symmetrization of rank $n$ tensors.}
\begin{gather}
  \mathcal{K}_{MN} = \frac{1}{8} \partial_M \mathcal{H}^{KL} \partial_N \mathcal{H}_{KL}
  - \frac{1}{4}\left( \partial_L - 2(\partial_L d)\right) \left( \mathcal{H}^{KL}\partial_K \mathcal{H}_{MN}\right)
  + 2\partial_M\partial_N d \nonumber \\
  - \frac{1}{2} \partial_{(M} \mathcal{H}^{KL} \partial_L \mathcal{H}_{N)K}
  + \frac{1}{2} \left( \partial_L - 2( \partial_L d) \right) \left( \mathcal{H}^{KL}\partial_{(M} \mathcal{H}_{N)K} + \mathcal{H}^K{}_{(M} \partial_K \mathcal{H}^L{}_{N)} \right)
  \,.
\end{gather}
However, this does not lead to the equations of motion for the generalized metric directly, because $\mathcal{H}^{MN}$ is a constrained field. To determine the proper projection that encodes the equations of motion, we have to remember that the generalized metric is O($D,D$) valued and must fulfill
\begin{equation}
  \mathcal{H}^{LM} \eta_{MN} \mathcal{H}^{KN} = \eta^{KL}\,.
\end{equation}
The variation of this constraint leads to
\begin{equation}\label{eqn:constrvariation}
  \delta \mathcal{H}^{LM} \mathcal{H}^K{}_M +
  \mathcal{H}^L{}_M \delta \mathcal{H}^{KM} = 0
\end{equation}
and after some relabeling of indices and using $\mathcal{H}^{ML} \mathcal{H}_{LN} = \delta^M_N$ one obtains
\begin{equation}
  \delta\mathcal{H}^{MN} = -\mathcal{H}^{MK}  
    \delta\mathcal{H}_{KL} \mathcal{H}^{LN}\,.
\end{equation}
As described in \cite{Hohm:2010pp,Hohm:2011si}, the most general variation $\delta\mathcal{H}^{MN}$ satisfying \eqref{eqn:constrvariation} can be written as
\begin{gather}
  \delta\mathcal{H}^{MN} =  
  P^{MK} \delta\mathcal{M}_{KL} \bar{P}^{LN} +
  \bar{P}^{MK} \delta\mathcal{M}_{KL} P^{LN}  \\ \label{eqn:projectorsPbarP}
\quad \text{with} \quad
  P^{MN} = \frac{1}{2}\left(\eta^{MN} - \mathcal{H}^{MN}\right)
\quad \text{and} \quad
  \bar{P}^{MN} = \frac{1}{2}\left(\eta^{MN} + \mathcal{H}^{MN}\right)\,,
\end{gather}
where $\delta \mathcal{M}_{MN}$ is now an arbitrary, unconstrained symmetric variation. Because this new variation is not subject to any constraints, it leads to 
\begin{equation}
  \delta S_\mathrm{DFT} = \int \mathrm{d}^{2D} X \mathcal{K}^{MN} \delta 
    \mathcal{H}_{MN} = \int \mathrm{d}^{2D} X \mathcal{R}_{MN} \delta \mathcal{M}^{MN}\,,
\end{equation}
where
\begin{equation}\label{eqn:genriccitensor}
  \mathcal{R}_{MN} = P_{MK} \mathcal{K}^{KL}
    \bar{P}_{LN} + \bar{P}_{MK} \mathcal{K}^{KL} P_{LN}
\end{equation}
is called the generalized Ricci tensor. Then the equation
\begin{equation}
  \mathcal{R}_{MN}=0
\end{equation}
is the equation of motion for the generalized metric. Because the generalized metric $\mathcal{H}^{MN}$ is symmetric, $\mathcal{K}_{MN}$ and $\mathcal{R}_{MN}$ are symmetric, too. The variation with respect to the generalized dilaton gives rise to
\begin{equation}
  \mathcal{R} = 0\,.
\end{equation}
An important difference between general relativity and DFT is that the generalized curvature scalar does not arise by contraction of the generalized Ricci tensors with $\eta^{MN}$. Both quantities are independent. 

The equations of motion inherit the covariant transformation behavior under generalized diffeomorphisms of the action. Thus, one obtains
\begin{equation}
  \delta_\xi \mathcal{R}_{MN} = \mathcal{L}_\xi \mathcal{R}_{MN} \quad \text{and} \quad
  \delta_\xi \mathcal{R} = \mathcal{L}_\xi \mathcal{R} = \xi^I \partial_I \mathcal{R}\,.
\end{equation}
Further, they are manifestly invariant under double Lorentz transformations.

\section{Flux formulation}\label{sec:covariantfluxes}
Besides the generalized metric formulation of DFT, which we have discussed in section~\ref{sec:dftsym}, there is the flux formulation \cite{Geissbuhler:2013uka,Aldazabal:2013sca}. Instead of the generalized metric $\mathcal{H}^{MN}$, the so called covariant fluxes
\begin{equation}\label{eqn:covfluxes}
  \mathcal{F}_{ABC} = 3 \Omega_{[ABC]} \quad \text{and} \quad
  \mathcal{F}_A = \Omega^B{}_{BA} + 2 E_A{}^I \partial_I d\,,
\end{equation}
where 
\begin{equation}\label{eqn:coeffanholo}
  \Omega_{ABC}=E_A{}^N\partial_N E_B{}^M E_{CM}
\end{equation}
denotes the coefficients of anholonomy, are considered as dynamical fields. They have to be compatible with the generalized diffeomorphisms presented in section~\ref{sec:dftsym}. Because the generalized Lie derivative acts on curved indices only, the covariant fluxes, which only possess flat indices, should transform like scalars. Let us explicitly check this property for $\mathcal{F}_{ABC}$. First, we provide an alternative way to express the covariant fluxes, namely in terms of the C-bracket
\begin{equation}\label{eqn:FABCCbracket}
  \mathcal{F}_{ABC} = \left[ E_A , E_B \right]_\mathrm{C}^L E_{CL}\,.
\end{equation}
To prove that this equation reproduces the definition \eqref{eqn:covfluxes}, we apply the definition of the C-bracket \eqref{eqn:Cbracket} and obtain
\begin{align}\label{eqn:omega_ABCexpanded}
  \mathcal{F}_{ABC} &=  E_A{}^N \partial_N
    E_B{}^L E_{CL} - \frac{1}{2} E_{AN} \partial^L 
    E_B{}^N E_{CL} - ( A \leftrightarrow B ) \nonumber \\
    &= \Omega_{ABC} + \frac{1}{2} \Omega_{CAB} - \Omega_{BAC} -
      \frac{1}{2} \Omega_{CBA} = \Omega_{ABC} + \Omega_{CAB} +
      \Omega_{BCA}\,.
\end{align}
In the last line we have used that the coefficients of anholonomy are antisymmetric with respect to their last two indices. This property is a consequence of 
\begin{equation}
  E_A{}^N \partial_N \big( E_B{}^M \eta_{ML}  E_C{}^L \big) = E_A{}^N \partial_N \eta_{BC} = 0
\end{equation}
and allows us to expand the antisymmetrization in the definition \eqref{eqn:covfluxes}
\begin{equation}\label{eqn:fluxescoeffanholo}
  \mathcal{F}_{ABC} = 3\Omega_{[ABC]} = \Omega_{ABC} + \Omega_{CAB} + \Omega_{BCA}\,,
\end{equation}
which indeed matches \eqref{eqn:omega_ABCexpanded}. Due to the closure of the gauge algebra under the strong constraint \eqref{eqn:strongconstraint}, the C-bracket transforms as a vector under generalized diffeomorphisms. To reproduce the generalized Lie derivative of the generalized metric \eqref{eqn:genLieH}, $E_A{}^I$ has to transform as a vector,
\begin{equation}\label{eqn:genLieEAM}
  \delta_\xi E^A{}_M = \mathcal{L}_\xi E^A{}_M = \xi^I \partial_I E^A{}_M + 
    (\partial_M \xi^I - \partial^I \xi_M) E^A{}_I\,,
\end{equation}
too. Hence, the contraction between the two vectors (C-bracket and the generalized vielbein) in \eqref{eqn:FABCCbracket} transforms like a scalar. For $\mathcal{F}_A$, one has to calculate $\delta_\xi \mathcal{F}_A$ explicitly. Taking \eqref{eqn:genLieEAM} and the strong constraint \eqref{eqn:strongconstraint} into account gives rise to
\begin{equation}
  \delta_\xi \mathcal{F}_A = \xi^I \partial_I \mathcal{F}_A\,,
\end{equation}
which proves that $\mathcal{F}_A$ is also scalar under generalized diffeomorphisms.

\subsection{Closure constraint}
It is possible to obtain the same results even if one replaces the strong constraint with the weaker closure constraint \cite{Geissbuhler:2013uka}
\begin{equation}\label{eqn:closureconst}
  \Delta_{\xi_1} \delta_{\xi_2} \cdot  = 0\,.
\end{equation}
Here, $\Delta_\xi \cdot$ denotes the failure of the quantity $\cdot$ to transform covariantly. It is defined as
\begin{equation}\label{eqn:failuregendiff}
  \Delta_\xi V = \delta_\xi V - \mathcal{L}_\xi V \,.
\end{equation}
In addition to \eqref{eqn:FABCCbracket}, another way to express $\mathcal{F}_{ABC}$ is in terms of the gauge transformation
\begin{equation}
  \mathcal{F}_{ABC} = \delta_{E_A} E_B{}^M E_{CM} = \mathcal{L}_{E_A} E_B{}^M E_{CM}\,.
\end{equation}
This identity follows directly from the connection
\begin{equation}
  [ \xi_1, \xi_2 ]_\mathrm{C}^I = \mathcal{L}_{\xi_1} \xi_2^I - \frac{1}{2} \partial^I ( \xi_1^J \xi_{2\,J} )
\end{equation}
between C-bracket and generalized Lie derivative. It allows us to substitute the requirement $\Delta_\xi \mathcal{F}_{ABC} = 0$ discussed in the last subsection with the closure constraint
\begin{equation}
  \Delta_{E_A} \mathcal{F}_{BCD} = \Delta_{E_A} \delta_{E_B} E_C{}^M E_{DM} = \mathcal{Z}_{ABCD} =  0
\end{equation}
for the generalized vielbein. In the same spirit, we calculate
\begin{equation}
  \mathcal{F}_A = 2 \delta_{E_A} d = 2 \mathcal{L}_{E_A} d = 2 E_A{}^I \partial_I d - \partial_I E_A{}^I = \Omega^B{}_{BA} - 2 E_A{}^I \partial_I d
\end{equation}
which allows us to identify the covariant transformation behavior of $\mathcal{F}_A$,
\begin{equation}\label{eqn:closuregendiltradi}
  \Delta_{E_A} \mathcal{F}_B = 2 \Delta_{E_A} \delta_{E_B} d = \mathcal{Z}_{AB} = 0\,,
\end{equation}
with the closure constraint for the generalized dilaton. Following \cite{Geissbuhler:2013uka}, we have introduced the two quantities
\begin{align}
  \mathcal{Z}_{ABCD} &= D_{[A} \mathcal{F}_{BCD]} - \frac{3}{4} \mathcal{F}_{[AB}{}^E \mathcal{F}_{CD]E} \\
  \mathcal{Z}_{AB}   &= D^C \mathcal{F}_{CAB} + 2 D_{[A} \mathcal{F}_{B]} - \mathcal{F}^C \mathcal{F}_{CAB}\,.
\end{align}
Further we use the abbreviation 
\begin{equation}
  D_A = E_A{}^I \partial_I
\end{equation}
which one denotes as flat derivative. Under the strong constraint $\mathcal{Z}_{AB}$ and $\mathcal{Z}_{ABCD}$ vanish. However, there are also cases where they vanish and still the strong constraint is violated. Simple, but non-trivial examples for these cases are generalized Scherk-Schwarz compactifications which we are going to discuss in chapter~\ref{chap:genScherkSchwarz} in detail. In the internal directions, their covariant fluxes are restricted by
\begin{equation}
  \mathcal{F}_{ABC} = \text{const.} \quad \text{and} \quad
  \mathcal{F}_A = 0 \,.
\end{equation}
In this case, $\mathcal{Z}_{AB}=0$ is automatically fulfilled and $\mathcal{Z}_{ABCD}=0$ can be identified with the Jacobi identity
\begin{equation}\label{eqn:jacobiid}
  \mathcal{Z}_{ABCD} = -\frac{1}{4} \big( \mathcal{F}_{AB}{}^E \mathcal{F}_{EC}{}^D + \mathcal{F}_{CA}{}^E \mathcal{F}_{EB}{}^D +
    \mathcal{F}_{BC}{}^E \mathcal{F}_{EA}{}^D \big)= 0\,.
\end{equation}

\subsection{Action and equations of motion}
The action of the flux formulation reads \cite{Geissbuhler:2013uka}
\begin{align}
  S_\mathrm{DFT} = \int d^{2D} X\, e^{-2d} &\big( \mathcal{F}_A \mathcal{F}_B S^{AB} + \frac{1}{4} \mathcal{F}_{ACD} \mathcal{F}_B{}^{CD} S^{AB} - \frac{1}{12} \mathcal{F}_{ABC} \mathcal{F}_{DEF} S^{AD} S^{BE} S^{CF} \nonumber \\ \label{eqn:Sdftfluxform}
    & \quad - \frac{1}{6} \mathcal{F}_{ABC} \mathcal{F}^{ABC} - \mathcal{F}_A \mathcal{F}^A \big)\,.
\end{align}
Its form is inspired by the scalar potential of the bosonic sector of half-maximal, electrically gauged supergravity \cite{Schon:2006kz}. The second line vanishes under the strong constraint. Expressing the generalized metric in terms of generalized vielbeins as $\mathcal{H}^{MN}= E_A{}^M S^{AB} E_B{}^N$ and applying the strong constraint, the first line is equivalent to the action in the generalized metric formulation \eqref{eqn:dftaction} up to boundary terms. Violating the strong constraint, \eqref{eqn:Sdftfluxform} contains additional terms in comparison with the action in generalized metric formulation. These terms were added by hand in order to make a connection to the scalar potential of half-maximal, gauged supergravities. Still, they are compatible with generalized diffeomorphisms after applying the closure constraint \eqref{eqn:closureconst}.

Due to the closure constraint, the action is manifestly invariant under generalized diffeomorphisms. But, in contrast to the generalized metric formulation it lacks a manifest double Lorentz invariance. Applying \eqref{eqn:localodxodsym} to the definition of the covariant fluxes \eqref{eqn:covfluxes} gives rise to
\begin{equation}
  \mathcal{F}_{ABC} \rightarrow T_A{}^D T_B{}^E T_C{}^F \mathcal{F}_{DEF} + 
    3 T_{[A}{}^D E_D{}^M \partial_M T_B{}^E T_{C]E}\,.
\end{equation}
The first term on the right hand side represents the covariant transformation behavior of a rank three tensor. But, the second term spoils covariance. According to \cite{Geissbuhler:2013uka}, an infinitesimal double Lorentz transformation of the full action \eqref{eqn:Sdftfluxform} yields
\begin{equation}
  \delta_\Lambda S_\mathrm{DFT} = \int d^{2D} X e^{-2d}\, \Lambda_A{}^C ( \eta^{AB} - S^{AB} ) \mathcal{Z}_{BC}
\end{equation}
which vanishes under the closure constraint due to $\mathcal{Z}_{AB} = 0$.

In order to derive the equations of motion, we calculate the variation of the flux formulation action with respect to the generalized vielbein giving rise to
\begin{equation}
  \delta S_\mathrm{DFT} = \int d^{2D} X e^{-2d} \,\mathcal{G}^{AB} \Delta_{AB}
    \quad \text{with} \quad
  \Delta_{AB} = \delta E_A{}^M E_{BM}\,.
\end{equation}
Due to the antisymmetry of $\Delta_{AB}$, which follows from
\begin{equation}
  \delta (E_A{}^M E_{BM} ) = \delta \eta_{AB} = 0 = \Delta_{AB} + \Delta_{BA}\,,
\end{equation}
only the antisymmetric part of $\mathcal{G}^{AB}$ contributes. After explicit calculations \cite{Geissbuhler:2013uka,Blumenhagen:2013zpa}, one obtains
\begin{equation}
   \mathcal{G}^{AB}= \mathcal{Z}^{AB} + 2 S^{C[A} D^{B]} \mathcal{F}_C + (\mathcal{F}_C- D_C)
   \breve{\mathcal{F}}^{C[AB]} +\breve{\mathcal{F}}^{CD[A}\, \mathcal{F}_{CD}{}^{B]}=0
\end{equation}
with
\begin{equation}
  \breve{\mathcal{F}}^{ABC}=\breve{S}^{ABCDEF}\,\mathcal{F}_{DEF} 
\end{equation}
and the projector
\begin{equation}
  \breve{S}^{ABCDEF} = \frac{1}{2} S^{AD}\, \eta^{BE}\, \eta^{CF} +
                       \frac{1}{2} \eta^{AD}\, S^{BE}\, \eta^{CF}+
                       \frac{1}{2} \eta^{AD}\, \eta^{BE}\, S^{CF}\\
                -\frac{1}{2} S^{AD}\, S^{BE}\, S^{CF}
\end{equation}
which fulfills $\breve{S}^2=1$. Applying the strong constraint, it is possible to express
\begin{equation}
  \mathcal{G}^{AB} = P^{[A}{}_C \bar{P}^{B]}{}_D \mathcal{K}_{AB}
\end{equation}
in terms of the symmetric tensor $\mathcal{K}_{AB}$, which we have introduced in section~\ref{sec:dfteom} to derive the equation of motion in the generalized metric formulation. For the projectors $P_{AB}$ and $\bar P_{AB}$ defined in \eqref{eqn:projectorsPbarP}, one is able to show \cite{Geissbuhler:2013uka} the equivalence
\begin{equation}
  P^{[A}{}_C \bar{P}^{B]}{}_D \mathcal{K}_{AB} = 0 \quad \Leftrightarrow \quad P^{(A}{}_C \bar{P}^{B)}{}_D \mathcal{K}_{AB} = 0\,.
\end{equation}
This allows us to identify the equations of motion of the flux and the generalized metric formulation
\begin{equation}
  \mathcal{G}^{AB} = 0 \quad \Leftrightarrow \quad \mathcal{R}^{AB} = 0
\end{equation}
under the strong constraint.

\subsection{\texorpdfstring{$H$-, $f$-, $Q$- and $R$-}{H-, f-, Q-, R-}flux}\label{sec:dftnongeofluxes}
There is an intriguing connection between the covariant fluxes $\mathcal{F}_{ABC}$ and geometric as well as non-geometric fluxes. As explained in section~\ref{sec:nongeombackgr}, geometric $H$- and $f$-fluxes arise e.g. in the scalar potential of SUGRA flux compactifications. In order to make such potentials invariant under T-duality transformations of the compact space, \cite{Shelton:2005cf} has introduced the additional, non-geometric $Q$- and $R$-fluxes. All these fluxes arise if one explicitly calculates the various components of $\mathcal{F}_{ABC}$. To this end, we start with the generalized vielbein \eqref{eqn:EAMgeneral} that possesses both the two-form $B_{ij}$ and the bi-vector $\beta^{ij}$. Its double Lorentz symmetry is not gauged fixed. Depending on the physical situation, one may gauge fix to a frame containing only the $B$-field, only the bivector $\beta^{ij}$, or some intermediate frame. For a gauge without independent $B$-field the covariant fluxes reduce to those identified in \cite{Andriot:2012wx,Andriot:2012an}. The generalized vielbein with the flat index lowered and the curved one raised reads
\begin{equation}
\label{eqn:paramUAM}
  E_A{}^M = \eta_{AB} E^B{}_N \eta^{NM} =
  \begin{pmatrix}
    e^a{}_i + e^a{}_j \beta^{jk} B_{ki} &
    e^a{}_j \beta^{ji} \\
    e_a{}^j B_{ji} & e_a{}^i
  \end{pmatrix}\,.
\end{equation}
Because the covariant fluxes $\mathcal{F}_{ABC}$ are totally antisymmetric, only four of the eight $D\!\times\! D\!\times\! D$ blocks they consist of are independent from each other. Each of these independent blocks, namely
\begin{equation}
  \mathcal{F}_{abc}\,, \quad \mathcal{F}^a{}_{bc}\,, \quad 
  \mathcal{F}^{ab}{}_c \quad \text{and} \quad \mathcal{F}^{abc}\,,
\end{equation}
will be evaluated. Combining these results, we obtain $H$-, $f$-, $Q$- and $R$-flux in flat indices.

We start with ${\mathcal F}_{abc}$ given in terms of
\begin{equation}\label{eqn:Fabcfromcoeffanholo}
  \mathcal{F}_{abc} = \Omega_{abc} + \Omega_{cab} +
  \Omega_{bca} = 3\Omega_{[abc]}\,.
\end{equation}
Putting \eqref{eqn:paramUAM} into \eqref{eqn:coeffanholo}, the relevant coefficients of anholonomy evaluate to
\begin{equation}
  \Omega_{abc}= e_a{}^i e_b{}^j e_c{}^k \big( \partial_i B_{jk} + B_{il} 
    \tilde{\partial}^l B_{jk} \big)\,.
\end{equation}
Combining this result with the antisymmetrization of $\Omega_{abc}$ in \eqref{eqn:Fabcfromcoeffanholo} gives rise to
\begin{equation}
  \mathcal{F}_{abc} = 3 e_a{}^i e_b{}^j e_c{}^k \big( 
    \partial_{[i} B_{jk]} - B_{l[i} \tilde{\partial}^l B_{jk]} \big) = H_{abc}\,.
\end{equation}
Choosing the explicit realization $\tilde \partial_i \cdot = 0$ of the strong constraint, this expression is equivalent to the $H$-flux in flat indices. In the next step, we calculate the three $\Omega_{ABC}$ components $\Omega^a{}_{bc}$, $\Omega_a{}^b{}_c$ and $\Omega_{ab}{}^c$ with two lowered indices and one raised index. They are given by the following expressions
\begin{align}\label{eqn:Omega^a_b_c}
  \Omega^a{}_{bc} &= e^a{}_i e_b{}^j e_c{}^k \big( \tilde{\partial}^i 
    B_{jk} + \beta^{il} \Omega_{ljk} \big) \nonumber  \\
  \Omega_a{}^b{}_c &= e_a{}^i \partial_i e^b{}_j e_c{}^j
    + e_a{}^i B_{ij} \tilde{\partial}^j e^b{}_k e_c{}^k
    + e_a{}^i e^b{}_j e_c{}^k \beta^{jl} \Omega_{ilk} \nonumber  \\
  \Omega_{ab}^c &= -\Omega_a{}^c{}_b \,.
\end{align}
With these three components, the covariant fluxes $\mathcal{F}^a{}_{bc}$ read
\begin{align}
  \mathcal{F}^a{}_{bc} &= \Omega^a_{[bc]}
  + \Omega_{[c}{}^a{}_{b]} + \Omega_{[bc]}{}^a = \Omega^a_{[bc]} + 
  2 \Omega_{[c}{}^a{}_{b]} \nonumber \\
  &= 2 \big( e_{[c}{}^i \partial_i e^a{}_j e_{b]}^j
    + e_{[c}{}^i B_{ij} \tilde{\partial}^j e^a{}_k e_{b]}{}^k 
    \big) + e^a{}_i e_b{}^j e_c{}^k \big( \tilde{\partial}^i 
    B_{jk} + \beta^{il} H_{ljk} \big)= f^a_{bc}\,.
\end{align}
They are equivalent to the geometric flux $f^a{}_{bc}$ in flat indices. This equivalence gets manifest, if a frame is chosen where $\tilde \partial^i \cdot = 0$ and $\beta^{ij}=0$ holds. Then $\mathcal{F}^a{}_{bc}$ reads
\begin{equation}
  \mathcal{F}^a{}_{bc} = 2 e_{[c}{}^i \partial_i e^a{}_j e_{b]}{}^j = f^a_{bc}\,,
\end{equation}
which is exactly the form given by e.g. \cite{Blumenhagen:2013hva}. In order to calculate $\mathcal {F}^{ab}_c$ we need the anholonomy coefficient's components
\begin{align}
  \Omega^{ab}_c &= e^a{}_i\tilde\partial^i e^b{}_j e_c{}^j + e^a{}_i e^b{}_j e_c{}^k ( \beta^{il}
    \Omega_l{}^j{}_k + \tilde\partial^i B_{lk} \beta^{jl} ) \nonumber \\
  \Omega_a{}^{bc} &= e_a{}^i e_j{}^b e_k{}^c \big( \partial_i \beta^{jk} + B_{il}\tilde\partial^l
    \beta^{jk} + \beta^{jl} \beta^{km} \Omega_{ilm} \big) \quad \text{and} \nonumber \\
  \Omega^a{}_b{}^c &= -\Omega^{ac}_b\,.
\end{align}
We combine them to
\begin{align}
  \mathcal{F}^{ab}_c &= \Omega^{[ab]}{}_c + \Omega_c{}^{[ab]} + \Omega^{[b}{}_c{}^{a]} 
  = 2 \Omega^{[ab]}{}_c + \Omega_c{}^{[ab]} = 2 e^{[a}{}_i \tilde\partial^i e^{b]}{}_j e_c{}^j 
    \nonumber \\ 
  &+ e_i{}^{[a} e_j{}^{b]} e_c{}^k \big( \partial_k \beta^{ij} + B_{kl}\tilde\partial^l \beta^{ij} +
    2 \tilde\partial^i B_{lk} \beta^{jl} - \beta^{li} \left[ 2 \Omega_l{}^j{}_k  +
    \beta^{jn} \Omega_{kln} \right]  \big) = Q_c^{ab}
\end{align}
which is equivalent to the $Q$-flux in flat indices. In the frame $\tilde \partial^i \cdot = 0$ and $B_{ij}=0$, this expression simplifies to
\begin{equation}
  \mathcal{F}^{ab}{}_c = e_i{}^a e_j{}^b e_c{}^k \big( \partial_k \beta^{ij} - 
    \beta^{l[i} f^{j]}_{kl} \big) =Q^{ab}_c
\end{equation}
and thus is equivalent to the $Q$-flux defined in e.g. \cite{Andriot:2013xca}. Finally, we have
\begin{align}
  \Omega^{abc} &= e^a{}_i e^b{}_j e^c{}_k \big(\tilde\partial^i \beta^{jk} + \beta^{il} 
    \Omega_l{}^{jk} + \tilde\partial^i B_{ml} \beta^{lj} \beta^{km} \big)\,,
\end{align}
which gives rise to
\begin{align}
  \mathcal{F}^{abc} = 3 \Omega^{[abc]} = e^a{}_i e^b{}_j e^c{}_k 3 \big(&
    \tilde\partial^{[i} \beta^{jk]} - \beta^{i[l} \partial_l \beta^{jk]} \nonumber\\ \label{eqn:Rflux}
  &+ B_{ln} \tilde\partial^n \beta^{[jk} \beta^{i]l}  + \beta^{l[k} \tilde\partial^i B_{lm} \beta^{j]m} + \frac{1}{3} \beta^{il} \beta^{jm} \beta^{kn} \mathcal{F}_{lmn}\big)
\end{align}
and is equivalent to the $R$-flux in flat indices. To see this, we use the frame $\tilde \partial^i \cdot =0$ and $B_{ij}=0$ in which \eqref{eqn:Rflux} reads
\begin{equation}
  \mathcal{F}^{abc} = e^a{}_i e^b{}_j e^c{}_k 3 \beta^{[il} \partial_l \beta^{jk]} = R^{abc}\,.
\end{equation}
This expression is equivalent to the $R$-flux defined in e.g.\cite{Andriot:2012an}. All these results agree with the ones presented in \cite{Geissbuhler:2013uka,Blumenhagen:2013hva} and show that the covariant fluxes are indeed a generalization of the fluxes known from the SUGRA effective action \eqref{eqn:nsnsaction}.

\chapter[Generalized Scherk-Schwarz compactification]{Generalized Scherk-Schwarz\\compactification}\label{chap:genScherkSchwarz}
While constructing backgrounds for closed string theory, a major challenge is to find non-trivial solutions for the background field equations. The NS/NS sector of these equations can be derived by varying the DFT action \eqref{eqn:dftaction} with respect to the generalized metric and the generalized dilaton. The resulting partial differential equations are involved, and in general it is impossible to solve them directly. One way to overcome this problem is to start with known supergravity solutions, like NS 5-branes or orthogonal intersections of them \cite{Hassler:2013wsa}. Here, we use another technique, called generalized Scherk-Schwarz compactification. It is the generalization of Scherk-Schwarz compactification \cite{Scherk:1978ta,Scherk:1979zr}, used for dimensional reduction of supergravity theories, to the doubled space of DFT. This compactification gives rise to a lower dimensional effective action whose equations of motion are easier to handle than the ones of full DFT. The action describes the bosonic sector of a half-maximal, electrically gauged supergravity. It is equipped with a scalar potential which severely restricts the vacua of the effective theory.

If the compactification ansatz is consistent, vacua of the effective theory can be uplifted to solutions of the DFT field equations. A consistent Scherk-Schwarz compactification has to possess enough isometries \cite{Scherk:1978ta,Scherk:1979zr,Pons:2003ka}. E.g., for a consistent dimensional reduction of a theory on a $n$-dimensional space one needs $n$ isometries. We show in section~\ref{sec:consistentcomp} that a consistent generalized Scherk-Schwarz compactification has to possess $2n$ isometries. Half of them are with respect to the coordinates $x^i$ and the other half is with respect to the dual coordinates $\tilde x_i$. In this case, the arrows in the diagram
\begin{equation}\label{eqn:consistentcomp}
  \begin{tikzpicture}[>=stealth',node distance=4em]
    \node (SDFT) {$S_\mathrm{DFT}$};
    \node (Seff) [right of=SDFT, xshift=20em] {$S_\mathrm{eff}$};
    \node (eomeff) [below of=Seff] {field equations};
    \node (soleff) [below of=eomeff] {solution};
    \node (eom) [below of=SDFT] {background field equations};
    \node (sol) [below of=eom] {background};
    \draw[->] (SDFT) -- (Seff) node[midway, above] {consistent compactification ansatz};
    \draw[->] (Seff) -- (eomeff) node[midway, above, anchor=west] {$\delta S_\mathrm{eff} = 0$};
    \draw[->] (eomeff) -- (soleff) node[midway, above, anchor=west] {solve};
    \draw[->] (soleff) -- (sol) node[midway, below] {uplift} ;
    \draw[->,dashed] (SDFT) -- (eom) node[midway, below, anchor=east] {$\delta S_\mathrm{DFT} = 0$};
    \draw[->,dashed] (eom) -- (sol) node[midway, below, anchor=east] {solve (involved)};
  \end{tikzpicture}
\end{equation}
commute.

In the last chapter we already discussed the dashed path in this diagram. Now, we will follow the path marked by the solid black lines. The following sections describe the way from $S_\mathrm{DFT}$ to the solution of the effective field theory's equations of motion. To this end, we start from a generalized Kaluza-Klein ansatz in section~\ref{sec:KKansatz} and restrict it to a group manifold to obtain a generalized Scherk-Schwarz ansatz. After discussing the properties of this ansatz, we drive the effective theory in section~\ref{sec:gaugedgravity}. Further, we explain how moduli of the compactification obtain masses due to non-vanishing background fluxes. These background fluxes can not be switched on arbitrarily but are severely restricted by the embedding tensor. Section~\ref{sec:examplecso202} presents an explicit example of a generalized Scherk-Schwarz compactification giving rise to an effective theory with CSO($2,0,2$) as gauge group. Based on the insights we gain from this example, we discuss conceptual issues of generalized Scherk-Schwarz compactifications. They motivate the derivation of double field theory on group manifolds in the next two chapters.

\section{Generalized Kaluza-Klein ansatz}\label{sec:KKansatz}
In a compactification one distinguishes between internal (compactified) and external (uncompactified) directions. Here, we assume that we have $D-n$ external and $n$ internal dimensions. To make this situation manifest, we split the $2D$ components of the vector $X^{\mathcal M}=\begin{pmatrix} \tilde x_i & x^i \end{pmatrix}$ into
\begin{equation}\label{eqn:compactindexstruct}
  X^{\hat M}= \begin{pmatrix} \tilde x_\mu & x^\mu  & Y^M \end{pmatrix} =    
    \begin{pmatrix} \mathds{X} & \mathds{Y} \end{pmatrix}\,, 
    \quad\text{where}\quad \mu=0,\dots, D-n-1
\end{equation}
counts the external directions and $Y^M$ is a vector in the internal doubled space. In this convention, the O($D,D$) invariant metric \eqref{eqn:etaMN} reads 
\begin{equation}
  \eta_{\hat M\hat N}=\begin{pmatrix}
    0 & \delta^\mu_\nu & 0 \\
    \delta_\mu^\nu & 0 & 0 \\
    0 & 0 & \eta_{MN} 
  \end{pmatrix}\,, \qquad
  \eta^{\hat M \hat N}=\begin{pmatrix}
    0 & \delta_\mu^\nu & 0\\
    \delta^\mu_\nu & 0 & 0\\
    0 & 0 & \eta^{MN} 
  \end{pmatrix}
\end{equation}
and the flat generalized metric is defined as
\begin{equation}
  S_{\hat A\hat B}=\begin{pmatrix}
    \eta^{ab} & 0 & 0 \\
    0 & \eta_{ab} & 0 \\
    0 & 0 & S_{AB}
  \end{pmatrix}\,, \qquad
  S^{\hat A\hat B}=\begin{pmatrix}
    \eta_{ab} & 0 & 0 \\
    0 & \eta^{ab} & 0 \\
    0 & 0 & S^{AB}
  \end{pmatrix}\,.
\end{equation} 
For flat indices we have adopted the index structure \eqref{eqn:compactindexstruct} of the curved ones. To obtain the curved version of the generalized metric, one introduces the gauged fixed generalized vielbein \cite{Geissbuhler:2011mx,Aldazabal:2013sca,Hohm:2013nja}
\begin{equation}\label{eqn:vielbeinKKansatz}
  E^{\hat A}{}_{\hat M} = \begin{pmatrix}
    e_\alpha{}^\mu & - e_\alpha{}^\rho C_{\mu\rho}
      & - e_\alpha{}^\rho A_{M\rho} \\
    0 & e^\alpha{}_\mu & 0 \\
    0 & E^A{}_L A^L{}_\mu & 
    E^A{}_M
  \end{pmatrix} \quad \text{with} \quad C_{\mu\nu} = B_{\mu\nu} + 
  \frac{1}{2} A^L{}_\mu A_{L\nu} \,.
\end{equation}
Its constituents are 
\begin{itemize}
  \item the $(D-n)$-dimensional vielbein $e^\alpha{}_\mu(\mathds{X})$ of the external space and
  \item the corresponding $B$-field $B_{\mu\nu}(\mathds{X})$,
  \item the $D-n$ $2n$-dimensional, covariant vectors $A_{M\mu}(\mathds{X},\mathds{Y})$ and
  \item the O($n,n$) valued vielbein $E^A{}_M(\mathds{X},\mathds{Y})$.
\end{itemize}
They form the field content of the effective theory which arises after the compactification. Altogether, they completely parameterize the $D^2$ degrees of freedom of the totally gauge fixed generalized vielbein \eqref{eqn:EAMfixed}. This special form of the generalized vielbein is called generalized Kaluza-Klein ansatz \cite{Hohm:2013nja}. We use the convention of the flux formulation. Thus, $E^{\hat A}{}_{\hat M}$ has to be O($D,D$) valued and must satisfy \eqref{eqn:vielbeinodd}. This is the case if and only if
\begin{equation}
  e_\alpha{}^\mu \eta^{\alpha\beta} e_\beta{}^\nu = 
    \eta^{\mu\nu} \quad \text{and} \quad
    E^A{}_M \eta_{AB} E^B{}_N = \eta^{MN}
\end{equation}
hold.

At the first glance, the presented parameterization of $E^{\hat A}{}_{\hat M}$ seems quite arbitrary. Its motivation becomes clear if one calculates the generalized Lie derivative
\begin{equation}\label{eqn:genlieKKansatz}
  \mathcal{L}_\xi E^{\hat A}{}_M =
  \left\{ \begin{aligned}
    \delta_\xi e^\alpha{}_\mu &= L_\xi  e^\alpha{}_\mu \\
    \delta_\xi B_{\mu\nu} &= L_\xi B_{\mu\nu} + \left(\partial_\mu \tilde \xi_\nu - 
    \partial_\nu \tilde \xi_\mu \right) + \partial_{[\mu} \Lambda_M A^M{}_{\nu]}\\
    \delta_\xi A_{M\mu} &= L_\xi A_{M\mu} + \mathcal{L}_\Lambda A_\mu{}^M + \partial_\mu \Lambda_M \\
    \delta_\xi E^A{}_M &= L_\xi E_A{}^M + \mathcal{L}_\Lambda E^A{}_M 
  \end{aligned} \right.
\end{equation}
with the parameter
\begin{equation}
  \xi^{\hat M} = \begin{pmatrix} \tilde \xi_\mu(x^\nu) & \xi^\mu(x^\nu) & \Lambda^M(x^\nu, \mathds{Y}) \end{pmatrix}\,.
\end{equation}
In the $D-n$ extended space time directions, there are no winding modes. Thus in these directions, the strong constraint \eqref{eqn:strongconstraint} is trivially solved by $\tilde \partial^\mu = 0$ and the partial derivative in doubled coordinates reduces to $\partial^{\hat M} = \begin{pmatrix} \partial_\mu & 0 & \partial^M \end{pmatrix}$. $L_\xi$ denotes the $(D-n)$-dimensional Lie derivative which acts on curved indices of the uncompactified directions only, e.g.
\begin{equation}
  L_\xi e^\alpha{}_\mu = \xi^\nu \partial_\nu e^\alpha{}_\mu + \partial_\mu \xi^\nu e^\alpha{}_\nu
    \quad \text{or} \quad
  L_\xi E^A{}_M = \xi^\mu \partial_\mu E^A{}_M\,,
\end{equation}
and $\mathcal{L}_\xi$ represents the generalized Lie derivative in the compactified space. According to \eqref{eqn:genlieKKansatz}, the parameters of $E^{\hat A}{}_{\hat M}$ transform covariantly with respect to both diffeomorphisms of the external and generalized diffeomorphisms of the internal space. This distinguished transformation behavior is the defining property of the generalized Kaluza-Klein ansatz \eqref{eqn:vielbeinKKansatz}. There are still additional terms besides $L_\xi \cdot$ and $\mathcal{L}_\xi \cdot$. They correspond to the gauge symmetries of the effective theory. We discuss them in detail in section~\ref{sec:gaugedgravity}.

\section{Generalized Scherk-Schwarz ansatz}\label{sec:scherkschwarz}
The Kaluza-Klein ansatz is still very general. In order to perform explicit calculations, we restrict it by splitting the generalized vielbein of the internal space 
\begin{equation}\label{eqn:twistofgenvielbein}
  E^A{}_M(X) = 
    \widehat E^A{}_N(\mathds{X}) U^N{}_M(\mathds{Y})
  \quad \text{and the vector} \quad
  A_{M\mu}(X) =
    \widehat A_{N\mu}(\mathds{X}) U^N{}_M(\mathds{Y})
\end{equation}
into $\mathds{X}$ and $\mathds{Y}$ dependent parts. Quantities which only depend on $\mathds{X}$ are marked with a hat. The tensor $U^N{}_M$ is called twist and plays a central role in Scherk-Schwarz compactifications. Note that the splitting \eqref{eqn:twistofgenvielbein} can also be written in the compact form
\begin{equation}\label{eqn:twistscherkschw}
  E^{\hat A}{}_{\hat M}(X) = 
    \widehat E^{\hat A}{}_{\hat N}(\mathds{X}) 
    U^{\hat N}{}_{\hat M}(\mathds{Y}) 
  \quad \text{with} \quad
  U^{\hat N}{}_{\hat M} = 
  \begin{pmatrix}
    \delta^\mu_\nu & 0 & 0 \\
      0 & \delta_\mu^\nu & 0 \\
      0 & 0 & U^N{}_M
  \end{pmatrix}\,.
\end{equation}
As previously emphasized, the generalized vielbein $E^{\hat A}{}_{\hat M}$ is O($D,D$) valued. The untwisted generalized vielbein $\hat E^{\hat A}{}_{\hat M}$ possesses this property by construction. Hence the twist $U^{\hat N}{}_{\hat M}$ also has to be O($D,D$) valued, which is the case if, and only if, $U^N{}_M$ is O($n,n$) valued.

\subsection{Twisted generalized diffeomorphisms}\label{sec:twistedgendiff}
The twist also affects the parameter of the generalized Lie derivative
\begin{equation}
  \label{eqn:ssansatzxi}
  \xi^{\hat M} = \widehat \xi^{\hat N} U_{\hat N}{}^{\hat M} =
  \begin{pmatrix} \tilde \xi_\mu & \xi^\mu & \Lambda^M \end{pmatrix}
    \quad \text{with} \quad
  \Lambda^M = \widehat{\Lambda}^N U_N{}^M\,.
\end{equation}
We use this parameter to calculate the generalized Lie derivative
\begin{align}
  \mathcal{L}_\xi V_{\hat M} & = \xi^{\hat P} \partial_{\hat P} V_{\hat M} + 
    \big( \partial_{\hat M} \xi^{\hat P} - \partial^{\hat P} \xi_{\hat M} \big)
    V_{\hat P} \nonumber \\
    &= \mathcal{L}_{\widehat \xi} \widehat{V}_{\hat I} U^{\hat I}{}_{\hat M} +
    \widehat{\xi}^{\hat L} \widehat{V}_{\hat N} \big( 
      U_{\hat L}{}^{\hat P} \partial_{\hat P} U^{\hat N}{}_{\hat M} +
      \partial_{\hat M} U_{\hat L}{}^{\hat P} U^{\hat N}{}_{\hat P} - 
      U^{\hat N}{}_{\hat P} \partial^{\hat P} U_{\hat L\hat M} \big) 
    \nonumber \\
    &= \big( \mathcal{L}_{\widehat \xi} \widehat{V}_{\hat I} 
      + \widehat{\xi}^{\hat L} \widehat{V}^{\hat N}
        \left[ \Omega_{\hat L\hat N \hat I} + \Omega_{\hat I\hat L\hat N} -
        \Omega_{\hat N\hat L\hat I} \right] \big) U^{\hat I}{}_{\hat M}
    \nonumber \\ \label{eqn:twistedgenLie}
    &= \big( \mathcal{L}_{\widehat \xi} \widehat V_{\hat I} +
    \mathcal{F}_{\hat I\hat N\hat L} \widehat \xi^{\hat N} \widehat V^{\hat L} \big)
      U^{\hat I}{}_{\hat M}
\end{align}
of the twisted vector $V_{\hat M} = \widehat{V}_{\hat N} U^{\hat N}{}_{\hat M}$. Analogous to the flux formulation of DFT, the abbreviations
\begin{equation}\label{eqn:maptwistflux}
  \Omega_{\hat I\hat J\hat K} = U_{\hat I}{}^{\hat M} \partial_{\hat M} U_{\hat J}{}^{\hat N} U_{\hat K\hat N}
    \quad \text{and} \quad
  \mathcal{F}_{\hat I\hat J\hat K} = 3 \Omega_{[\hat I\hat J\hat K]}
\end{equation}
are used. Due to the structure of the twist \eqref{eqn:twistscherkschw}, $\mathcal{F}_{\hat I\hat J\hat K}$ vanishes for all indices labeling external directions. Its non-vanishing components are linked to the covariant fluxes introduced in \eqref{eqn:fluxescoeffanholo} in section~\ref{sec:covariantfluxes} by
\begin{equation}\label{eqn:fluxflatcurved}
  \mathcal{F}_{ABC} = \widehat{E}_{A}{}^{I}
    \widehat{E}_{B}{}^{J} \widehat{E}_{C}{}^{K} \mathcal{F}_{IJK}\,.
\end{equation}
Hence in the following we will also call $\mathcal{F}_{\hat I\hat J\hat K}$ covariant fluxes.

An important consistency constraint the reduction ansatz has to fulfill is the closure of generalized diffeomorphisms mediated by \eqref{eqn:twistedgenLie}. Assuming that the strong constraint holds for the internal and external space, closure is automatically guaranteed. In this case, the relation
\begin{equation}\label{eqn:closuretwistedgenlie}
  \big[ \mathcal{L}_{\xi_1}, \mathcal{L}_{\xi_2} ] V_{\hat M} = \mathcal{L}_{\xi_{12}} V_{\hat M}
\end{equation}
holds with the resulting parameter
\begin{equation}
  \xi_{12}^{\hat M} = \big( [ \widehat{\xi}_1, \widehat{\xi}_2 ]_\mathrm{C}^{\hat I} + 
  \mathcal{F}^{\hat I}{}_{\hat J \hat K} \widehat{\xi}_1^{\hat J} \widehat{\xi}_1^{\hat K} \big)
    U_{\hat I}{}^{\hat M} = [\xi_1, \xi_2]_\mathrm{C}^{\hat M}
\end{equation}
where the explicit expression for the C-bracket arises from the identity
\begin{equation}
  [\widehat{\xi}_1, \widehat{\xi}_2 ]_\mathrm{C}^{\hat M} = \frac{1}{2} (
    \mathcal{L}_{\widehat{\xi}_1} \widehat{\xi}_2^{\hat M} - \mathcal{L}_{\widehat{\xi}_2} \widehat{\xi}_1^{\hat M} )
\end{equation}
and \eqref{eqn:twistedgenLie}. Further, one can check the closure constraint \eqref{eqn:closureconst}
\begin{equation}
  \Delta_{E_{\hat A}} \delta_{E_{\hat B}} E_{\hat C}{}^{\hat I} E_{\hat D\hat I} = \mathcal{Z}_{\hat A\hat B\hat C\hat D} = D_{[\hat A} \mathcal{F}_{\hat B\hat C\hat D]} - \frac{3}{4} \mathcal{F}_{[\hat A\hat B}{}^{\hat E} \mathcal{F}_{\hat C\hat D]\hat E} = 0\,.
\end{equation}
In the internal directions it yields
\begin{equation}
  \partial_{[I} \mathcal{F}_{JKL]}- \frac{3}{4} \mathcal{F}_{[IJ}{}^M \mathcal{F}_{KL]}{}^M = 0
\end{equation}
after applying \eqref{eqn:fluxflatcurved}. In the following we assume that the covariant fluxes are constant 
\begin{equation}
  \mathcal{F}_{IJK} = \text{const.}
\end{equation}
and so we are left with the Jacobi identity
\begin{equation}
\label{eqn:quadraticc}
  \mathcal{F}_{LMN}\mathcal{F}^{L}{}_{IK} +
  \mathcal{F}_{LIM}\mathcal{F}^{L}{}_{NK} +
  \mathcal{F}_{LNI}\mathcal{F}^{L}{}_{MK} = 0
    \quad \text{or} \quad 
  \mathcal{F}_{L[MN} \mathcal{F}^{L}{}_{I]K} = 0\,, 
\end{equation}
taking the total antisymmetry $\mathcal{F}_{NML} = \mathcal{F}_{[MNL]}$ into account. In the external directions, the strong constraint is implemented trivially by vanishing winding modes $\tilde \partial^\mu \cdot = 0$. Thus, all contributions to $\mathcal{Z}_{\hat A\hat B\hat C\hat D}$ in the external directions vanish, too. These constraints are sufficient for the closure of generalized diffeomorphisms stated in \eqref{eqn:closuretwistedgenlie} \cite{Aldazabal:2013sca,Grana:2012rr}.

Further restrictions arise if one considers the generalized dilaton
\begin{equation}\label{eqn:genSSdilaton}
  d(X) = \widehat{d}(\mathds{X})
\end{equation}
which we assume to be constant in the internal directions. Its generalized Lie derivative reads
\begin{equation}
  \mathcal{L}_\xi d = \mathcal{L}_{\widehat{\xi}} \widehat{d} - \frac{1}{2} \partial_{\hat M} U_{\hat N}{}^{\hat M}\,.
\end{equation}
The last term would introduce a $\mathds{Y}$ dependence which we ruled out by the ansatz \eqref{eqn:genSSdilaton}. Hence, this term has to vanish and we obtain the additional constraint
\begin{equation}\label{eqn:OmegaIJI=0}
  \partial_{\hat M} U_{\hat N}{}^{\hat M} = 0
    \quad \text{or equally} \quad
  \Omega_{\hat I\hat J}{}^{\hat I} = 0\,.
\end{equation}
Applying it, it is trivial to check that
\begin{equation}
  [\mathcal{L}_{\xi_1}, \mathcal{L}_{\xi_2}] d = \mathcal{L}_{[\xi_1, \xi_2]_\mathrm{C}} d
\end{equation}
closes due to the strong constraint in the external directions.

Although not necessary for a closing gauge algebra, it is desirable to check whether the strong constraint holds in the internal directions. If this is the case, the twist $U^M{}_N$ fulfills
\begin{equation}\label{eqn:strongconsttwist}
  \partial_{\hat M} U_{\hat I}{}^{\hat J} \partial^{\hat M} U_{\hat L}{}^{\hat K} = 0\,.
\end{equation}
It would be convenient to check this constraint directly at the level of the covariant fluxes. Trying to find an explicit result, one encounters the problem of inverting the map \eqref{eqn:maptwistflux}. Still, one can show that \eqref{eqn:strongconsttwist} annihilates the contraction
\begin{equation}\label{eqn:strongconstfluxes}
  \mathcal{F}_{\hat M\hat N\hat L} \mathcal{F}^{\hat M\hat N\hat L} = 3 \Omega_{\hat M\hat N\hat L}\Omega^{\hat M\hat N\hat L} + 6 \Omega_{\hat M\hat N\hat L}\Omega^{\hat L\hat M\hat N}
  =  3 \partial_{\hat M} U_{\hat N}{}^{\hat L} \partial^{\hat M} U^{\hat N}{}_{\hat L} - 
  6 \partial_M U_N{}^L \partial_L U^{NM} = 0\,.
\end{equation}
To see that the second term in the second line vanishes, we use 
\begin{equation}
  \partial_M \partial_L \left( U_N{}^L U^{NM} \right) = 0 =
    \partial_M U_N{}^L \partial_L U^{NM} \quad \text{with} \quad
    \partial_M U_N{}^M = 0\,.
\end{equation}
With this result one is able to identify covariant fluxes which definitely violate the strong constraint in the internal directions.

Combining \eqref{eqn:twistedgenLie} with \eqref{eqn:genlieKKansatz}, one gets the generalized Lie derivatives
\begin{align}
  \label{eqn:gendiffeomorphAMmu}
  {\mathcal L}_\xi A_{M\mu} &= L_\xi A_{M\mu} + \mathcal{F}_{IJK} \widehat{\Lambda}^J 
    \widehat{A}^K{}_\mu U^I{}_M + \partial_\mu \widehat{\Lambda}_M  \quad \text{and} \\
  \label{eqn:gendiffeomorphEAM}
  {\mathcal L}_\xi E^A{}_M &= L_\xi E^A{}_M +
    \mathcal{F}_{IJK} \widehat{\Lambda}^J \widehat{E}^{AK} U^I{}_M
\end{align}
for the twisted fields. It is obvious that both $A_{M\mu}$ and $E^A{}_M$ transform under generalized diffeomorphisms with non-vanishing $\Lambda^M$ as non-abelian vector fields.

\subsection{Isometries}
An isometry is defined as a shift of the coordinates $X^{\hat J} \rightarrow X^{\hat J} - K^{\hat J}$ which does not change the generalized metric. Using the generalized Lie derivative, which generates such coordinate shifts, yields
\begin{equation}\label{eqn:killinggenmetric}
  {\mathcal L}_{K} {\mathcal H}_{\hat M\hat N} = 0\,.
\end{equation}
The transformation parameter $K^{\hat J}$ is called Killing vector. This equation generalizes the Killing equation known from conventional geometry to the doubled space. In total we need $2n$ independent isometries for a consistent compactification. They are mediated by the Killing vectors $K_I{}^{\hat J}$ where $I$ is running from $1,\dots,2n$. Further, they act on the internal space only. Thus all Killing vectors are independent of the external directions $\mathds{X}$ and their external components vanish. Condition \eqref{eqn:killinggenmetric} is equivalent to
\begin{equation}\label{eqn:killinggenvielbein}
  {\mathcal L}_{K_I} E^{\hat A}{}_{\hat M} = 0 \quad \Leftrightarrow \quad
  {\mathcal L}_{K_I} {\mathcal H}_{\hat M\hat N} = 2 \big( {\mathcal L}_{K_I} E^{\hat A}{}_{(\hat M}
    \big) S_{AB} E^{\hat B}{}_{\hat N)} = 0
\end{equation}
which allows us to use the generalized vielbein $E^{\hat A}{}_{\hat M}$ to define Killing vectors.

Checking how a Killing vector acts on a twisted vector $V_{\hat M} = \widehat{V}_{\hat N} U^{\hat N}{}_{\hat M}$ gives rise to
\begin{equation}
  \mathcal{L}_K V^{\hat M} = \widehat{V}^{\hat N} \big[ K^{\hat P} \partial_{\hat P} U_{\hat N}{}^{\hat M} +( \partial^{\hat M} K_{\hat P}  - \partial_{\hat P} K^{\hat M} ) U_{\hat N}{\hat P} \big] =
    \widehat{V}^{\hat N} \mathcal{L}_K U_{\underline{\hat N}}{}^{\hat M}\,.
\end{equation}
In the last step, we have introduced underlined curved indices, e.g. $\underline{\hat N}$, which are not affected by the generalized Lie derivative. Because the generalized vielbein is a twisted vector, \eqref{eqn:killinggenvielbein} implies
\begin{equation}\label{eqn:killingeqtwist}
  \mathcal{L}_{K_I} U_{\underline{\hat N}}{}^{\hat M} = 0 \quad \text{and} \quad
  \mathcal{L}_{K_I} V^{\hat M} = 0\,.
\end{equation}
Keeping in mind that the generalized Lie derivative \eqref{eqn:twistedgenLie} of a twisted vector is again a twisted vector, we obtain
\begin{equation}
  [\mathcal{L}_{K_I}, \mathcal{L}_\xi] V_{\hat M} = 0 \,.
\end{equation}
Thus, isometries commute with twisted generalized diffeomorphisms. The Killing vectors form the algebra
\begin{equation}
  \mathcal{L}_{K_I} K_{\underline{J}}{}^{\hat K} = \tilde{\mathcal F}_{IJ}{}^L K_L{}^{\hat K}
\end{equation}
with the structure coefficients
\begin{equation}\label{eqn:structurekilling}
  \tilde{\mathcal F}_{IJ}{}^K = 
    K_I{}^N \partial_N K_J{}^M K^K{}_M +
    K^{KN} \partial_N K_I{}^M K_{JM} + 
    K_J{}^N \partial_N K^K{}_M K_I{}^M\,.
\end{equation}
Here $K^I{}_J$ denotes the inverse transpose of $K_I{}^J$ and thus $K_I{}^L K^J{}_L = \delta^J_I$ holds. In contrast to the twist $U_I{}^J$, $K_I{}^J$ is in general not an O($n,n$) element. Its first index \underline{cannot} be raised or lowered with $\eta_{MN}$ or $\eta^{MN}$, respectively. Like in the last section, additional restrictions arise due to the generalized dilaton. It is invariant under a generalized Lie derivative with a Killing vector as parameter. Hence, we find
\begin{equation}\label{eqn:invgendilaton}
  \mathcal{L}_{K_I} d = K_I{}^J \partial_J d - \frac{1}{2} \partial_J K_I{}^J =
    \frac{1}{2} \partial_J K_I{}^J = 0 \quad \rightarrow \quad \partial_J K_I{}^J = 0
\end{equation}
analogous to \eqref{eqn:OmegaIJI=0}.

\subsection{Consistent compactification}\label{sec:consistentcomp}
Like indicated in \eqref{eqn:consistentcomp}, a consistent compactification ansatz is used twofold. First, it gives rise to the effective action $S_\mathrm{eff}$, whose Lagrangian is independent of the internal coordinates. Afterwards solutions of the effective action's field equations are uplifted with the ansatz to solutions of the full DFT equations of motion. As we discuss in this section, both steps depend on the existence of $2n$ linear independent Killing vectors.

In combination with the additional constraint \eqref{eqn:invgendilaton}, they guarantee that the Lagrangian of the effective action does not depend on the internal coordinates. To see this, consider the action of a Killing vector $K_I$ on the Lagrangian defining DFT. It is a scalar density and transforms as 
\begin{equation}\label{eqn:genlieLdft2}
  \delta_{K_I} (e^{-2d} \mathcal{R}) = \partial_J K_I{}^J e^{-2d} \mathcal{R} + K_I{}^J \partial_J (e^{-2} \mathcal{R})  = K_I{}^J \partial_J (e^{-2d} \mathcal{R}) = 0
\end{equation}
after using \eqref{eqn:invgendilaton} to drop the term with the partial derivative acting on the Killing vector. Because $K_I{}^J$ consists of $2n$ linearly independent vector fields, from this equation we conclude 
\begin{equation}
  \partial_J (e^{-2d} \mathcal{R}) = 0\,.
\end{equation}
In other words, the Lagrangian is independent of the internal coordinates $\mathds{Y}$.

Next, consider a generalized Ricci tensor which vanishes
\begin{equation}
  \mathcal{R}_{\hat M\hat N}(X_0) = 0
\end{equation}
at a fixed position $X_0$ in the internal space. As a function of the generalized metric and dilaton, it inherits their Killing equation
\begin{equation}
  \delta_{K_I} \mathcal{R}_{\hat M\hat N} = 0\,.
\end{equation}
Further, it transforms under generalized diffeomorphisms as a tensor. Thus, we are able to identify
\begin{equation}
  \delta_{K_I} \mathcal{R}_{\hat M\hat N} = \mathcal{L}_{K_I} \mathcal{R}_{\hat M\hat N} = 0\,.
\end{equation}
At the point $X_0$ where $\mathcal{R}_{\hat M \hat N}$ vanishes, this equation reads
\begin{equation}
  \left.\mathcal{L}_{K_I} \mathcal{R}_{\hat M\hat N}\right|_{X_0} = \left.K_I{}^J \partial_J \mathcal{R}_{\hat M\hat N}\right|_{X_0} = 0 \,.
\end{equation}
Using that there are $2n$ linear independent Killing vectors, we find
\begin{equation}
  \left.\partial_I \mathcal{R}_{\hat M\hat N}\right|_{X_0} = 0\,.
\end{equation}
Thus, in an infinitesimally small neighbourhood around $X_0$, $\mathcal{R}_{\hat M\hat N}$ is constant and vanishes. If we repeat this procedure successively to different points in this neighbourhood, one is able to cover the complete internal space which is simply connected. We conclude that, assuming $2n$ linear independent Killing vectors, if the generalized Ricci tensor vanishes at one point in the internal space, it also vanishes at all other points therein.

This result is useful in constructing solutions to the $2 D$-dimensional equations of motion
\begin{equation}
  \delta S_\mathrm{DFT} = \int d^{2 D} X e^{-2 d} \mathcal{R}_{\hat M\hat N} \delta \mathcal{H}^{\hat M\hat N} = 0\,.
\end{equation}
In order to solve them, it is sufficient to find a solution in the external directions
\begin{equation}
  0 = \int d^{2(D-n)} X e^{-2 d} \left. \mathcal{R}_{\hat M\hat N}\right|_{X_0} \delta \mathcal{H}^{\hat M\hat N} = \int d^{2(D-n)} X e^{-2 d} \widehat{\mathcal{R}}_{\hat M\hat N} \delta \widehat{\mathcal{H}}^{\hat M\hat N} = \delta S_\mathrm{eff}
\end{equation}
with
\begin{equation}
  \widehat{\mathcal{R}}_{\hat M\hat N} = U^{\hat I}{}_{\hat M} \mathcal{R}_{\hat I\hat J} U^{\hat J}{}_{\hat N}\,.
\end{equation}
Such a solution arises from the field equations of the effective theory. Hence, one is able to uplift solutions of the effective theory to the full $2 D$-dimensional DFT if there exist $2n$ linear independent Killing vectors.

\section{Effective theory}\label{sec:gaugedgravity}
In the last section, we have proved that a consistent Scherk-Schwarz ansatz leads to an effective action $S_\mathrm{eff}$ which is independent of the internal coordinates. The effective theory is a gauge theory. Its gauge transformations are mediated by generalized diffeomorphisms with the parameters
\begin{equation}
  \Lambda^{\hat M} = \begin{pmatrix} 0 & 0 & \widehat{\Lambda}^N U_N{}^M \end{pmatrix}\,.
\end{equation}
It is convenient to express the effective theory in terms of quantities which transform covariantly under these transformations. To this end, we introduce the gauge covariant derivative
\begin{equation}\label{eqn:covderiv}
  D_\mu = \partial_\mu - \mathcal{L}_{A^M{}_\mu}
\end{equation}
with the defining property
\begin{equation}
  \Delta_\Lambda (D_\mu V) = 0\,,
\end{equation}
calculating the failure \eqref{eqn:failuregendiff} of $D_\mu V$ for a generic O($n,n$) tensor $V$ to transform covariantly under generalized diffeomorphisms parameterized by $\Lambda$. Applied to the generalized metric $\mathcal{H}_{MN}$ of the internal space, it gives rise to
\begin{align}
  D_\mu \mathcal{H}_{MN} &= U^I{}_M \widehat{D}_\mu \widehat{\mathcal H}_{IJ} U^J{}_N \quad
  \text{with} \nonumber \\
  \widehat{D}_\mu \widehat{\mathcal H}_{MN} &= \partial_\mu 
    \widehat{\mathcal H}_{MN}
  - \mathcal{F}_{MJ}{}^I \widehat{A}^J{}_\mu \widehat{\mathcal H}_{IN}
  - \mathcal{F}_{NJ}{}^I \widehat{A}^J{}_\mu 
    \widehat{\mathcal H}_{MI}\,.
\end{align}
The field strength of the gauge field $A_\mu{}^M$ is defined in analogy with Yang-Mills theory by setting
\begin{align}\label{eqn:F_mu_nu^M}
  F^M{}_{\mu\nu} &= 2 \partial_{[\mu} A^M{}_{\nu]} - [ A_\mu, A_\nu ]^M_\mathrm{C} = 
    \widehat{F}^N{}_{\mu\nu} U_N{}^M \quad \text{with} \nonumber\\
  \widehat{F}^M{}_{\mu\nu} &= 2 \partial_{[\mu}
  \widehat{A}^M{}_{\nu]} -
    \mathcal{F}^M{}_{NL} \widehat{A}^N{}_\mu
    \widehat{A}^L{}_\nu
\end{align}
and describes how two covariant derivatives commute
\begin{equation}
  [D_\mu, D_\nu] = - \mathcal{L}_{F^M{}_{\mu\nu}}\,.
\end{equation}
In generalized Kaluza-Klein compactifications, $F^M{}_{\mu\nu}$ normally does not transform covariantly under gauge transformations \cite{Hohm:2013nja}. This problem is fixed by adding the partial derivative of a 2-form gauge potential to the field strength defined in \eqref{eqn:F_mu_nu^M}. It compensates for the wrong transformation behavior and gives rise to the so called tensor hierarchy. Due to the restrictions introduces by the generalized Scherk-Schwarz ansatz, the failure
\begin{equation}
  \Delta_\Lambda F^M{}_{\mu\nu}  = \delta_\Lambda F^M{}_{\mu\nu} - \mathcal{L}_\Lambda
    F^M{}_{\mu\nu} = \partial^M ( \partial_{[\mu} \Lambda^N A_{\nu]N} ) 
    = \partial^M (\partial_{[\mu} \widehat{\Lambda}^M \widehat{A}_{\nu] N}) = 0
\end{equation}
of the field strength to transform covariantly vanishes. Thus, we use \eqref{eqn:F_mu_nu^M} without any modifications. Further, note that the Bianchi identity
\begin{equation}
  D_{[\mu} F^M{}_{\nu\rho]} = 0 
\end{equation}
is fulfilled for $F^M{}_{\mu\nu}$. The canonical field strength for the $B$-field is extended by a Chern-Simons term
\begin{equation}
  \widehat{G}_{\mu\nu\rho} = 
  3\partial_{[\mu} B_{\nu\rho]} + 3\partial_{[\mu} 
    \widehat{A}^M{}_\nu
    \widehat{A}_{M \rho ]} - \mathcal{F}_{MNL} 
    \widehat{A}^M{}_\mu
    \widehat{A}^N{}_\nu
    \widehat{A}^L{}_\rho
\end{equation}
in order to be invariant under gauge transformations. It fulfills the Bianchi identity
\begin{equation}
  \partial_{[\mu} G_{\nu\rho\lambda]} = 0\,.
\end{equation}

To derive the effective action in terms of these quantities, it is convenient to start from the DFT flux formulation \eqref{eqn:Sdftfluxform} and calculate the covariant fluxes $\mathcal{F}_{\hat A\hat B\hat C}$ and $\mathcal{F}_{\hat A}$. Their non-vanishing components read\cite{Geissbuhler:2011mx,Hohm:2013nja}
\begin{align}
  \mathcal{F}_{abc} &= e_a{}^\mu e_b{}^\nu e_c{}^\rho \widehat{G}_{\mu\nu\rho} &
  \mathcal{F}_{ab}{}^c &= 2 e_{[a}{}^\mu \partial_\mu e_{b]}{}^\nu e^c{}_\nu = f_{ab}^c \nonumber \\
  \mathcal{F}_{abC} &= -e_a{}^\mu e_b{}^\nu \widehat{E}_{CM} \widehat{F}^M{}_{\mu\nu} &
  \mathcal{F}_{aBC} &= e_a{}^\mu \widehat{D}_\mu \widehat{E}_B{}^M \widehat{E}_{CM} \nonumber \\
  \mathcal{F}_{ABC} &= 3 \Omega_{[ABC]} &
  \mathcal{F}_{a}   &= f^{b}_{ab} + 2 e_a{}^\mu \partial_\mu \phi\,. \label{eqn:fluxcomponents}
\end{align}
Plugging them into \eqref{eqn:Sdftfluxform} and switching to curved indices, the effective action
\begin{gather}
  S_\mathrm{eff} = \int \mathrm{d}x^{(D-n)} \sqrt{-g} e^{-2\phi} \Bigl( R + 
    4 \partial_\mu\phi \partial^\mu\phi -\frac{1}{12}  \widehat{G}_{\mu\nu\rho} \widehat{G}^{\mu\nu\rho} \nonumber \\
    \label{eqn:ddimeffaction} \qquad
    -\frac{1}{4} \widehat{\mathcal H}_{MN} \widehat{F}^{M\mu\nu} \widehat{F}^N{}_{\mu\nu}
    +\frac{1}{8} \widehat{D}_\mu \widehat{\mathcal H}_{MN} \widehat{D}^\mu \widehat{\mathcal H}^{MN}
    -\widehat{V}\Bigr)
\end{gather}
\cite{Geissbuhler:2011mx,Hohm:2013nja} arises. Here, $R$ denotes the standard scalar curvature in the external directions. As discussed in section~\ref{sec:consistentcomp}, the Lagrange density of DFT is constant in the internal directions if there are $2n$ linear independent Killing vectors. We assume this case and solve the integral in these directions. It gives rise to a global factor, which is neglected in \eqref{eqn:ddimeffaction}. The resulting action is equivalent to the one presented by \cite{Aldazabal:2011nj}. Its scalar potential $\hat V$ reads
\begin{equation}\label{eqn:scalarpotential}
  \widehat V = -\frac{1}{4} \mathcal{F}_I{}^{KL}
    \mathcal{F}_{JKL} \widehat{\mathcal H}^{IJ} +
    \frac{1}{12} \mathcal{F}_{IKM}
    \mathcal{F}_{JLN} \widehat{\mathcal H}^{IJ} \widehat{\mathcal H}^{KL} 
    \widehat{\mathcal H}^{MN}\,.
\end{equation}
Note that we assume the strong constraint to hold for the complete doubled space. Therefore, we do not have a $\mathcal{F}_{IJK} \mathcal{F}^{IJK}$ term in the scalar potential. As discussed in section~\ref{sec:twistedgendiff}, it vanishes after imposing the strong constraint for the internal space. There is a straightforward geometric interpretation of the scalar potential \cite{Hohm:2013nja}: It is equivalent to the internal space's generalized scalar curvature, namely
\begin{equation}
  \widehat{V} = - \mathcal{R}(d,\mathcal{H}_{MN})\,.
\end{equation}
To avoid overloading our notation, we drop the hats on all quantities of the effective theory in the following.

\subsection{Vacua}\label{sec:vacua}
Since we have performed a consistent compactification, each solution of the effective action is also a solution of the DFT we started with. At this point it helps to note that \eqref{eqn:ddimeffaction} represents the bosonic sector of a half-maximal, electrically gauged supergravity. So in order to find vacua, we do not have to explicitly solve its full field equations. Instead, it is sufficient to look for minima of the scalar potential. A necessary condition for a local minimum is
\begin{equation}
  \frac{\partial V}{\partial \mathcal{H}_{MN}} = 0\,.
\end{equation}
Again, one has to keep in mind that the generalized metric $\mathcal{H}_{IJ}$ of the internal space is O($n,n$) valued and symmetric. Thus, one has to apply the projection
\begin{equation}\label{eqn:vacuumeqs}
  0 = P_{MK} \mathcal{K}^{KL}
    \bar{P}_{LN} + \bar{P}_{MK} \mathcal{K}^{KL} P_{LN}
  \quad\text{with}\quad 
  \mathcal{K}^{MN} = \frac{\partial V}{\partial \mathcal{H}_{MN}}
\end{equation}
to obtain $n^2$ linear independent equations fixing valid vacua. In order to solve these equations, we first calculate
\begin{equation}\label{eqn:kappamnfull}
  \mathcal{K}^{MN} = \frac{1}{4}\left( -\mathcal{F}^{MKL} 
    \mathcal{F}^N{}_{KL} + \mathcal{F}^M{}_{IK} \mathcal{F}^N_{JL}
    \mathcal{H}^{IJ} \mathcal{H}^{KL} \right)\,.
\end{equation}
It is evaluated for $\bar{\mathcal H}^{MN}$, the vacuum expectation value of $\mathcal{H}^{MN}$. We express this value in terms of the vacuum generalized vielbein
\begin{equation}\label{eqn:vacuumgenvielbein}
  \bar{\mathcal H}^{MN} = \bar{E}_A{}^M S^{AB} \bar{E}_B{}^N\,.
\end{equation}
In the following, flat and curved indices will be related by means of this background frame field. Applying this prescription to the indices of \eqref{eqn:kappamnfull}, one obtains
\begin{equation}\label{eqn:confluxesintern} 
  \mathcal{K}^{AB} = \frac{1}{4}\big( \mathcal{F}^A{}_{CD} \eta^{DE}
    \mathcal{F}^B{}_{EF} \eta^{FC} - \mathcal{F}^A{}_{CD} S^{DE}
    \mathcal{F}^B{}_{EF} S^{FC} \big)\,.
\end{equation}
A further simplification is achieved by changing to flat, $GL(n)\times GL(n)$ indices which were introduced in section~\ref{sec:doublelorentz}. Expanding \eqref{eqn:confluxesintern}, in terms of these indices yields
\begin{equation}\label{eqn:KAB}
  \mathcal{K}^{\bar A\bar B} = \mathcal{F}^{\bar A}{}_{\bar c d} \mathcal{F}^{\bar B \bar c d} 
\end{equation}
where we use the parameterization
\begin{equation}
  \mathcal{F}_{\bar A\bar B\bar C} =
    \begin{pmatrix}
      \mathcal{F}_{\bar A bc} & \mathcal{F}_{\bar A b\bar c} \\
      \mathcal{F}_{\bar A \bar bc} & \mathcal{F}_{\bar A \bar b\bar c} 
    \end{pmatrix}
\end{equation}
of the covariant fluxes. The projectors $P_{MK}$ and $\bar{P}_{LN}$ needed to calculate \eqref{eqn:vacuumeqs} take the simple form
\begin{equation}
  P_{\bar A\bar B} = \frac{1}{2}\left( \eta_{\bar A\bar B} -
    \delta_{\bar A\bar B} \right) = \begin{pmatrix}
      0 & 0 \\ 0 & -\eta_{\bar a\bar b} \end{pmatrix} 
\quad \text{and} \quad
  \bar{P}_{\bar A\bar B} = \frac{1}{2}\left( \eta_{\bar A\bar B} + 
    \delta_{\bar A\bar B} \right) = \begin{pmatrix}
      \eta_{ab} & 0 \\ 0 & 0 \end{pmatrix} \;,
\end{equation}
in barred, flat indices. They allow to write \eqref{eqn:vacuumeqs} as
\begin{equation}
  0 = - \begin{pmatrix}
    0 & \mathcal{K}_{a \bar b} \\
    \mathcal{K}_{\bar a b} & 0
  \end{pmatrix}\,.
\end{equation}
Further $\mathcal{K}_{\bar a b}=\mathcal{K}_{a \bar b}$ holds. This identity follows immediately from \eqref{eqn:KAB}. Thus, we obtain the $n^2$ independent equations
\begin{equation}\label{eqn:fluxeseom}
  0 = \mathcal{F}_{a\bar c d} \mathcal{F}_{\bar b}{}^{\bar c d}
\end{equation}
as a necessary condition for a local minimum of the scalar potential and therewith a vacuum of the effective theory. An obvious solution is
\begin{equation}
  \mathcal{F}_{a\bar b \bar c} = 0 \quad \text{and} \quad \mathcal{F}_{a b\bar c} = 0
\end{equation}
which leaves only $\mathcal{F}_{abc}$ and $\mathcal{F}_{\bar a\bar b\bar c}$ as non-vanishing components of the covariant fluxes.

We conclude this section with an important remark: Note that the scalar potential \eqref{eqn:scalarpotential} includes flux from the NS/NS sector only. To obtain the full scalar potential one has to add R/R fluxes, too. So even if \eqref{eqn:fluxeseom} is violated, there is a good chance to fix this situation by switching on additional R/R fluxes.

\subsection{Spectrum}\label{sec:spectrum}
We already discussed the necessary condition \eqref{eqn:vacuumeqs} for a local minimum of the scalar potential. To formulate a sufficient condition, we have to calculate the Hesse matrix
\begin{equation}  
  H_{\alpha\beta} = \frac{\partial^2 V}{\phi_\alpha \phi_\beta}
    \quad \text{with} \quad
  \alpha, \beta = 1, \cdots, n^2
\end{equation}
and check whether it is positive definite for the vacuum generalized vielbein $\bar{\mathcal H}^{IJ}$. At this point, we introduce $n^2$ auxiliary scalar fields $\phi_\alpha$. They account for the fact that the generalized vielbein $\mathcal{H}^{MN}$ in the internal space is O($n,n$) valued and thus not all of its entries correspond to physical degrees of freedom. The unconstrained, physical degrees of freedom are represented by the fields $\phi_\alpha$. They correspond to the moduli discussed in section~\ref{sec:introcomp}. With this convention, the Hesse matrix reads
\begin{equation}\label{eqn:secondvarV}
  H_{\alpha\beta} = \frac{\partial^2 V}{\partial \mathcal{H}_{IJ} \partial \mathcal{H}_{KL}}
    \frac{\partial \mathcal{H}^{IJ}}{\partial \phi_\alpha}
    \frac{\partial \mathcal{H}^{KL}}{\partial \phi_\beta}
      + \frac{\partial V}{\partial \mathcal{H}_{KL}}
    \frac{\partial^2 \mathcal{H}^{KL}}{ \partial \phi_\alpha \, 
    \partial \phi_\beta} \,.
\end{equation}
With
\begin{equation}\label{eqn:(talpha)IJ}
  \frac{\partial^2 V}{\partial \mathcal{H}_{IJ} \partial \mathcal{H}_{KL}} =
    \frac{1}{2} {\mathcal F}_{IKM}
    {\mathcal F}_{JLN} {\mathcal H}^{MN}
  \quad \text{and the abbreviation} \quad
  \left( t_\alpha \right){}^{IJ} = 
    \frac{\partial {\mathcal H}^{IJ}}{\partial \phi_\alpha}\,,
\end{equation}
we obtain
\begin{equation}\label{eqn:secondvarV2}
  H_{\alpha\beta} =  \frac{1}{2} {\mathcal F}_{IKM}
    {\mathcal F}_{JLN} {\mathcal H}^{MN} \left( t_\alpha 
    \right){}^{IJ} \left( t_\beta
    \right){}^{KL} + \mathcal{K}_{KL} \frac{\partial}{\partial \phi_\alpha} 
    \left( t_\beta \right){}^{KL} \,.
\end{equation}
The matrix $\left( t_\alpha \right){}_I{}^J$ is an element of the Lie algebra $\mathfrak{o}(n,n)$ and is constraint by
\begin{equation}\label{eqn:talphacurved}
  \left( t_\alpha\right)_I{}^K \eta_{KJ} + \left( t_\alpha \right)_J{}^K \eta_{IK} = 0\,.
\end{equation}

For the following calculations, it is convenient to switch from curved to flat indices. We do so with the vacuum generalized frame field, e.g.
\begin{equation}
  \mathcal{H}^{IJ} = {\bar E}_A{}^I \mathcal{H}^{AB} {\bar E}_B{}^J\,.
\end{equation}
In general $\mathcal{H}^{AB}$ is not equivalent to $S^{AB}$ used in the flux formulation. Only for the vacuum they are. It is straightforward to check that $\left( t_\alpha \right)_I{}^J$ transforms in the canonical way
\begin{equation}
  \left( t_\alpha \right)_A{}^B = \left( t_\alpha \right)_I{}^J {\bar E}_A{}^I {\bar E}^B{}_J\,.
\end{equation}
Further, the constraint \eqref{eqn:talphacurved} can be written as
\begin{equation}
  \left( t_\alpha\right)_A{}^C \eta_{CB} + \left( t_\alpha \right)_B{}^C \eta_{AC} = 0
\end{equation}
and its solution is given by
\begin{equation}\label{eqn:talphaAB}
  \sqrt{2} \left( t_{CD} \right)_A{}^B = S^{B}{}_{[C} \eta_{D] A} \quad \text{with} \quad
  \alpha=\begin{pmatrix} C & D \end{pmatrix}\,.
\end{equation}
A normalization factor was introduced to guarantee that
\begin{equation}\label{eqn:(talpha)ABnormalization}
  \left(t_\alpha \right)_A{}^B \left( t_\beta\right)_B{}^A = \delta_{\alpha\beta}
\end{equation}
holds. For $C < D$, \eqref{eqn:talphaAB} produces the adjoint representation of the Lie algebra $\mathfrak{o}(n,n)$. After switching to bared flat indices, $n^2$ of the generators $\left(t_\alpha\right)_{\bar A}{}^{\bar B}$ are symmetric, the others are antisymmetric. We drop the antisymmetric ones, because the generalized metric is symmetric and so are its generators. With these generators at hand, the generalized metric can be expressed by the exponential map
\begin{equation}\label{eqn:genmetricfluctuation}
  \mathcal{H}^{AB} = \exp \big( \left(t_\alpha \right)^{AB} \phi^\alpha \big) = 
    S^{AB} + (t_\alpha)^{AB} \phi^\alpha + \frac{1}{2} (t_\alpha)^{AC} S_{CD} 
    (t_\beta)^{DB} \phi^\alpha \phi^\beta + \dots\ .
\end{equation}
To simplify the notation we apply Einstein sum convention to the indices $\alpha, \beta, \cdots$ of the auxiliary fields. In the vacuum all $\phi_\alpha$ vanish and according to \eqref{eqn:genmetricfluctuation} the generalized metric equals $S^{AB}$. Back in curved indices this gives rise to the vacuum generalized metric
\begin{equation}
  \bar{\mathcal H}^{MN} = \mathcal{H}^{MN}(\phi_\alpha=0)\,.
\end{equation}
With the parameterization of the generalized metric in \eqref{eqn:genmetricfluctuation}, one obtains
\begin{equation}
  \left. \frac{\partial^2 \mathcal{H}^{AB}}{\partial \phi_\alpha \, \partial \phi_\beta} 
    \right|_{\phi_\gamma = 0} 
  = \begin{cases}
        \left( t_\alpha \right)^{AC} S_{CD} \left( t_\beta \right)^{DB} & 
          \text{for } \alpha \le \beta \\
        \left( t_\beta \right)^{AC} S_{CD} \left( t_\alpha \right)^{DB} &
          \text{otherwise}
    \end{cases}
\end{equation}
and is able to express the Hesse matrix \eqref{eqn:secondvarV2} as
\begin{equation}\label{eqn:massmatrix}
  H_{\alpha\beta} = \Big( \frac{1}{2} {\mathcal F}_{ACE}
    {\mathcal F}_{BDF} S^{EF} + {\mathcal K}_{AD} S_{BC} \Big) 
    \left( t_\alpha \right){}^{AB} \left( t_\beta \right){}^{CD}\,.
\end{equation}

In order to identify massive scalar excitations, we diagonalize the Hesse matrix. Because $H_{\alpha\beta}$ is symmetric, this is always possible and leads to $n^2$ eigenvalues $\lambda_i$ and the corresponding orthonormal eigenvectors $v_i$ with the components $v_i{}_\beta$. To diagonalize, we rotate the generators $\left( t_\alpha \right){}^{AB}$ by defining
\begin{equation}\label{eqn:rotatedgenh}
  (t_i ){}^{AB} = v_i{}^\beta \left( t_\beta \right){}^{AB}\,.
\end{equation}
The generalized metric $\mathcal{H}^{AB}$ in \eqref{eqn:genmetricfluctuation} has to be invariant under this rotation. Thus one also has to rotate the scalar fields
\begin{equation}
  \phi_i = v_i{}^\beta \phi_\beta\,. 
\end{equation}
By plugging the rotated generators \eqref{eqn:rotatedgenh} into the expression for the mass matrix \eqref{eqn:massmatrix}, one obtains the requested diagonal form
\begin{equation}\label{eqn:hessediag}
  H_{ij} = v_i{}^\alpha v_j{}^\beta H_{\alpha\beta} = \diag ( \lambda_i )\,.
\end{equation}
After inserting the expansion for the flat generalized metric \eqref{eqn:genmetricfluctuation} into the scalar potential, we find
\begin{equation}
  V(\phi_k) = V(0) + \frac{1}{2} H_{i j} (0)  \phi^i \phi^j + \mathcal{O}(\phi^3)
\end{equation}
where the argument $0$ is an abbreviation for $\phi_k = 0$. There is no linear term in this expansion, because 
\begin{equation}
  \left. \frac{\partial V}{\partial \phi_i}\right|_{\phi_k = 0} = 0
\end{equation}
holds already due to the necessary condition for a minimum of the scalar potential, which we discussed in the last section. Note that the second term in \eqref{eqn:massmatrix} in general does not vanish for the vacuum. Only $\mathcal{R}_{AB}$, a projection of $\mathcal{K}_{AB}$, vanishes due to the equations of motion. We are now able to state the sufficient condition for a minimum of the scalar potential: All eigenvalues of $H_{ij}(0)$ has to be greater than zero,
\begin{equation}
  \lambda_i(0) > 0 \quad \forall\, i \,.
\end{equation}
Moduli $\phi_i$ with $\lambda_i(0) = 0$ are called flat directions and these with $\lambda_i(0)< 0$ are called tachyonic. 

Considering the kinetic term
\begin{equation}
  D_\mu \mathcal{H}_{MN} D^\mu \mathcal{H}^{MN} = \partial_\mu \phi_i \partial^\mu \phi^i 
\end{equation}
in the effective action \eqref{eqn:ddimeffaction} for a vanishing gauge field $A_{M\mu}$, gives rise to the quadratic action
\begin{equation}
  S_\phi^{(2)} = \frac{1}{8} \int dx^{(D-n)} \sqrt{-g} e^{-2\phi} \left( \partial_\mu \phi_i
    \partial^\mu \phi^i - 4 \lambda_i (\phi^i)^2 \right)\,.
\end{equation}
It allows us to identify $2 \sqrt{\lambda_i} = m_i$ as the mass of the scalar field $\phi_i$. Thus, the eigenvalues $\lambda_i$ have to be positive or zero in order to avoid tachyons. If all of them are positive, all moduli are stabilized.

\subsection{Embedding tensor}\label{sec:solutionsconstr}
The covariant fluxes $\mathcal{F}_{IJK}$ which fix a generalized Scherk-Schwarz ansatz are closely related to the embedding tensor $\Theta_I{}^\alpha$ of gauged supergravities. It describe a subset of the global O($n,n$) symmetry in the compact directions, which is promoted to a gauge symmetry in the effective theory. Comparing the formalism reviewed in \cite{Samtleben:2008pe} and the one shown here, we find the connection
\begin{equation}\label{eqn:embeddingtensor}
  \mathcal{F}_{IJ}{}^K = \Theta_I{}^\alpha 
    (t_\alpha)_J{}^K = \left( X_I \right)_J{}^K \,.
\end{equation}
Here, $t_\alpha$ denote the different O($n,n$) generators and $\left(t_\alpha\right)_J{}^K$ is their representation with respect to $2n$-dimensional vectors. There are two constraints one imposes on the embedding tensor: a linear and a quadratic constraint. Following \cite{Dibitetto:2012rk}, we now discuss them for a $n=3$-dimensional internal space. In this case the number of compact dimensions is large enough to find non-trivial solutions and at the same time still small enough to completely classify all relevant solutions with an appropriate effort.

For $n=3$, the $X_I$ in \eqref{eqn:embeddingtensor} describe six different O($3,3$) generators labeled by $I=1,\dots,6$. Group-theoretically, $\left(X_I\right)_J{}^K$ lives in the tensor product
\begin{equation}\label{eqn:decomptensorprod}
  \mathbf{6} \otimes \mathbf{15} = \mathbf{6} \oplus \overline{\mathbf{10}} \oplus \mathbf{10} \oplus \mathbf{64} \,.
\end{equation}
The first factor in this product is the vector representation of SO($3,3$) and the second is the adjoint representation of the same group. A linear constraint projects out distinguished irreps. Here, the irreps $\mathbf{6}$ and $\mathbf{64}$ of the tensor product decomposition \eqref{eqn:decomptensorprod} are projected out. They have to be absent because the covariant fluxes are totally antisymmetric. This implies that the remaining irreps $\overline{\mathbf{10}} \oplus \mathbf{10}$ represent the $6\cdot5\cdot4/ 3! = 20$ independent components of $\mathcal{F}_{IJK}$ in six dimensions.

Following the reasoning in \cite{Dibitetto:2012rk}, one can express $\left(X_I\right)_J{}^K$ also as irreps of SL($4$), which is isomorphic to SO($3,3$). In this case the decomposition \eqref{eqn:decomptensorprod} does not change. To distinguish between the two different groups, one introduces fundamental SL($4$) indices $p,q,r = 1,\dots,4$. The generators $\left(X_I\right)_J{}^K$ can be expressed in terms of these indices as
\begin{equation}\label{eqn:decompfluxes} 
  \left(X_{mn}\right)_p{}^q = \frac{1}{2} \delta^q_{[m} M_{n]p} - 
    \frac{1}{4} \varepsilon_{mnpr}\tilde M^{rq}\,,
\end{equation}
where $M_{np}$ and $\tilde M^{rq}$ are symmetric matrices and $\varepsilon$ denotes the Levi-Civita symbol. The matrices $M_{np}$ and $\tilde M^{rq}$ have $4\cdot 5/2=10$ independent components each and match the irreps $\overline{\mathbf{10}}$ and $\mathbf{10}$ in \eqref{eqn:decomptensorprod}. The indices $mn$ in $\left(X_{mn}\right)_p{}^q$ are antisymmetric and label the $6=4\cdot 3/2$ independent components of the SL($4$) irrep $\mathbf{6}$. The dual representation carries raised, antisymmetric indices. Both are connected by
\begin{equation}
  X_{mn} = \frac{1}{2} \varepsilon_{mnpq} X^{pq}\,.
\end{equation}
Equation \eqref{eqn:decompfluxes}, embeds the relevant irreps $\overline{\mathbf{10}}\oplus\mathbf{10}$ into the product $\mathbf{6}\otimes\mathbf{15}$. The covariant fluxes as rank 3 tensor live formally in the product $\mathbf{6}\otimes\mathbf{6}\otimes\mathbf{6}$. Hence, one has to embed $\left(X_{mn}\right)_p{}^q$ into this product which is done by
\begin{equation}\label{eqn:doubleindices}
  \left(X_{mn}\right)_{pq}{}^{rs} \ = \ 
    2 \left(X_{mn}\right)_{[p}{}^{[r} \delta^{s]}_{q]}\,.
\end{equation}
The irrep $\mathrm{6}$ of SL($4$) is linked to the $\mathrm{6}$ of SO($3,3$) by the 't Hooft symbols $(G_I)^{mn}$. For $n=3$, they read
\begin{align}
  \left( G^1 \right)^{mn} &= \begin{pmatrix}
    0 & -1 & 0 & 0 \\
    1 & 0 & 0 & 0 \\
    0 & 0 & 0 & 0 \\
    0 & 0 & 0 & 0
  \end{pmatrix} &
  \left( G^2 \right)^{mn} &= \begin{pmatrix}
    0 & 0 & -1 & 0 \\
    0 & 0 & 0 & 0 \\
    1 & 0 & 0 & 0 \\
    0 & 0 & 0 & 0
  \end{pmatrix} &
  \left( G^3 \right)^{mn} &= \begin{pmatrix}
    0 & 0 & 0 & -1 \\
    0 & 0 & 0 & 0 \\
    0 & 0 & 0 & 0 \\
    1 & 0 & 0 & 0
  \end{pmatrix} \nonumber \\
  \left( G_1 \right)^{mn} &= \begin{pmatrix}
    0 & 0 & 0 & 0 \\
    0 & 0 & 0 & 0 \\
    0 & 0 & 0 & -1 \\
    0 & 0 & 1 & 0
  \end{pmatrix} &
  \left( G_2 \right)^{mn} &= \begin{pmatrix}
    0 & 0 & 0 & 0 \\
    0 & 0 & 0 & 1 \\
    0 & 0 & 0 & 0 \\
    0 & -1 & 0 & 0
  \end{pmatrix} &
  \left( G_3 \right)^{mn} &= \begin{pmatrix}
    0 & 0 & 0 & 0 \\
    0 & 0 & -1 & 0 \\
    0 & 1 & 0 & 0 \\
    0 & 0 & 0 & 0
  \end{pmatrix}
\end{align}
and fulfill the identities
\begin{align}
  \left( G_I \right)_{mn} \left( G_J \right)^{mn} &= 2\eta_{IJ}\,, \\
  \left( G_I \right)_{mp} \left( G_J \right)^{pn} +
  \left( G_J \right)_{mp} \left( G_I \right)^{pn} &= -\delta^n_m \eta_{IJ}\,, \\
  \left( G_I \right)_{mp} \left( G_J \right)^{pq} \left( G_K \right)_{qr}
  \left( G_L \right)^{rs} \left( G_M \right)_{st} \left( G_N \right)^{tn}
  &= \delta^n_m \varepsilon_{IJKLMN}\,.
\end{align}
With them, we finally obtain the covariant fluxes
\begin{equation}\label{eqn:SL(4)toSO(3,3)}
  \mathcal{F}_{IJK} = \left( X_{mn} \right)_{pq}{}^{rs}
    \left( G_I \right)^{mn} \left( G_J \right)^{pq} \left( G_K \right)_{rs}
\end{equation}
in their familiar form.

In order to find vacua of the effective theory, we have to evaluate the condition \eqref{eqn:fluxeseom}. To this end, we need the covariant fluxes in flat indices
\begin{equation}\label{eqn:fluxesflatvacuum}
  \mathcal{F}_{ABC} = \bar{E}_A{}^I \bar{E}_B{}^J \bar{E}_A{}^K \mathcal{F}_{IJK}
\end{equation}
which arise after the contraction with the vacuum generalized vielbein $\bar{E}_A{}^I$. At this point, it is essential to note that neither the fluxes $\mathcal{F}_{IJK}$ nor the vacuum generalized vielbein fix a physical background. Only their contraction $\mathcal{F}_{ABC}$ does. Hence, without loss of generality we are able to fix
\begin{equation}\label{eqn:defvacuumvielbein}
  \bar{E}_A{}^I := \delta_A^I 
\end{equation}
and identify the components of the covariant fluxes in flat and curved indices. Other choices would be possible too, but they would make explicit calculations more complicated.

\begin{table}[t!]
\centering
\begin{tabular}{| c | r r r r | r r r r | c | c |}
\hline
\textrm{ID} & \multicolumn{4}{|c|}{$\diag M_{mn}/\,\cos\alpha\,$} & \multicolumn{4}{|c|}{$\diag \tilde{M}^{mn}/\,\sin\alpha\,$} & range of $\alpha$ & gauging \\[1mm]
\hline \hline \rowcolor{fillcolor}
$1$ & $1$ & $1$ & $1$ & $1$ & 
      $1$ & $1$ & $1$ & $1$ & $-\frac{\pi}{4}\,<\,\alpha\,\le\,\frac{\pi}{4}$ & $\left\{%
\setlength{\arraycolsep}{2pt}\begin{array}{cc}\textrm{SO}($4$)\ , & \alpha\,\ne\,\frac{\pi}{4}\ ,\\ \textrm{SO}(3)\ , & \alpha\,=\,\frac{\pi}{4}\ .\end{array}\right.$\\[4mm]
\hline
$2$ & $1$ & $1$ & $1$ & $-1$ & 
      $1$ & $1$ & $1$ & $-1$ & $-\frac{\pi}{4}\,<\,\alpha\,\le\,\frac{\pi}{4}$ & SO($3,1$)\\[1mm]
\hline
$3$ & $1$ & $1$ & $-1$ & $-1$ &
      $1$ & $1$ & $-1$ & $-1$ & $-\frac{\pi}{4}\,<\,\alpha\,\le\,\frac{\pi}{4}$ & $\left\{%
\setlength{\arraycolsep}{2pt}\begin{array}{cc}\textrm{SO}($2,2$)\ , & \alpha\,\ne\,\frac{\pi}{4}\ ,\\ \textrm{SO}(2,1)\ , & \alpha\,=\,\frac{\pi}{4}\ .\end{array}\right.$\\[2mm]
\hline \hline \rowcolor{fillcolor}
$4$ & $1$ & $1$ & $1$ & $0$ & 
      $0$ & $0$ & $0$ & $1$ & $-\frac{\pi}{2}\,<\,\alpha\,<\,\frac{\pi}{2}$ & ISO($3$)\\[1mm]
\hline
$5$ & $1$ & $1$ & $-1$ & $0$ & 
      $0$ & $0$ & $0$  & $1$ & $-\frac{\pi}{2}\,<\,\alpha\,<\,\frac{\pi}{2}$ & ISO($2,1$)\\[1mm]
\hline \hline \rowcolor{fillcolor}
$6$ & $1$ & $1$ & $0$ & $0$ & 
      $0$ & $0$ & $1$ & $1$ & $-\frac{\pi}{4}\,<\,\alpha\,\le\,\frac{\pi}{4}$ & $\left\{%
\setlength{\arraycolsep}{2pt}\begin{array}{cc}\textrm{CSO}(2,0,2)\ , & \alpha\,\ne\,\frac{\pi}{4}\ ,\\ \mathfrak{f}_{1}\quad(\textrm{Solv}_{6}) \ , & \alpha\,=\,\frac{\pi}{4}\ .\end{array}\right.$\\[4mm]
\hline
$7$ & $1$ & $1$ & $0$ & $0$ & 
      $0$ & $0$ & $1$,& $-1$ & $-\frac{\pi}{2}\,<\,\alpha\,<\,\frac{\pi}{2}$ & $\left\{%
\setlength{\arraycolsep}{2pt}\begin{array}{cc}\textrm{CSO}(2,0,2)\ , & |\alpha|\,<\,\frac{\pi}{4}\ ,\\ \textrm{CSO}(1,1,2)\ , & |\alpha|\,>\,\frac{\pi}{4}\ ,\\ \mathfrak{g}_{0}\quad(\textrm{Solv}_{6}) \ , & |\alpha|\,=\,\frac{\pi}{4}\ .\end{array}\right.$\\[4mm]
\hline \rowcolor{fillcolor}
$8$ & $1$ & $1$ & $0$ & $0$ & 
      $0$ & $0$ & $0$ & $1$ & $-\frac{\pi}{2}\,<\,\alpha\,<\,\frac{\pi}{2}$ & $\mathfrak{h}_{1}\quad(\textrm{Solv}_{6})$\\[1mm]
\hline
$9$ & $1$ & $-1$ & $0$ & $0$ & 
      $0$ & $0$  & $1$ & $-1$ & $-\frac{\pi}{4}\,<\,\alpha\,\le\,\frac{\pi}{4}$ & $\left\{%
\setlength{\arraycolsep}{2pt}\begin{array}{cc}\textrm{CSO}(1,1,2)\ , & \alpha\,\ne\,\frac{\pi}{4}\ ,\\ \mathfrak{f}_{2}\quad(\textrm{Solv}_{6}) \ , & \alpha\,=\,\frac{\pi}{4}\ .\end{array}\right.$\\[4mm]
\hline
$10$ & $1$ & $-1$ & $0$ & $0$ & 
       $0$ & $0$  & $0$ & $1$ & $-\frac{\pi}{2}\,<\,\alpha\,<\,\frac{\pi}{2}$ & $\mathfrak{h}_{2}\quad(\textrm{Solv}_{6})$\\[1mm]
\hline \hline \rowcolor{fillcolor}
$11$ & $1$ & $0$ & $0$ & $0$ & 
       $0$ & $0$ & $0$ & $1$ &
$-\frac{\pi}{4}\,<\,\alpha\,\le\,\frac{\pi}{4}$ &
$\left\{%
\setlength{\arraycolsep}{2pt}\begin{array}{cc}\mathfrak{l}\quad(\textrm{Nil}_{6}(3)\,)\ , & \alpha\,\ne\,0\ ,\\
\textrm{CSO}(1,0,3)\ , &
\alpha\,=\,0\ .\end{array}\right.$\\[4mm]
\hline \hline \rowcolor{fillcolor}
$12$ & $0$ & $0$ & $0$ & $0$ & 
       $0$ & $0$ & $0$ & $0$ &
$\alpha = 0$ & 
$\textrm{U}(1)^6$ \\
\hline
\end{tabular}
\caption{Solutions of the embedding tensor for half-maximal, electrically gauged supergravity in $n=3$ dimensions from \cite{Dibitetto:2012rk}. All shaded entries give rise to compact background spaces. Besides semi-simple Lie groups, serveral solvable Lie groups apprear. Details about $\mathfrak{f}_{1}$, $\mathfrak{f}_{2}$, $\mathfrak{g}_{0}$, $\mathfrak{h}_{1}$ and $\mathfrak{h}_{2}$ can be found in the appendinx of \cite{Dibitetto:2012rk}. CSO($p,q,r$) and esspecially CSO($2,0,2$) are discussed in section~\ref{sec:examplecso202}.} \label{tab:solembedding}
\end{table}
Next, we study solutions of the Jacobi identity \eqref{eqn:quadraticc} for the covariant fluxes. Instead of using their SO($3,3$) representation, we switch to the SL($4$) representation \eqref{eqn:doubleindices} and obtain \cite{Dibitetto:2012rk}
\begin{equation}\label{eqn:quadraticcMMtilde}
  M_{mp} \tilde M^{pn} = \frac{1}{4} \delta^m_n M_{qp} \tilde M^{pq}
\end{equation}
as an equivalent SL($4$) version of \eqref{eqn:quadraticc}. This constraint is not linear but quadratic and thus it is called quadratic constraint. Because $M_{np}$ is symmetric, it can always be diagonalized by a SO($4$) transformation. This group is the maximal compact subgroup of SL($4$) and it is, up to a discrete $\mathds{Z}_2$, isomorphic to SO($3$)$\times$SO($3$), the maximal compact subgroup of SO($3,3$). Hence it is always possible to diagonalize $M_{np}$ by a double Lorentz transformation applied to the covariant fluxes. When $M_{np}$ is diagonal, $\tilde M_{rq}$ has to be diagonal, too. Otherwise the constraint \eqref{eqn:quadraticcMMtilde} is violated. In this case one can identify the components
\begin{equation}\label{eqn:mapMMtildefluxes}
  \diag M_{mn} = \begin{pmatrix} H_{123} & Q^{23}_1 & Q^{31}_2 
    & Q^{12}_3 \end{pmatrix} \quad \text{and} \quad
  \diag \tilde{M}_{mn} = \begin{pmatrix} R^{123} & f^1_{23} & f^2_{31} 
    & f^3_{12} \end{pmatrix}
\end{equation}
by applying \eqref{eqn:decompfluxes}, \eqref{eqn:doubleindices}, \eqref{eqn:SL(4)toSO(3,3)} and the mapping between the covariant fluxes $\mathcal{F}_{ABC}$ and the $H$-, $f$-, $Q$- and $R$-flux derived in section~\ref{sec:covariantfluxes}. Table~\ref{tab:solembedding} summarizes the twelve different solutions for \eqref{eqn:quadraticcMMtilde} found in \cite{Dibitetto:2012rk}. Each of them has a real parameter $\alpha$.

Performing a compactification, we are only interested in solutions which give rise to compact geometries in the internal space. The necessary condition for a minimum of the scalar potential \eqref{eqn:fluxeseom} further restricts these solutions by
\begin{align}
  \left( H_{123} - Q^{23}_1 \right)^2 - 
    \left( Q^{31}_2 - Q^{12}_3 \right)^2 &=
    \left( R^{123} - f^1_{23} \right)^2 -
    \left( f^2_{31} - f^3_{12} \right)^2 
    \\
  \left( H_{123} - Q^{31}_2 \right)^2 -
    \left( Q^{12}_3 -  Q^{23}_1 \right)^2 &=
    \left( R^{123} - f^2_{31} \right)^2 -
    \left( f^3_{12} - f^1_{23} \right)^2 \\
  \left( H_{123} - Q^{12}_3 \right)^2 -
    \left( Q^{23}_1 - Q^{31}_2 \right)^2 &=
    \left( R^{123} - f^3_{12} \right)^2 -
    \left( f^1_{23} - f^2_{31} \right)^2 \,.
\end{align}
Only the solutions 1, 6, 12 and 4 with $\alpha=\pm \pi/4$ are compatible with these constraints. Note that neither 4 nor 6 are a priori compact spaces. E.g., ISO($3$) represents the isometries of the three dimensional euclidean space which is extended. It is generated by three independent translations and rotations, respectively. The translations are responsible for the non-compactness. Thus, points connected to each other by translations with are the elements of a lattice are identified. This procedure gives rise to a compact, internal space. However, it also puts severe restrictions on the rotation generators. We will discuss them in the next section in detail. According to \eqref{eqn:strongconstfluxes}, the strong constraint restricts the fluxes by
\begin{equation}\label{eqn:fluxessc}
  H_{123} R^{123} + Q^{23}_1 f^1_{23} + Q^{31}_2 f^2_{31} + 
    Q^{12}_3 f^3_{12} = M_{qp} {\tilde M}^{pq} = 0
\end{equation}
and in conjunction with the Jacobi identity \eqref{eqn:quadraticcMMtilde} gives rise to
\begin{equation}
  H_{123} R^{123} = 0 \qquad  
  Q^{23}_1 f^1_{23} = 0 \qquad
  Q^{31}_2 f^2_{31} = 0 \qquad 
  Q^{12}_3 f^3_{12} = 0\,.
\end{equation}
Solutions 4, 6 and 12 fulfill these equations. In general solution 1 violates the strong constraint. Only for for $\alpha=0$ it fulfills \eqref{eqn:fluxessc}. Further, it is possible to derive solutions 4, 6 and 12 from 1 by a procedure which is called group contraction \cite{Wigner:1953}. For the explicit contraction from SO($4$) to ISO($3$) see, e.g. \cite{Subag:2012}. Hence in the set of all different compact solutions of the embedding tensor presented here, solution 1 is distinguished. For $\alpha=0$, it corresponds to the $S^3$ with $H$-flux background known from supergravity compactifications.

\section{\texorpdfstring{CSO($2,0,2$)}{CSO(2,0,2)} compactifications}\label{sec:examplecso202}
To give an explicit example of a generalized Scherk-Schwarz compactification, we discuss the solution 6 of table~\ref{tab:solembedding} which gives rise to an effective theory with CSO$(2,0,2)$ gauging. Doing so will highlight some fundamental problems of DFT while dealing with spaces which are not T-dual to geometric ones. In general the group CSO($p,q,r$) arises from a group contraction of SO($p,q+r$) which preserves a SO($p,q$) subalgebra \cite{deRoo:2006ms}.

\subsection{Geometry of internal space}
First, we change to appropriate coordinates which allow to formulate the Scherk-Schwarz ansatz in a compact form. To this end, we apply the O($2n$) transformation
\begin{equation}
  T_{\tilde M}{}^N = \frac{1}{\sqrt{2}} \begin{pmatrix}
    \sqrt{2} & 0 & 0 & 0        & 0 & 0 \\
    0        & 0 & 0 & \sqrt{2} & 0 & 0 \\
    0        & 1 & 0 & 0        & 1 & 0 \\
    0        & 0 & 1 & 0        & 0 & 1 \\
    0        &-1 & 0 & 0        & 1 & 0 \\
    0        & 0 &-1 & 0        & 0 & 1
  \end{pmatrix} = T^{\tilde M}{}_N 
\end{equation}
to the covariant fluxes $\mathcal{F}_{IJK}$ with the non-vanishing components
\begin{equation}\label{eqn:solutionfluxes}
  H_{123} = Q^{23}_1 = H\,, \quad  Q^{31}_2 = Q^{12}_3 = 0\,, \quad
  R^{123} = f^1_{23} = 0  \quad \text{and} \quad f^2_{31} = f^3_{12} = f\,.
\end{equation}
Afterwards, we identify the components of the resulting fluxes
\begin{equation}
  \mathcal{F}_{\tilde I\tilde J}{}^{\tilde K} = T_{\tilde I}{}^L T_{\tilde J}{}^M T^{\tilde K}{}_N \mathcal{F}_{LM}{}^N
\end{equation}
with the five generators
\begin{equation}
  t_0 = \left(\mathcal{F}_2\right)_{\tilde J}{}^{\tilde K}\,,\quad
  t_1 = \left(\mathcal{F}_3\right)_{\tilde J}{}^{\tilde K}\,,\quad 
  t_2 = \left(\mathcal{F}_4\right)_{\tilde J}{}^{\tilde K}\,,\quad 
  s_1 = \left(\mathcal{F}_5\right)_{\tilde J}{}^{\tilde K} \quad \text{and} \quad
  s_2 = \left(\mathcal{F}_6\right)_{\tilde J}{}^{\tilde K}\,.
\end{equation}
They satisfy the $\mathfrak{cso}(2,0,2)$ algebra with the non-vanishing commutator relations
\begin{align}
  [ t_0, t_1 ] &= (- f - H) t_2 &
  [ t_0, t_2] &= (f + H) t_1 \nonumber \\
  [ t_0, s_1] &= (-f + H) s_2 &
  [ t_0, s_2] &= (f - H) s_1 \,.
\end{align}
Further the algebra exhibits a central charge, which we will not discuss here.
\begin{figure}[t!]
  \centering
  \tikzstyle link=[postaction={decorate,decoration={markings,mark=at position .85 with {%
    \node[name=pt1,point] {};}}}]
  \tikzstyle arrow=[postaction={decorate,decoration={markings,mark=at position .80 with {%
    \arrow{>}; \node[anchor=north] {$x$};}}}]
  \tikzstyle zero=[postaction={decorate,decoration={markings, mark=at position .75 with {%
    \node[anchor=center] {\tikz \draw (0,0.3) -- (0,-0.3);};}}}]
  \begin{tikzpicture}
    \draw[zero,arrow, link] (0,0) ellipse (2cm and 0.8cm);
    \draw[bend angle=25,bend left,thick] (pt1) to (2,1);
    \begin{scope}[shift={(2,1)},xslant=0.5]
      \draw[fill=fillcolor] (0,0) rectangle (4,2);
      \draw[fill=fillcolor] (0.2,0.2) rectangle (4.2,2.2);
      \draw[->] (-0.2,-0.2) -- (-0.2,1) node[anchor=south east] {$y^{\bar 1}$};
      \draw[->] (-0.2,-0.2) -- (1,-0.2) node[anchor=north west] {$y^{\bar 2}$};
      \draw[->] (0.4,0.4) -- (0.4,1.6) node[anchor=south west] {$y^1$};
      \draw[->] (0.4,0.4) -- (1.6,0.4) node[anchor=south west] {$y^2$};
    \end{scope}
  \end{tikzpicture}
  \caption{Geometry of the internal space used for the generalized Scherk-Schwarz compactification giving rise to an effective theory with the gauge group CSO($2,0,2$).}\label{fig:csogeometry}
\end{figure}
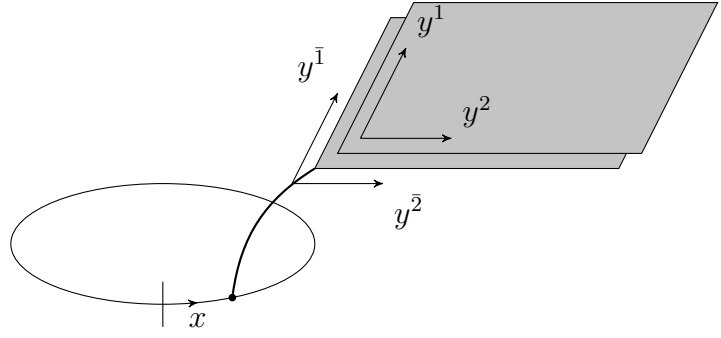
Exponentiating these generators, two two-dimensional euclidean planes $\mathds{R}^2$ arise from $t_1$, $t_2$ and $s_1$, $s_2$, respectively. Coordinates on these planes are $y^1$, $y^2$ and $y^{\bar 1}$, $y^{\bar 2}$. The remaining generator $t_0$ gives rise to a circle $S^1$ with the coordinate $x\in[0,2 \pi)$. Combining these coordinates, we obtain
\begin{equation}\label{eqn:adaptedcoordinates}
  X^{\tilde M} = \begin{pmatrix} \tilde x & x & y^1 & y^2 & y^{\bar 1} & y^{\bar 2} \end{pmatrix}\,.
\end{equation}
Figure~\ref{fig:csogeometry} visualizes the resulting geometry of the internal space $\mathcal{M}$ used for the compactification. It can be described in terms of the fibration
\begin{equation}
  \mathds{R}^4 \,\hookrightarrow\, {\mathcal M}
    \,\rightarrow\,\mathrm{S}^1 \,.
\end{equation}
Expressed in the coordinates \eqref{eqn:adaptedcoordinates}, the twist of the generalized Scherk-Schwarz compactification reads
\begin{align}\label{eqn:twistcso202}
  U_{\tilde N}{}^{\tilde M}(x) &= \exp ( t_0 x) \\ 
  &= \begin{pmatrix}
    1 & 0 & 0 & 0 & 0 & 0 \\
    0 & 1 & 0 & 0 & 0 & 0 \\
    0 & 0 & \cos(x[f+H]) & \sin(x[f+H]) & 0 & 0   \\
    0 & 0 & -\sin(x[f+H]) & \cos(x[f+H]) & 0 & 0  \\
    0 & 0 & 0 & 0 & \cos(x[f-H]) & \sin(x[f-H])   \\ 
    0 & 0 & 0 & 0 & -\sin(x[f-H]) & \cos(x[f-H])
  \end{pmatrix}\,. \nonumber
\end{align}

In its current form, the internal space $\mathcal{M}$ is extended. To make it compact, we identify points in the fiber by a four-dimensional lattice
\begin{equation}
  \Gamma=\big\{\sum\limits_{a=1}^4 \vec v_a z_a |\, \forall\, z_a \in \mathds{Z}\big\}
\end{equation}
which is fixed by four vectors $\vec v_a$. Applied on the fiber, it gives rise to
\begin{equation}
  \mathds{R}^4\, /\, \Gamma = \mathrm{T}^4
\end{equation}
and finally to
\begin{equation}\label{eqn:torusfibr}
  \mathrm{T}^4 \,\hookrightarrow\, {\mathcal M}
    \,\rightarrow\,\mathrm{S}^1 \,.
\end{equation}
However, this identification does not work for arbitrary values of $f$ and $H$. They are restricted by the monodromy
\begin{equation}
  M_{\tilde N}{}^{\tilde M} = U_{\tilde N}{}^{\tilde M}(2\pi) = \begin{pmatrix} \mathbf{1}_{2\times2} & 0 \\
    0 & M
  \end{pmatrix}
\end{equation}
which arises after one complete cycle around the base circle. Only if the SO($4$) matrix $M$ yields two integers for the quantities \cite{Blumenhagen:2013fgp}
\begin{align}
  \Tr M &= 4 \cos(2\pi f) \cos(2\pi H) \quad \text{and} \\
  \chi(M) &= \det( \mathbf{1} - M ) = 4 \big( \cos(2 \pi H) - \cos(2 \pi f) \big)
\end{align}
the space $\mathcal{M}$ in \eqref{eqn:torusfibr} is globally well defined. These two conditions quantize the covariant fluxes \eqref{eqn:solutionfluxes}. The monodromy $M$ is only compatible with lattices $\Gamma$ on which it acts as automorphism
\begin{equation}\label{eqn:automorphismlattice}
  \Gamma = M \Gamma\,.
\end{equation}
This condition severely restricts the number of compatible lattices for a pair of quantized fluxes $H$ and $f$. The vectors $\vec v_a$ which span the lattice form the vacuum generalized vielbein of the fiber. Thus, we are able to rewrite \eqref{eqn:automorphismlattice} as
\begin{equation}
  \bar E_A{}^{\tilde N} M_{\tilde N}{}^{\tilde M} \bar E^B{}_{\tilde M} \in \mathrm{O}(3,3,\mathds{Z})\,.
\end{equation}
By parameterizing the vacuum generalized vielbein in the same way as the monodromy,
\begin{equation}
  \bar E_{A}{}^{\tilde M} = \begin{pmatrix} \mathbf{1}_{2\times2} & 0 \\
    0 & E
  \end{pmatrix}\,,
\end{equation}
it yields the constraint
\begin{equation}\label{eqn:EME-1}
  \bar E M \bar E^{-1} \in \mathrm{O}(2,2,\mathds{Z})\,.
\end{equation}

\subsection{Monodromy and T-duality}
Looking for explicit solutions of \eqref{eqn:EME-1}, we note that $M$ is a SO($2,2$) element. Thus, it can be decomposed into
\begin{equation}\label{eqn:decompSL2xSL2}
  M = \begin{pmatrix}
    M_\tau & 0 \\
    0 & M_\tau^{-T}
  \end{pmatrix}
  T \begin{pmatrix}
    M_\rho & 0 \\
    0 & M_\rho^{-T}
  \end{pmatrix} T^{-1}
  \quad \text{with} \quad 
 T=\begin{pmatrix}
    1 & 0 & 0 & 0 \\
    0 & 0 & 0 & 1 \\
    0 & 0 & 1 & 0 \\
    0 & 1 & 0 & 0
  \end{pmatrix}=T^{-1}=T^T
\end{equation}
and $M_\tau$, $M_\rho$ denoting two independent SL($2$) matrices. We interpret $\tau$ as the complex structure and $\rho$ as the Kähler parameter of the fiber torus. SL($2$) transformations act on these two parameters as 
\begin{equation}\label{eqn:actiontau&rho}
  \tau \rightarrow \frac{a \tau + b}{c \tau + d} \qquad M_\tau =
    \begin{pmatrix} a & b \\ c & d \end{pmatrix}
  \quad \text{and} \quad
  \rho \rightarrow  \frac{a' \rho +  b'}{c' \rho + d'} \qquad M_\rho =
    \begin{pmatrix} a' & b' \\ c' & d' \end{pmatrix}\;, 
\end{equation}
respectively. To fill out the complete T-duality group in two dimensions, which is isomorphic to \cite{Giveon:1994fu}
\begin{equation}\label{eqn:decompO(2,2)}
  \mathrm{O}(2,2,\mathds{Z}) \cong \mathrm{SL}_\tau(2,\mathds{Z}) \times \mathrm{SL}_\rho(2,\mathds{Z}) 
  \times \mathds{Z}_2^{\tau\leftrightarrow\rho} \times \mathds{Z}_2\,,
\end{equation}
we need two additional discrete transformations. One of them swapping $\tau$ and $\rho$. However, they do not occur in the monodromy of the space which we discuss in this section. SL($2$) elements are classified by their conjugacy classes. In total there are three different classes, which are discriminated by the traces
\begin{equation}
  \left| \Tr N \right| < 2 \quad \text{elliptic} \qquad 
  \left| \Tr N \right| = 2 \quad \text{parabolic} \qquad \text{and} \qquad
  \left| \Tr N \right| > 2 \quad \text{hyperbolic}
\end{equation}
of the corresponding SL($2$) element $N$. For
\begin{equation}
  M_\tau = \begin{pmatrix}
    \cos 2\pi f & \sin 2\pi f \\
   -\sin 2\pi f & \cos 2\pi f
  \end{pmatrix} \quad \text{and} \quad
  M_\rho = \begin{pmatrix}
    \cos 2\pi H & \sin 2\pi H \\
   -\sin 2\pi H & \cos 2\pi H
  \end{pmatrix}\,,
\end{equation}
we obtain
\begin{equation}
  | \Tr M_\tau | = 2 |\cos( 2\pi f)| \le 2
  \quad \text{and} \quad
  | \Tr M_\rho | = 2 |\cos( 2\pi H)| \le 2 
\end{equation}
which renders them either elliptic or parabolic. We are interested in the following combinations:
\begin{itemize}
  \item A {\it Single elliptic space} with $f\neq0$, $H=0$ is a geometric space with $f$-flux, only. T-duality along all fiber directions maps it to itself. T-duality along only one fiber direction exchanges $\tau$ and $\rho$ and gives rise to
  \item a {\it Single elliptic space} with $f=0$, $H\neq0$ which is a non-geometric space with $H$-flux and $Q$-flux. Again, T-duality along all fiber directions maps it to itself.
  \item A {\it Double elliptic space} with $f\neq0$, $H\neq0$ is a genuinely non-geometric background. In contrast to the two cases we discussed so far, it can not be mapped to a geometric background by any T-duality transformation. Hence, the double elliptic space cannot be handled with standard supergravity. It needs a full DFT description.
\end{itemize}
Double elliptic spaces are most interesting, because they go beyond the framework of supergravity. Still, they can be described in terms of a consistent CFT \cite{Dabholkar:2002sy,Condeescu:2012sp,Condeescu:2013yma} and were discussed by \cite{Hohm:2013bwa} in the context of large generalized diffeomorphisms in DFT.

To obtain the flux quantization conditions for a double elliptic setup, we note that similarity transformations like \eqref{eqn:EME-1} leave the trace invariant. Thus, \eqref{eqn:EME-1} gives rise to the constraints
\begin{equation}
  \Tr M_\tau \in \{ -1, 0, 1\} \quad \text{and} \quad \Tr M_\rho \in \{-1, 0, 1\}\,.
\end{equation}
They are solved by the three elliptic elements of SL($2,\mathds{Z}$) which read \cite{Dabholkar:2002sy}
\begin{equation}
  M_3 = \begin{pmatrix} 0 & 1 \\ -1 & -1 \end{pmatrix}\,\quad 
  M_4 = \begin{pmatrix} 0 & 1 \\ -1 & 0 \end{pmatrix}\, \quad
  M_6 = \begin{pmatrix} 1 & 1 \\ -1 & 0 \end{pmatrix}
\end{equation}
and give rise to the quantized fluxes
\begin{equation}\label{eqn:quantizedfluxes}
  f \in \big\{ \frac{1}{3}, \frac{1}{4}, \frac{1}{6} \big\} \quad \text{and} \quad
  H \in \big\{ \frac{1}{3}, \frac{1}{4}, \frac{1}{6} \big\}\,.
\end{equation}
These values represent irreducible crystallographic rotations in two dimensions. With the decomposition
\begin{equation}
  \tau=\tau_\compliR + i\tau_\compliI \quad \text{and} \quad
  \rho=\rho_\compliR + i\rho_\compliI
\end{equation}
of the complex structure and the Kähler parameter, we can parameterize
\begin{equation}
  E_\tau = \frac{1}{\sqrt{\tau_\compliI}} \begin{pmatrix}
    1 & 0 \\
    \tau_\compliR & \tau_\compliI
  \end{pmatrix} \quad \text{and} \quad
  E_\rho = \frac{1}{\sqrt{\rho_\compliI}} \begin{pmatrix}
    1 & 0 \\
    \rho_\compliR & \rho_\compliI
  \end{pmatrix}\,, 
\end{equation}
respectively. From
\begin{equation}
  \bar E_\tau M_\tau \bar E_\tau^{-1} = M_3
\end{equation}
we obtain the vacuum expectation value of the complex structure
\begin{equation}
  \bar \tau = -\frac{1}{2} + i \frac{\sqrt{3}}{2} \quad \text{for} \quad f=\frac{1}{3} \,.
\end{equation}
The same value we obtain for $f=1/6$, whereas $f=1/4$ gives rise to $\bar\tau = i$. For the vacuum expectation value $\bar \rho$ of $\rho$, the calculation proceeds in exactly the same way.

\subsection{Spectrum}
We now calculate the spectrum arising from the CSO($2,0,2$) compactification. Following the reasoning in section~\ref{sec:spectrum}, one finds in general four massive excitations and five massless ones. To assign them to deformations of the vacuum, we consider the generalized metric
\begin{equation}
  \mathcal{H}^{MN} = \begin{pmatrix}
    \mathbf{1}_{2\times 2} & 0 \\
    0 & \mathcal{H}
  \end{pmatrix}
\end{equation}
of the internal space with an explicit split between base and fiber coordinates. Again, we are only interested in the fiber part
\begin{equation}
  \mathcal{H} = \begin{pmatrix}
    g - B g^{-1} B & B g \\
    -g B & g^{-1}
  \end{pmatrix}
\end{equation}
which is parameterized in terms of the metric
\begin{equation}\label{eqn:metricbrhotau}
  g = \frac{\rho_\compliI}{\tau_\compliI} \begin{pmatrix}
    1 & \tau_\compliR \\
    \tau_\compliR & \tau_\compliI^2 + \tau_\compliR^2
  \end{pmatrix} \quad \text{and the $B$-field} \quad
  B = \begin{pmatrix}
    0 & \rho_\compliR \\
    -\rho_\compliR & 0
  \end{pmatrix}\,.
\end{equation}
According to \eqref{eqn:(talpha)IJ}, we define the O($2,2$) generators \begin{align}
  t_1 &= \left.\frac{\partial \mathcal{H}}{\partial 2 \tau_\compliI}\right|_{\bar \tau,\bar \rho}
  = \frac{1}{2} \begin{pmatrix}
    -1 & 0 & 0 & 0 \\
    0  & 1 & 0 & 0 \\
    0  & 0 & 1 & 0 \\
    0  & 0 & 0 & - 1
  \end{pmatrix} &
  t_2 &= \left.\frac{\partial \mathcal{H}}{\partial 2 \tau_\compliR}\right|_{\bar \tau,\bar \rho}
  = \frac{1}{2} \begin{pmatrix}
    0 & 1 & 0 & 0 \\
    1 & 0 & 0 & 0 \\
    0 & 0 & 0 & -1 \\
    0 & 0 & -1& 0 
  \end{pmatrix} \nonumber \\
  t_3 &= \left.\frac{\partial \mathcal{H}}{\partial 2 \rho_\compliI}\right|_{\bar \tau,\bar \rho}
  = \frac{1}{2} \begin{pmatrix}
    1 & 0 & 0 & 0 \\
    0  & 1 & 0 & 0 \\
    0  & 0 & -1 & 0 \\
    0  & 0 & 0 & - 1
  \end{pmatrix} &
  t_4 &= \left.\frac{\partial \mathcal{H}}{\partial 2 \rho_\compliR}\right|_{\bar \tau,\bar \rho}
  = \frac{1}{2} \begin{pmatrix}
    0 & 0 & 0 & 1 \\
    0 & 0 & -1 & 0 \\
    0 & -1 & 0 & 0 \\
    1& 0 & 0 & 0
  \end{pmatrix}
\end{align}
which are evaluated for
\begin{equation}
  \bar \tau = i \quad \text{and} \quad \bar \rho = i\,.
\end{equation}
This background complex structure and Kähler parameter correspond to the vacuum generalized vielbein $\bar{E}_A{}^M=\delta_A^M$ defined in \eqref{eqn:defvacuumvielbein}. It is straightforward to check that the generators fulfill the normalization condition \eqref{eqn:(talpha)ABnormalization}, namely
\begin{equation}
  \Tr t_i t_j = \delta_{ij}\,.
\end{equation}
Calculating the Hesse matrix \eqref{eqn:hessediag} for the basis spanned by $t_1$, \ldots, $t_4$ gives rise to
\begin{equation}
  H_{ij} = \diag (\lambda_i) \quad \text{with} \quad
  \lambda_1 = f^2 \,, \quad
  \lambda_2 = f^2 \,, \quad
  \lambda_3 = H^2 \,, \quad \text{and} \quad
  \lambda_4 = H^2\,.
\end{equation}
Thus, we find two different masses
\begin{equation}
  m(\tau) = 2 |f| \quad \text{and} \quad m(\rho) = 2 |H|
\end{equation}
for the complex structure and the Kähler parameter of the fiber. Taking into account the quantization conditions for the covariant fluxes \eqref{eqn:quantizedfluxes}, mass scales are comparable to the string scale. Hence any dynamics in these scalars is frozen out in the low energy effective theory. They are stabilized to their vacuum expectation values $\bar \tau$ and $\bar \rho$ which we calculated in the last subsection. This demonstrates the strength of genuinely non-geometric backgrounds to stabilize combinations of moduli at tree-level with could not be stabilized in a geometric compactification.

\subsection{Killing vectors}
Because the twist $U_{\tilde N}{}^{\tilde M}$ in \eqref{eqn:twistcso202} depends on the coordinate $x$ only, five of the six Killing vectors we need to obtain a consistent compactification are trivial. E.g., they can be chosen to be
\begin{equation}
  K_{\tilde N}{}^{\tilde M} = \delta_{\tilde N}{}^{\tilde M} \quad \text{for} \quad 
    \tilde N \in \{1,\,3,\,\dots,\,6 \}\,.
\end{equation}
For the remaining $K_2$, the situation is a bit more challenging. Fixing its second component
\begin{equation}
  K_2{}^2 = 1
\end{equation}
yields
\begin{equation}
  K_2{}^{\tilde P} \partial_{\tilde P} U_{\tilde N}{}^{\tilde M} = U_{\tilde N}{}^{\tilde P}
    \left(t_0\right)_{\tilde P}{}^{\tilde M}\,.
\end{equation}
With this result, the Killing condition \eqref{eqn:killingeqtwist} gives rise to
\begin{equation}
  \mathcal{L}_{K_I} U_{\underline{\hat N}}{}^{\hat M} =
    \big( \left( t_0 \right)_{\tilde P}{}^{\tilde M} + \partial^{\tilde M} K_{2\tilde P} - 
      \partial_{\tilde P} K_2^{\tilde M} \big) U_{\tilde N}{}^{\tilde P} = 0
\end{equation}
and is solved by fixing the remaining components of $K_2$ according to
\begin{equation}
  (K_2)^{\tilde M} =  \frac{1}{2} X^{\tilde N} \left(t_0\right)_{\tilde N}{}^{\tilde M}
    \quad \text{for} \quad
  \tilde M \in \{1,\,3,\,\dots,\,6 \}\,.
\end{equation}
Thus, the complete, non-trivial Killing vector reads
\begin{equation}
  K_2{}^{\tilde M} = \begin{pmatrix} 0 & 1 & -\frac{1}{2} (f + H) y^2 & \frac{1}{2} (f + H) y^1 &
    -\frac{1}{2} (f - H) y^{\bar 2} & \frac{1}{2} (f - H) y^{\bar 1} \end{pmatrix}\,.
\end{equation}
It is instructive to transform it back to the standard DFT coordinates
\begin{equation}
  X^M = \begin{pmatrix} \tilde x_1 & \tilde x_2 & \tilde x_3 & x^1 & x^2 & x^3 \end{pmatrix}
\end{equation}
giving rise to
\begin{equation}
  (K_2)^M =
  \begin{pmatrix}
    0 & -\frac{1}{2}( H x^3 + f \tilde x^3) & \frac{1}{2}( H x^2 + f \tilde x^2) & 1 & 
      -\frac{1}{2}( f x^3 + H \tilde x^3) & \frac{1}{2}( f x^2 + H \tilde x^2 )
  \end{pmatrix}\,.
\end{equation}
For the double elliptic case $H\ne 0$, $f\ne 0$, this Killing vector violates the strong constraint. In this case it mediates diffeomorphisms, $B$- and $\beta$-transformations at the same time. The algebra of infinitesimal transformations along the Killing vectors still closes. After introducing an additional normalization factor $2$, it gives rise to the structure coefficients
\begin{equation}\label{eqn:tildeFIJK}
  \tilde{\mathcal F}_{IJK} = - \mathcal{F}_{IJK}\,.
\end{equation}

\section{Open questions}\label{sec:questiongenSS}
Generalized Scherk-Schwarz compactifications essentially depend on the twist $U_I{}^J$ and the corresponding Killing vectors $K_I{}^J$. Starting from some fixed covariant fluxes $\mathcal{F}_{IJK}$, which arise from the embedding tensor formalism, there is no explicit prescription to construct both of them. Of course, one could try to solve the first order partial differential equation arising from the definition
\begin{equation}\label{eqn:PDEtwist}
  \mathcal{F}_{IJK} = 3 U_{[I}{}^M \partial_M U_J{}^N U_{K]N}\,.
\end{equation}
In the context of EFT, \cite{Hohm:2014qga} follows this approach to obtain twists for a limited set of different fluxes. But in general, it is not even clear whether well defined twists exists for all covariant fluxes suggested by the embedding tensor. From this perspective, generalized Scherk-Schwarz compactifications are far less well understood than their geometric counter parts.

For the traditional Scherk-Schwarz compactification, the twist and its Killing vectors arise as left- and right-invariant Maurer-Cartan forms on a group manifold \cite{Kaloper:1999yr}. As we will see in the next chapter, it is straightforward to construct them explicitly from a group element. Such an element arises from exponentiating the generators which form the embedding tensor. This procedure is not applicable to DFT, because it generates twists which are in general GL($2n$) valued and depend on all coordinates. Thus, they violate the strong constraint even for geometric compactifications.

Intuitively, it seems that solving \eqref{eqn:PDEtwist} gets more difficult the more covariant fluxes are non-vanishing. Looking at the solutions of the embedding tensor in table~\ref{tab:solembedding}, the most difficult one corresponds to the compact, semi-simple Lie algebra of orbit 1. In comparison to all other compact solutions, it shares the least properties with a torus. Switching off fluxes, the solutions become more toroidal, like the CSO($2,0,2$) example in the last section demonstrates. Originally, DFT was developed starting from a torus as background. Following this observation, we will derive in the next two chapters a version of DFT on a compact, semi-simple Lie group. Finally chapter~\ref{chap:fluxform} will link this new theory with generalized Scherk-Schwarz compactifications. We see that it will eliminate all the ambiguities discussed in this section.

\chapter{DFT on group manifolds}\label{chap:groupdft}
Motivated by the results of the last chapter, we will now derive DFT from scratch on a group manifold \cite{Blumenhagen:2014gva}. Due to their isometries, these manifolds have the same local properties at each point. In general, the isometries are non-abelian, but they include also the torus with abelian isometries. Group manifolds are closely related to Scherk-Schwarz compactifications and also well suited to study various properties of doubled geometries \cite{Dabholkar:2005ve,Hull:2009sg}. On the worldsheet, they give rise to an exactly solvable background described by a Wess-Zumino-Witten model (WZW) \cite{Witten:1983ar} in the large level limit $(k\gg 1)$. Employing the occurring current algebras, we derive a cubic action and the corresponding gauge transformations from CSFT. Just like in DFT, we find that one also has to impose a weak and a strong constraint, which however take a different form.

Before starting with the actual CSFT calculations, we review the relevant features of the WZW model and its current algebra in section~\ref{sec:worldsheet}. Furthermore, we give a representation for two- and three-point correlators involving these currents in terms of scalar functions on a group manifold in the limit of large level $k$. Section~\ref{sec:CSFTaction&gauge} presents the derivation of the action and its gauge transformations to cubic order in CSFT. We close this chapter with an explicit example, the $S^3$ with $H$-flux in section~\ref{sec:examplesu(2)}. It corresponds to the semisimple solution 1 of the embedding tensor in table~\ref{tab:solembedding}.

\section{Worldsheet theory}\label{sec:worldsheet}
In the following, we discuss the basic properties of the WZW model and its current algebra. Doing so we set up the notation for the rest of the chapter. For a more detailed review of WZW models, we refer to e.g. \cite{Walton:1999xc} or the appendix of \cite{Schulz:2011ye}. Additionally, we show how the various representations of a semisimple Lie algebra can be expressed in terms of scalar functions on a group manifold. We use this result to express two- and three-point correlators and show that they fulfill the Knizhnik-Zamolodchikov equation \cite{Knizhnik:1984nr}. Finally, we provide the two- and three-point off-shell amplitude for Ka\v{c}-Moody primary fields.

\subsection{Wess-Zumino-Witten model}\label{sec:wzwmodel}
A string propagating on a group manifold of a semisimple Lie group $G$ is described by the non-linear sigma model
\begin{equation}\label{eqn:WZWaction}
  S = \frac{1}{4\pi\alpha'} \int_{\partial M} \mathcal{K}(\omega_\gamma, \star \omega_\gamma) + S_\mathrm{WZ}
\end{equation}
on the worldsheet two-sphere $S^2=\partial M$. Note that its prefactor does not match the common choice $-k/(8\pi)$, but is convenient for comparing \eqref{eqn:WZWaction} with a non-linear sigma model given in terms of a metric and an asymmetric two-form field. We will compensate this uncommon choice in the definition of the Killing metric \eqref{eqn:killingform}. The action presented here is exactly the same as the one presented in \cite{Walton:1999xc}.

Let us explain the notation used in \eqref{eqn:WZWaction} in more detail. As usual, $\star$ denotes the Hodge dual and $\omega_\gamma$ is the left-invariant Maurer-Cartan form\footnote{We could also use the right-invariant Maurer-Cartan form and would obtain the same results. But in the literature it is common to use the left-invariant one.}. The function $\gamma(\sigma)$, which appears as subscript of $\omega_\gamma$, maps each point of $S^2$ to an element of the group $G$. In this way the string worldsheet is embedded into the target space. In order to fix a certain group element $\gamma \in G$, one needs $n$ different parameters $x^i$ where $i=1,\dots, n$. Infinitesimal changes of them at a fixed $\gamma$ create the tangent space $T_\gamma G$ of the group manifold. At the identity, $T_e G$ is identified with the Lie algebra $\mathfrak{g}$ associated to $G$. The tangent space at an arbitrary group element $T_\gamma$ is mapped to $\mathfrak{g}$ by the left- or right-invariant Maurer-Cartan form
\begin{equation}\label{eqn:maurer-cartan}
  \omega_\gamma = \gamma^{-1} d \gamma = \gamma^{-1} \partial_i \gamma \,d x^i \quad \text{or} \quad
  \bar\omega_\gamma = d \gamma \gamma^{-1} = \partial_i \gamma \gamma^{-1} \, d x^i \quad  \text{with} \quad
  \partial_i = \frac{\partial}{\partial x^i}\,.
\end{equation}
They arise if $\gamma$ is assumed to act as a left or right translation of $G$. Both of them take values in the Lie algebra $\mathfrak{g}$. Two elements of this algebra are contracted to a scalar by the symmetric, bilinear Killing form\footnote{We use the common convention that the length square of the longest root in the root system of $\mathfrak{g}$ is normalized to 2.}
\begin{equation}\label{eqn:killingform}
  \mathcal{K}(x, y) = - \frac{\alpha' k}{2} \frac{\Tr (\adj_x \adj_y )}{2 h^\vee}\,,
    \quad \text{with} \quad
  x,\,y\in \mathfrak{g}
\end{equation}
where $\adj_x$ is the adjoint representation of $x$ and $h^\vee$ denotes the dual Coxeter number of $\mathfrak{g}$. The generalization of this equation to $n$-forms is straightforward: One has to insert a wedge product $\wedge$ between $\adj_x$ and $\adj_y$. With these definitions at hand, one is able to expand \eqref{eqn:WZWaction} as
\begin{equation}\label{eqn:WZWactionmetric}
  S = \frac{1}{4\pi\alpha'} \int_{\partial M} g_{ij} \, d x^i \wedge \star d x^j + S_\mathrm{WZ} 
    \quad \text{with} \quad 
  g_{ij} = \mathcal{K}(\gamma^{-1}\partial_i \gamma, \gamma^{-1}\partial_j \gamma)
\end{equation}
where $g_{ij}$ is the target space metric of the group manifold. The parameters $x^i$ parameterizing the elements of the group $G$ are equivalent to coordinates on the manifold. They are related to the word-sheet coordinates $\sigma^\alpha$ by the mapping $x^i(\sigma^a)$ giving rise to $d x^i = \partial_\alpha x^i d\sigma^\alpha$.

Since the metric part \eqref{eqn:WZWactionmetric} of the action $S$ alone spoils local conformal symmetry, one has to add the topological Wess-Zumino term
\begin{gather}\label{eqn:SWZ}
  S_\mathrm{WZ} = \frac{1}{12 \pi \alpha'} \int_M  \mathcal{K}\left(\omega_\gamma, 
    [ \omega_\gamma, \omega_\gamma ]\right) = \frac{1}{2 \pi \alpha'} \int_M H \\
\intertext{with the 3-form flux}\label{eqn:WZWHflux}
  H = \frac{1}{3!} H_{ijk} \, d x^i \wedge d x^j \wedge d x^k
    \quad \text{and} \quad
  H_{ijk} = \mathcal{K}\left(\gamma^{-1} \partial_i \gamma, [\gamma^{-1} \partial_j \gamma, \gamma^{-1} \partial_k \gamma]\right)\,.
\end{gather}
Here, the $H$-flux is the field strength associated to the massless, antisymmetric Kalb-Ramond field $B_{ij}$. Both are linked via the relation
\begin{equation}
  H = d B \quad \text{with} \quad 
  B = \frac{1}{2!} B_{ij} d x^i \wedge d x^j 
  \quad \text{and} \quad
  H_{ijk} = 3 \partial_{[i} B_{jk]}\,.
\end{equation}
Of course, a physically meaningful sigma model only depends on the worldsheet $\partial M$ and not on its extension to the three-dimensional space $M$. Thus, physics has to be independent of the specific choice for $M$. For $G$ being a compact semisimple Lie groups with non-trivial homotopy $\pi_3(G)=\mathds{Z}$, this is only the case if $S_\mathrm{WZ}$ is an integer multiple of $2\pi$ \cite{Witten:1983tw}. Thereby, the $H$-flux of a compact background is quantized.

The variation of the action with respect to the $G$-valued field $\gamma$ gives rise to the equation of motion
\begin{equation}\label{eqn:eomwzwmodel}
  \partial_\alpha (\gamma^{-1} \partial^\alpha \gamma) + \frac{1}{2} \epsilon_{\alpha\beta} \partial^\alpha (\gamma^{-1} \partial^\beta \gamma) = 0\,.
\end{equation}
It is interesting to note that the second term in this equation originates from the Wess-Zumino term in the action. By fixing the word sheet metric to $h^{z\bar z} = 2$, $h^{zz}=h^{\bar z\bar z}=0$ and writing out the components of the totally antisymmetric tensor $\epsilon_{\alpha\beta}$ with $\epsilon_{z \bar z}=1$, one obtains
\begin{equation}
  \partial (\gamma^{-1} \bar \partial \gamma) = 0 \,.
\end{equation}
Now, one can directly read off the anti-chiral Noether current
\begin{equation}\label{eqn:antichiralcurrent}
  \bar j(\bar z) = - \frac{2}{\alpha'} \gamma^{-1} \bar \partial \gamma
\end{equation}
from the equation of motion. Note that without the second term in \eqref{eqn:eomwzwmodel} we would not be able to find this current. To obtain the chiral current, we apply complex conjugation to \eqref{eqn:antichiralcurrent} and substitute $\gamma$ by $\gamma^{-1}$ afterwards. This procedure yields
\begin{equation}\label{eqn:chiralcurrent}
  j(z) = \frac{2}{\alpha'} \partial \gamma \gamma^{-1}\,.
\end{equation}
To motivate the normalization of these currents, consider the infinitesimal transformations
\begin{equation}\label{eqn:deltaxigamma}
  \delta_\xi \gamma(z,\bar z) = \xi(z) \gamma(z,\bar z) 
    \quad \text{and} \quad
  \delta_{\bar \xi} \gamma(z,\bar z) = - \gamma(z,\bar z) \bar \xi(\bar z)
\end{equation}
of the field $\gamma$. Here, $\xi(z)$ and $\bar \xi(\bar z)$ are the Lie algebra valued parameters of the transformations. It is sufficient to discuss the chiral part $\xi(z)$ only. Applying \eqref{eqn:deltaxigamma} to the action $S$, we obtain 
\begin{equation}
  \delta_\xi S = - \frac{1}{2\pi i} \oint_0 dz\, \mathcal{K}(\xi(z), j(z))
\end{equation}
where $\oint_w dz$ denotes a closed contour integral around the point $w$. We have chosen the normalization factor of $j_a$ in \eqref{eqn:chiralcurrent} to obtain precisely the factor $1/(2\pi i)$ in this expression. With $\delta S$ one can  compute small changes
\begin{equation}\label{eqn:deltaxi<x>}
  \delta_\xi \langle X \rangle = \langle \delta_\xi S X \rangle = \frac{1}{2\pi i} \oint_0 dz \langle \mathcal{K}(\xi(z), j(z)) X \rangle
\end{equation}
of an arbitrary expectation value
\begin{equation}
  \langle X \rangle = \frac{\int [d \gamma]\, X e^{-S[\gamma]}}{\int [d \gamma]\, e^{-S[\gamma]}}
\end{equation}
in the Euclidean path integral. 

As a brief interlude, let us discuss the $n = \dim\mathfrak{g}$ generators $t_a$ of the Lie algebra $\mathfrak{g}$. They form a basis of the adjoint representation. We define the symmetric tensor
\begin{equation}\label{eqn:defetaab}
  \eta_{ab} = \mathcal{K}(t_a, t_b) = - \frac{\alpha' k}{2} \frac{\Tr(t_a t_b)}{2 x_\lambda} = - \frac{1}{2 h^\vee}
    f_{ad}{}^c f_{bc}{}^d \,.
\end{equation}
In the last step we have expressed the generators in terms of the structure coefficients of the Lie algebra appearing in the commutation relation\footnote{There are different conventions. Some use an additional $i$ in front of the structure coefficients. We stick to the convention in \cite{Schulz:2011ye} without $i$.}
\begin{equation}\label{eqn:liealgebra}
  [t_a, t_b] = \sqrt{\frac{2}{\alpha' k}} f_{ab}{}^c\, t_c = F_{ab}{}^c\, t_c
    \quad \text{with} \quad
  F_{ab}{}^c := \sqrt{\frac{2}{\alpha' k}}\, f_{ab}{}^c  \,. 
\end{equation}
For later convenience, we have defined the rescaled structure coefficients $F_{ab}{}^c$. Note that it is always possible to choose the generators $t_a$ of a semisimple Lie algebra $\mathfrak{g}$ in a way that $\eta_{ab}$ is a diagonal matrix with entries $\pm 1$. Thus, $\eta_{ab}$ is completely specified by its signature. A compact Lie group $G$  has a Lie algebra with a negative definite Killing form, e.g. the signature of $\eta_{ab}$ is $(-, \dots,-)$. In combination with its inverse $\eta^{ab}$, $\eta_{ab}$ is used to raise and lower flat indices $a,b,\dots$. 

Coming back, the chiral current \eqref{eqn:chiralcurrent} can be written in terms of the generators $t_a$ as
\begin{equation}
  j(z) = t^a j_a(z) \quad \text{with} \quad j_a(z) = \mathcal{K}(t_a, j(z)) \,.
\end{equation}
In this form, the infinitesimal transformation $\delta_\xi$ of the chiral current reads
\begin{equation}\label{eqn:deltaxij}
  \delta_\xi  j_b(z) = F_{ab}{}^c\, j_c(z)\, \xi^a(z)  + \frac{2}{\alpha'} \eta_{ab} \partial \xi^a(z) 
    \quad \text{with} \quad
  \xi_a(z) = \mathcal{K}(t_a, \xi(z))\,.
\end{equation}
Plugging this into \eqref{eqn:deltaxi<x>}, one obtains the Ward identity
\begin{equation}
  \delta_\xi \langle j_b(z) \rangle = \frac{1}{2\pi i} \oint d w \langle j_a(w) j_b(z) \rangle \xi^a(w) 
    = F_{ab}{}^c\, \langle j_c(z) \rangle \xi^a(z) + \frac{2}{\alpha'} \eta_{ab} \partial \xi^a(z)
\end{equation}
allowing to read off the OPE
\begin{equation}\label{eqn:opejj}
  j_a(z) j_b(w) = \frac{F_{ab}{}^c\, j_c(w)}{z-w} - \frac{2}{\alpha'} \frac{\eta_{ab}}{(z-w)^2} + \dots
\end{equation}
of the chiral currents. The analogous algebra holds for the anti-chiral current $\bar j(\bar z)$. Normally one would expect the level $k$ in front of the flat metric $\eta_{ab}$ instead of $-\alpha'/2$. Here, $k$ is hidden in the rescaled structure coefficients $F_{ab}{}^c$. For this reason, the OPE \eqref{eqn:opejj} corresponds to the usual form of the Ka\v{c}-Moody algebra at level $k$. Applying the same procedure to the transformation in \eqref{eqn:deltaxigamma}, we get the OPE
\begin{equation}\label{eqn:opejg}
  j_a(z) \gamma(w,\bar w) = \frac{t_a \gamma(w,\bar w)}{z-w} + \cdots
\end{equation}
defining a Ka\v{c}-Moody primary. Introducing the mode expansion
\begin{equation}\label{eqn:modeexpja}
  j_a(z) = \sum\limits_n j_{a,n} \, z^{-n-1}\,,
\end{equation}
the OPE \eqref{eqn:opejj} is equivalent to the Ka\v{c}-Moody algebra  
\begin{equation}\label{eqn:algebrajj} 
  [j_{a,m}, j_{b,n}] = F_{ab}{}^c\, j_{c,m+n} - \frac{2}{\alpha'}\, m\, \eta_{ab}\, \delta_{m+n} \,.
\end{equation}

\subsection{Geometric representation}\label{sec:representation}
In the following we will show that there exist highest weight representations of a semisimple Lie algebra in terms of scalar functions defined on the group manifold. For that purpose, let us first change from the abstract notation with Maurer-Cartan forms to a more explicit one by introducing vielbeins. Expressing $\omega_\gamma$ in \eqref{eqn:maurer-cartan} in terms of the generators $t_a$, we obtain
\begin{equation}\label{eqn:vielbein}
  \omega_\gamma = t_a \, e^a{}_i \, d x^i \quad \text{with the vielbein} \quad
    e^a{}_i = \mathcal{K}(t^a, \gamma^{-1} \partial_i \gamma)\,.
\end{equation}
It carries two different kinds of indices: flat ones are labeled by $a,b,c,\cdots$ and curved ones by $i,j,k,\cdots$. Flat indices are raised and lowered with the metric $\eta_{ab}$, whereas for curved indices we use the target space metric $g_{ij}$ in \eqref{eqn:WZWactionmetric}, which in terms of the vielbein reads
\begin{equation}
  g_{ij} = \eta_{ab}\, e^a{}_i\, e^b{}_j \,.
\end{equation}
Moreover, $e_a{}^i$ denotes the inverse transposed of $e^a{}_i$ and the $H$-flux defined in \eqref{eqn:WZWHflux} can be written as
\begin{equation}\label{eqn:WZWHfluxexplicit}
  H_{ijk} = e^a{}_i\, e^b{}_j\, e^c{}_k\, F_{abc}\, .
\end{equation}
Introducing the flat derivative
\begin{equation}\label{eqn:flatderivative}
  D_a = e_a{}^i \partial_i
\end{equation}
the commutator of two of them satisfies
\begin{equation}\label{eqn:[Da,Db]}
  [D_a, D_b] = F_{ab}{}^c D_c \, ,
\end{equation}
with
\begin{equation}\label{eqn:fabcfromvielbein}
  F_{ab}{}^c = 2 e_{[a}{}^i \partial_i e_{b]}{}^j e^c{}_j = 2 D_{[a} e_{b]}{}^i e^c{}_i\,.
\end{equation}
Thus, we found a representation of the generators $t_a$ in terms of the differential operators $D_a$ acting on functions defined on a patch of the group manifold. We will see that these functions include all highest weight  representations of the Lie algebra.

Flat derivatives are mainly used under volume integrals with the volume element $d^n x \sqrt{|g|}$ where $g$ denotes the determinant of the target space metric $g_{ij}$. In this case, one finds
\begin{equation}\label{eqn:inttotalderiv}
  \int d^n x \, \sqrt{|g|} D_a v= \int d^n x \, \partial_i(\sqrt{|g|} e_a{}^i v)\,,
\end{equation}
where $v$ is an arbitrary scalar function depending on the target space coordinates $x^i$. Thus, the right hand side reduces to a boundary term which we always assume to vanish. Then one can perform integration by parts
\begin{equation}
  \int d^n x \, \sqrt{|g|} (D_a v) w = - \int d^n x \, \sqrt{|g|} v (D_a w)\,.
\end{equation}
Note that \eqref{eqn:inttotalderiv} is not restricted to semisimple Lie algebras, but is much more general and always holds if
\begin{equation}\label{eqn:unimodular}
  F_{ab}{}^b = 0 \quad \text{or equivalently} \quad \Tr \adj_x = 0 \quad \forall x \in \mathfrak{g}
\end{equation}
is fulfilled. Lie algebras with this property are called unimodular. 

The well known procedure of building highest weight representations also carries over to the flat derivatives discussed above. Take e.g. the group $SU(2)$ parameterized by Hopf coordinates $x^i = \begin{pmatrix} \eta^1 & \eta^2 & \eta^3 \end{pmatrix}$ with $0\le\eta^1<\pi/2$ and $0\le\eta^{2,3}<2\pi$. A detailed derivation of the vielbeins for this group is presented in section~\ref{sec:examplesu(2)}. Here we are only interested in the flat derivatives
\begin{align}
  \tilde D_3 &= - \sqrt{\frac{\alpha' k}{2}} D_3 = - \frac{i}{\sqrt{2}}\bigl( \partial_2 + \partial_3 \bigr) \quad \text{and} \\
  \tilde D_\pm &= - \sqrt{\frac{\alpha' k}{2}} ( \pm i D_1 - D_2 ) \nonumber\\ 
    &= - \frac{i e^{\pm i(\eta^2+\eta^3)}}{\sqrt{2} \sin(2 \eta^1)} \left[ \pm i \sin(2 \eta^1) \,\partial_1 + 2 \sin^2(\eta^1)\, \partial_2 - 2 \cos^2(\eta^1)\, \partial_3 \right]\,.
\end{align}
We look for eigenfunctions of $\tilde D_3$ which are annihilated by $\tilde D_+$. A short calculation shows that this is the case for
\begin{equation}
  y_\lambda(x^i) = C_\lambda ( \sin \eta^1 )^{\sqrt{2} \lambda} e^{i \sqrt{2} \lambda \eta^3}
\end{equation}
where $C_\lambda$ denote normalization constants fixed by the requirement
\begin{equation}\label{eqn:normalizationy}
  \int d^n x\, \sqrt{|g|} y_\lambda^* y_\lambda^{} = |C_\lambda|^2 4 \pi^2 (\alpha' k)^{3/2} \int\limits_0^{\pi/2} d \eta^1 \, 
  \cos(\eta^1) \sin(\eta^1)^{1 + 2 \sqrt{2} \lambda} = |C_\lambda|^2
  \frac{2 \pi^2 (\alpha' k)^{3/2}}{\sqrt{2} \lambda + 1} = 1\, ,
\end{equation}
which is only possible if $\sqrt{2} \lambda + 1 > 0$. Furthermore, we know from $\mathfrak{su}(2)$ representation theory that $\lambda$ is an element of the 1-dimensional weight lattice $\Lambda = \mathds{Z}/\sqrt{2}$. Therefore, $\lambda$ has to be an element of $\mathds{N}_0/\sqrt{2}$ in order to allow the normalization \eqref{eqn:normalizationy}. Starting from these highest weight states, one can construct the full $\mathfrak{su}(2)$ representation by acting with $\tilde D_-$ on $y_\lambda$. We denote the resulting functions according to their $\tilde D_3$ eigenvalues as
\begin{equation}
  y_{\lambda q} = C_{\lambda q} (\tilde D_-)^{(\lambda - q)/\sqrt{2}} y_\lambda \quad \text{with} \quad
  \tilde D_3 y_{\lambda q} = q\, y_{\lambda q}\quad \text{and} \quad
  q=-\lambda, -\lambda + \sqrt{2}, \dots, \lambda \,.
\end{equation}
Some of these functions are listed in section~\ref{sec:examplesu(2)}. According to the integral
\begin{equation}
  \int d^n x \sqrt{|g|}\, y_{\lambda_1 q_1}^*\, y_{\lambda_2 q_2}^{} = \delta_{\lambda_1 \lambda_2} \delta_{q_1 q_2}\,,
\end{equation}
which fixes the normalization constants $C_{\lambda q}$, they are orthonormal. It is straightforward to generalize this procedure for other compact semisimple Lie algebras. In this case $\lambda$ and $q$ are not just scalars, but vectors of dimension $r = \rank \mathfrak{g}$.

For non-compact Lie algebras, the structure becomes more involved: First, one has to consider lowest weight states in addition to the highest weight states discussed so far. These are states annihilated by all negative simple roots. A representation is build by acting with all negative simple roots on highest weight states $v_\lambda$ and with all positive simple roots on lowest weight states $v_{-\lambda}$. In contrast to a compact Lie algebra, this process does not terminate. Thus, there is an infinite tower of states for each highest and lowest weight. A simple example for a non-compact Lie algebra is $\mathfrak{sl}(2)$. Its representations are discussed in the context of the $SL(2)$ WZW model in \cite{Maldacena:2000hw}. 

\subsection{Correlation functions}\label{sec_twothreepointkm}
In order to perform the CSFT calculations in the next section, we need to know the correlation functions $\langle \gamma_1(z_1) \dots \gamma_n(z_n)\rangle$ of Ka\v{c}-Moody primary fields whose OPE we already defined in \eqref{eqn:opejg}. These correlation functions have to fulfill the Knizhnik-Zamolodchikov equation \cite{Knizhnik:1984nr}
\begin{equation}\label{eqn:KZeq}
  \left(\partial_{z_i} + \frac{2}{\alpha'} \frac{k}{k + h^\vee} \sum\limits_{i\ne j} \frac{\eta^{ab}\, t_a^{(i)} \otimes t_b^{(j)}}{z_i - z_j} \right) \langle \gamma_1(z_1) \dots \gamma_n(z_n)\rangle = 0 \,,
\end{equation}
where the notation $t_a^{(i)}$ indicates that the generator $t_a$ acts on the $i$th field $\gamma_i(z_i)$. The chiral energy momentum tensor is given by the Sugawara construction as
\begin{equation}\label{eqn:T(z)}
  T(z) = -\frac{\alpha'}{2} \frac{k}{2(k + h^\vee)} :\eta^{ab} j_a(z) j_b(z): \,.
\end{equation}
Again, the uncommon factors in the Knizhnik-Zamolodchikov equation and the energy momentum tensor are due to the normalization we performed in section~\ref{sec:worldsheet}. With the OPE of the chiral currents $j_a(z)$ in \eqref{eqn:opejj}, it is straightforward to calculate
\begin{align}
  T(z) j_a(w) &= \frac{j_a(w)}{(z-w)^2} + \frac{\partial_w j_a(w)}{z-w} + \dots 
    \quad \text{and} \quad \\
    T(z) T(w) &= \frac{c}{2 (z-w)^4} + \frac{2\, T(w)}{(z-w)^2} + \frac{\partial_w T(w)}{z-w} + \dots
\end{align}
with the central charge
\begin{equation}\label{eqn:centralcharge}
  c  = \frac{k n}{k + h^\vee}
    \quad \text{and} \quad n = \dim \mathfrak{g}\,.
\end{equation}
Combining \eqref{eqn:opejg} and \eqref{eqn:T(z)}, one can compute the OPE
\begin{equation}\label{eqn:opeTg}
  T(z) \gamma(w) = \frac{h}{(z-w)^2} \gamma(w) + \frac{\partial_w \gamma(w)}{z-w}  + \dots 
    \quad \text{with} \quad
  h = -\frac{\alpha' k}{4(k + h^\vee)}t_a t^a\,.
\end{equation}
For $\gamma(z)$ to be a Ka\v{c}-Moody and a Virasoro primary, it needs to be an eigenstate of the Lie algebra's quadratic Casimir operator $\eta^{ab} t_a t_b$.

The CSFT calculation in this chapter will be performed up to cubic order so that we need to know only the two-point and three-point correlation functions. Recall that for Virasoro primaries, these are completely determined up to some structure constants. We introduce a Fourier-type expansion of a Ka\v{c}-Moody primary
\begin{equation}
\label{gammaexpand}
  \gamma(z) = \sum\limits_{\lambda, q} c_{\lambda q}\, \phi_{\lambda q}(z, x^i)
\end{equation}
in terms of the Virasoro primaries $\phi_{\lambda q}(z, x^i)$ with constant coefficients $c_{\lambda q}$. Due to the linearity of the correlation functions, it is sufficient to know the correlation functions of $\phi_{\lambda q}$. As mentioned above, these are fixed by conformal symmetry as
\begin{align}\label{eqn:conformal2point}
  \langle \phi_{\lambda_1 q_1}(z_1) \phi_{\lambda_2 q_2}(z_2) \rangle &=
    \frac{d_{\lambda_1 q_1 \, \lambda_ 2 q_2}
    \delta_{h_{\lambda_1} h_{\lambda_2}}}{z_{12}^{2 h_{\lambda_1}}} \quad
    \text{with} \quad z_{12} = z_1 - z_2\, ,\\
  \label{eqn:conformal3point}
  \langle \phi_{\lambda_1 q_1}(z_1) \phi_{\lambda_2 q_2}(z_2) \phi_{\lambda_3 q_3}(z_3) \rangle &=
    \frac{C_{\lambda_1 q_1\,\lambda_2 q_2\,\lambda_3 q_3}}{z_{12}^{h_{\lambda_1}+h_{\lambda_2}-h_{\lambda_3}}
    z_{23}^{h_{\lambda_2}+h_{\lambda_3}-h_{\lambda_1}} z_{13}^{h_{\lambda_1}+h_{\lambda_3}-h_{\lambda_2}}}\,.
\end{align}
In these equations, $h_\lambda$ denotes the conformal weight of $\phi_{\lambda q}$ as written in \eqref{eqn:opeTg}. Note that it is independent of $q$. 

Finally, we apply the Knizhnik-Zamolodchikov equation \eqref{eqn:KZeq} to fix the constants $d_{\lambda_1 q_1 \,\lambda_2 q_2}$ and $C_{\lambda_1 q_1\, \lambda_2 q_2\, \lambda_3 q_3}$ in \eqref{eqn:conformal2point} and
\eqref{eqn:conformal3point}. To do so, we realize that the functions $y_{\lambda q}(x^i)$ we introduced in the last section are eigenstates of $L_0$. Therefore, a natural candidate for the two-point structure constants is
\begin{equation}
  d_{\lambda_1 q_1 \, \lambda_2 q_2} = \int d^n x \sqrt{|g|}\, y_{\lambda_1 q_1}^*\, y_{\lambda_2 q_2}^{} = \delta_{\lambda_1 \lambda_2} \delta_{q_1 q_2}\,.
\end{equation}
It automatically implies the delta function $\delta_{h_{\lambda_1} h_{\lambda_2}}$ in \eqref{eqn:conformal2point} by its $\delta_{\lambda_1 \lambda_2}$ part. We now show that this choice is compatible with the Knizhnik-Zamolodchikov equation. Plugging the correlation function into \eqref{eqn:KZeq} gives rise to
\begin{equation}
  h_{\lambda_1} d_{\lambda_1 q_1\, \lambda_2 q_2} - \frac{\alpha'}{2}
  \frac{k}{2(k + h^\vee)} \int d^n x\, \sqrt{|g|} \, D_a y_{\lambda_1
    q_1}^*\,  D^a y_{\lambda_2 q_2}^{} = 0 \,,
\end{equation}
where we used that the differential operators $D_a$ form a representation of the Lie algebra generators $t_a$. Now, we perform integration by parts, pull the constant factor in front of the integral into the integrand and obtain
\begin{equation}
  h_{\lambda_1} d_{\lambda_1 q_1\, \lambda_2 q_2} - \int d^n x \sqrt{|g|}\; L_0\, y_{\lambda_1 q_1}^*\, y_{\lambda_2 q_2}^{} = 0\,.
\end{equation}
Recalling the eigenvalue equation $L_0\, y_{\lambda q} = h_\lambda\, y_{\lambda
  q}$, one immediately sees that the Knizhnik-Zamolodchikov equation is indeed fulfilled. A similar calculation proves that in order to fulfill
\eqref{eqn:KZeq} for the three-point correlation function \eqref{eqn:conformal3point}, we have to set
\begin{equation}\label{eqn:Cconformal3point}
  C_{\lambda_1 q_1\,\lambda_2 q_2\,\lambda_3 q_3} = \int d^n x\, \sqrt{|g|}\, y_{\lambda_1 q_1}^*\, y_{\lambda_2 q_2}^{ }\, y_{\lambda_3 q_3}^{ }\,.
\end{equation}

Let us discuss how the usual toroidal case fits into this scheme. A torus corresponds to an abelian group manifold with $F_{ab}{}^c = 0$ and a coordinate independent vielbein $e_a{}^i$. Applied to the torus metric $g_{ij}=\delta_{ij}$, it gives rise to the flat metric $\eta_{ab}= e_a{}^i g_{ij} e_b{}^j$. Plugging these quantities in \eqref{eqn:algebrajj} and introducing the abelian currents 
\begin{equation}\label{eqn:alpha}
  \alpha_{i,m} = - i \sqrt{\frac{\alpha'}{2}} e^a{}_i\, j_{a,m}\,,
\end{equation}
we obtain the same current algebra
\begin{equation}
  [\alpha_{i, m}, \alpha_{j,n}] = m\, g_{ij}\, \delta_{m+n}
\end{equation}
as used for the derivation of DFT on a torus in \cite{Hull:2009mi}. To reproduce the zero modes $\alpha_{i,0}$, we perform the substitution $j_{a,0} \rightarrow D_a$ giving rise to
\begin{equation}
  \alpha_{i,0} = - i \sqrt{\frac{\alpha'}{2}} \partial_i \,.
\end{equation}
Finally, the Virasoro zero mode reads
\begin{equation}
  L_0 = -\frac{\alpha'}{4} \eta^{ab} \sum_n : j_{a,n}\, j_{b,-n}: = N +
  \frac{1}{2} g^{ij}\, \partial_i \, \partial_j
    \quad \text{with} \quad
  N = \sum\limits_{n>0} g^{ij}\, \alpha_{i,n} \alpha_{j,-n}\,.
\end{equation}

Note that the operator $D_a D^a$ is the Laplace operator on the group manifold. As we have seen above, the functions $y_{\lambda q}$ are its eigenfunctions. Consider now flat space where we find
\begin{equation}
  y_k(x^i) = \frac{1}{\sqrt{2 \pi}} e^{i k_i x^i}
\end{equation}
as eigenfunctions of the Laplace operator. The corresponding expansion \eqref{gammaexpand} is nothing else than a Fourier expansion. According to \eqref{eqn:Cconformal3point}, the constants in the three-point correlation function read
\begin{equation}
  C_{k_1 \, k_2 \, k_3} = \delta_{- k_1 + k_2 + k_3}\,.
\end{equation}
Physically, this reflects  momentum conservation in a scattering process with two incoming particles (momentum $k_2$ and $k_3$) and one outgoing particle (momentum $k_1$). Switching to the $SU(2)$ example discussed in section~\ref{sec:examplesu(2)}, one obtains \cite{WenAvery85}
\begin{equation}
  C_{\lambda_1 q_1\,\lambda_2 q_2\,\lambda_3 q_3} = \langle j_1 q_1 | j_2 q_2 \, j_3 q_3 \rangle
\end{equation}
with $\langle j_1 q_1 | j_2 q_2 \, j_3 q_3 \rangle$ denoting the Clebsch-Gordan coefficients. In contrast to flat space, the corresponding scattering process is not ruled by momentum conservation but by angular momentum conservation.

\subsection{CSFT off-shell amplitudes}\label{sec:fundamentalamp}
In the previous subsection we considered only the chiral primaries $\phi_{\lambda q}(z)$. Now, we take also their anti-chiral counterparts $\bar \phi(\bar z)_{\bar \lambda \bar q}$ into account. In order to keep the notation as simple as possible, we introduce the following abbreviations:
\begin{equation}
  R = \begin{pmatrix} \lambda q & \bar\lambda \bar q \end{pmatrix} \quad \text{and} \quad \phi_R(z, \bar z) = \phi_{\lambda q}(z) \bar\phi_{\bar \lambda \bar q}(\bar z)\,.
\end{equation} 
For the WZW model in section~\ref{sec:wzwmodel}, the anti-chiral current $\bar j_a(\bar z)$ is governed by the same Ka\v{c}-Moody algebra as the chiral one. 

Analogous to \eqref{eqn:flatderivative} and \eqref{eqn:vielbein}, we introduce a flat derivative $D_{\bar a}$ defined in terms of the vielbein
\begin{equation}\label{eqn:vielbeinbared}
  e^{\bar a}{}_{\bar i} = \mathcal{K}(t^a, \partial_{\bar i} \gamma\,  \gamma^{-1})
    \quad \text{as} \quad
  D_{\bar a} = e_{\bar a}{}^{\bar i} \partial_{\bar i}\,.
\end{equation}
In order to distinguish between the chiral and the anti-chiral part, it is convenient to use bared indices so that e.g. the commutator of two flat, bared derivatives is written as
\begin{equation}\label{eqn:[Dbara,Dbarb]}
  [ D_{\bar a}, D_{\bar b} ] = F_{\bar a\bar b}{}^{\bar c} D_{\bar c}\,.
\end{equation}
In the left/right symmetric WZW model corresponding to a geometric background, the bared and unbared structure coefficients are related by
\begin{equation}\label{eqn:signfandfbar}
  F_{\bar a\bar b}{}^{\bar c} = - F_{ab}{}^c \,.
\end{equation}
However in general, we want to treat them as independent quantities. The derivative in \eqref{eqn:vielbeinbared} acts on the right-moving coordinates $x^{\bar i}$ only. Combining these $n$ right-moving coordinates with the $n$ left-moving ones, we obtain a doubled space parameterized by the $2n$ coordinates $X^I = \begin{pmatrix} x^i & x^{\bar i} \end{pmatrix}$. It is straightforward to generalize the structure constants $d_{\lambda_1 q_1\,\lambda_2 q_2}$ and $C_{\lambda_1 q_1\,\lambda_2 q_2\,\lambda_3 q_3}$ to the combination of the chiral and anti-chiral fields $\phi_R$. In doing so, we obtain
\begin{align}
  d_{R_1\,R_2} &= \int d^{2n} X \sqrt{|H|}\, Y_{R_1}^*\, Y_{R_2}^{} = \delta_{R_1 R_2}
    \quad \text{and} \\
  C_{R_1\,R_2\,R_3} & = \int d^{2n} X \sqrt{|H|}\, Y_{R_1}^*\, Y_{R_2}^{}\, Y_{R_3}^{}
\end{align}
with
\begin{equation}\label{eqn:metricH}
  Y_R(X^I) = y_{\lambda q}(x^i)\, y_{\bar \lambda \bar q}(x^{\bar i})
    \quad \text{and} \quad
  H_{IJ} = \begin{pmatrix} 
      g_{ij} & 0 \\
      0 & g_{\bar i\bar j} 
  \end{pmatrix} \,.
\end{equation}

As we will see, all expressions arising in the CSFT calculation in the next section can be eventually reduced to two different off-shell amplitudes of the primaries $\phi_R$. In the vertex notation \cite{Zwiebach:1992ie,Taylor:2003gn}, these amplitudes read
\begin{align}\label{eqn:basic2point}
  \langle {\cal R}_{12} | \phi_{R_1} \rangle_1 | \phi_{R_2} \rangle_2 &= \lim_{z_i \to 0}\langle 
    I \circ \phi_{R_1}(z_1, \bar z_1)\;
    \phi_{R_2}(z_2,\bar z_2) \rangle \quad \text{and} \\
  \label{eqn:basic3point}
  \langle {\cal V}_3 | \phi_{R_1}\rangle_1 |\phi_{R_2}\rangle_2 |\phi_{R_3}\rangle_3 &= \lim_{z_i \to 0}\langle 
    I\circ f_1 \circ \phi_{R_1}(z_1,\bar z_1)\; f_2 \circ \phi_{R_2}(z_2,\bar z_2)
    \; f_3 \circ \phi_{R_3}(z_3,\bar z_3) \rangle\,,
\end{align}
where $\langle {\cal R}_{12} |$ denote the so-called reflector state and $\langle {\cal V}_3 |$ the three-point vertex. Moreover, $I$ is the BPZ conjugation defined as
\begin{equation}
  I(z) = \frac{1}{z} \quad \text{and} \quad
  I\circ \phi_R(z,\bar z) = z^{-2 h_R} \bar z^{-2 \bar h_R} \phi_R(I(z), \bar I(\bar z))\,.
\end{equation}
Further, 
\begin{equation}
  f_i(z_i)=z_{0\,i} + \rho_i z_i + \mathcal{O}(z_i^2)= z
\end{equation}
is a conformal mapping between the local coordinates $z_i$ around the $i$-th puncture of the sphere $S^2$ and a common, uniformizing coordinate $z$. We fix the punctures to $(z_{0\,1}, z_{0\,2}, z_{0\,3}) = (\infty, 0, 1)$. The parameter $\rho_i$ appearing in $f_i$ is called mapping radius \cite{Belopolsky:1994sk}. We will comment on its significance later. Note that for  Virasoro primaries, like $\phi_R$, a conformal transformation acts as
\begin{equation}
  f_i \circ \phi_R(z_i, \bar z_i) = \left(\frac{d f_i}{d z_i}\right)^{h_{R_i}}
    \left(\frac{d \bar f_i}{d \bar z_i}\right)^{\bar h_{R_i}} \phi_R(f_i(z_i),
    \bar f_i(\bar z_i))\,.
\end{equation}
 
An important consistency condition of CSFT is that all primaries have to be level matched ($h_R = \bar h_R$). In this case, the off-shell amplitudes take the simple form
\begin{align}
  \langle {\cal R}_{12} | \phi_{R_1} \rangle_1 | \phi_{R_2} \rangle_2 &= d_{R_1\,R_2}  \quad \text{and} 
    \quad \label{eqn:basicR12}\\
  \langle {\cal V}_3 | \phi_{R_1}\rangle_1 |\phi_{R_2}\rangle_2 |\phi_{R_3}\rangle_3 &= |\rho_1|^{2 h_{R_1}} 
    |\rho_2|^{2 h_{R_2}} |\rho_3|^{2 h_{R_3}} \, C_{R_1\,R_2\,R_3} \label{eqn:basicV3}\,.
\end{align}
Now, we have introduced all necessary tools to perform the CSFT calculations in the next section.

\section{Effective theory}\label{sec:CSFTaction&gauge}
After having worked out the details of the worldsheet theory, the corresponding CFT correlation functions and off-shell amplitudes, we derive the low energy effective theory in this section. It is called DFT${}_\mathrm{WZW}$ \cite{Blumenhagen:2014gva} because it originates from a Wess-Zumino-Witten model. We start with introducing the string fields describing a massless closed string state on a group manifold and the parameter for its gauge transformations. Then, from CSFT we derive the DFT${}_\mathrm{WZW}$ action and its gauge transformations up to cubic order. After introducing a modified version of the strong constraint, we simplify the results by applying the same field redefinitions as in \cite{Hull:2009mi}. Interestingly, the form of the strong constraint differs from the one imposed on traditional DFT. Finally, we calculate the gauge algebra (C-bracket).

In the following, we will work in the large level $k$ limit corresponding to the large volume limit of the group manifold. Therefore, many of the quantities we will compute receive higher order corrections in $k^{-1}$ which are closely linked to $\alpha'$ corrections.

\subsection{String fields}\label{sec:stringfields}
The starting point for the CSFT calculations are two string fields $|\Psi\rangle$ and $|\Lambda\rangle$. They are level matched and in Siegel gauge \cite{Siegel:1988yz}. Thus they are annihilated by
\begin{equation}\label{eqn:levelmatching}
  L_0 - \bar L_0
    \quad \text{and} \quad
  b_0^- = b_0 - \bar b_0\,.
\end{equation}
The former has ghost number two and the latter has ghost number one. A general string field consists of fields corresponding to all order Ka\v{c}-Moody modes acting on the Ka\v{c}-Moody ground states $|\phi_R\rangle$. Recall that for toroidal DFT, one restricts the string field to just the lowest lying massless oscillation modes acting on the Kaluza-Klein (momentum) and winding ground states. Since in this case there does not exist a regime for the radius such that all these states are lighter than the first excited oscillation mode, this is not a low energy truncation of the theory. However, the strong constraint prohibits simultaneous winding and momentum excitations in the same direction. In this sense, the torus can always be chosen in a way permitting a consistent low energy truncation.

For the WZW model the situation is similar. Analogous to the toroidal case, we first remove all massive string excitations from the string field. Then, we recall the explicit Sugawara form of the Virasoro operator
\begin{equation}\label{eqn:Lm}
  L_m = -\frac{\alpha'}{4} \Bigl( 1 - h^\vee k^{-1} \Bigr) \eta^{ab} \sum\limits_n :j_{a,n-m}\; j_{b,-n}: + \mathcal{O}(k^{-3})
\end{equation}
where we have expanded the prefactor as
\begin{equation}
 - \frac{\alpha'}{2} \frac{k}{2(k + h^\vee)} = -\frac{\alpha'}{4} ( 1
 - h^\vee k^{-1} + \cdots )
\end{equation}
and have taken into account that the chiral currents $j_a$ and $j_b$ include a normalization factor $k^{-1/2}$. Thus, we find exactly the order $\mathcal{O}(k^{-3})$ stated in \eqref{eqn:Lm}. Consider e.g. the state $j_{a,-1} \, j_{\bar b,-1}\, c_1 \bar c_1 |\phi_R\rangle$. It is still present in the truncated string field and its energy $E$ is given by
\begin{align}
  2 (L_0 + \bar L_0) j_{a,-1} j_{\bar b,-1} c_1 \bar c_1 |\phi_R\rangle &= \alpha' E j_{a,-1} j_{\bar b,-1} c_1 \bar c_1 |\phi_R\rangle \quad \text{with} \nonumber \\
  E &= \frac{1}{\alpha' k} \big( 1 - h^\vee k^{-1} \big) \big( c_2 (\lambda) + c_2(\bar \lambda) \big) + \mathcal{O}(k^{-3}) \label{hannover96}
\end{align}
where $c_2(\lambda)$ denotes the quadratic Casimir of the representation with the highest weight $\lambda$. Now, for a fixed ground state in the representation $\lambda$, one can always choose the level $k$ large enough so that the energy in \eqref{hannover96} is much smaller than one. For fixed level $k$, there exist always ground states with an energy much larger than one\footnote{For instance for $SU(2)_k$, there are finitely many highest
  weight representations with conformal dimension $h=\frac{l(l+2)}{4(k+2)}$ 
  with $0\le l\le k$. The state carrying  highest energy is $l=k$ with $h=k/4$.}. This is the same behavior as for the toroidal case, but only after one applies the strong constraint there. Thus, the truncated string field is given by
\begin{align}
  |\Psi\rangle &= \sum\limits_R \Bigl[ {\textstyle\frac{\alpha'}{4}} \epsilon^{a\bar
      b}(R)\, j_{a, -1}\, \bar j_{\bar b,-1}\, c_1\bar c_1 + e(R)\, c_1 c_{-1}
    + \bar e(R)\,\bar c_1 \bar c_{-1} + \nonumber \\ &
    \phantom{aaaaaaaaaaaaaaaa}
  {\textstyle \frac{\alpha'}{2}} \bigl( f^a(R)\, c_0^+c_1\, j_{a,-1} + f^{\bar b}(R)\,
  c_0^+\bar c_1\, \bar j_{\bar b,-1} \bigr) \Bigr] |\phi_R\rangle \label{eqn:stringfield}
\end{align}
and for the gauge parameters the corresponding string field is
\begin{equation}
\label{gaugestringfield}
  |\Lambda \rangle = \sum\limits_R \Bigl[ {\textstyle\frac{1}{2}} \lambda^a(R) j_{a,-1}
    c_1 - {\textstyle \frac{1}{2}} \lambda^{\bar b}(R)\, \bar j_{\bar b,-1}\, \bar c_1 +
    \mu(R)\, c_0^+ \Bigr] |\phi_R\rangle \, 
\end{equation}
with $c_0^\pm = \frac{1}{2} ( c_0 \pm \bar c_0 )$. The fields $\epsilon^{a\bar b}(R)$, $e(R)$ etc. can be considered as fluctuations around the WZW background. In contrast to the toroidal case \cite{Hull:2009mi}, in \eqref{eqn:stringfield} one does not sum over winding and momentum modes but over the different representations $R = \begin{pmatrix} \lambda q & \bar\lambda\bar q\end{pmatrix}$. 

\subsection{Weak constraint}\label{sec:weakconstr}
Now, let us derive the consequences of the level-matching constraint \eqref{eqn:levelmatching} in more detail. This will guide us to the DFT$_{\rm WZW}$ generalization of the weak and strong constraint. For that purpose, let us take a closer look at a component of the string field, like e.g. $e(R)$. We assume that the group manifold $G$ is simply-connected so that the functions $Y_{R}(X)$ introduced in section~\ref{sec:fundamentalamp} form a basis for the Hilbert space spanned by the states $|\phi_R\rangle$. Hence, we are able to substitute $e(R)$ by a field
\begin{equation}\label{eqn:modeexpansion}
  e(X) = \sum\limits_R e(R)\, Y_{R}(X)
\end{equation}
on the doubled space. For this field, the level matching constraint \eqref{eqn:levelmatching} translates into
\begin{equation}\label{eqn:weakconstr}
  \left( D_a D^a - D_{\bar a} D^{\bar a} \right) e = 0\,.
\end{equation}
It has not only to hold for $e$, but for all physical fields $e,\,\bar e,\,\epsilon^{a\bar b},\,f^a,\,f^{\bar b}$ and the gauge parameters $\lambda^a,\,\lambda^{\bar  b},\,\mu$. Denoting them as $\cdot$, we obtain
\begin{equation}
   \label{eqn:weakconstflat}
   \left( D_a D^a - D_{\bar a} D^{\bar a} \right) \cdot = 0 \,.
\end{equation}
In this notation, the level matching closely resembles the weak constraint of toroidal DFT. However, it is given in flat and not in curved indices so that for a proper comparison, we have to transform it into curved ones. To this end, we employ the identities
\begin{equation}
  \Omega_b{}^{ba} = - \Omega_b{}^{ab} + \partial_i g^{ij} e^a{}_j
    \quad \text{with the coefficients of anholonomy} \quad
  \Omega_{ab}{}^c = e_a{}^i \partial_i e_b{}^j e^c{}_j
\end{equation}
and 
\begin{equation}
  F_{ab}{}^b = 0 = 2\Omega_{[ab]}{}^b = \Omega_{ab}{}^b - \Omega_{ba}{}^b
    \quad \Rightarrow \quad
  \Omega_{ab}{}^b = \Omega_{ba}{}^b\,,
\end{equation}
which follows from unimodularity of the Lie algebra $\mathfrak{g}$, as required in \eqref{eqn:unimodular}. Moreover, for a constant dilaton $\phi$ one gets
\begin{equation}\label{eqn:Omegaabb}
  2 D^a d = \Omega^a{}_b{}^b\,,
    \quad \text{where} \quad
  d = \phi - \frac{1}{2} \log \sqrt{|H|}
\end{equation}
is the generalized dilaton of DFT. Remember that $H$ denotes the determinant of the metric on the doubled space defined in \eqref{eqn:metricH}. Combining these results, we obtain the relation
\begin{equation}
  \Omega_b{}^{ba} = - 2 D^a d + \partial_i g^{ij} e^a{}_j
\end{equation}
which yields
\begin{equation}
  D_a D^a \cdot = (\Omega_b{}^{ba} D_a + g^{ij} \partial_i \partial_j) \cdot =
   ( - 2\partial_i d\, \partial^i + \partial_i \partial^i ) \cdot \,.
\end{equation}
The analogous relation holds for bared indices, as well. Thus, the curved indices version of \eqref{eqn:weakconstflat} reads 
\begin{equation}\label{eqn:weakconstcurved}
  ( \partial_I \partial^I - 2\, \partial_I d\, \partial^I ) \cdot = 0
\end{equation}
with
\begin{equation}
  \partial_I = \begin{pmatrix} \partial_i & \partial_{\bar i} \end{pmatrix}
    \quad \text{and} \quad
  \partial^I = \begin{pmatrix} \partial^i & \partial^{\bar i} \end{pmatrix}\,.
\end{equation}
In DFT, the weak constraint consists of the first term in \eqref{eqn:weakconstcurved} only. The second one is absent there. This observation is consistent with our result: If we restrict the background to be a torus, $|H|$ is constant. Then, according to \eqref{eqn:Omegaabb}, $d$ also is constant. Thus, the second term vanishes and we reproduce the familiar result.

Applying \eqref{eqn:weakconstflat} to a product of two elementary objects we arrive at the strong constraint
\begin{equation}\label{eqn:strongconst}
  D_a f\, D^a g - D_{\bar a} f \, D^{\bar a} g = 0\,.
\end{equation}

\subsection{Action and gauge transformations}
In CSFT, the tree-level action is given by \cite{Zwiebach:1992ie,Hull:2009mi}
\begin{equation}\label{eqn:treelevelaction}
  (2\kappa^2) S= \frac{2}{\alpha'} \bigl( \{\Psi, Q \Psi\} + \frac{1}{3} \{\Psi, \Psi, \Psi\}_0 +
    \frac{1}{3\cdot4} \{\Psi,\Psi,\Psi,\Psi\}_0 + \dots \bigr)
\end{equation}
where $\psi$ denotes the string field \eqref{eqn:stringfield}. It is a sum over infinitely many string vertices $\{\cdot,\,\dots\,,\cdot\}_0$, also called string functions, evaluated at the genus zero worldsheet $S^2$. As in \cite{Hull:2009mi}, here we will evaluate only vertices up to order three. The fourth order term is already very challenging as it involves an integral over a region in $\mathds C$, whose boundary is not analytically known \cite{Belopolsky:1994bj,Moeller:2004yy}. First we calculate the quadratic order and then discuss the appearance of Ward identities which are going to be used along the line of \cite{Rastelli:2000iu} to calculate the cubic order. This will give the simplest interactions among the components of the string field.
 
Besides the action \eqref{eqn:treelevelaction}, CSFT admits to calculate gauge transformations of the action, too. They read
\begin{equation}
\label{eqn:treelevelgauge}
  \delta_\Lambda \Psi = Q\Lambda + [\Lambda, \Psi]_0 + \frac{1}{2!} [\Lambda, \Lambda, \Psi]_0 + \dots
\end{equation}
and are parameterized by $\Lambda$, the ghost number one string field introduced in \eqref{gaugestringfield}. Here, the string product $[\cdot, \cdot]_0$ appears. It is connected to the string function by the identity
\begin{equation}
  [B_1, \dots, B_n]_0 = \sum\limits_s |\phi_s \rangle \{\phi_s^c, B_1, \dots, B_n\}_0\,.
\end{equation}
The string fields $\phi_s^c$ are called conjugate fields of $\phi_s$. Since for CSFT on the torus, the CFT is free, it is straightforward to obtain the conjugate fields. However, on group manifolds, the worldsheet theory is in general interacting so that the notion of conjugate fields becomes more involved. We will tackle this problem while discussing the gauge transformations at quadratic order.

\subsubsection{Quadratic order}\label{sec:freetheory}
Let us start discussing the leading order CSFT action
\begin{equation}\label{eqn:2pointstringfunc}
  \{\Psi, Q \Psi\} = \langle\Psi| c_0^- Q | \Psi\rangle\,
\end{equation}
with the BRST operator given by
\begin{equation}
  Q = \sum\limits_m \bigl( :c_{-m} L_m: + \frac{1}{2}:c_{-m} L^{gh}_{m}:
  \bigr) + \text{anti-chiral}\,.
\end{equation}
In a theory free from conformal anomalies, the BRST operator has to be nilpotent. This is only the case if the central charge $c_\mathrm{gh}=-26$ of the ghost system cancels the one of the bosons. Thus, we have to add $26-n$ extended directions to the $n$ compact ones. Furthermore, for finite level $k$, the external space has to have a negative curvature. A classical example would be e.g.
\begin{equation}\label{eqn:ads3s3}
  \mathrm{AdS}_3 \times \mathrm{S}^3 \times \mathds{R}^{20}
\end{equation}
where the $n=3$ directions we consider in this chapter parameterize the $\mathrm{S}^3$.

We know the exact definition of $L_m$ and $L^{gh}_m$ in terms of the modes $j_{a m}$, $c_m$ and $b_m$, but for most purposes we only need to employ the commutator
\begin{equation}\label{eqn:algebraLphi}
  [L_m, \phi_n] = \big((h - 1) m - n\big)
\end{equation}
between a Virasoro generator and a primary field $\phi$ of conformal weight $h$ and similarly for the ghost contribution $L^{gh}_m$. 

As we have already defined in \eqref{eqn:basicR12}, a convenient way to express the expectation value \eqref{eqn:2pointstringfunc} is in terms of the reflector state $\langle {\cal R}_{12}|$, namely
\begin{equation}\label{eqn:actionR12}
  \langle \Psi | c_0^- Q | \Psi\rangle = \langle {\cal R}_{12} | \Psi \rangle_1 c_0^{-(2)} Q^{(2)} | \Psi\rangle_2\,.
\end{equation}
Then, we can use the identities \cite{Zwiebach:1992ie}
\begin{equation}\label{eqn:wardR12}
  \langle {\cal R}_{12} | c_m^{(1)} + c_{-m}^{(2)} = 0 \quad\text{and}\quad
  \langle {\cal R}_{12} | j_{a,m}^{(1)} + j_{a,-m}^{(2)} = 0 
\end{equation}
to move operators from one side of the reflector to the other. As \eqref{eqn:actionR12} is bilinear, one can treat each term in the string field \eqref{eqn:stringfield} separately. To continue, we use the following algorithm: On each side of the reflector state we move operators annihilating the primary $|\phi_R\rangle$ or the ghost vacuum to the right by using the commutation relations \eqref{eqn:algebraLphi} and \eqref{eqn:algebrajj}. This procedure is called normal ordering. It is performed in such a way that the Virasoro generators are transported directly to the primary field in each slot of the reflector state. Only $L_0$ and $L_{-1}$ survive this procedure. According to \eqref{eqn:Lm}, one can replace $L_0$ and $L_{-1}$ by
\begin{align}
  L_0 |\phi_R\rangle &= -\frac{\alpha'}{4} (1 - h^\vee k^{-1} + \dots )
  \eta^{ab}\, j_{a,0}\, j_{b,0} |\phi_R\rangle \nonumber \\
    L_{-1} |\phi_R\rangle &= -\frac{\alpha'}{2} (1 - h^\vee k^{-1} + \dots) \eta^{ab}\, j_{a,-1}\, j_{b,0} |\phi_R\rangle\,
    \label{eqn:L0L1alphapexpansion}
\end{align}
for large $k$. Afterwards, we perform normal ordering again until only zero modes or creation operators are left over. All operators acting on the first part of $\langle {\cal R}_{12}|$ are moved to the second one utilizing the identities \eqref{eqn:wardR12}. We establish normal ordering and so, finally, only zero modes are left over.

Just to give an impression, one of the many terms of the resulting expression is
\begin{equation}
  \{\Psi, Q \Psi\} = \dots + \frac{\alpha'}{2} \sum\limits_{R_1,\, R_2} \bar e(R_1)\, e(R_2)\, \eta^{ab}\,
    \langle {\cal R}_{12}| \phi_{R_1}\rangle_1 c_{-1}\bar c_{-1} c_0 c_1 \bar c_1\,
    j_{a,0}\, j_{b,0} |\phi_{R_2}\rangle_2 + \dots \,.
\end{equation}
To get rid of the ghost zero modes $c_{-1}$, $c_0$ and $c_1$, we apply the ghost overlap\footnote{We use the convention of \cite{Hull:2009mi} which  differs by a sign from earlier works like \cite{Zwiebach:1992ie}.}
\begin{equation}\label{eqn:ghostoverlap}
  \langle \phi_{R_1} | c_{-1} c_0 c_1 \bar c_{-1} \bar c_0 \bar c_1 |
  \phi_{R_2} \rangle = 2 \delta_{R_1\,R_2}\, .
\end{equation}
Recalling the two-point amplitude \eqref{eqn:basicR12} and combining it with the substitution 
\begin{equation}
  j_{a,0} |\phi_R\rangle = t_a |\phi_R\rangle \qquad \text{and} \qquad
    t_a \rightarrow D_a\,,
\end{equation}
we obtain the final result
\begin{equation}
 (2\kappa^2)S= \dots + \frac{\alpha'}{2} \int d^{2n} X \,\sqrt{|H|} \, \bar e\,  D_a D^a e + \dots \,.
\end{equation}
After a tedious computation, at leading order $\mathcal{O}(k^{-1})$ the complete quadratic action reads
\begin{align}
  (2\kappa^2) S^{(2,-1)} &= \int d^{2n}\, \sqrt{|H|}\, \Bigl[ {\textstyle \frac{1}{4}} \epsilon_{ab} \square \epsilon^{ab} + 2\, \bar e \,\square e - f_a\, f^a - f_{\bar b}\, f^{\bar b} \nonumber \\
  &\qquad\qquad - f_a ( D_{\bar b} e^{a\bar b} - 2 D^a \bar e) + f_{\bar b} ( D_a e^{a\bar b} + 2 D^{\bar b} e) \Bigr] \label{eqn:S2-1}
\end{align}
where the generalized Laplace operator is defined as
\begin{equation}
  \square = \frac{1}{2} \left( D_a D^a + D_{\bar a} D^{\bar a} \right)\, .
\end{equation}
Let us make a couple of comments:
\begin{itemize}
  \item Note that we assumed the auxiliary fields $f_a$ and $f_{\bar a}$ to be proportional to $k^{-1/2}$, as otherwise we would also find additional terms in \eqref{eqn:S2-1}. This situation is in total accordance with toroidal DFT, where the auxiliary fields are also weighted by an additional factor $\sqrt{\alpha'}$.
  \item On the torus, the vielbeins $e_a{}^i$ and $e_{\bar a}{}^{\bar i}$ are independent of the coordinates $X^I$, so that one can simply substitute the flat indices in \eqref{eqn:S2-1} by curved ones. In this way, one exactly reproduces the result derived in \cite{Hull:2009mi}.
  \item Even though \eqref{eqn:S2-1} looks like the action of toroidal DFT, there is a substantial difference: The derivatives appearing there do not commute.
\end{itemize}
At subleading orders in $k^{-1}$ the differences become even more striking. Recall that such corrections have the interpretation of $\alpha'$ corrections. Whereas for the toroidal case such corrections are absent in the CFT action at quadratic order, for the WZW model there exist a good deal of them. Thus, all quantities on the worldsheet receive corrections. This is already reflected in \eqref{eqn:L0L1alphapexpansion}, where the Virasoro generators $L_0$ and $L_{-1}$ receive corrections in all orders of $k^{-1}$.

\vspace{0.2cm}
Now, we come to the evaluation of the gauge transformation \eqref{eqn:treelevelgauge} at leading order, which involves the conjugate fields $\phi_s$. These are defined by the relation
\begin{equation}\label{eqn:defconjugatestate}
  \{\phi_s^c, \phi_{s'}\}_0 = \langle \phi_s^c| c_0^- |\phi_{s'} \rangle =
  \langle {\cal R}_{12} | \phi_s^c\rangle_1 c_0^{-(2)} |\phi_{s'}\rangle_2 = \delta_{s s'}\,.
\end{equation}
Since $j_{a,-1}$ and $j_{\bar b,-1}$ are the only creation operators appearing in the massless string fields, it is sufficient to know the conjugate field of $\phi_s = j_{a,-1} |\phi_R\rangle$ with $s=\begin{pmatrix} a & R \end{pmatrix}$ and its anti-chiral counterpart. A first guess for this conjugate field is $\phi_s^c = j^{a}_{-1} |\phi_R\rangle$, which  is along the lines of the abelian case. Evaluating \eqref{eqn:defconjugatestate}, we obtain
\begin{equation}\label{eqn:defconjsf}
  \langle {\cal R}_{12} | j_{-1}^{a\,(1)} |\phi_{R_1}\rangle_1\, j_{b,-1}^{(2)} |\phi_{R_2}\rangle_2 =
    -F^a{}_b{}^c \, \langle {\cal R}_{12} | \phi_{R_1}\rangle_1\, j_{c,0}^{(2)} |\phi_{R_2}\rangle_2 
    + \frac{2}{\alpha'}\, \delta^a_b\, \delta_{R_1\,R_2} \,.
\end{equation}
We realize that, even though the second term on the right hand side looks quite good, the first one spoils everything. We can get rid of this term by defining the conjugate field as
\begin{equation}\label{eqn:conjsf}
  \phi_s^c = \frac{\alpha'}{2} \Bigl( 1 - {\textstyle \frac{2 h^\vee}{k}} \Bigr)
    \Bigl( j^a_{-1} + {\textstyle \frac{\alpha'}{2}}\, F^{abc}\, j_{b,0}\, j_{c,-1} \Bigr)\, .
\end{equation}
Indeed, after some algebra and using \eqref{eqn:defetaab}, up to order $k^{-3/2}$, this ansatz gives rise to the desired result 
\begin{equation}\label{eqn:conjja-1}
  \langle {\cal R}_{12} | \phi_{s_1}^c \rangle_1 j_{b,-1}^{(2)} |\phi_{R_2}\rangle_2 =
    \delta^a_b\, \delta_{R_1\,R_2}  + \mathcal{O}(k^{-3/2})\,,
\end{equation}
which  is an improvement in comparison to our first guess. There it was only satisfied up to the order $k^{-1/2}$. In general, one has to determine the conjugate fields order by order in inverse powers of $k$. However, for all orders we are considering here, \eqref{eqn:conjja-1} is sufficient.

Now, we have collected all ingredients to calculate the gauge transformation
\begin{equation}
  \delta_\Lambda \Psi = \sum\limits_s |\phi_s \rangle \{\phi_s^c, Q \Lambda \}_0
\end{equation}
using the same techniques as for computing the CSFT action. In the end, at leading order $\mathcal{O}(k^{-1})$ we obtain the gauge transformations
\begin{align}
  \delta_\Lambda \epsilon_{a\bar b} &= D_a \lambda_{\bar b} + D_{\bar b} \lambda_a &
  \delta_\Lambda e      &= \mu - \frac{1}{2} D_a \lambda^a &
  \delta_\Lambda f_a    &= D_a \mu - \frac{1}{2} \square \lambda_a \nonumber \\
  &&
  \delta_\Lambda \bar e  &= \mu + \frac{1}{2} D_{\bar b} \lambda^{\bar b} &
  \delta_\Lambda f_{\bar b}  &= D_{\bar b}\, \mu + \frac{1}{2} \square \lambda_{\bar b}\,.
\end{align}
These and the quadratic action \eqref{eqn:S2-1} possess the $\mathds{Z}_2$ symmetry
\begin{align}
  \epsilon_{a\bar b} \,\, &\leftrightarrow \,\, \epsilon_{\bar b a} &
  e \,\, &\leftrightarrow \,\, - \bar e &
  f_a \,\, &\leftrightarrow \,\, - f_{\bar a} \nonumber \\
  D_a \,\, &\leftrightarrow \,\, D_{\bar a} &
  \lambda_a \,\, &\leftrightarrow \,\, \lambda_{\bar a} &
  \mu \,\, &\leftrightarrow \,\, -\mu\,, \label{eqn:Z2sym}
\end{align}
which is a direct consequence of vanishing (anti-)commutators between chiral and anti-chiral operators in the theory. 

\subsubsection{Cubic order}\label{sec:cubicorder}
We now compute the string function 
\begin{equation}
\label{stringfuncthree}
  \{\Psi, \Psi, \Psi\} = \langle {\cal V}_3 | \Psi \rangle_1 |\Psi \rangle_2 |\Psi \rangle_3 \,,
\end{equation}
which forms the cubic part of the tree-level action \eqref{eqn:treelevelaction}. Like proposed in \cite{Rastelli:2000iu}, we apply Ward identities to evaluate this expression. Even though \cite{Rastelli:2000iu} considers open string field theory, our CSFT computations are similar.

From the discussion in section~\ref{sec:worldsheet}, we know that each mode $j_{a,n}$ of the current $j_a(z)$ is a symmetry generator of our theory. Hence, the variation
\begin{equation}\label{eqn:jsymm}
  \delta_\varepsilon \langle I \circ f_1 \circ V_1\; f_2 \circ V_2\; f_3 \circ V_3 \rangle =
  \oint \frac{d z}{2\pi i} \langle \varepsilon(z) j_a(z) I \circ f_1 \circ V_1\; f_2 \circ V_2\; f_3 \circ V_3 \rangle = 0
\end{equation}
has to vanish for arbitrary vertex operators $V_i$. In the vertex notation $\langle V_3|$, introduced in \eqref{eqn:basic3point}, this expression translates into \cite{Rastelli:2000iu}
\begin{equation}\label{eqn:vertexrulegen}
  \sum\limits_{i=1}^3 \oint_{\mathcal{C}_i} \frac{dz}{2\pi i}\, \langle {\cal V}_3| \varepsilon(z)\, j_a(z) = 0\,.
\end{equation}
Here, we do not explicitly write the right hand side of the equation, because it holds for arbitrary vertex operators $V_i$. The integral in \eqref{eqn:jsymm} receives only contributions around the punctures introduced by these vertex operators. Each puncture is enclosed by the contour $\mathcal{C}_i$\,. To pull the integration directly in front of the corresponding vertex operator, one has to change the integration variable from $z$ to $z_i = f^{-1}_i(z)$. Since $j_a(z)$ has conformal weight one, this transformation gives rise to
\begin{equation}
  dz\, \varepsilon(z)\, j_a(z) = dz_i\, \frac{d z}{d z_i} \left( \frac{d z_i}{d z} \right)^1 \varepsilon(f_i(z_i))\, j_a(z_i) = dz_i\, \varepsilon_i(z_i)\, j_a(z_i) 
\end{equation}
with $\varepsilon_i(z_i) = \varepsilon(f_i(z_i))$. Thus, for \eqref{eqn:vertexrulegen} we obtain
\begin{equation}
  \sum\limits_{i=1}^3 \oint_{\mathcal{C}_i} \frac{d z_i}{2\pi i}\, \langle {\cal
    V}_3| \varepsilon_i(z_i)\, j_a(z_i) = 0 \,.
\end{equation}
The functions $z=f_i(z_i)$ map the local coordinates around the punctures at $z_{0\,i}=\{\infty, 0, 1\}$ to a common, uniformizing coordinate system $z$.

\begin{figure}[t]
  \centering
  \begin{tikzpicture}
    \begin{scope}[x=2.05cm,y=2.05cm]
      \node[at={(0.5,0)},anchor=center]{\includegraphics{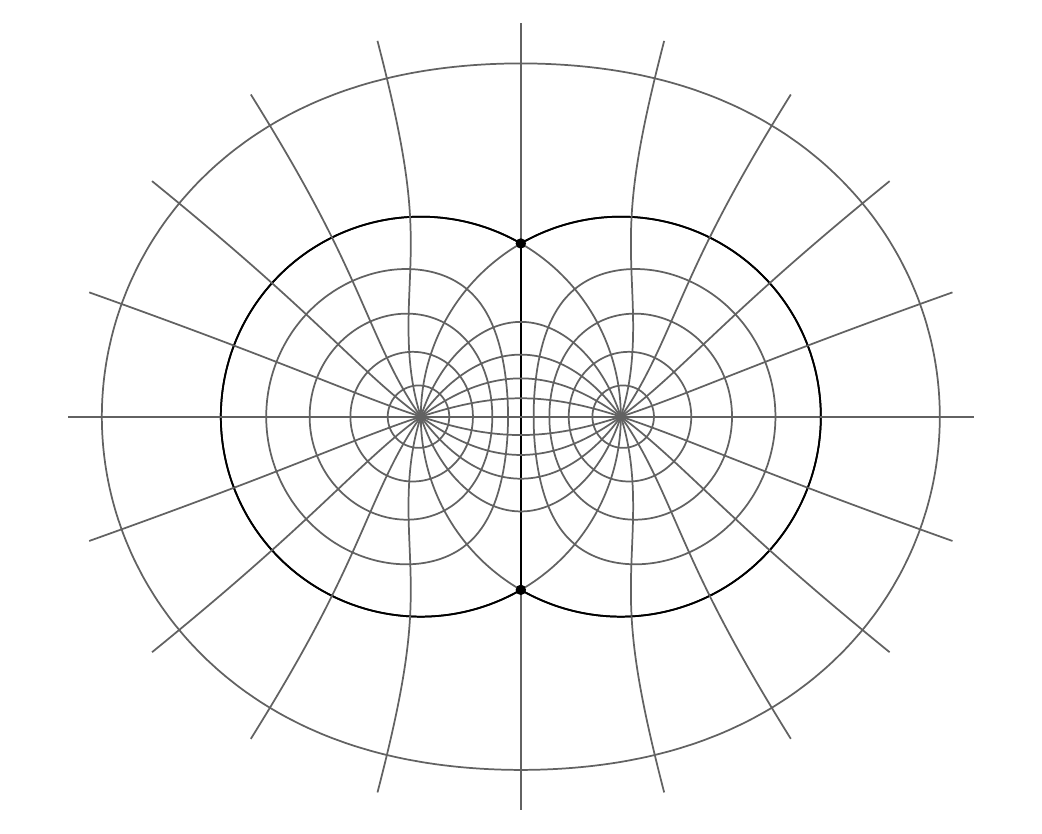}};
      \draw[->] (-2, 0) -- (3 , 0);
      \draw[->] (0, -2) -- (0  ,2);
      \draw (1,0.2em) -- (1,-0.2em);
      \node[at={(1,0)},anchor=north,name=z03] {$z_{0\,3}$};
      \node[at={(0,0)},anchor=north west,name=z02] {$z_{0\,2}$};
      \draw (2.2,1.7) -- ++(0, -1em) node[anchor=south west] {$z$} -- ++(1em,0);
      \node[at={(0.5,+0.866025)},point,name=pt1] {};
      \node[at={(0.5,-0.866025)},point] {};
    \end{scope}
    \begin{scope}[shift={(-4.5,3)},scale=1.5]
      \draw (0,0) circle (1);
      \draw[plotgray] (0,0) circle (0.8);
      \foreach \deg in {0,18,...,360} {\draw[plotgray] (\deg:0.6111111) -- (\deg:1);};
      \node[at={(0,1)},point,name=pt1z1] {};
      \node[at={(0,-1)},point] {};
      \draw[->] (-1.2,0) -- (1.2,0);
      \draw[->] (0,-1.2,0) -- (0,1.2);
      \draw (-1,1.2) -- ++(0, -1em) node[anchor=south east] {$z_1$} -- ++(-1em,0);
    \end{scope}
    \begin{scope}[shift={(-7.5, 0)},scale=1.5]
      \draw (0,0) circle (1);
      \foreach \r in {0.2, 0.4, 0.6, 0.8} {\draw[plotgray] (0,0) circle (\r);};
      \foreach \deg in {0,18,...,360} {\draw[plotgray] (0,0) -- (\deg:1);};
      \node[at={(0,1)},point] {};
      \node[at={(0,-1)},point] {};
      \draw[->] (-1.2,0) -- (1.2,0);
      \draw[->] (0,-1.2,0) -- (0,1.2);
      \draw (-1,1.2) -- ++(0, -1em) node[anchor=south east] {$z_2$} -- ++(-1em,0);
    \end{scope}
    \begin{scope}[shift={(-4.5,-3)},scale=1.5]
      \draw (0,0) circle (1);
      \foreach \r in {0.2, 0.4, 0.6, 0.8} {\draw[plotgray] (0,0) circle (\r);};
      \foreach \deg in {0,18,...,360} {\draw[plotgray] (0,0) -- (\deg:1);};
      \node[at={(0,1)},point] {};
      \node[at={(0,-1)},point] {};
      \draw[->] (-1.2,0) -- (1.2,0);
      \draw[->] (0,-1.2,0) -- (0,1.2);
      \draw (-1,1.2) -- ++(0, -1em) node[anchor=south east] {$z_3$} -- ++(-1em,0);
    \end{scope}
    \draw[->] (-7.5,0) to[bend left=20] node[midway,anchor=south] {$f_2$} (z02.north west);
    \draw[->] (-4.5,-3) -- (z03.north) node[pos=0.35,anchor=south] {$f_3$};
    \draw[->] (pt1z1.center) to[bend left=20] node[pos=0.35,anchor=south] {$f_1$} (pt1.center);
  \end{tikzpicture}
  \caption{Relation between the local coordinates $z_i$ and the uniformizing coordinate $z$. The small black dots mark the zeros of the quadratic differential $\phi(z)$.}\label{fig:uniformizingcoord}
\end{figure}
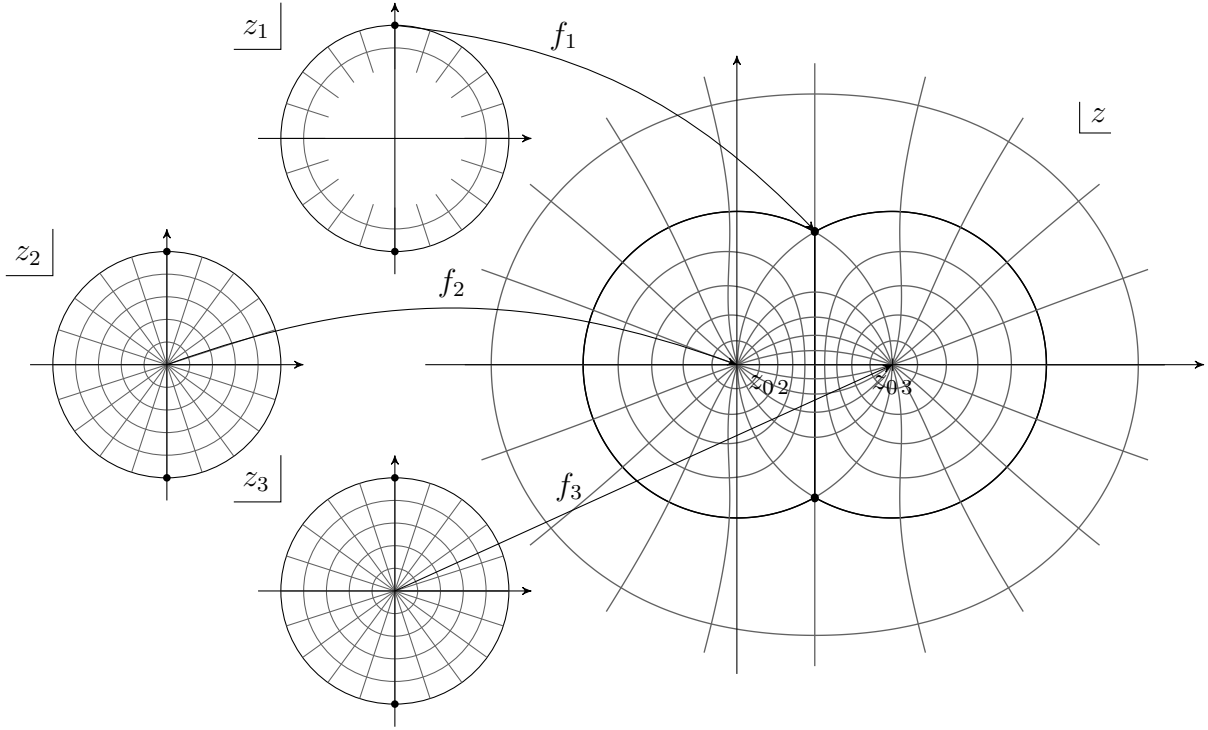

In doing so, they decompose the three-punctured sphere into three disc domains like depicted in figure~\ref{fig:uniformizingcoord}. This decomposition is governed by the quadratic differential \cite{Witten:1985cc,Sonoda:1989wa}
\begin{equation}
  \varphi(z) = \phi(z) (d z)^2 \quad \text{with} \quad 
    \phi(z) = - \frac{1}{(z-1)^2} - \frac{1}{z^2} + \frac{1}{z (z+1)}\,.
\end{equation}
Local coordinates around a puncture have to reproduce $\varphi(z)$ in the corresponding ring domain. To this end, the functions $f_i(z_i)$ have to fulfill
\begin{equation}\label{eqn:localcoords}
  \frac{d f_i}{d z_i} = \sqrt{\phi(z_i)}\,.
\end{equation}
By expanding the left and right hand side of this equation into a Laurent series around $z_{0\,2}=0$, it is 
straightforward to show that
\begin{equation}
  f_2(z_2) = \frac{(\sqrt{3} - i) \bigl[ (i + z)^{2/3} + (i - z)^{2/3} \bigr]}{%
    (\sqrt{3} + i) (i + z)^{2/3} + 2 i(i - z)^{2/3}}
\end{equation}
is a solution of \eqref{eqn:localcoords}. A Taylor expansion of $f_2(z_2)$ around $z_{0\,2}$ gives rise to
\begin{equation}
  f_2(z_2) = -\frac{4}{3\sqrt{3}} z_2 - \frac{8}{27} z_2^2 + \frac{4}{81\sqrt{3}} z_2^3 + \frac{16}{243} z_2^4 - \frac{52}{2187\sqrt{3}} z_2^5 + \cdots
\end{equation}
from which we read of the coefficients 
\begin{equation}
  \rho = -\frac{4}{3\sqrt{3}}\,, \quad
  d_1 = -\frac{1}{2} \,, \quad
  d_2 = -\frac{1}{16} \,, \quad
  d_3 = \frac{3}{16} \,, \quad
  d_4 = \frac{13}{256} \,, \quad \dots
\end{equation}
in the general expansion
\begin{equation}
  f_2(z_2) = z_{0\,2} + \rho z_2 + d_1 (\rho z_2)^2 + d_2 (\rho z_2)^3 + d_2 (\rho z_2)^4 + \dots \,.
\end{equation}
The local coordinates for the remaining punctures arise from the Möbius transformations
\begin{equation}
  z \rightarrow \frac{1}{1 - z} \quad \text{and} \quad
  z \rightarrow 1 - \frac{1}{z}
\end{equation}
which permute the punctures of the sphere.

Choosing $\varepsilon(z) = \rho z^{-1}$ and utilizing the mode expansion of the chiral current $j_a(z_i)$ in \eqref{eqn:modeexpja}, we obtain the Ward identity
\begin{equation}\label{eqn:wardj-1}
  0 = \langle {\cal V}_3| \bigl( \rho \, j_{a,0}^{(1)} - \rho^2 \,j_{a,1}^{(1)} +j_{a,-1}^{(2)}  - \rho d_1\, j_{a, 0}^{(2)} +  \rho^2 d_1^2 \, j_{a, 1}^{(2)}
    - \rho^2\, j_{a,1}^{(3)}  + \dots \bigr)\,.
\end{equation}
A similar argument holds for the $c$-ghosts, which are Virasoro primaries of conformal weight $-1$. Thus, the main difference is the transformation behavior of
\begin{align}
  dz\, \phi(z)\, c(z) & = d z_i \frac{d z}{d z_i} \left(\frac{d z_i}{d z}\right)^{-1} \phi(f(z_i))\, c(z_i) =
    d z_i\, \phi_i(z_i)\, c(z_i) 
\end{align}
with $\phi_i(z_i) = ( f'(z_i) )^{-2} \, \phi( f (z_i) )$. Again, for the specific choices
\begin{equation}
  \phi(z) = \frac{1}{(1-z) z^2} \quad \text{and} \quad
  \phi(z) = \frac{(z-2)\rho}{2(z-1)z^3}\,
\end{equation}
the two Ward identities
\begin{align}
  0 &= \langle {\cal V}_3| \bigl( \rho c_1^{(1)}+ c_0^{(2)} + \rho( 1 + 2 d_1)\, c_1^{(2)} - \rho\, c_1^{(3)} \bigr) 
    \label{eqn:wardc0} \\[0.2cm]
  0 &= \big\langle {\cal V}_3\big| \big(  -\frac{\rho^2}{2}\, c_1^{(1)} + c_{-1}^{(2)} +
  \frac{\rho}{2} (1 + 2 d_1)\, c_0^{(2)} + \frac{\rho^2}{2}( 1 + 2 d_1 - 4 d_1^2
  + 6 d_2 )\, c_{1}^{(2)} -\frac{\rho^2}{2}\, c_1^{(3)}  \big)
    \label{eqn:wardc-1}
\end{align}
follow. For bared operators, analogous Ward identities hold.

Equipped with these Ward identities, we can now proceed and compute the string function \eqref{stringfuncthree}. Like for the quadratic term, we again use the bilinearity of the string function and obtain $5^3=125$ different terms to calculate. Considering their symmetries, it is sufficient to calculate only $35$ different terms and weight them with the corresponding combinatoric prefactors.

To evaluate each of these 35 remaining string functions, we apply the following algorithm: First we use one of the Ward identities \eqref{eqn:wardj-1}, \eqref{eqn:wardc0} or \eqref{eqn:wardc-1} to remove the corresponding operator from the second slot of $\langle {\cal V}_3|$. Afterwards we establish normal ordering of all slots and remove terms where annihilation operators hit the primaries. We repeat this procedure until slot two of $\langle {\cal V}_3|$ contains the operators $c_1$, $\bar c_1$ and $j_{a, 0}$ only. Now, we rotate the vertex according to the rule
\begin{equation}
  \langle {\cal V}_3| V_1\rangle_1\, V_2\rangle_2\, V_3\rangle_3 = (-)^{V_1(V_2 +
    V_3)} \langle {\cal V}_3| V_2 \rangle_1\, V_3 \rangle_2\, V_1 \rangle_3
\end{equation}
and start over again by applying the Ward identities and normal ordering. Then we rotate again and we continue until all slots of $\langle {\cal V}_3|$ contain $c_1$, $\bar c_1$, $j_{a, 0}$ and $\bar j_{\bar a, 0}$ only. Finally, we apply the ghost overlap \eqref{eqn:ghostoverlap} giving rise to the substitution rule
\begin{equation}
  \langle {\cal V}_3 | c_1^{(1)} \bar c_1^{(1)} c_1^{(2)} \bar c_1^{(2)}
  c_1^{(3)} \bar c_1^{(3)} = \frac{2}{|\rho|^6} \langle  {\cal V}_3|
\end{equation}
where the $|\rho|^6$ term in the denominator arises because we have 6 ghosts with conformal weight $-1$. It is canceled completely by the $|\rho|^6$ due to the successive application of the Ward identities. After all these steps, only the fundamental three-point off-shell amplitudes \eqref{eqn:basicV3} are left over. Writing them in terms of an integral over the doubled space, we have to take care of the $|\rho|^{2 h_i}$ factors in \eqref{eqn:basicV3}. However, they can be expressed as
\begin{equation}
  |\rho|^{2 h_R} = |\rho|^{-\frac{\alpha'}{2 (k + h^\vee)} \square} = 1 - \frac{\alpha'}{2} \ln |\rho| \square + \dots  = 1 + \mathcal{O}(k^{-1})\,
\end{equation}
and therefore at leading order do not give any contribution to the action\footnote{Even though the algorithm presented here is straightforward, the calculations are lengthy and cumbersome. For that purpose we developed a Mathematica package that was inspired to some extent by Lambda \cite{Ekstrand:2010bp}, a package to evaluate operator product expansions in vertex algebras. It also extensively uses MathGR \cite{Wang:2013mea} to simplify tensor expressions.}. Finally, at leading order $O(k^{-1})$, the cubic part of the action can be
expressed as
\begin{align}
  (2\kappa^2) S^{(3,-1)} &= \int d^{2n} X \sqrt{|H|} \, \biggl[ -\frac{1}{8} \epsilon_{a\bar b} \Bigl( 
    - D_c \epsilon^{c \bar b}\, D_{\bar d} \epsilon^{a\bar d} - D_c
    \epsilon^{c \bar d} \, D_{\bar d} 
    \epsilon^{a\bar b} -  2 D^a \epsilon_{c \bar d}\, D^{\bar b} \epsilon^{c\bar d} \nonumber\\
  & \hspace{11em} + 2 D^a \epsilon_{c\bar d} \, D^{\bar d} \epsilon^{c\bar b} + 
    2 D^c \epsilon^{ad} \, D^{\bar b} \epsilon_{c \bar d} \Bigr) \nonumber\\
  & \quad - \frac{1}{4} \epsilon_{a\bar b} \Bigl( 
    F^a{}_{cd}\, \epsilon^{c\bar e}\,  D_{\bar e} \epsilon^{d \bar b} + 
    F^{\bar b}{}_{\bar c \bar d}\, \epsilon^{e \bar c}\, D_e \epsilon^{a \bar d}\Bigr) 
    -\frac{1}{12} F_{ace} \, F_{\bar d\bar b\bar f} \, \epsilon^{a\bar b}\, \epsilon^{c\bar d}\, 
    \epsilon^{e\bar f} \label{eqn:newcubic} \\
  & \quad + \frac{1}{2} \epsilon_{a \bar b}\, f^a f^{\bar b}
    -\frac{1}{2} f_a f^a \, \bar e +  \frac{1}{2} f_{\bar a} f^{\bar a}\, e \nonumber\\
  & \quad -\frac{1}{8} \epsilon_{a\bar b}\, \Bigl( D^a D^{\bar b} e\; \bar e -
    D^a e \, D^{\bar b} \bar e
    -D^{\bar b} e\,  D^a \bar e + e\, D^a D^{\bar b} \bar e \Bigr) \nonumber \\
  & \quad -\frac{1}{4} f^a \Bigl( 2 \epsilon_{a\bar b}\, D^{\bar b} \bar e + 
    D^{\bar b} \epsilon_{a\bar b}\; \bar e \Bigr)
    +\frac{1}{4} f^a \Bigl( D_a e \; \bar e - e\,  D_a \bar e \Bigr) \nonumber\\
  & \quad -\frac{1}{4} f^{\bar b} \Bigl( 2 \epsilon_{a \bar b}\,  D^a e + D^a \epsilon_{a\bar b}\; e\Bigr)
    +\frac{1}{4} f^{\bar b} \Bigl( D_{\bar b} e\; \bar e - e\, D_{\bar b} \bar e \Bigr) \biggr]\,.
\end{align}
Like already observed for the quadratic action \eqref{eqn:S2-1}, large parts of it resemble the original result obtained by Hull and Zwiebach. However, there are also additional terms \eqref{eqn:newcubic}, linear and quadratic in the structure coefficients $F_{abc}$. On the abelian torus they vanish and then the action \eqref{eqn:newcubic} reduces to the one derived in \cite{Hull:2009mi}. Whereas in toroidal DFT, there are kinetic terms in the action only, one of the additional terms in \eqref{eqn:newcubic} represents a potential
\begin{equation}
  V = -\frac{1}{12}\, F_{ace}\, F_{\bar b\bar d\bar f}\, \epsilon^{a\bar b}\, \epsilon^{c\bar d}\, \epsilon^{e\bar f} 
\end{equation}
for the fluctuations $\epsilon_{a\bar b}$.

In order to evaluate the gauge transformation in cubic order, we again use the conjugate string field $\phi_s^c$ from section~\ref{sec:freetheory}. It allows to express the string product
\begin{equation}
  [\Psi, \Lambda]_0 = \sum\limits_s |\phi_s\rangle \{\phi_s^c, \Psi, \Lambda\}_0
\end{equation}
in terms of string functions, which we compute like those appearing in the action. One finally obtains for the gauge variations of the fluctuations
\begin{align}\label{eqn:deltalambdaepsilon}
  \delta_\lambda \epsilon_{a \bar b} &= - \frac{1}{4} \Bigl( 
    \lambda^c\, D_a \epsilon_{c\bar b} - D_a \lambda^c\; \epsilon_{c\bar b} + \lambda_a\, D^c \epsilon_{c \bar b}
    + 2 D^c \lambda_a\; \epsilon_{c\bar b} - \lambda_c\, D^c \epsilon_{a\bar b} 
    - 2 \lambda_c\, D^c \epsilon_{a\bar b} \Bigr) \nonumber \\
  & \quad - \frac{1}{4}\Bigl( \lambda_a\, D_{\bar b} \bar e - D_{\bar b} \lambda_a\; \bar e \Bigr)
    + \frac{1}{2}\lambda_a\, f_{\bar b} + \frac{1}{2} F_{ac}{}^d\, \lambda^c\, \epsilon_{d \bar b} \\
  \delta_\lambda e &= -\frac{1}{4} f^a\, \lambda_a + \frac{1}{8} e\, D^a \lambda_a + \frac{1}{4} \lambda_a\, D^a e \\
  \delta_\lambda \bar e &= \frac{1}{16} \bar e\, D^a \lambda_a + \frac{1}{8} \lambda_a\, D^a \bar e\,. 
\end{align}
The corresponding ones for $\lambda_{\bar a}$ arise after applying the $\mathds{Z}_2$ symmetry \eqref{eqn:Z2sym}. Here, we are not interested in the gauge transformations of the auxiliary fields $f_a$ and $f_{\bar a}$, because they are eliminated by their equations of motion in the next subsection anyway. A $\mu$-type gauge transformation acts as
\begin{equation}\label{eqn:deltamu}
  \delta_\mu \epsilon_{a\bar b} = 0\,, \quad
  \delta_\mu e = - \frac{3}{8} \mu e
    \quad \text{and} \quad
  \delta_\mu \bar e = \frac{3}{8} \mu \bar e \,.
\end{equation}

\subsection{Simplifying action and gauge transformations}\label{sec:simpactiongauge}
Following \cite{Hull:2009mi,Hohm:2014xsa}, we simplify the action by first fixing the $\mu$ gauge in such a way that
\begin{equation}
  e = d \quad \text{and} \quad \bar e = -d \,.
\end{equation}
Afterwards, we redefine the fields
\begin{equation}\label{eqn:fieldredefepsilond}
  \epsilon_{a\bar b}' = \epsilon_{a\bar b} + \epsilon_{a\bar b}\, d\,, \quad
  d' = d + \frac{1}{32} \epsilon_{a\bar b}\, \epsilon^{a\bar b} \,
\end{equation}
and the gauge parameter
\begin{equation}\label{eqn:fieldredefgaugepara}
  \lambda_a' = \lambda_a + \frac{3}{4}\lambda_a\, d - \frac{1}{4}
  \lambda^{\bar b} \, \epsilon_{a \bar b}\,. 
\end{equation}
Let us briefly discuss how the level matching condition works for these redefined fields. We know that the unprimed fields in \eqref{eqn:fieldredefepsilond} have to satisfy the weak constraint \eqref{eqn:weakconstflat}. Since the primed ones contain products of unprimed fields, they do not automatically satisfy it. However, requiring also the strong constraint \eqref{eqn:strongconst} guarantees that the primed fields still do it. Therefore, already at the level of this field redefinition the strong constraint is necessary.

Now, plugging the redefined quantities into the quadratic and cubic gauge transformations and removing all contributions that are not linear in the parameter $\lambda$ or the fields, we obtain
\begin{align}
  \delta_{\lambda} \epsilon_{a\bar b} =& D_{\bar b} \lambda_a  + \frac{1}{2}\Bigl( 
    D_a \lambda^c\, \epsilon_{c\bar b} - D^c \lambda_a\; \epsilon_{c\bar b} + \lambda_c \,D^c \epsilon_{a \bar b}
    + F_{ac}{}^d\, \lambda^c\, \epsilon_{d\bar b} \Bigr) + \nonumber \\ 
  &  D_a \lambda_{\bar b} + \frac{1}{2}\Bigl( D_{\bar b} \lambda^{\bar c} \,\epsilon_{a\bar c} - 
    D^{\bar c} \lambda_{\bar b}\; \epsilon_{a \bar c} + \lambda_{\bar c} \,D^{\bar c} \epsilon_{a \bar b}
    + F_{\bar b\bar c}{}^{\bar d}\, \lambda^{\bar c}\, \epsilon_{a \bar d}
    \Bigr) \nonumber \\
    \delta_\lambda d =& -\frac{1}{4} D_a \lambda^a + \frac{1}{2} \lambda_a\, D^a d - 
    \frac{1}{4} D_{\bar a} \lambda^{\bar a} + \frac{1}{2} \lambda_{\bar a}\,
    D^{\bar a} d \,. \label{eqn:gaugetrafo}
\end{align}
For simplicity of the notation, we dropped the prime. Except for the flux term, they have the same form as the gauge transformations of toroidal DFT. 

As already mentioned above, it is convenient to simplify the action by eliminating the auxiliary fields $f_a$ and $f_{\bar a}$. To this end, we solve their equations of motion up to quadratic order in the remaining fields, yielding
\begin{align}
  f^a & = -\frac{1}{2} D_{\bar b} \epsilon^{a \bar b} - D^a d
    +\frac{1}{2}\Bigl( \epsilon^{a \bar b} \, D_{\bar b} d + d\, D^a d\Bigr) 
    +\frac{1}{8}\Bigl(D^c \epsilon_{c\bar b}\; \epsilon^{a\bar b} - d\, D_{\bar b} \epsilon^{a \bar b} \Bigr) \\
  f^{\bar b} &=  \frac{1}{2} D_{a} \epsilon^{a \bar b} + D^{\bar b} d
    -\frac{1}{2}\Bigl( \epsilon^{a \bar b}\, D_{a} d + d\, D^{\bar b} d\Bigr) 
    -\frac{1}{8}\Bigl(D^c \epsilon_{a \bar c}\; \epsilon^{a\bar b} - d\, D_{a} \epsilon^{a \bar b} \Bigr)\,.
\end{align}
Furthermore, we apply the field redefinitions \eqref{eqn:fieldredefepsilond}, which we already used to simplify the gauge transformations. Finally, we obtain
\begin{align}
  (2 \kappa^2)&S = \int d^{2n} X\sqrt{|H|}\, \biggl[ 
      \frac{1}{4} \epsilon_{a\bar b} \,\square \epsilon^{a\bar b} + \frac{1}{4} (D^{\bar b} \epsilon_{a\bar b})^2
      + \frac{1}{4} (D^a \epsilon_{a \bar b})^2 - 2 d\, D^a D^{\bar b} \epsilon_{a\bar b} 
      - 4 d\, \square d \nonumber \\[0.1cm]
    & + \frac{1}{4} \epsilon_{a\bar b} \Bigl( D^a \epsilon_{c\bar d}\, D^{\bar b} \epsilon^{c\bar d} - 
      D^a \epsilon_{c \bar d}\, D^{\bar d} \epsilon^{c\bar b} - D^c \epsilon^{a\bar d}\, D^{\bar b} 
      \epsilon_{c\bar d} \Bigr) \nonumber \\[0.1cm]
    & - \frac{1}{4} \epsilon_{a\bar b} \Bigl( F^{ac}{}_d\, D^{\bar e} \epsilon^{d\bar b} \;\epsilon_{c\bar e} 
      + F^{\bar b\bar c}{}_{\bar d}\, D^e \epsilon^{a\bar d}\; \epsilon_{e\bar c} \Bigr)
      -\frac{1}{12} F^{ace}\, F^{\bar b\bar d\bar f}\, \epsilon_{a\bar b}\, \epsilon_{c\bar d}\,
      \epsilon_{e\bar f} \nonumber \\[0.1cm]
    & +\frac{1}{2} d \Bigl( (D^a \epsilon_{a\bar b})^2 + (D^{\bar b} \epsilon_{a\bar b})^2 + 
      \frac{1}{2} (D_c \epsilon_{a\bar b})^2 + \frac{1}{2} (D_{\bar c} \epsilon_{a\bar b})^2 +
      2 \epsilon^{a\bar b} ( D_a D^c \epsilon_{c\bar b} + D_{\bar b} D^{\bar c} \epsilon_{a\bar c} )
      \Bigr) \nonumber \\[0.1cm]
    & +4 \epsilon_{a\bar b}\, d\, D^a D^{\bar b} d + 4 d^2\, \square d \biggr]   \label{eqn:action}
\end{align}
where we defined
\begin{equation}
  (D^{\bar b} \epsilon_{a\bar b})^2=
   (D^{\bar b} \epsilon_{a\bar b})(D_{\bar c} \epsilon^{a\bar c})\, .
\end{equation}
Thus, we have derived the leading order form of the DFT$_{\rm WZW}$ action, which reduces to the usual DFT action for a flat torus. Still, it contains additional terms which go beyond the traditional formulation. First, the derivatives $D_a$ are non-commuting and, second, the fluxes $F_{abc}$ appear explicitly.

\subsection{C-bracket}
Let us finally analyze the gauge algebra of the theory, which arises from the gauge transformations \eqref{eqn:gaugetrafo}. In CSFT, at cubic order the commutator of two gauge transformations $\delta_{\Lambda_1}$ and $\delta_{\Lambda_2}$ gives another one parameterized by
\begin{equation}
  \Lambda_{12} = [\Lambda_2, \Lambda_1]_0 \,.
\end{equation}
Using the techniques presented in section~\ref{sec:cubicorder}, it is straightforward to evaluate this expression and one obtains
\begin{align}
  \lambda_{12\,a} &= -\frac{1}{2} \lambda_1^b\, D_b \lambda_{2,a}
    + \frac{1}{4} \big( \lambda_{1,b} \, D_a \lambda_2^b + \lambda_{1, a}\, D_b \lambda_2^b
    - \lambda_1^{\bar b}\, D_{\bar b} \lambda_{2,a} + \lambda_{2,a} \,\mu_1 + 
    F_{abc} \,\lambda_1^b\, \lambda_2^c \big) \nonumber \\
    & \quad - \frac{1}{8} \lambda_{2, a}\, D_{\bar b} \lambda_1^{\bar b} \, -\, (1 \leftrightarrow 2)\,.
\end{align}
Due to the $\mathds{Z}_2$ symmetry \eqref{eqn:Z2sym}, the equation for the $\lambda_{12\,\bar a}$ has exactly the same form. Note that these commutators hold before the field redefinition of the gauge parameter \eqref{eqn:fieldredefgaugepara} is applied. As explained in section~3.1 of \cite{Hohm:2014xsa} after the field redefinition we have to adapt $\lambda_{12\,a}$ according to
\begin{equation}
  \lambda'_{12,a} = \lambda_{12,a} + \Bigl[ 
    \frac{1}{4}\big( D_{\bar b} \lambda_{1,a}\; \lambda_2^{\bar b}
      + D_a \lambda_{1,\bar b}\, \lambda_2^{\bar b} \big)
    + \frac{3}{16}\big( D_{\bar b} \lambda_1^{\bar b}\; \lambda_{2,a} + D_b \lambda_1^b\; \lambda_{2,a} \big)
    - (1 \leftrightarrow 2) \Bigr]\, .
\end{equation}
In addition, we have to set
\begin{equation}
  \mu = \frac{1}{4} D_a \lambda^a - \frac{1}{4} D_{\bar a} \lambda^{\bar a}\,
\end{equation}
which takes into account the $\mu$ gauge fixing performed in the last subsection. After removing all terms which are not linear in $\lambda_1$, $\lambda_2$ or in both, we obtain the result
\begin{equation}
  \label{eqn:Cbracketlambda12}
  \lambda'_{12,a} = -\frac{1}{2} \big(\lambda_1^b D_b + \lambda_1^{\bar b}\, D_{\bar b}\big) \lambda_{2,a}
    + \frac{1}{4} \big( \lambda_{1,b}\,  D_a \lambda_2^b - \lambda_{1,\bar
      b}\,  D_a \lambda_2^{\bar b} 
    - F_{abc}\, \lambda_1^b\, \lambda_2^c\big) - (1 \leftrightarrow 2)\,.
\end{equation}
For the bared parameter, we obtain by the same procedure
\begin{equation}
  \label{eqn:Cbracketlambda12bar}
  \lambda'_{12,\bar a} = -\frac{1}{2} \big(\lambda_1^b D_b +
  \lambda_1^{\bar b}\,  D_{\bar b} \big) \lambda_{2,\bar a}
    - \frac{1}{4} \big( \lambda_{1,b}\, D_{\bar a} \lambda_2^b - \lambda_{1,\bar b}\, D_{\bar a} \lambda_2^{\bar b} 
    - F_{\bar a\bar b\bar c}\, \lambda_1^{\bar b}\, \lambda_2^{\bar c}\big) - (1 \leftrightarrow 2)\,.
\end{equation}
At linear order, $\lambda$ is equivalent to $\lambda'$ and therefore $\lambda$ can be substituted by $\lambda'$ on the right hand side of these two equations.

\section{\texorpdfstring{$S^3$}{Three-sphere} with \texorpdfstring{$H$}{H}-flux}\label{sec:examplesu(2)}
We close this chapter with a simple toy model, the group manifold $SU(2)$. It corresponds to a $S^3$ with $H$-flux and is part of the duality orbit 1 in table~\ref{tab:solembedding}. On this background, we compute all relevant quantities discussed throughout this chapter. We start with the generators
\begin{equation}
  t_a = \frac{1}{\sqrt{\alpha' k}} \sigma_a \quad \text{with} \quad a = 1, 2, 3
\end{equation}
in the fundamental representation. Here, $\sigma_a$ denotes the Pauli-matrices
\begin{equation}
  \sigma_1 = \begin{pmatrix} 0 & 1 \\ 1 & 0 \end{pmatrix}\,, \quad
  \sigma_2 = \begin{pmatrix} 0 & -i \\ i & 0\end{pmatrix}\,, \quad
  \sigma_3 = \begin{pmatrix} 1 & 0 \\ 0 & -1\end{pmatrix} \quad \text{and} \quad
  \sigma_0 = \begin{pmatrix} 1 & 0 \\ 0 & 1 \end{pmatrix}\,.
\end{equation}
The normalization of the generators is chosen in such a way that, according to \eqref{eqn:defetaab}, they give rise to the Killing metric
\begin{equation}
  \eta_{ab} = - \frac{\alpha' k}{2} \frac{\Tr(t_a t_b)}{2 x_f} = \diag (-1 , -1 , -1)
    \quad \text{with} \quad x_f = \frac{1}{2}
\end{equation}
denoting the Dynkin index of the fundamental representation. Each group element
\begin{equation}
  g =  y^0 t_0 - i y^a t_a
\end{equation}
is parameterized in terms of four coordinates $y^i$ which have to fulfill
\begin{equation}
  (y^1)^2 + (y^2)^2 + (y^3)^2 + (y^4)^2 = R^2 = k \alpha'\,.
\end{equation}
Doing so they describe the embedding of a three-sphere $S^3$ with radius $R = \sqrt{k \alpha'}$ into the four dimensional, euclidean space $\mathds{R}^4$. To parameterize the sphere, we choose Hopf coordinates $x^i = \begin{pmatrix} \eta^1 & \eta^2 & \eta^3 \end{pmatrix}$ with
\begin{align}
  y^0 &= \sqrt{\alpha' k} \cos \eta^2 \cos \eta^1 &
  y^1 &= \sqrt{\alpha' k} \sin \eta^2 \cos \eta^1 \\
  y^2 &= \sqrt{\alpha' k} \cos \eta^3 \sin \eta^1 &
  y^3 &= \sqrt{\alpha' k} \sin \eta^3 \sin \eta^1 \,.
\end{align}

\begin{table}[b]
  \centering
  \begin{tabular}{l|cccc}
    $y_{\lambda q}$ & $\lambda = 0$ & $\lambda = 1/\sqrt{2}$ & $\lambda=\sqrt{2}$ & $\cdots$ \\
    \hline\hline
    $\vdots$ & & & &  $\iddots$ \\
    $q=\sqrt{2}$ & & & 
      $\displaystyle \frac{\sqrt{3} e^{i 2 \eta^3} \sin^2 \eta^1}{\sqrt{2} \pi (\alpha'k)^{3/4}}$ & $\cdots$ \\
    $q=\displaystyle \frac{1}{\sqrt{2}}$ & &
      $ \displaystyle \frac{ e^{i \eta^3} \sin \eta^1}{\pi (\alpha'k)^{3/4}}$ &
      -- & $\cdots$ \\
      $q=0$ & 0 & -- & $\displaystyle - \frac{ \sqrt{3} e^{i (\eta^3 - \eta^2)} \cos \eta^1 \sin \eta^1}{\pi (\alpha'k)^{3/4}}$ & $\cdots$ \\
    $q=\displaystyle -\frac{1}{\sqrt{2}}$ & & 
      $-\displaystyle\frac{e^{-i\eta^2} \cos\eta^1}{\pi (\alpha'k)^{3/4}}$ & -- & $\cdots$ \\
    $q=-\sqrt{2}$ & & &
      $\displaystyle \frac{\sqrt{3} e^{-i 2 \eta^2} \cos^2 \eta^1}{\sqrt{2} \pi (\alpha' k)^{3/4}}$ & $\cdots$ \\
      $\vdots$ & & & & $\ddots$
  \end{tabular}
  \caption{Scalar functions $y_{\lambda q}$ which implement the SU($2$) highest weight representations for $\lambda\in\{0,\,1/\sqrt{2},\,\sqrt{2}\}$ on $S^3$. The general procedure to obtain these functions is outlined in section~\ref{sec:representation}.}\label{tab:ylambdaqsu(2)}
\end{table}
After this preparation, we apply \eqref{eqn:vielbein} and \eqref{eqn:vielbeinbared} to obtain the vielbeins
\begin{align}
  e^a{}_i &= -i \sqrt{k \alpha'}
    \begin{pmatrix}
      0 & \cos^2 \eta^1 & -\sin^2 \eta^1 \\
      \cos\eta^{23}_- & \sin\eta^1 \cos\eta^1 \sin\eta^{23}_- & 
      -\sin\eta^1 \cos\eta^1\sin\eta^{23}_-  \\
      -\sin\eta^{23}_- & \sin\eta^1 \cos\eta^1 \cos\eta^{23}_- &
      \sin\eta^1 \cos\eta^1 \cos\eta^{23}_-
    \end{pmatrix} 
  \quad \text{and} \\
  e^{\bar a}{}_{\bar i} &=  -i \sqrt{k \alpha'}
    \begin{pmatrix}
      0 & \cos^2 \eta^1 & \sin^2 \eta^1 \\
      \cos\eta^{23}_+ & \sin\eta^1 \cos\eta^1 \sin\eta^{23}_+ & 
      -\sin\eta^1 \cos\eta^1\sin\eta^{23}_+  \\
      \sin\eta^{23}_+ & - \sin\eta^1 \cos\eta^1 \cos\eta^{23}_+ &
      \sin\eta^1 \cos\eta^1 \cos\eta^{23}_+
    \end{pmatrix} 
\end{align}
with the abbreviation $\eta^{23}_\pm = \eta^2 \pm \eta^3$. They give rise to the structure coefficients \eqref{eqn:fabcfromvielbein}
\begin{equation}\label{eqn:SU2Fabc}
  F_{abc} = \frac{2 i}{\sqrt{\alpha' k}} \epsilon_{abc}
    \quad \text{and} \quad
  F_{\bar a\bar b\bar c} = \frac{2 i}{\sqrt{\alpha' k}} \epsilon_{abc}\,,
\end{equation}
where $\epsilon_{abc}$ denotes the totally antisymmetric tensor in three dimensions with $\epsilon_{123}=1$. As expected for a geometric background, they fulfill the relation $F_{abc} = - F_{\bar a\bar b\bar c}$. The target space metric obtained from the vielbein $e^a{}_i$ reads
\begin{equation}
  g_{ij} = \alpha' k \diag ( 1, \cos^2 \eta^1, \sin^2 \eta^1 )\,.
\end{equation}
By construction, it belongs to a $S^3$ with radius $R = \sqrt{\alpha' k}$ parameterized by the Hopf coordinates. With the structure coefficients \eqref{eqn:SU2Fabc}, \eqref{eqn:WZWHflux} and \eqref{eqn:WZWHfluxexplicit}, we calculate the 3-form
\begin{equation}
  H = 2\alpha' k\, \sin \eta^1 \cos \eta^1 d\eta^1 \wedge d\eta^2 \wedge d\eta^3 \,.
\end{equation}
As a consistency check we evaluation the quantization condition
\begin{equation}
  \frac{1}{2\pi \alpha'} \int_{S^3} H = \frac{k}{\pi} \int\limits_0^{2\pi} d\eta^2 
    \int\limits_0^{2\pi} d\eta^3 \int\limits_0^{\pi/2} d\eta^1\,\sin \eta^1 \cos\eta^1 =
      2 \pi k
\end{equation}
for the $H$-flux. It reproduces the quantization condition $k\in\mathds{N}$ for the level on compact group manifolds. Following the prescription presented in section~\ref{sec:representation}, one obtains the orthonormal functions $y_{\lambda q}$ listed in table~\ref{tab:ylambdaqsu(2)}. They form an orthonormal basis for the primaries of the WZW model.

\chapter{Generalized metric formulation}\label{chap:genmetric}
The traditional formulation of DFT we presented in chapter~\ref{chap:DFTreview} relies on doubled objects like e.g. the generalized metric $\mathcal{H}^{IJ}$. These doubled quantities can be understood as a very effective tool to organize the fields and gauge parameters appearing in the CSFT calculations. They are used to rewrite the action and its gauge transformations in a compact form. Further, they make the underlying structure of the theory manifest and allow to extend the cubic CSFT results to all orders in the fields. So far, it is not clear what the corresponding doubled objects in DFT${}_\mathrm{WZW}$ are. Thus we dedicate this chapter to find them and to rewrite the action and gauge transformations derived in the last chapter. The result, which is derived step by step through the sections~\ref{sec:notation}--\ref{sec:actiongenmetric}, is a generalized metric formulation of our theory \cite{Blumenhagen:2015zma}. We present its equations of motion and discuss its symmetries. Besides the invariance under generalized diffeomorphisms, DFT${}_\mathrm{WZW}$ possesses a $2D$-diffeomorphism invariance which is absent in the traditional version. Further, the closure of the gauge algebra is checked in section~\ref{sec:cbracket}. It depends on two constraints: Whereas the fluctuations have to fulfill the modified strong constraint discussed in the last chapter, the background structure coefficients have to fulfill a Jacobi identity only. At last, we show the equivalence of our theory and original DFT under an additional constraint, the extended strong constraint.

\section{Notation and convention}\label{sec:notation}
Instead of directly using the action \eqref{eqn:action}, its gauge transformations \eqref{eqn:gaugetrafo}, and the C-bracket \eqref{eqn:Cbracketlambda12}/\eqref{eqn:Cbracketlambda12bar} derived in the last chapter, it is convenient to perform the field redefinition
\begin{equation}\label{eqn:fieldredef}
  \epsilon^{a\bar b} \rightarrow -2 \epsilon^{a\bar b} \,, \quad
  \lambda^a \rightarrow 2 \lambda^a \quad \text{and} \quad
  \lambda^{\bar a} \rightarrow 2 \lambda^{\bar a} \,,
\end{equation}
which gives rise to
\begin{align}
    (2 \kappa^2)S &= \int d^{2n} X\sqrt{|H|}\, \Big[ 
      \epsilon_{a\bar b} \,\square \epsilon^{a\bar b} + (D^{\bar b} \epsilon_{a\bar b})^2
      +  (D^a \epsilon_{a \bar b})^2 + 4 \tilde d\, D^a D^{\bar b} \epsilon_{a\bar b} 
      - 4 \tilde d\, \square \tilde d \nonumber \\[0.1cm]
    & -2 \epsilon_{a\bar b} \bigl( D^a \epsilon_{c\bar d}\, D^{\bar b} \epsilon^{c\bar d} - 
      D^a \epsilon_{c \bar d}\, D^{\bar d} \epsilon^{c\bar b} - D^c \epsilon^{a\bar d}\, D^{\bar b} 
      \epsilon_{c\bar d} \bigr) \nonumber \\[0.1cm]
    & +2 \epsilon_{a\bar b} \bigl( F^{ac}{}_d\, D^{\bar e} \epsilon^{d\bar b} \;\epsilon_{c\bar e} 
      + F^{\bar b\bar c}{}_{\bar d}\, D^e \epsilon^{a\bar d}\; \epsilon_{e\bar c} \bigr)
      + \frac{2}{3} F^{ace}\, F^{\bar b\bar d\bar f}\, \epsilon_{a\bar b}\, \epsilon_{c\bar d}\,
      \epsilon_{e\bar f} \nonumber \\[0.1cm]
    & -\tilde d \bigl( 2 (D^a \epsilon_{a\bar b})^2 + 2 (D^{\bar b} \epsilon_{a\bar b})^2 + 
      (D_c \epsilon_{a\bar b})^2 + (D_{\bar c} \epsilon_{a\bar b})^2 + 4 \epsilon^{a\bar b}
      ( D_a D^c \epsilon_{c\bar b} + D_{\bar b} D^{\bar c} \epsilon_{a\bar c} )
      \bigr) \nonumber \\[0.1cm]
    & -8 \epsilon_{a\bar b}\, \tilde d\, D^a D^{\bar b} \tilde d + 4 {\tilde d}^2\, \square \tilde d \Big]\,,
    \label{eqn:action2}
\end{align}
the corresponding gauge transformations
\begin{align}
  \delta_{\lambda} \epsilon^{a\bar b} =& - D^{\bar b} \lambda^a + 
    D^a \lambda_c \epsilon^{c\bar b} - D_c \lambda^a \epsilon^{c\bar b} + \lambda^c D_c \epsilon^{a \bar b}
    + F^a{}_{cd}\, \lambda^c \epsilon^{d\bar b}  + \nonumber \\ 
  & -D^a \lambda^{\bar b} + D^{\bar b} \lambda_{\bar c} \epsilon^{a\bar c} - 
    D_{\bar c} \lambda^{\bar b} \epsilon^{a \bar c} + \lambda^{\bar c} D_{\bar c} \epsilon^{a \bar b}
    + F^{\bar b}{}_{\bar c \bar d} \lambda^{\bar c}\, \epsilon^{a \bar d} \nonumber \\
    \delta_\lambda \tilde d =& -\frac{1}{2} D_a \lambda^a + \lambda_a\, D^a \tilde d - 
    \frac{1}{2} D_{\bar a} \lambda^{\bar a} + \lambda_{\bar a}\,
    D^{\bar a} {\tilde d} \label{eqn:gaugetrafo2}
\end{align}
and the C-bracket
\begin{align}
  \lambda_{12,a} &= -\lambda_1^b D_b \lambda_{2,a} - \lambda_1^{\bar b}\, D_{\bar b} \lambda_{2,a} 
    + \frac{1}{2} \big( \lambda_{1,b}\,  D_a \lambda_2^b - \lambda_{1,\bar
      b}\,  D_a \lambda_2^{\bar b} 
    - F_{abc}\, \lambda_1^b\, \lambda_2^c\big) - (1 \leftrightarrow 2) \nonumber \\
  \lambda_{12,\bar a} &= - \lambda_1^b D_b \lambda_{2,\bar a} - \lambda_1^{\bar b}\,  D_{\bar b} 
    \lambda_{2,\bar a} - \frac{1}{2} \big( \lambda_{1,b}\, D_{\bar a} \lambda_2^b -
    \lambda_{1,\bar b}\, D_{\bar a} \lambda_2^{\bar b} 
    - F_{\bar a\bar b\bar c}\, \lambda_1^{\bar b}\, \lambda_2^{\bar c}\big) - (1 \leftrightarrow 2)\,.
  \label{eqn:cbracket2}
\end{align}
Applying the field redefinition \eqref{eqn:fieldredef} helps to get rid of a $1/2$ factor which would otherwise arise between DFT and DFT${}_\mathrm{WZW}$ results. Further, we refine the notation of the generalized dilaton to allow a clear distinction between background fields and fluctuations: From now on, $\tilde d$ denotes fluctuations of the generalized dilaton
\begin{equation}\label{eqn:splitdilaton}
  d = \bar d + \tilde d = -\frac{1}{2} \log \sqrt{|H|} + \tilde d
\end{equation}
which combines the background field $\bar d$ and the fluctuations.

One of the main objectives in this chapter is to rewrite \eqref{eqn:action2}-\eqref{eqn:cbracket2} in terms of doubled objects like they are common in the generalized metric formulation of DFT. To this end, we now define successively the required doubled quantities. As mentioned in section~\ref{sec:fundamentalamp}, the left- and right-mover coordinates on the group manifold can be combined into
\begin{equation}
  X^I = \begin{pmatrix}x^i & x^{\bar i}\end{pmatrix}\,.
\end{equation}
Besides curved doubled indices like $I$, $J$, $K$, \ldots, there are also flat ones denotes by $A$, $B$, $C$, \ldots\,\footnote{Note that according to the notation introduced in section~\ref{sec:doublelorentz}, we should decorate these indices with a bar to stress their GL($D$)$\times$GL($D$) structure. However, to avoid overloading our notation we drop these bars.}. Both of them are connected by the generalized vielbein
\begin{equation}\label{eqn:EAI}
  E_A{}^I = \begin{pmatrix}
    e_a{}^i & 0 \\
    0 & e_{\bar a}{}^{\bar i}
  \end{pmatrix} \quad \text{and its inverse transposed} \quad
  E^A{}_I = \begin{pmatrix}
    e^a{}_i & 0 \\
    0 & e^{\bar a}{}_{\bar i}
  \end{pmatrix}\,.
\end{equation}
At this point it is important to note that in contrast to the flux formulation of DFT, this generalized vielbein is not restricted to be O($D,D$) valued. Flat indices are raised a lowered with the $\eta$ metric
\begin{equation}\label{eqn:etaAB}
  \eta_{AB} = 2 \begin{pmatrix}
    \eta_{ab} & 0 \\
    0 & -\eta_{\bar a\bar b}
  \end{pmatrix} \quad \text{and its inverse} \quad
  \eta^{AB} = \frac{1}{2} \begin{pmatrix}
    \eta^{ab} & 0 \\
    0 & -\eta^{\bar a\bar b}
  \end{pmatrix}\,.
\end{equation}
Their curved versions read
\begin{equation}
  \eta_{IJ} = E^A{}_I \eta_{AB} E^B{}_J = 2 \begin{pmatrix}
    g_{ij} & 0 \\
    0 & -g^{\bar i\bar j}
  \end{pmatrix} \quad \text{and} \quad
  \eta^{IJ} = E_A{}^I \eta^{AB} E_B{}^J = \frac{1}{2} \begin{pmatrix}
    g^{ij} & 0 \\
    0 & -g^{\bar i\bar j}
  \end{pmatrix}\,.
\end{equation}
Because the generalized vielbein $E_A{}^I$ is not O($D,D$) valued, $\eta$ in curved indices is not constant. This is one of the main differences between DFT${}_\mathrm{WZW}$ and DFT. It has far-reaching consequences for the whole structure of the theory. E.g., it is not possible anymore to pull $\eta_{IJ}$ in and out of the doubled partial derivative
\begin{equation}
  \partial_I = \begin{pmatrix} \partial_i  & \partial_{\bar i} \end{pmatrix}\,.
\end{equation}
As we are going to see in section~\ref{sec:covderiv}, this problem is fixed by introducing a covariant derivative. In addition to the partial derivative, one defines the flat derivative
\begin{equation}\label{eqn:flatderivdoubled}
  D_A = E_A{}^I \partial_I = \begin{pmatrix} D_a & D_{\bar a} \end{pmatrix}
\end{equation}
analogous to the flux formulation. Further, we define the doubled structure coefficients in terms of the commutator
\begin{equation}
  D_A = E_A{}^I \partial_I
    \quad \text{with} \quad
  [D_A, D_B] = F_{AB}{}^C D_C\,.
\end{equation}
With \eqref{eqn:flatderivdoubled}, \eqref{eqn:[Dbara,Dbarb]} and \eqref{eqn:[Da,Db]}, this relation gives rise to
\begin{equation}
\label{eqn:conventFABC}
  F_{AB}{}^C = \begin{cases}
    F_{ab}{}^c & \\
    F_{\bar a\bar b}{}^{\bar c} & \\
    0 & \text{otherwise\,.}
  \end{cases}
\end{equation}
We also define the doubled parameter of a gauge transformation as
\begin{equation}\label{eqn:lambda^A}
  \lambda^A = \begin{pmatrix} \lambda^a & \lambda^{\bar a} \end{pmatrix}\,.
\end{equation}
The objects $D_A$, $F_{AB}{}^C$ and $\xi^A$ are considered as fundamental, meaning their bared and unbared components do not receive additional minus signs or prefactors. From these quantities all others are derived by raising or lowering the doubled indices with the $\eta$ metric. A simple example is
\begin{equation}\label{eqn:lambda_A}
  \lambda_A = \lambda^B \eta_{BA} = \begin{pmatrix} 2 \lambda_a & -2 \lambda_{\bar a} \end{pmatrix}\,.
\end{equation}

Finally, one needs a doubled object holding the fluctuations $\epsilon^{a\bar b}$. For that purpose, let us consider first the symmetric tensor $\mathcal{H}^{AB}$ which leaves $\eta$ invariant
\begin{equation}\label{eqn:HABOdd}
  \mathcal{H}^{AC} \eta_{CD} \mathcal{H}^{DB} = \eta^{AB}\,.
\end{equation}
An example for such a tensor is 
\begin{equation}\label{eqn:SAB}
  S_{AB} = 2 \begin{pmatrix}
    \eta_{ab} & 0 \\
    0 & \eta_{\bar a\bar b}
  \end{pmatrix} \quad \text{and its inverse} \quad
  S^{AB} = \frac{1}{2} \begin{pmatrix}
    \eta^{ab} & 0 \\
    0 & \eta^{\bar a\bar b}
  \end{pmatrix}\,.
\end{equation}
A small, symmetric perturbation of it which is compatible with \eqref{eqn:HABOdd} is denoted as $\epsilon^{AB}$. It is straightforward to check that $\epsilon^{AB}$ has to satisfy the relation
\begin{equation}\label{eqn:O(D,D)gens}
  \epsilon^{AC} \eta_{CD} S^{DB} + S^{AC}\eta_{CD} \epsilon^{DB} +
  \mathcal{O}(\epsilon^2) = 0 \,.
\end{equation}
The most general, symmetric solution for this equation reads
\begin{equation}
  \epsilon^{AB} = \begin{pmatrix}
    0 & - \epsilon^{a\bar b} \\
    - \epsilon^{\bar a b} & 0
  \end{pmatrix} \quad \text{with} \quad
  \epsilon^{a \bar b} = (\epsilon^T)^{\bar b a}\,.
\end{equation}
Its $D^2$ different entries $\epsilon^{a\bar b}$ allow us to express any $\mathcal{H}^{AB}$ in terms of the series expansion
\begin{equation}\label{eqn:HAB}
  \mathcal{H}^{AB} = S^{AB} + \epsilon^{AB} + \frac{1}{2} \epsilon^{AC}\, S_{CD}\, \epsilon^{DB} + \dots = \exp ( \epsilon^{AB} )\,.
\end{equation}
This equation has exactly the same structure as \eqref{eqn:genmetricfluctuation}, which governs fluctuations of the generalized metric around a vacuum of a generalized Scherk-Schwarz compactification. Hence, we make the following identifications:
\begin{enumerate}
  \item We call $\mathcal{H}^{AB}$ generalized metric.
  \item We associate the symmetric, doubled tensor $\epsilon^{AB}$ to the small fluctuations $\epsilon^{a\bar b}$ which appear in the action \eqref{eqn:action} and its gauge transformations \eqref{eqn:gaugetrafo2}.
  \item We denote $S^{AB}$ as flat vacuum generalized metric.
\end{enumerate}
For later considerations, one needs to expand the generalized metric \eqref{eqn:HAB} into components
\begin{equation}\label{eqn:HABexpansion}
\mathcal{H}^{AB} = \begin{pmatrix}
  \frac{1}{2}\eta^{ab} + \epsilon^{a\bar{c}} \eta_{\bar{c}\bar{d}} \epsilon^{\bar{d}b} & \epsilon^{a\bar{b}} + \frac{2}{3} \epsilon^{a\bar{c}} \eta_{\bar{c}\bar{d}} \epsilon^{\bar{d}e} \eta_{ef} \epsilon^{f\bar{b}} \\ \epsilon^{\bar{a}b} + \frac{2}{3} \epsilon^{\bar{a}c} \eta_{cd} \epsilon^{d\bar{e}} \eta_{\bar{e}\bar{f}} \epsilon^{\bar{f}b} & \frac{1}{2} \eta^{\bar{a}\bar{b}} + \epsilon^{\bar{a}c} \eta_{cd} \epsilon^{d\bar{b}}
\end{pmatrix} + \mathcal{O}(\epsilon^4)
\end{equation}
up to cubic order in the fields. 

Finally $H_{IJ}$, whose determinant $H$ is used in \eqref{eqn:action}, is defined as the curved version
\begin{equation}
  H_{IJ}=E^A{}_I S_{AB} E^B{}_J = 2 \begin{pmatrix}
    g_{ij} & 0 \\
    0 & g_{\bar i\bar j}
  \end{pmatrix}
\end{equation}
of $S_{AB}$. In comparison to \eqref{eqn:metricH}, this quantity possesses an additional prefactor 2. To keep the action integral \eqref{eqn:action2} invariant, one has to perform the compensating change of variables $X^I \rightarrow X^I / \sqrt{2}$.

\section{Covariant derivative}\label{sec:covderiv}
As it is going to become obvious in the next sections, the covariant derivative
\begin{equation}\label{eqn:flatcovderv}
  \nabla_A V^B = D_A V^B + \frac{1}{3} F^B{}_{AC} V^C \quad \text{and} \quad
  \nabla_A V_B = D_A V_B + \frac{1}{3} F_{BA}{}^C V_C
\end{equation}
plays a central role in the generalized metric formulation of DFT${}_\mathrm{WZW}$. Here, we discuss its properties. Note that this covariant derivative is \underline{not} the O($D,D$) covariant derivative of DFT \cite{Hohm:2010xe,Hohm:2011si,Aldazabal:2013sca}. Still it fulfills similar compatibility conditions:
\begin{itemize}
  \item Compatibility with the frame requires
    \begin{equation}\label{eqn:compframe}
      \nabla_A E_B{}^I = 0\,.
    \end{equation}
    Here the covariant derivative acts on a tensor with both flat and curved indices. Thus, we have to extend its definition
    \begin{equation}
      \nabla_A E_B{}^I = D_A E_B{}^I + \frac{1}{3} F_{BA}{}^C E_C{}^I + E_A{}^K \Gamma_{KJ}{}^I E_B{}^J = 0
    \end{equation}
    by the curved connection $\Gamma_{IK}{}^J$. Due to \eqref{eqn:compframe}, it is completely determined
    \begin{equation}\label{eqn:GammaIJK}
      \Gamma_{IJ}{}^K = - E^A{}_I E^B{}_J E_C{}^K\frac{1}{3} \bigl( 2 \Omega_{AB}{}^C + \Omega_{BA}{}^C \bigr)
      = -\frac{1}{3} \bigl(2 \Omega_{IJ}{}^K + \Omega_{JI}{}^K \bigr)
    \end{equation}
    in terms of the coefficients of anholonomy $\Omega_{ABC} = D_A E_B{}^I E_{CI}$ and the vielbein $E_A{}^I$.
  \item Compatibility with the invariant metric
    \begin{equation}\label{eqn:compinvmetric}
      \nabla_A \eta_{BC} = D_A \eta_{BC} + F_{BA}{}^D \eta_{DC} + F_{CA}{}^D \eta_{BD} = F_{BAC} + F_{CAB} = 0
    \end{equation}
    is fulfilled due to the total antisymmetry of $F_{ABC}$, a direct consequence of the total antisymmetry of its components $f_{abc}$ and $f_{\bar a\bar b\bar c}$. Split into bared and unbared indices, the non-trivial contributions of \eqref{eqn:compinvmetric} read
    \begin{equation}\label{eqn:compinvmetriccomp}
      f_{bac} + f_{cab} = 0 \quad \text{and} \quad - f_{\bar b \bar a \bar c} - f_{\bar  c\bar a \bar b} = 0\,.
    \end{equation}
  \item Compatibility with the background metric
    \begin{equation}\label{eqn:compbgmetric}
      \nabla_A S_{BC} = D_A S_{BC} + F_{BA}{}^D S_{DC} + F_{CA}{}^D S_{BD} = 
        F_{BA}{}^D S_{DC} + F_{CA}{}^D S_{BD} = 0
    \end{equation}
    is checked along the same lines as for  $\eta$. The only difference is a plus sign instead of a minus sign in the bared part of \eqref{eqn:compinvmetriccomp}.
  \item Compatibility with integration by parts
    \begin{equation}\label{eqn:compibp}
      \int d^{2 n} X \,e^{-2 \bar d}\, U \,\nabla_M V^M =
        -\int d^{2 n} X \,e^{-2 \bar d}\, \nabla_M U\; V^M
    \end{equation}
    fixes the trace
    \begin{equation}\label{eqn:GammaJIJ}
      \Gamma_{JI}{}^J = \Gamma_I = -2 \partial_I \bar d
    \end{equation}
    of the curved connection. Employing the relation between curved and flat connection \eqref{eqn:GammaIJK}, unimodularity
    \begin{equation}
      F_{AB}{}^B = \Omega_{AB}{}^B - \Omega_{BA}{}^B = 0 \qquad \Leftrightarrow \qquad
        \Omega_{AB}{}^B = \Omega_{BA}{}^B \label{eqn:omegaABB=omegaBAB}
    \end{equation}
    and \eqref{eqn:Omegaabb}, linking $\Omega_{AB}{}^{B}$ with the flat derivative of $\bar d$, we finally obtain
    \begin{equation}\label{eqn:condibp}
      \Gamma_I = - E^A{}_I \Omega_{AB}{}^B = - 2 E^A_I D_A \bar d = -2 \partial_I \bar d\,.
    \end{equation}
    This proves compatibility with integration by parts.
\end{itemize}

To calculate the curvature and the torsion of the covariant derivative, we evaluate the commutator
\begin{equation}\label{eqn:defcurv&torsion}
  [\nabla_A,\nabla_B] V_C = R_{ABC}{}^D V_D - T^D{}_{AB} \nabla_D V_C \,,
\end{equation}
giving rise to the skew-symmetric torsion
\begin{equation}
  T^A{}_{BC} = -\frac{1}{3} F^A{}_{BC}
\end{equation}
and the Riemann curvature
\begin{equation}\label{eqn:RABCD}
  R_{ABC}{}^D = \frac{2}{9} F_{AB}{}^E F_{EC}{}^D\,.
\end{equation}
In calculating the Riemann curvature, we used the Jacobi identity
\begin{equation}
  F_{AB}{}^E F_{EC}{}^D + F_{CA}{}^E F_{EB}{}^D +
    F_{BC}{}^E F_{EA}{}^D = 0
\end{equation}
for the structure coefficients $F_{AB}{}^C$. Note that both the curvature and the torsion, do not vanish.

The covariant derivative is not completely fixed by the four compatibility constraints \eqref{eqn:compframe}, \eqref{eqn:compinvmetric}, \eqref{eqn:compbgmetric} and \eqref{eqn:compibp}. E.g., one can check that the one parameter family
\begin{equation}
  \nabla_A^t V^B = D_A V^B + t F^B{}_{AC} V^C \quad \text{and} \quad
  \nabla_A^t V_B = D_A V_B + t F_{BA}{}^C V_C \,,
\end{equation}
also fulfills all of them. A short calculation gives rise to its torsion
\begin{equation}
  T^A{}_{BC} = (2 t - 1) F^A{}_{BC}
\end{equation}
and its curvature
\begin{equation}
  R_{ABC}{}^D = (t - t^2) F_{AB}{}^E F_{EC}{}^D\,.
\end{equation}
For $t=1/2$, the torsion free Levi-Civita connection on the group manifold arises. Choosing $t=0$ or $t=1$, the curvature vanishes and one obtains two flat connections. These choices correspond to the left- or right-invariant trivialization of the tangent bundle \cite{Ferreira:2010}. In this context, it is especially remarkable that DFT${}_\mathrm{WZW}$ requires neither of the three canonical choices, but instead $t=1/3$.

Finally, we have to check how the covariant derivative acts on the generalized dilaton $d$. Its background part $\bar d$ is covariantly constant and the fluctuations $\tilde d$ transform like a scalar. Thus, we are able to identify
\begin{equation} \label{eqn:nablaAd}
  \nabla_A d = D_A \tilde d\,.
\end{equation}

\section{Strong constraint}
The level matching or weak constraint \eqref{eqn:weakconstr} can be easily rewritten
\begin{equation}
  \eta^{AB} D_A D_B \, \cdot = D_A D^A \, \cdot = 0
\end{equation}
in terms of the doubled, flat derivatives introduced in section~\ref{sec:notation}. Further, one can express this constraint with the covariant derivative introduced in the last section as
\begin{equation}
  \nabla_A \nabla^A \cdot = D_A D^A \cdot + \frac{1}{3} F^A{}_{AC} D^C \cdot = D_A D^A \cdot\,.
\end{equation}
At this point, it is inevitable to treat all arguments $\cdot$ the constraint is applied to as scalars. Further, the constraint only applies to quantities in flat indices. Otherwise, one would obtain incorrect results. Keeping this convention in mind and using the frame compatibility of the covariant derivative \eqref{eqn:compframe} yields the curved version of the weak constraint
\begin{equation}
  \nabla_I \partial^I \cdot = 0\,.
\end{equation}
After applying \eqref{eqn:GammaJIJ}, we obtain
\begin{equation}
  ( \partial_I \partial^I - 2 \partial_I d \partial^I ) \cdot = 0 
\end{equation}
which matches \eqref{eqn:weakconstcurved} discussed in section~\ref{sec:weakconstr}.

If $\cdot$ is not only interpreted as a placeholder for fields and parameters of gauge transformations but arbitrary products of them, the weak constraint becomes the strong constraint. A direct consequence of this additional restriction is
\begin{equation}
  D_A D^A ( f g) = ( D_A D^A f ) g + ( D_A D^A g ) f + 2 D_A f D^A g = 0
\end{equation}
which is equivalent to
\begin{equation}\label{eqn:sc}
  D_A f D^A g = 0 \quad \text{or} \quad \partial_I f \partial^I g = 0
\end{equation}
in curved indices. As already discussed in the last chapter, the weak constraint is not sufficient for a consistent formulation of DFT${}_\mathrm{WZW}$ up to cubic order in the fields. One has to impose the strong constraint.

\section{Gauge transformations}\label{sec:gaugetrafogenmetric}
Now, we rewrite the gauge transformations \eqref{eqn:gaugetrafo2} in terms of a generalized Lie derivative acting on the generalized metric $\mathcal{H}^{AB}$ and the generalized dilaton $d$. If the gauge algebra closes, the commutator of two such generalized Lie derivatives gives rise to another generalized Lie derivative whose parameter results from the C-bracket. We obtain this bracket by rewriting \eqref{eqn:cbracket2} in terms of doubled objects. Finally we show that the strong constraint for fluctuations and the Jacobi identity for the background's structure coefficients are sufficient for the gauge algebra to close.

\subsection{Generalized Lie derivative}
Guided by the flux formulation of toroidal DFT \cite{Geissbuhler:2013uka,Aldazabal:2013sca}, let us define the generalized Lie derivative of DFT$_{\rm WZW}$ as
\begin{equation}
  \mathcal{L}_\lambda V^{A} =  \lambda^B D_B V^{A} + \big( D^A \lambda_B -
  D_B \lambda^A \big)\, V^{B} + F^A{}_{BC} \lambda^B V^C \label{eqn:genLieVAflat}\,.
\end{equation}
Objects transforming like $\delta_\lambda V^A= \mathcal{L}_\lambda V^{A}$ are called generalized vectors. The generalized Lie derivative extends to tensors in the usual way so that e.g. the generalized Lie derivative of  $\epsilon^{AB}$ reads
\begin{align}
  \mathcal{L}_\lambda \epsilon^{AB} &=  \lambda^C D_C \epsilon^{AB} + ( D^A \lambda_C - D_C \lambda^A ) \epsilon^{CB} + \nonumber \\ 
  & ( D^B \lambda_C - D_C \lambda^B ) \epsilon^{AC} + F^A{}_{CD} \lambda^C \epsilon^{DB} + F^B{}_{CD} \lambda^C \epsilon^{AD} \label{eqn:genLieepsilonAB}\,.
\end{align}
Moreover, it leaves $\eta^{AB}$ invariant
\begin{equation}
  \mathcal{L}_\lambda \eta^{AB} = 0
\end{equation}
and for a closed gauge parameter it acts trivially, i.e. 
\begin{equation}
\label{liederivtrivial}
  \mathcal{L}_{D^A \chi} V^{B} = 0\,,
\end{equation}
after applying the strong constraint \eqref{eqn:sc}. Note that the gauge transformations \eqref{eqn:gaugetrafo2} affect fluctuations only. They are trivial 
\begin{equation}\label{eqn:trafobackground} 
  \delta_\lambda S^{AB} = 0 \,
\end{equation}
for the background metric. A straightforward computation shows that the gauge transformation of $\epsilon^{AB}$ can be expressed in terms of the generalized Lie derivative as
\begin{equation}\label{eqn:deltaepsilongenLie}
  \delta_\lambda \epsilon^{AB} = \mathcal{L}_\lambda S^{AB} + \mathcal{L}_\lambda \epsilon^{AB} + \mathcal{L}_\lambda S^{(A}{}_C S^{B)}{}_D\,  \epsilon^{CD}\,.
\end{equation}
With \eqref{eqn:trafobackground}, one can evaluate the gauge transformation
of the generalized metric
\begin{align}
  \delta_\lambda \mathcal{H}^{AB} &= \delta_\lambda \epsilon^{AB} + \frac{1}{2} \delta_\lambda \epsilon^{AC} S_{CD} \epsilon^{DB} + \frac{1}{2} \epsilon^{AC} S_{CD} \delta_\lambda e^{DB} + \mathcal{O}(\epsilon^2) \nonumber \\
  &= \mathcal{L}_\lambda S^{AB} + \mathcal{L}_\lambda \epsilon^{AB} + \mathcal{L}_\lambda S^{(A}{}_C S^{B)}_D \epsilon^{CD} + \epsilon^{C(A} S_{CD} \mathcal{L}_\lambda S^{B)D} + \mathcal{O}(\epsilon^2) \nonumber \\
  &= \mathcal{L}_\lambda S^{AB} + \mathcal{L}_\lambda \epsilon^{AB} + \mathcal{O}(\epsilon^2)
   = \mathcal{L}_\lambda \mathcal{H}^{AB} + \mathcal{O}(\epsilon^2)\,.
\end{align}
Being equivalent to \eqref{eqn:O(D,D)gens}, we have applied the identity
\begin{equation}
  S^A{}_C \, \epsilon^{CB} = - S^B{}_C \,\epsilon^{CA}
\end{equation}
in the step from the second to the third line. In a similar vein, the gauge transformation of the generalized dilaton fluctuations $\tilde d$
\begin{equation}
  \delta_\lambda \tilde d = \mathcal{L}_\lambda \tilde d \quad \text{with} \quad
  \mathcal{L}_\lambda \tilde d = \lambda^A D_A \tilde d - \frac{1}{2} D_A \lambda^A
\end{equation}
can be expressed by using the generalized Lie derivative for a density. In summary, we obtain the very compact notation for the gauge transformations
\begin{equation}\label{eqn:deltaHABdeltatilded}
  \delta_\lambda \mathcal{H}^{AB} = \mathcal{L}_\lambda \mathcal{H}^{AB}
    \quad \text{and} \quad
  \delta_\lambda \tilde d = \mathcal{L}_\lambda \tilde d \,.
\end{equation}

It is convenient to express the  generalized Lie derivative \eqref{eqn:genLieVAflat} in terms of the covariant derivative, which was introduced in section~\ref{sec:covderiv}, as
\begin{equation}\label{eqn:genLieVA}
  \mathcal{L}_{\lambda} V^A = \lambda^C \nabla_C V^A + (\nabla^A \lambda_C - \nabla_C \lambda^A) V^C\,.
\end{equation}
After this rewriting, \eqref{eqn:deltaHABdeltatilded} gives rise to the gauge transformations in their final form
\begin{align}
  \delta_\xi \mathcal{H}^{AB} = \mathcal{L}_\xi \mathcal{H}^{AB} &= \lambda^C \nabla_C \mathcal{H}^{AB} + (\nabla^A \lambda_C - \nabla_C \lambda^A) \mathcal{H}^{CB} + (\nabla^B \lambda_C - \nabla_C \lambda^B) \mathcal{H}^{AC} \nonumber \\
  \delta_\xi \tilde d = \mathcal{L}_\xi \tilde d & = \xi^A D_A \tilde d - \frac{1}{2} D_A \xi^A \,. \label{eqn:gendiffHAB&tilded}
\end{align}
Plugging in the expansion of the generalized metric \eqref{eqn:HABexpansion}, one recovers the gauge transformations \eqref{eqn:gaugetrafo2} up to additional terms which are not linear in the field or the gauge parameter. 

\subsection{C-bracket}
Using the conventions \eqref{eqn:lambda^A}, \eqref{eqn:lambda_A} and \eqref{eqn:flatderivdoubled}, one can also write the C-bracket \eqref{eqn:cbracket2} in double index notation, where it takes the compact form
\begin{equation}
  \lambda_{12}^A = - \lambda_1^B\, D_B \lambda_2^A
    + \frac{1}{2} \lambda_1^B\, D^A \lambda_{2,B} 
    - \frac{1}{2} F^A{}_{BC} \,\lambda_1^B\, \lambda_2^C - (1\leftrightarrow 2)\,.
\end{equation}
Analogous to the convention in DFT, we denote
\begin{equation}\label{eqn:cbracket}
  [\lambda_1, \lambda_2]^A_\mathrm{C} := -\lambda_{12}^A=
\lambda_1^B D_B \lambda_2^A -
    \frac{1}{2} \lambda_1^B D^A \lambda_{2\,B} + \frac{1}{2} F^A{}_{BC} \lambda_1^B \lambda_2^C - (1\leftrightarrow 2)
\end{equation}
as C-bracket of DFT${}_\mathrm{WZW}$. It differs in the third term from the expression presented for toroidal backgrounds in \cite{Hull:2009zb}. Furthermore, the derivatives appearing in \eqref{eqn:cbracket} do not commute. On the torus, they do.

Like the generalized Lie derivative, the C-bracket can also be expressed in terms of the generalized covariant derivative \eqref{eqn:flatcovderv} as
\begin{equation}\label{eqn:cbracketcov}
  [\lambda_1, \lambda_2]_C^A = \lambda_1^B \,\nabla_B \lambda_2^A - \frac{1}{2} \lambda_1^B\, \nabla^A\, \lambda_{2\,B} - (1 \leftrightarrow 2)\,.
\end{equation}
With this result, we are also able to calculate the generalized torsion of $\nabla_A$. Like for DFT, it is defined as the difference between the C-bracket with covariant and partial derivatives. In our case, this leads to
\begin{equation}
  [\lambda_1, \lambda_2]_\mathrm{C}^I - \mathcal{T}^I{}_{JK} \lambda_1^J \lambda_2^K = \lambda_1^J \partial_J \lambda_2^I - \frac{1}{2}\lambda_1^J \partial^I \lambda_{2\,J} - (1 \leftrightarrow 2)
\end{equation}
with 
\begin{equation}
  [\lambda^B_1, \lambda^C_2]_\mathrm{C}^I=[E^B{}_J \lambda^J_1,
    E^C{}_K\lambda^K_2]_C^A\, E_A{}^I\,.
\end{equation}
Evaluating this expression by using the compatibility with the frame results in the non-vanishing generalized torsion
\begin{equation}\label{eqn:gentorsion}
  \mathcal{T}^I{}_{JK} = 2  \Gamma_{[JK]}{}^I + \Gamma^I{}_{[JK]}  = -\frac{1}{3} \bigl( 2 \Omega_{[JK]}{}^I + 2\Omega^I{}_{[JK]} + \Omega_{[J}{}^I{}_{K]} \bigr) = - \Omega^I{}_{[JK]}\,.
\end{equation}
Thus in contrast to the covariant derivative of toroidal DFT, the generalized torsion of the covariant derivative of DFT$_{\rm WZW}$ does not vanish.

\subsection{Closure of gauge algebra}\label{sec:cbracket}
We now check the closure of the gauge algebra. There are two different ways to prove closure which are completely equivalent. First, one can compute the Jacobiator
\begin{equation}
  J(\lambda_1, \lambda_2, \lambda_2) = [\lambda_1, [\lambda_2, \lambda_3]_\mathrm{C}]_\mathrm{C} +
    [\lambda_3, [\lambda_1, \lambda_2]_\mathrm{C}]_\mathrm{C} +
    [\lambda_2, [\lambda_3, \lambda_1]_\mathrm{C}]_\mathrm{C}
\end{equation}
and impose that it vanishes up to terms parameterizing trivial gauge transformations. According to \eqref{liederivtrivial}, then the constraint
\begin{equation}
  \mathcal{L}_{J(\lambda_1, \lambda_2, \lambda_3)} V^A = 0
\end{equation}
has to hold. Alternatively, one can show that the commutator of two generalized Lie derivatives closes in the sense that
\begin{equation}\label{eqn:closurecond}
  \mathcal{L}_{[\lambda_1, \lambda_2]_\mathrm{C}} V^A = (\mathcal{L}_{\lambda_1} \mathcal{L}_{\lambda_2} 
    - \mathcal{L}_{\lambda_2} \mathcal{L}_{\lambda_1}) V^A\,.
\end{equation}
Here, we will show this second property of the generalized C-bracket.

Evaluating the condition \eqref{eqn:closurecond}, one eventually arrives at the expression 
\begin{align}
  \mathcal{L}_{[\lambda_1,\lambda_2]_\mathrm{C}} V^A = (\mathcal{L}_{\lambda_1} & \mathcal{L}_{\lambda_2} 
    - \mathcal{L}_{\lambda_2} \mathcal{L}_{\lambda_1}) V^A \nonumber\\
    & - F_{BC}{}^E F_{ED}{}^A - F_{DB}{}^E F_{EC}{}^A - F_{CD}{}^E F_{EB}{}^A \,,
\end{align}
where the second line vanishes due to the Jacobi identity \eqref{eqn:jacobiid}. Let us emphasize that this closure result goes beyond what one would expect from the CSFT construction. A priori CSFT at cubic order only forces the $V^A$ independent part of \eqref{eqn:closurecond} to hold \cite{Hohm:2014xsa}. For all terms depending on $V^A$, there are in general corrections and closure is only guaranteed on-shell. However, here we do not face any of these problems. Moreover, for the closure of the usual DFT algebra, the strong constraint was essential for the fluctuations and the background, whereas here one only needs the Jacobi-identity for the background fluxes.

\section{Action}\label{sec:actiongenmetric}
In this section, we rewrite the action \eqref{eqn:action} in terms of the generalized metric and the generalized dilaton. The guiding principle is inspired by the results for the gauge transformations and the C-bracket discussed in the last sections: in the expressions known from traditional DFT, one has to substitute partial derivatives by covariant derivatives \eqref{eqn:flatcovderv}. Taking into account the original DFT action in the generalized metric formulation \cite{Hohm:2010pp} and following this principle, the action should read
\begin{align}
  S = \int d^{2n} X e^{-2d} \Big( & \frac{1}{8} \mathcal{H}^{CD} \nabla_C \mathcal{H}_{AB} \nabla_D \mathcal{H}^{AB} -\frac{1}{2} \mathcal{H}^{AB} \nabla_{B} \mathcal{H}^{CD} \nabla_D \mathcal{H}_{AC} \nonumber \\
  \label{eqn:actiongenmetricnaive}
  & - 2 \nabla_A d \nabla_B \mathcal{H}^{AB} + 4 \mathcal{H}^{AB} \nabla_A d \nabla_B d \Big)\,.
\end{align}
Subsequently, we prove that it indeed reproduces \eqref{eqn:action} up to cubic terms in the fields and a missing term which has to be added to \eqref{eqn:actiongenmetricnaive}. To keep this straightforward though  cumbersome calculation as traceable as possible, we begin with terms containing two flat derivatives like e.g.
\begin{equation}\label{eqn:HDHDH}
  e^{-2 d} \frac{1}{8} \mathcal{H}^{CD} D_C \mathcal{H}_{AB} D_D \mathcal{H}^{AB}\,.
\end{equation}
We further simplify the calculation by first considering the term
\begin{equation}
  \frac{1}{8} S^{CD} D_C \mathcal{H}_{AB} D_D \mathcal{H}^{AB},
\end{equation}
which gives rise to
\begin{align} \label{eqn:SDHDH1}
  \frac{1}{8} S^{CD} D_C \mathcal{H}_{AB} D_D \mathcal{H}^{AB} &= -\frac{1}{2} \Big( D_c \epsilon_{a\bar{b}} D^c \epsilon^{a\bar{b}} + D_{\bar{c}} \epsilon_{a\bar{b}} D^{\bar{c}} \epsilon^{a\bar{b}} \Big) + \mathcal{O}(\epsilon^4) \nonumber \\
  &= \epsilon_{a\bar{b}}\Box \epsilon^{a\bar{b}} - \epsilon_{a\bar{b}} D_c \tilde{d} D^c \epsilon^{a\bar{b}} - \epsilon_{a\bar{b}} D_{\bar{c}} \tilde{d} D^{\bar{c}} \epsilon^{a\bar{b}} + \mathcal{O}(\epsilon^4)\,,
\end{align}
after plugging in the components of $S^{AB}$ and $\mathcal{H}^{AB}$, according to \eqref{eqn:SAB} and \eqref{eqn:HABexpansion}. From the first to the second line in \eqref{eqn:SDHDH1}, we have performed integration by parts by applying the rule
\begin{equation}\label{eqn:ibp}
  \int d^{2n} X e^{-2d} u D_a v = - \int d^{2n} \sqrt{|H|} e^{-2 \tilde d} ( -2 u D_a \tilde d  + D_a u ) v = \int d^{2n} X e^{-2d} (2 u D_a \tilde d   - D_a u) v \,.
\end{equation}
It automatically arises if one remembers the splitting of the generalized dilaton \eqref{eqn:splitdilaton} into the background part $\bar d$ and the fluctuations $\tilde d$ around this background. Performing integration by parts again and dropping all terms in quartic order in the fields, we obtain
\begin{equation}\label{eqn:SDHDH2}
  \frac{1}{8} S^{CD} D_C \mathcal{H}_{AB} D_D \mathcal{H}^{AB} = \epsilon_{a\bar{b}} \Box \epsilon^{a\bar{b}} + \tilde{d} \big( D_c \epsilon_{a\bar{b}} \big)^2 + \tilde{d} \big( D_{\bar{c}} \epsilon_{a\bar{b}} \big)^2 + 2 \tilde{d}\, \epsilon_{a\bar{b}} \Box \epsilon^{a\bar{b}}  + \mathcal{O}(\epsilon^4) + \mathcal{O}(\tilde d^2 \epsilon^2)\,.
\end{equation}
Now, it is straightforward to read off the remaining terms of \eqref{eqn:HDHDH}, namely
\begin{equation}
  e^{-2 d} \frac{1}{8} \mathcal{H}^{CD} D_C \mathcal{H}_{AB} D_D \mathcal{H}^{AB} = \sqrt{H} \Big[ \epsilon_{a\bar{b}} \Box \epsilon^{a\bar{b}} -2 \epsilon^{c\bar{d}} D_c \epsilon_{a\bar{b}} D_{\bar{d}} \epsilon^{a\bar{b}} + \tilde{d} \big( D_c \epsilon_{a\bar{b}} \big)^2 + \tilde{d} \big( D_{\bar{c}} \epsilon_{a\bar{b}} \big)^2 \Big]\,.
\end{equation}
Here and in the following, $\mathcal{O}(\dots)$ is suppressed for brevity. The last term in \eqref{eqn:SDHDH2} cancels against a term arising in the expansion of
\begin{equation}
  e^{-2 d} = \sqrt{|H|} \big(1 - 2 \tilde d + 2 \tilde d^2 + \dots \big)\,.
\end{equation}
Next, we turn to the term
\begin{equation}
  -\frac{1}{2} \mathcal{H}^{AB} D_{B} \mathcal{H}^{CD} D_D \mathcal{H}_{AC}
\end{equation}
for which the calculations are more cumbersome. Using the commutation relations for flat derivatives
\begin{equation}
  [ D_a, D_b ] = F_{ab}{}^c D_c\, ,\quad [ D_{\bar a}, D_{\bar b} ] = F_{\bar{a}\bar{b}}{}^{\bar c} D_{\bar c}
\end{equation}
and performing integration by parts, we finally obtain the result
\begin{align}
  - e^{-2 d} \frac{1}{2} \mathcal{H}^{AB} D_{B} \mathcal{H}^{CD} D_D \mathcal{H}_{AC} &= \sqrt{|H|} \Big[ \big( D^a e_{a\bar{b}} \big)^2 + \big( D^{\bar{b}} e_{a\bar{b}} \big)^2 \nonumber \\ 
  &\,- \big( F_d{}^{ac} D_c \epsilon^{d\bar{b}}  \epsilon_{a\bar{b}}\,+ F_{\bar d}{}^{\bar a\bar c} D_{\bar{c}} \epsilon^{b\bar{d}} \epsilon_{b\bar{a}} \big) \big( 1-2\tilde{d} \big) \nonumber \\ 
  &\,+ 2 \tilde{d}\, \epsilon^{a\bar{b}} D_a D^c \epsilon_{c\bar{b}} - 2 \tilde{d} D_a D^c \epsilon^{a\bar{b}} \epsilon_{c\bar{b}} - 2 \tilde{d} D^c \epsilon^{a\bar{b}}  D_a \epsilon_{c\bar{b}} \nonumber \\ 
  &\,+2 \tilde d\, \epsilon^{a\bar{b}} D_{\bar{b}} D^{\bar{c}} \epsilon_{a\bar{c}} - 2 \tilde{d} D_{\bar{b}} D^{\bar{c}} \epsilon^{a\bar{b}} \epsilon_{a\bar{c}} - 2 \tilde{d} D^{\bar{c}} \epsilon^{a\bar{b}} D_{\bar{b}} \epsilon_{a\bar{c}} \nonumber \\ 
  &\,+ 2 \epsilon_{a\bar{b}} \big( D^a \epsilon_{c\bar{d}} D^{\bar{d}} \epsilon^{c\bar{b}} + D^c \epsilon^{a\bar{d}} D^{\bar{b}} \epsilon_{c\bar{d}} \big) \Big].
\end{align}
All remaining terms in the action \eqref{eqn:actiongenmetricnaive} contain covariant derivatives acting on the generalized dilaton. Thus, one has to remember \eqref{eqn:nablaAd}, specifying the action of the covariant derivative on $d$. In combination with the expansion \eqref{eqn:HABexpansion} of $\mathcal{H}^{AB}$, it gives rise to
\begin{equation}
  4 \mathcal{H}^{AB} D_A \tilde d D_B \tilde d = 2 D_a \tilde d D^a \tilde d +  2 D_{\bar a} \tilde d D^{\bar a} \tilde d + 8 \epsilon^{a\bar b} D_a \tilde d D_{\bar b} \tilde d\,.
\end{equation}
Taking into account the prefactor $e^{-2d}$, we obtain
\begin{equation}
  e^{-2 d} 4 \mathcal{H}^{AB} D_A \tilde{d} D_B \tilde{d} = \sqrt{|H|} \big[ -4\tilde{d} \Box \tilde{d} + 8 \epsilon^{a\bar{b}} D_a \tilde{d} D_{\bar{b}} \tilde{d}  + 4 \tilde{d}^2 \Box \tilde{d} \big]
\end{equation}
where we have applied the relation
\begin{equation}
  \sqrt{H} 4 \tilde{d}^2 \Box d = \sqrt{H} \big(-4\tilde{d} D_a \tilde{d} D^a \tilde{d} - 4\tilde{d} D_{\bar{a}} \tilde{d} D^{\bar{a}} \tilde{d}\;\big)\,.
\end{equation}
The last term in \eqref{eqn:actiongenmetricnaive}, which contains two flat derivatives, gives rise to
\begin{align}
  - e^{-2d} 2 D_A \tilde{d} D_B \mathcal{H}^{AB} &= \sqrt{|H|} \Big[ 4 \tilde{d} D_a D_{\bar{b}} \epsilon^{a\bar{b}} - 8\epsilon^{a\bar{b}} D_a \tilde{d} D_{\bar{b}} \tilde{d} - 8\tilde{d} \epsilon^{a\bar{b}} D_a D_{\bar{b}} \tilde{d} \nonumber \\ 
  &\,+ 2 \tilde{d} \big( D^a \epsilon_{a\bar{b}} \big)^2 + 2 \tilde{d} \big( D^{\bar{b}} \epsilon_{a\bar{b}} \big)^2 + 2 \tilde{d} \epsilon^{a\bar{b}} D_a D^c \epsilon_{a\bar{b}} + 2 \tilde{d} \epsilon^{a\bar{b}} D_{\bar{b}} D^{\bar{c}} \epsilon_{a\bar{c}} \nonumber \\ 
  &\,+2 \tilde{d} D^c \epsilon^{a\bar{b}} D_a \epsilon_{c\bar{b}} + 2 \tilde{d} D^{\bar{c}} \epsilon^{a\bar{b}}  D_{\bar{b}} \epsilon_{a\bar{c}} + 2 \tilde{d}  D_a D^c \epsilon^{a\bar{b}} \epsilon_{c\bar{b}} \nonumber \\
  &\,+ 2 \tilde{d} \big( D_{\bar{b}} D^{\bar{c}} \epsilon^{a\bar{b}} \big) \epsilon_{a\bar{c}}\Big]\,.
\end{align}
Now, we are done with all terms required for the abelian case $F_{ABC}=0$. Hence, it is a convenient check of the results obtained so far to write down the complete abelian action
\begin{align}
  \left. S\right|_{F_{ABC}=0} &= \int d^{2 D}X \sqrt{|H|} \Big[ \epsilon_{a\bar{b}} \Box \epsilon^{a\bar{b}} +  \big( D^a e_{a\bar{b}} \big)^2 + \big( D^{\bar{b}} e_{a\bar{b}} \big)^2  + 4\tilde{d} D_a D_{\bar{b}} \epsilon^{a\bar{b}} -4\tilde{d} \Box \tilde{d} \nonumber \\ & -2 \epsilon_{a\bar{b}} \big( D_a \epsilon_{c\bar{d}} D_{\bar{b}} \epsilon^{c\bar{d}} - D^a \epsilon_{c\bar{d}} D^{\bar{d}} \epsilon^{c\bar{b}} - D^c \epsilon^{a\bar{d}} D^{\bar{b}} \epsilon_{c\bar{d}} \big) \nonumber \\ & + \tilde{d} \Big( 2 \big( D^a e_{a\bar{b}} \big)^2  + 2 \big( D^{\bar{b}} e_{a\bar{b}} \big)^2 + \big( D_c \epsilon_{a\bar{b}} \big)^2 + 
  \big( D_{\bar{c}} \epsilon_{a\bar{b}} \big)^2 + 4 \epsilon^{a\bar{b}} \big( D_a D^c \epsilon_{a\bar{b}} + D_{\bar{b}} D^{\bar{c}} \epsilon_{a\bar{c}} \big) \Big) \nonumber \\ & + 4\tilde{d}^2 \Box \tilde{d} - 8\tilde{d} \epsilon^{a\bar{b}} D_a D_{\bar{b}} \tilde{d} \Big]\,.
\end{align}
It indeed matches with the action \eqref{eqn:action} after dropping all terms depending on the structure coefficients $F_{abc}$ and $F_{\bar a\bar b\bar c}$. 

Let us now take into account these terms so that we have to consider the full covariant derivative instead of only using its flat derivative part. We start with
\begin{align}
- 2 \nabla_A d \nabla_B \mathcal{H}^{AB} &= - D_A \tilde{d} \Big( D_B \mathcal{H}^{AB} + \frac{1}{3} \big( {F^A}_{BC} \mathcal{H}^{CB} + {F^B}_{BC} \mathcal{H}^{AC} \big) \Big) \nonumber \\ 
  &= - D_A \tilde{d} D_B \mathcal{H}^{AB}
\end{align}
where the second term in the first line vanishes due to the total antisymmetry of $F_{ABC}$ and the symmetry of $\mathcal{H}^{AB}$. The third term is zero due to the unimodularity condition
\begin{equation}
  F^A{}_{AB} = 0\,,
\end{equation}
which the structure coefficients have to fulfill according to \eqref{eqn:unimodular}. At this point, we come to the more challenging part
\begin{equation}\label{eqn:Htermscov}
  \frac{1}{8} \mathcal{H}^{CD} \nabla_C \mathcal{H}_{AB} \nabla_D \mathcal{H}^{AB} -\frac{1}{2} \mathcal{H}^{AB} \nabla_{B} \mathcal{H}^{CD} \nabla_D \mathcal{H}_{AC}\,.
\end{equation}
In the subsequent computation, we ignore all terms which contain more than one flat derivative, because we already discussed these contributions above. The first part of \eqref{eqn:Htermscov} gives rise to
\begin{align}
  \frac{1}{8} \mathcal{H}^{CD} \nabla_C & \mathcal{H}_{AB} \nabla_D \mathcal{H}^{AB} = \frac{1}{12} \mathcal{H}^{CD} D_C \mathcal{H}_{AB} {F^{A}}_{DE} \mathcal{H}^{EB} + \frac{1}{12} \mathcal{H}^{CD} {F_{AC}}^{F} \mathcal{H}_{FB} D_D \mathcal{H}^{AB} \nonumber \\ & +\frac{1}{36} \mathcal{H}^{CD} {F_{AC}}^{F} \mathcal{H}_{FB} {F^{A}}_{DE} \mathcal{H}^{EB} + \frac{1}{36} \mathcal{H}^{CD} {F_{AC}}^{F} \mathcal{H}_{FB} {F^{B}}_{DE} \mathcal{H}^{AE} + \mathcal{O}(D^2) \label{eqn:HCHCH}
\end{align}
where the second term on the right hand side is equivalent to
\begin{equation}
  \mathcal{H}^{CD} {F_{AC}}^{F} \mathcal{H}_{FB} D_D \mathcal{H}^{AB} = \mathcal{H}^{CD} {F^{A}}_{DE} \mathcal{H}^{EB} D_C \mathcal{H}_{AB}
\end{equation}
after using the symmetry of $\mathcal{H}^{AB}$ and relabeling the indices. For the fourth term, we use the total antisymmetry of the structure coefficients, yielding
\begin{equation}
    \mathcal{H}^{CD} {F_{AC}}^{F} \mathcal{H}_{FB} {F^{B}}_{DE} \mathcal{H}^{AE} = - F_{ACE} F_{BDF} \mathcal{H}^{AB} \mathcal{H}^{CD} \mathcal{H}^{EF} \,.
\end{equation}
Applying these two substitutions, \eqref{eqn:HCHCH} simplifies to
\begin{align}
  \frac{1}{8} \mathcal{H}^{CD} \nabla_C \mathcal{H}_{AB} \nabla_D \mathcal{H}^{AB} &\,= \frac{1}{6} \mathcal{H}^{CD} D_C \mathcal{H}_{AB} {F^{A}}_{DE} \mathcal{H}^{EB} +\frac{1}{36} \mathcal{H}^{CD} {F_{AC}}^{F} {F^{A}}_{DE} \mathcal{H}_{FB} \mathcal{H}^{EB} \nonumber \\ &\quad\; - \frac{1}{36} F_{ACE} F_{BDF} \mathcal{H}^{AB} \mathcal{H}^{CD} \mathcal{H}^{EF} + \mathcal{O}(D^2)\,.
\end{align}
For the second part of \eqref{eqn:HCHCH}, we obtain in a similar fashion
\begin{align}
  -\frac{1}{4} \mathcal{H}^{AB} \nabla_{B} \mathcal{H}^{CD} \nabla_D \mathcal{H}_{AC} 
&\,= \frac{1}{6} \mathcal{H}^{CD} D_C \mathcal{H}_{AB} {F^{A}}_{DE} \mathcal{H}^{EB} - \frac{1}{12} \mathcal{H}^{AB} D_B \mathcal{H}^{CD} {F_{CD}}^E \mathcal{H}_{AE} \nonumber \\ &\;-\frac{1}{12} \mathcal{H}^{AB} {F^D}_{BE} \mathcal{H}^{CE} D_D \mathcal{H}_{AC} - \frac{1}{36} F_{ACE} F_{BDF} \mathcal{H}^{AB} \mathcal{H}^{CD} \mathcal{H}^{EF}  \nonumber \\ &\;+\frac{1}{36} \mathcal{H}^{CD} {F_{AC}}^{F} {F^{A}}_{DE} \mathcal{H}_{FB} \mathcal{H}^{EB} + \mathcal{O}(D^2)\,.
\end{align}
After combining these results, we finally get
\begin{align}\label{eqn:Ftermsaction}
  e^{-2 d} \Big[ \frac{1}{8} \mathcal{H}^{CD} &\nabla_C \mathcal{H}_{AB} \nabla_D \mathcal{H}^{AB} -\frac{1}{2} \mathcal{H}^{AB} \nabla_{B} \mathcal{H}^{CD} \nabla_D \mathcal{H}_{AC}\Big] \nonumber\\ &\,= \sqrt{|H|} \Big[ 2 \epsilon_{a\bar{b}} \big( {F^{ac}}_d D^{\bar{e}} \epsilon^{d\bar{b}} \epsilon_{c\bar{e}} + {F^{\bar{b}\bar{c}}}_{\bar{d}} D^{e} \epsilon^{a\bar{d}} \epsilon_{e\bar{c}} \big) + \frac{2}{3} F_{ace} F_{\bar{b}\bar{d}\bar{f}} \epsilon^{a\bar{b}} \epsilon^{c\bar{d}} \epsilon^{e\bar{f}} \nonumber \\ &\;-\frac{1}{6} \big( F^a{}_{cd} F_b{}^{cd} \epsilon_{a\bar e} \epsilon^{b \bar e} + F^{\bar a}{}_{\bar c\bar d} F_{\bar b}{}^{\bar c\bar d} \epsilon_{e \bar a} \epsilon^{e \bar b} \big) \big( 1 - 2\tilde{d} \big) + \mathcal{O}(D^2) \Big]\,.
\end{align}
The first line on the right hand side exactly reproduces the structure coefficients dependent terms in the cubic action \eqref{eqn:action}, but the second line has to be canceled to successfully reproduce the action. Achieving this is done by adding the term
\begin{equation}
  \frac{1}{6} F_{ACE} F_{BDF} \mathcal{H}^{AB} \eta^{CD} \eta^{EF} + V_0 =
    \frac{1}{6} \big( F^a{}_{cd} F_b{}^{cd} \epsilon_{a\bar e} \epsilon^{b \bar e} +
      F^{\bar a}{}_{\bar c\bar d} F_{\bar b}{}^{\bar c\bar d} \epsilon_{e \bar a} \epsilon^{e \bar b}
      \big)
\end{equation}
with
\begin{align}
  V_0 &= - \frac{1}{6} F_{ACE} F_{BDF} S^{AB} S^{CD} S^{EF} \nonumber \\
      &= - \frac{1}{4} F_{ACE} F_{BDF} S^{AB} \eta^{CD} \eta^{EF} +
         \frac{1}{12} F_{ACE} F_{BDF} S^{AB} S^{CD} S^{EF}
\end{align}
to the naive action \eqref{eqn:actiongenmetricnaive}. To obtain the second line in the expression for $V_0$, we applied the identity 
\begin{equation}
  F_{ACE} F_{BDF} S^{CD} S^{EF} =  F_{ACE} F_{BDF} \eta^{CD} \eta^{EF}
\end{equation}
which holds due to the strict separation of bared and unbared structure coefficients \eqref{eqn:conventFABC}. Substituting the structure coefficients $F_{ABC}$ by the covariant fluxes $\mathcal{F}_{ABC}$, $V_0$ matches the vacuum expectation value of the scalar potential \eqref{eqn:scalarpotential}, which we derived in section~\ref{sec:gaugedgravity}. Even though we do not impose the strong constraint on the background, $V_0$ lacks the $1/6 F_{ABC} F^{ABC}$ term suggested by the flux formulation. If we consider the full $2D$-dimensional doubled space time instead of only its $2n$-dimensional compact subspace, $V_0$ has to vanish for a background giving rise to a well defined CFT. As outlined in section~\ref{sec:missingFABCFABCterm}, otherwise the combined central charge of the ghost system and the bosons would not vanish. We close this section with the complete action of DFT${}_\mathrm{WZW}$
\begin{align}
  S = & \int d^{2D} X e^{-2d} \Big(  \frac{1}{8} \mathcal{H}^{CD} \nabla_C \mathcal{H}_{AB} \nabla_D \mathcal{H}^{AB} -\frac{1}{2} \mathcal{H}^{AB} \nabla_{B} \mathcal{H}^{CD} \nabla_D \mathcal{H}_{AC} \nonumber \\
  & - 2 \nabla_A d \nabla_B \mathcal{H}^{AB} + 4 \mathcal{H}^{AB} \nabla_A d \nabla_B d + \frac{1}{6} F_{ACD} F_B{}^{CD} \mathcal{H}^{AB} \Big) \label{eqn:actiongenmetric}
\end{align}
in the generalized metric formulation for the full, $2D$-dimensional doubled space time. For obtaining the action in curved indices, one has to remember the vielbein compatibility condition \eqref{eqn:compframe} of the covariant derivative. Due to this condition it is legitimate to simply substitute all flat indices with curved ones.

\section{Equations of motion}\label{sec:eom}
After deriving the full action of DFT${}_\mathrm{WZW}$ in the last section, we now discuss its equations of motion. It is convenient to split them into two independent parts. First, we present the variation of the action \eqref{eqn:actiongenmetric} with respect to the generalized dilaton $d$ in subsection \ref{sec:gencurvature}. It gives rise to the generalized curvature scalar $\mathcal{R}$. Furthermore, we show how the action can be rewritten in terms of this scalar. In the second step, we perform the variation with respect to the generalized metric $\mathcal{H}^{AB}$ in subsection \ref{sec:genriccitensor}. Just as in the generalized metric formulation of DFT \cite{Hohm:2010pp}, we have to apply an appropriate projection, taking into account the O($D,D$) property of the generalized metric, to obtain the generalized Ricci tensor $\mathcal{R}_{IJ}$.

\subsection{Generalized curvature scalar}\label{sec:gencurvature}
Following \cite{Hohm:2010pp}, we define the generalized scalar curvature $\mathcal{R}$ of DFT${}_\mathrm{WZW}$ using the variation of the action \eqref{eqn:actiongenmetric}
\begin{equation}\label{eqn:variationd}
  \delta S = - 2 \int d^{2D} X\, e^{-2d}\, \mathcal{R}\, \delta d
\end{equation}
with respect to the generalized dilaton $d$. A straightforward calculation gives rise to
\begin{align}
  \mathcal{R} &=  4 \mathcal{H}^{AB} \nabla_A \nabla_B d - \nabla_A \nabla_B \mathcal{H}^{AB} - 4 \mathcal{H}^{AB} \nabla_A d\, \nabla_B d + 4 \nabla_A d \,\nabla_B \mathcal{H}^{AB} \nonumber \\ 
  &\,+ \frac{1}{8} \mathcal{H}^{CD} \nabla_C \mathcal{H}_{AB} \nabla_D \mathcal{H}^{AB} - \frac{1}{2} \mathcal{H}^{AB} \nabla_B \mathcal{H}^{CD} \nabla_D \mathcal{H}_{AC} +\frac{1}{6} F_{ACD} F_B{}^{CD} \mathcal{H}^{AB}\,. \label{eqn:gencurvature}
\end{align}
In order to prove the invariance of the action \eqref{eqn:actiongenmetric} under generalized diffeomorphisms in the next section, it is convenient to express it in the form
\begin{equation}\label{eqn:actiongenricci}
  S = \int d^{2D} X\, e^{-2 d}\, \mathcal{R}\,.
\end{equation}
To this end, we rewrite \eqref{eqn:actiongenmetric} as
\begin{equation}
  S = \int d^{2D} X\, e^{-2d}\, \mathcal{R} + \int d^{2D} X \,\sqrt{|H|}\, D_A \big[ e^{-2\tilde{d}} ( \nabla_B \mathcal{H}^{AB} - 4 \mathcal{H}^{AB} \nabla_B d ) \big]\,,
\end{equation}
where the last term is a vanishing boundary term. Due to the compatibility of the covariant derivative with the generalized vielbein, it is trivial to express the generalized scalar curvature in curved instead of flat indices. One only has to relabel the indices to obtain the curved version. The generalized dilaton part of the equations of motion reads
\begin{equation}
  \mathcal{R} =  0\,.
\end{equation}

\subsection{Generalized Ricci tensor}\label{sec:genriccitensor}
Now, we calculate the variation of the action \eqref{eqn:actiongenmetric} with respect to the generalized metric $\mathcal{H}^{AB}$. By analogy with \eqref{eqn:variationd}, we consider
\begin{equation}
  \delta S = \int d^{2 D}X\, e^{-2d}\, \delta \mathcal{H}^{AB}\, \mathcal{K}_{AB}\,.
\end{equation}
As discussed in \cite{Hohm:2010pp}, the variation $\delta \mathcal{H}^{AB}$ is symmetric and thus it is sufficient to study the symmetric part of $\mathcal{K}_{AB}$ only. Performing the variation explicitly and afterwards symmetrizing $\mathcal{K}_{AB}$ gives rise to
\begin{align}
  \mathcal{K}_{AB} &= \frac{1}{8} \nabla_A \mathcal{H}_{CD} \nabla_B \mathcal{H}^{CD} - \frac{1}{4} \big[ \nabla_C - 2 (\nabla_C d) \big] \mathcal{H}^{CD} \nabla_D \mathcal{H}_{AB} + 2 \nabla_{(A} \nabla_{B)} d \nonumber \\ &\,- \nabla_{(A} \mathcal{H}^{CD} \nabla_D \mathcal{H}_{B)C} + \big[ \nabla_D - 2 (\nabla_D d) \big] \big[ \mathcal{H}^{CD} \nabla_{(A} \mathcal{H}_{B)C} + {\mathcal{H}^C}_{(A} \nabla_C {\mathcal{H}^D}_{B)} \big]  \nonumber \\ &\,+ \frac{1}{6} F_{ACD} F_B{}^{CD}\,. \label{eqn:Ktensor}
\end{align}
Furthermore, the O($D,D$) constraint
\begin{equation}
  \mathcal{H}^{AC} \eta_{CD} \mathcal{H}^{DB} = \eta^{AB}
\end{equation}
has to be preserved under the variation \cite{Hohm:2010pp}. This implies that only a certain projection of $\mathcal{K}_{AB}$ gives rise to the equations of motion. Hence, it is necessary to introduce the projection operators
\begin{equation}
P_{AB} = \frac{1}{2} \big( \eta_{AB} - S_{AB} \big),\quad\text{and}\quad\bar{P}_{AB} = \frac{1}{2} \big( \eta_{AB} + S_{AB} \big)
\end{equation}
which are used to define the generalized Ricci tensor
\begin{equation}\label{eqn:projectiongenricci}
  \mathcal{R}_{AB} = 2 {P_{(A}}^C {\bar{P}_{B)}}^{\;\;D} \mathcal{K}_{CD}\,.
\end{equation}
This projection is exactly the same as the one we applied in the sections~\ref{sec:dfteom} and \ref{sec:vacua}. It cancels the term in the last line of \eqref{eqn:Ktensor}. Thus, we find a generalized Ricci tensor whose structure matches the one of toroidal DFT. However, all partial derivatives have to be replaced with covariant ones.

\section{Local symmetries}\label{sec:symmetries}
The CSFT derivation of DFT${}_\mathrm{WZW}$ in chapter~\ref{chap:groupdft} was challenging. Even if all calculations were performed with much care, there is still a small chance that some prefactors are wrong or some terms are missing completely. The recasting of the gauge transformations and the action in the sections \ref{sec:gaugetrafogenmetric} and \ref{sec:actiongenmetric} is a first indication that everything went well: All the different terms with bared and unbared indices integrate nicely into doubled objects. However, a much more important consistency check is the invariance of the action \eqref{eqn:actiongenmetric} under the gauge transformations \eqref{eqn:gendiffHAB&tilded}. If all previous calculations were performed correctly, the CSFT framework guarantees this invariance up to cubic order in the fields. As we show in subsection \ref{sec:gendiffinv}, it even holds to all higher orders introduced by the generalized metric formulation. Besides generalized diffeomorphism invariance, the action is also manifestly invariant under 2D-diffeomorphisms, as we prove in subsection \ref{sec:2Ddiff}.

\subsection{Generalized diffeomorphisms}\label{sec:gendiffinv}
It does not matter whether one proves the invariance under gauge transformations for the action \eqref{eqn:actiongenmetric} or \eqref{eqn:actiongenricci}. Both only differ by a vanishing total derivative. We choose the latter one, with the generalized curvature scalar $\mathcal{R}$. Proving its invariance requires two steps: First, we show that $\mathcal{R}$ transforms as a scalar under generalized diffeomorphisms. Second, we consider the remaining term $e^{-2d}$ and show that it transforms as a weight +1 scalar density.

In order to show that the generalized curvature \eqref{eqn:gencurvature} is a scalar under generalized diffeomorphisms, we have to compare its transformation behavior under gauge transformations with the results we expect from generalized diffeomorphisms mediated by the generalized Lie derivative. The failure of a quantity $V$ to transform covariantly under generalized diffeomorphisms reads
\begin{equation}\label{eqn:Deltaxigendiff}
  \Delta_\xi V = \delta_\xi V - \mathcal{L}_\xi V,
\end{equation}
where $\mathcal{L}_\xi$ is the generalized Lie derivative \eqref{eqn:genLieVA} and $\delta_\xi$ denotes the gauge transformations \eqref{eqn:gendiffHAB&tilded} discussed in section \ref{sec:gaugetrafogenmetric}. From the definition \eqref{eqn:Deltaxigendiff}, it is obvious that
\begin{equation}
  \Delta_\xi \mathcal{H}^{AB} = 0 \quad \text{and} \quad \Delta_\xi \tilde d = 0
\end{equation}
hold. Furthermore, $\Delta_\xi$  is linear and fulfills the product rule
\begin{equation}
  \Delta_\xi \big( V W \big) = \big( \Delta_\xi V \big) W + V \big( \Delta_\xi W \big)\,.
\end{equation}
Note that the gauge transformations $\delta_\xi$ act on the fields $\mathcal{H}^{AB}$ and $\tilde d$ only, whereas the generalized Lie derivative $\mathcal{L}_\xi$ acts on the full tensorial structure. As an instructive example take e.g.
\begin{equation}
  \Delta_\xi \big( D_A \mathcal{H}^{BC} \big) = \delta_\xi \big( D_A \mathcal{H}^{BC} \big) - \mathcal{L}_\xi \big( D_A \mathcal{H}^{BC} \big) = D_A \big( \mathcal{L}_\xi \mathcal{H}^{BC} \big) - \mathcal{L}_\xi \big( D_A \mathcal{H}^{BC} \big)\,.
\end{equation}

We now calculate $\Delta_\xi$ for all sub-terms appearing in the generalized curvature scalar \eqref{eqn:gencurvature}. Finally, we combine these results using the product rule and the linearity of $\Delta_\xi$ to compute $\Delta_\xi \mathcal{R}$. We begin with
\begin{equation}
  \Delta_\xi \big( \nabla_A d \big) = \Delta_\xi \big( D_A \tilde{d} \big) = - \frac{1}{2} D_A \big( D_D \xi^D \big)
\end{equation}
and since
\begin{equation}
  \mathcal{H}^{MN} \nabla_M \nabla_N d =  H^{MN} D_M D_N \tilde{d}
\end{equation}
holds, we only need to consider
\begin{equation}
\Delta_\xi \big( D_A D_B \tilde{d} \big) =  \big( D_A D_B \xi^D \big) D_D \tilde{d} - \frac{1}{2} D_A D_B \big( D_D \xi^D \big) + {F_{BD}}^{C} \big( D_A \xi^D \big) \big( D_C \tilde{d} \big).
\end{equation}
Furthermore, we obtain
\begin{align}
\Delta_\xi \big( \nabla_A \mathcal{H}^{BC} \big) &= 2 D_A D^{(B} \xi_D \mathcal{H}^{C)D} - 2 D_A D_D \xi^{(B} \mathcal{H}^{C)D} + \frac{2}{3} {F^{(B}}_{AE} H^{C)D} \Big( D^E \xi_D - D_D \xi^E \Big) \nonumber \\ &+\frac{4}{3} {F^{(B}}_{DE} \mathcal{H}^{C)E} D_A \xi^D + \frac{2}{3} {F^{(B}}_{DE} \mathcal{H}^{C)E} D^D \xi_A \nonumber \\ &+\frac{2}{3} {F^D}_{AE} \big( D_D \xi^{(B} \big) \mathcal{H}^{C)E} - \frac{2}{3} {F^D}_{AE} \big( D^{(B} \xi_D \big) H^{C)E}
\end{align}
and
\begin{align}
\Delta_\xi & \big( \nabla_A \nabla_B \mathcal{H}^{AB} \big) = \frac{2}{3} F_{ACE} {F^{E}}_{BD} \xi^C D^B \mathcal{H}^{AD} + \frac{4}{3} F_{ACE} {F^{E}}_{BD} \mathcal{H}^{AB} D^C \xi^D - \frac{1}{3} F_{ACE} \mathcal{H}^{AB} D_B D^C \xi^E \nonumber \\ &\,+\frac{2}{3} F_{ACE} \xi^A D^C D_F \mathcal{H}^{EF} + \frac{10}{3} F_{ACE} \mathcal{H}^{AB} D^C D_B \xi^E + 2 F_{ACE} D^A \xi^C D_D \mathcal{H}^{DE} \nonumber \\ &\,+F_{ACE} D^A \mathcal{H}^{DE} D_D \xi^C - \frac{2}{3} F_{ACE} \xi^A D_D D^C \mathcal{H}^{DE} - D_A D_B \xi^C D_C \mathcal{H}^{AB} \nonumber \\ &\,-2D_A \mathcal{H}^{AB} D_C D_B \xi^C - 2 \mathcal{H}^{AB} D_C D_A D_B \xi^C +\frac{2}{27} F_{ACE} F_{BDF} F^{EDF} \mathcal{H}^{BC} \xi^A\,.
\end{align}
On the right hand side, we have canceled all terms of the form
\begin{equation}
  \label{eqn:strongconstflux}
    F_{ABC} \big(D^B \cdot \big) \big(D^C \cdot \big) = \big( D^B \cdot \big) \big( [D_A,D_B] \cdot \big) = 0\,.
\end{equation}
They vanish due to the strong constraint \eqref{eqn:strongconst}. Combining these results, we are finally able to calculate $\Delta_\xi$ of the naive generalized Ricci scalar \eqref{eqn:gencurvature} without the $1/6 F_{ACD} F_B{}^{CD} \mathcal{H}^{AB}$ term. It is denoted as $\tilde{\mathcal R}$ and its failure to transform as a scalar under generalized diffeomorphisms reads
\begin{align}
  \Delta_\xi \mathcal{\tilde{R}} &= \frac{1}{6} \Big(\frac{1}{3} F_{AFH} F_{CGI} {F_{E}}^{HI} \eta_{BD} - \frac{1}{3} F_{ACH} F_{EFI} {F_{G}}^{HI} \eta_{BD}  - F_{ABH} F_{CDF} {F_{EG}}^H \Big) \mathcal{H}^{BC} \mathcal{H}^{DE} \mathcal{H}^{FG} \xi^A \nonumber \\ &\,+\frac{1}{3}F_{ACD} {F^{CD}}_E \mathcal{H}^{AB} D_B \xi^E + \frac{1}{6} F_{ACD} {F^{CD}}_E \mathcal{H}^{AB} D^E \xi_B \nonumber \\ &\,+\frac{1}{6} \Big( F_{IAG} {F^G}_{CD} + F_{CIG} {F^G}_{AD} + F_{ACG} {F^G}_{ID} \Big) \mathcal{H}^{BC} \mathcal{H}^{DE} \xi^A D_E \mathcal{H}^{FI} \nonumber \\ &\,+F_{ACD} D^A \xi_B D^C \mathcal{H}^{BD} - \frac{1}{2} F_{ACD} D^A \mathcal{H}^{BD} D_B \xi^C + F_{ACD} \mathcal{H}^{AB} D^D D^C \xi_B \nonumber \\ &\,-\frac{1}{2} F_{ACD} \mathcal{H}^{AB} \mathcal{H}^{EF} D_F \xi^D D^C \mathcal{H}_{BE} +\frac{1}{2} F_{ACD} \mathcal{H}^{AB} \mathcal{H}^{EF} D^C \xi_E D^D \mathcal{H}_{BF} \label{eqn:deltaxiRlong}\,.
\end{align}
Here, we ordered the terms according to the number of derivatives. All terms with three flat derivatives vanish in the same way as they do for toroidal DFT \cite{Hohm:2010pp}. The third line of \eqref{eqn:deltaxiRlong} vanishes due to the Jacobi identity
\begin{equation}
  F_{AB}{}^E F_{EC}{}^D + F_{CA}{}^E F_{EB}{}^D + F_{BC}{}^E F_{EA}{}^D = 0\,.
\end{equation}
Additionally, one is able to rewrite the first line as
\begin{equation}
\frac{1}{18} \mathcal{H}^{AB} \xi^G \Big( {F_{EA}}^P \mathcal{H}_{PF} + {F_{FA}}^P \mathcal{H}_{PE} \Big) \Big( {F_B}^{J(E} F_{GHJ} + {F_G}^{J(E} F_{HBJ} + {F_H}^{J(E} F_{BGJ} \Big) \mathcal{H}^{F)H}\,,
\end{equation} 
showing that it is zero due to the Jacobi identity, too. Simplifying the remaining terms in \eqref{eqn:deltaxiRlong}, we make use of the O($D,D$) property
\begin{equation}
  \mathcal{H}_{AB} \mathcal{H}^{BC} = \delta_A{}^C \quad \text{and following from it} \quad
  D_D \mathcal{H}_{AB} \mathcal{H}^{BC} = - \mathcal{H}_{AB} D_D \mathcal{H}^{BC}
\end{equation}
which gives rise to
\begin{align}
\Delta_\xi \mathcal{\tilde{R}} &= \frac{1}{3}F_{ACD} {F^{CD}}_E \mathcal{H}^{AB} D_B \xi^E + \frac{1}{6} F_{ACD} {F^{CD}}_E \mathcal{H}^{AB} D^E \xi_B \\ &\,+\frac{1}{2} F_{ACD} D^A \xi_B D^C \mathcal{H}^{BD} + F_{ACD} \mathcal{H}^{AB} D^D D^C \xi_B \nonumber\,.
\end{align}
Due to the antisymmetry of the structure coefficients, we identify
\begin{align}
F_{ACD} \mathcal{H}^{AB} D^D D^C \xi_B &= \frac{1}{2} F_{ACD} \mathcal{H}^{AB} \big[D^D,D^C\big] \xi_B = \frac{1}{2} F_{ACD} \mathcal{H}^{AB} {F^{DC}}_E D^E \xi_B \nonumber \\ &= -\frac{1}{2} F_{ACD} \mathcal{H}^{AB} {F^{CD}}_E D^E \xi_B
\end{align}
and obtain
\begin{equation}
  \Delta_\xi \mathcal{\tilde{R}} = \frac{1}{3}F_{ACD} {F^{CD}}_E \mathcal{H}^{AB} D_B \xi^E - \frac{1}{3} F_{ACD} {F^{CD}}_E \mathcal{H}^{AB} D^E \xi_B +\frac{1}{2} F_{ACD} D^A \xi_B D^C \mathcal{H}^{BD}\,.
\end{equation}
The last term vanishes under the strong constraint \eqref{eqn:strongconstflux} and 
\begin{equation}
  \Delta_\xi \mathcal{\tilde{R}} = \frac{1}{3} \mathcal{H}^{AB} F_{ACD} F_E{}^{CD} \big( D_B \xi^E - D^E \xi_B \big)
  \label{eqn:ricciwithout}
\end{equation}
remains. This non-vanishing failure of $\tilde{\mathcal R}$ to transform like a scalar should be canceled by the term
\begin{align}
  \frac{1}{6} F_{ACD} F_B{}^{CD} \mathcal{H}^{AB}
\end{align}
that we have not taken into account yet. Indeed, $\Delta_\xi$ applied on this term gives rise to
\begin{equation}
  \frac{1}{6} \Delta_\xi \big( F_{ACD} F_B{}^{CD} \mathcal{H}^{AB} \big) =
    -\frac{1}{3} \mathcal{H}^{AB} F_{ACD} F_E{}^{CD} \big( D_B \xi^E - D^E \xi_B \big)
\end{equation}
after remembering $\delta_\xi F_{ABC} = 0$ (gauge transformations act on fluctuations only, but not on background fields). Ultimately, we obtain the desired result
\begin{equation}
  \Delta_\xi \mathcal{R} = \Delta_\xi \tilde{\mathcal R} + \frac{1}{6} \Delta_\xi \big( F_{ACD}F_B{}^{CD} \mathcal{H}^{AB} \big) = 0
\end{equation}
which proves that the generalized curvature \eqref{eqn:gencurvature} is a scalar under generalized diffeomorphisms.

In addition to $\mathcal{R}$, we have to check the transformation behavior of the factor $e^{-2d}$ in the action \eqref{eqn:actiongenricci}. To this end, we first rewrite the generalized Lie derivative of the dilaton fluctuations \eqref{eqn:gendiffHAB&tilded} in terms of covariant derivatives
\begin{equation}\label{eqn:genLietildedcov}
  \mathcal{L}_\xi \tilde d = \xi^A \nabla_A \tilde d - \frac{1}{2} \nabla_A \xi^A
    = \xi^A D_A \tilde d - \frac{1}{2} D_A \xi^A - \frac{1}{6} F^A{}_{AB} \xi^B
\end{equation}
where the last term vanishes due to the unimodularity of the structure coefficients. Next, we consider
\begin{equation}
  \delta_\xi e^{-2 d} = -2 e^{-2 d} \delta_\xi d = -2 e^{-2 d} \mathcal{L}_\xi \tilde d
\end{equation}
where we take into account that the background field $\bar d$ is not affected by gauge transformations. With $\mathcal{L}_\xi \tilde d$ written in terms of covariant derivatives, it is trivial to switch to curved indices. Doing so and plugging in \eqref{eqn:genLietildedcov}, $\delta_\xi e^{-2 d}$ reads
\begin{align}
  \delta_\xi e^{-2 d} &= \xi^I \partial_I e^{-2 d} + e^{-2 d} ( \nabla_I \xi^I  + \xi^I 2 \partial_I \bar d ) = \xi^I \partial_I e^{-2 d} + e^{-2d} ( \nabla_I \xi^I  - \Gamma_{JI}{}^J \xi^I ) \nonumber \\
  &= \xi^I \partial_I e^{-2d} + e^{-2 d} \partial_I \xi^I 
\end{align}
after applying \eqref{eqn:GammaJIJ}. Thus, we see that $e^{-2 d}$ transforms like a scalar density with the weight +1 and the integral over the product $e^{-2 d} \mathcal{R}$, which is equivalent to the action, is invariant.

Besides the action, the generalized Lie derivative \eqref{eqn:genLieVA} transforms covariantly under generalized diffeomorphisms. Indirectly, this property has already been proven by showing the closure of the gauge algebra 
\begin{equation}\label{eqn:closuregaugealg}
  [\mathcal{L}_{\xi_1}, \mathcal{L}_{\xi_2}] V^A = \mathcal{L}_{[\xi_1,\xi_2]_\mathrm{C}} V^A
\end{equation}
in section~\ref{sec:gaugetrafogenmetric}. However to make it more explicit, we consider
\begin{equation}
  \Delta_\xi \mathcal{L}_\lambda V^A = \mathcal{L}_\xi ( \mathcal{L}_\lambda V^A ) - \mathcal{L}_{\mathcal{L}_\xi \lambda} V^A - \mathcal{L}_\lambda ( \mathcal{L}_\xi V^A) = 0\,.
\end{equation}
In combination with \eqref{eqn:closuregaugealg} it vanishes
\begin{equation}
  \Delta_\xi \mathcal{L}_\lambda V^A = \mathcal{L}_{[\xi,\lambda]_\mathrm{C}} V^A - 
  \mathcal{L}_{\mathcal{L}_\xi \lambda} V^A = 0
\end{equation}
after rewriting the C-bracket
\begin{equation}
  [\xi, \lambda]^A_\mathrm{C} = \mathcal{L}_\xi \lambda^A - \frac{1}{2} \nabla^A (\xi_B \lambda^B) 
\end{equation}
in terms of the generalized Lie derivative and the trivial gauge transformation $-1/2 \nabla^A (\xi_B \lambda^B)$.

\subsection{2D-diffeomorphisms}\label{sec:2Ddiff}
Besides the generalized diffeomorphisms discussed in the previous subsection, one can change the doubled coordinates of DFT${}_\mathrm{WZW}$ through the standard Lie derivative. This gives rise to infinitesimal $2D$-diffeomorphisms under which the action \eqref{eqn:actiongenricci} is even manifestly invariant. In order to prove this claim, we follow very similar steps as in the subsection \ref{sec:gendiffinv}. However, in this case we will not apply the strong constraint in any of the following steps.

Again, we start by introducing the failure
\begin{equation}
\Delta_\xi V = \delta_\xi V - L_\xi V
\end{equation}
of an arbitrary quantity $V$ to transform covariantly. Here, we use the standard Lie derivative $L_\xi$ instead of the generalized Lie derivative. The transformation behavior of the generalized vielbein $E_A{}^I$ and the generalized dilaton fluctuations $\tilde d$ is given by
\begin{align}
  \delta_\xi E_A{}^I &= L_\xi E_A{}^I = \xi^J \partial_J E_A{}^I - E_A{}^J \partial_J \xi^I \quad \text{and} \\
  \delta_\xi \tilde{d} &= L_\xi \tilde{d} = \xi^P \delta_P \tilde{d}\,.
\end{align}
From these two equations, we see that  $E_A{}^I$ transforms as a vector and $\tilde d$ as a scalar under 2D-diffeomorphisms. Next, we check the failure
\begin{equation}\label{eqn:DeltaxinablaIVJ}
  \Delta_\xi \big( \nabla_I V^J \big) = \Delta_\xi \big( \partial_I V^J \big) + \Delta_\xi \big( {\Gamma^J}_{IL} \big) V^L
\end{equation}
of the covariant derivative
\begin{equation}
  \nabla_I V^J = \partial_I V^J + \Gamma_{IK}{}^J V^K
\end{equation}
to transform as a covariant quantity. Being called a `covariant' derivative, this failure should vanish of course. We start by calculating the first term in \eqref{eqn:DeltaxinablaIVJ} and obtain
\begin{equation}
  \Delta_\xi \big( \partial_I V^J \big) = - V^K \partial_K \partial_I \xi^J\,.
\end{equation}
The second term is a bit more challenging. In order to evaluate it, we need the definition of the Christoffel symbols
\begin{equation}
  \Gamma_{IJ}{}^K = -\frac{1}{3} \bigl(2 \Omega_{IJ}{}^K + \Omega_{JI}{}^K \bigr),
\end{equation}
where $\Omega_{IJK}$ denotes the coefficients of anholonomy
\begin{equation}\label{eqn:OmegaIJK}
  \Omega_{IJK} = E^A{}_I E^B{}_J E^C{}_K \Omega_{ABC} = -\partial_I E^A{}_J E_{AK}\,
\end{equation}
in curved indices. With these definitions at hand, one obtains
\begin{equation}\label{eqn:DeltaxiGammaIJK}
  \Delta_\xi \Omega_{IJ}{}^K = - \partial_I \partial_J \xi^K
    \quad \text{and finally} \quad
  \Delta_\xi \Gamma_{IJ}{}^K = \partial_I \partial_J \xi^K\,.
\end{equation}
Thus, \eqref{eqn:DeltaxinablaIVJ} gives rise to the expected result
\begin{equation}
  \Delta_\xi \big( \nabla_I V^J \big) = -V^K \partial_K \partial_I \xi^J + V^K \partial_I \partial_K \xi^J = 0
\end{equation}
and $\nabla_I$ is indeed a covariant derivative under 2D-diffeomorphisms.

Even though we have shown the vanishing $\Delta_\xi$ of the covariant derivative applied on a vector this result generalizes to arbitrary tensors. Especially, the failures
\begin{equation}
  \Delta_\xi \big( \nabla_I \mathcal{H}^{JK} \big) = 0
    \quad \text{and} \quad
  \Delta_\xi \big( \nabla_I d \big) = \Delta_\xi \big( \partial_I \tilde{d} \big) = 0
\end{equation}
vanish. The last ingredients in the definition of the generalized curvature scalar \eqref{eqn:gencurvature} are the structure coefficients $F_{IJK}$. Fortunately, their failure to transform covariantly
\begin{equation}
  \Delta_\xi F_{IJ}{}^K = 2 \Omega_{[IJ]}{}^K = \partial_{[I} \partial_{J]} \xi^K = 0
\end{equation}
vanishes, too. Applying the linearity and the product rule of $\Delta_\xi$, we immediately obtain
\begin{equation}
  \Delta_\xi \big( e^{-2 \tilde d} \mathcal{R} \big) = 0, 
\end{equation}
which proves that the product $e^{-2 \tilde d} \mathcal{R}$ transforms as a scalar under 2D-diffeomorphisms. For the action \eqref{eqn:actiongenricci} to be invariant, the remaining factor $e^{-2 \bar d}$ has to transform as a weight +1 scalar density. Indeed, we have
\begin{equation}
  e^{-2\bar d} = \sqrt{|H|}\,,
\end{equation}
which exactly transforms in the right way. Hence, the DFT${}_\mathrm{WZW}$ action exhibits a manifest 2D-diffeomorphism invariance.

Containing covariant derivatives only, the generalized Lie derivative \eqref{eqn:genLieVA} transforms covariantly, too. Hence, it fulfills
\begin{equation}
  \Delta_\xi \mathcal{L}_\lambda V^A = 0\,.
\end{equation}
Rewriting this equation, we obtain
\begin{equation}
  \Delta_\xi \mathcal{L}_\lambda V^A = L_\xi (\mathcal{L}_\lambda V^A) - \mathcal{L}_{L_\xi \lambda} V^A -
    \mathcal{L}_\lambda (L_\xi V^A) = 0 
\end{equation}
giving rise to the algebra
\begin{equation}
  [L_\xi, \mathcal{L}_\lambda] V^A = \mathcal{L}_{L_\xi \lambda} V^A
\end{equation}
which links 2D-diffeomorphisms and generalized diffeomorphisms. Equipped with this algebra, our theory implements an extension of the DFT gauge algebra proposed by Cederwall \cite{Cederwall:2014kxa,Cederwall:2014opa}.

However, there are some important differences we would like to comment on. Cederwall considered a covariant derivative without torsion on an arbitrary, pseudo Riemannian manifold in order to define a generalized Lie derivative formally matching the one of DFT${}_\mathrm{WZW}$. Applying the Bianchi identity without torsion,
\begin{equation}
  R_{[IJK]}{}^L = 0\,,
\end{equation}
he shows that the gauge algebra closes. To this end, he applies a strong constraint which is manifestly covariant
\begin{equation}
  \nabla_I \nabla^I \cdot = 0
\end{equation}
and does not treat $\cdot$ as scalars like in DFT${}_\mathrm{WZW}$. We rather consider a torsionful covariant derivative on a group manifold, a very special case of a pseudo Riemannian manifold. Interestingly, the Bianchi identity with torsion
\begin{equation}
  R_{[IJK]}{}^L + \nabla_{[I} T^L{}_{JK]} - T^M{}_{[IJ} T^L{}_{K]M} = 
  \frac{2}{9} \big( F_{IJ}{}^M F_{MK}{}^L + F_{KI}{}^M F_{MJ}{}^L + F_{JK}{}^M F_{MI}{}^L \big) = 0
\end{equation}
reproduces on the group manifold the Jacobi identity which we use to show the closure of the DFT${}_\mathrm{WZW}$ gauge algebra and the invariance of the action under generalized diffeomorphisms. Thus, one is inclined to conjecture that the whole formalism presented here is not limited to a group manifold as background but could hold for arbitrary pseudo Riemannian manifolds.

\section{Relation to DFT}\label{sec:DFTWZWtoDFT}
We observed that the usual notions of DFT like a generalized Lie derivative, a C-bracket and the strong constraint undergo a natural generalization which encodes the background fields in an intricate way. Both the frame fields and the fluxes of the background appear in these objects, making the theory derived in the last two chapters explicitly background dependent.

The original DFT is claimed to be background independent so that the question arises how it is related to DFT${}_\mathrm{WZW}$. Assuming a geometric group manifold as a background, we study this question now. Based on the generalized metric formulation, we prove that under an additional constraint both theories can be identified. For that purpose, first we introduce a distinguished generalized vielbein in subsection \ref{sec:genvielbein}. Afterwards, we discuss an additional constraint that links the background fields with the fluctuations around it. We call it the extended strong constraint. As subsection \ref{sec:extendedstrongconst} shows, this constraint allows us to identify the covariant fluxes $\mathcal{F}_{ABC}$ of the DFT flux formulation \cite{Geissbuhler:2013uka,Aldazabal:2013sca,Hassler:2014sba} with the structure coefficients $F_{ABC}$ of the group manifold. Applying the extended strong constraint in the subsections~\ref{sec:=gauge} and \ref{sec:=action}, we prove the equivalence of the gauge transformations and the action in both theories. In this context, we will briefly discuss the background independence of DFT.

\subsection{Generalized vielbein}\label{sec:genvielbein}
The starting point for the following discussion is a background generalized vielbein $E_A{}^I$ fulfilling the strong constraint of DFT. In this subsection, we explicitly work with both O($D,D$) and GL($D$)$\times$GL($D$) flat indices. Thus, we apply the notation of section~\ref{sec:doublelorentz} which decorates the latter with a bar. Due to 2D-diffeomorphism invariance proven in section \ref{sec:2Ddiff}, one is not forced to parameterize $E_A{}^I$ with the left/right moving coordinates $x^i$/$x^{\bar i}$. Instead, we choose the momentum $x^i$ and winding $\tilde x_i$ coordinates which are common in the generalized metric formulation of DFT \cite{Hohm:2010pp}. They give rise to
\begin{equation}
  X^I= \begin{pmatrix} \tilde x_i & x^i \end{pmatrix} \,,\quad
  \partial_I= \begin{pmatrix} \tilde{\partial}^i & \partial_i \end{pmatrix}
    \quad \text{and} \quad
  \eta_{IJ}=\begin{pmatrix}
    0 & \delta^i_j \\
    \delta_i^j & 0
  \end{pmatrix}\,.
\end{equation}
A canonical choice for the vielbein in the DFT flux formulation \cite{Geissbuhler:2013uka,Aldazabal:2013sca,Hassler:2014sba} is\footnote{Compared to chapter~\ref{chap:DFTreview}, we flipped the sign of the $B$-field here. This sign is mere convention. It does not change e.g. the NS/NS action~\eqref{sec:loweneffaction}, because $B_{ij} \to - B_{ij}$ is a manifest symmetry of this action and its gauge transformations.}
\begin{equation}\label{eqn:EhatAI}
  E_A{}^I = \begin{pmatrix}
    e^a{}_i & 0 \\
    - e_a{}^j B_{ji} &
    e_a{}^i
  \end{pmatrix}\,.
\end{equation}
The strong constraint of DFT requires that it only depends on half of the coordinates. Without any loss of generality, we choose $E_A{}^I$ to depend on the momentum coordinates $x^i$. Remember that an unbared doubled index indicates that the $\eta$ metric
\begin{equation}
  \eta_{AB}=\begin{pmatrix}
    0 & \delta_b^a \\
    \delta^b_a & 0
  \end{pmatrix}
  \quad \text{and its inverse} \quad
  \eta^{AB}=\begin{pmatrix}
    0 & \delta_a^b \\
    \delta^a_b & 0
  \end{pmatrix}
\end{equation}
are used to lower and raise this index. In order to identify this representation of $\eta$ with the diagonal form \eqref{eqn:etaAB} common in DFT${}_\mathrm{WZW}$, we apply the coordinate independent transformation
\begin{equation}\label{eqn:MbarAB}
  M_{\bar A}{}^B = \begin{pmatrix}
    \eta_{ab} & \delta^b_a \\
    -\eta_{\bar a b} & \delta^b_{\bar a}
  \end{pmatrix}
    \quad \text{with} \quad
    M_{\bar A}{}^C M_{\bar B}{}^D \eta_{CD} = \eta_{\bar A\bar B}\,.
\end{equation}
For the background metric, it yields
\begin{equation}
    M_{\bar A}{}^C M_{\bar B}{}^D S_{CD} = S_{\bar A\bar B}
      \quad \text{with} \quad
    S_{AB} = \begin{pmatrix}
      \eta^{ab} & 0 \\
      0 & \eta_{ab}
    \end{pmatrix}\,.
\end{equation}
Switching to curved indices, $S^{AB}$ gives rise to the generalized metric
\begin{equation}
  H^{IJ} = E_A{}^I S^{AB} E_{\hat B}{}^J = 
    \begin{pmatrix} g_{ij} - B_{ik} g^{kl} B_{lj} & B_{ik} g^{kj} \\
      - g^{ik} B_{kj} & g^{ij}
    \end{pmatrix}\,.
\end{equation}
It is important to note that the canonical generalized vielbein \eqref{eqn:EAI} of DFT${}_\mathrm{WZW}$ is not an O($D,D$) element, because it gives rise to different representations of the $\eta$ metric in flat and curved indices, namely
\begin{equation}
  E_{\bar A}{}^I \eta^{\bar A\bar B} E_{\bar B}{}^J = \eta^{IJ} = 2 \begin{pmatrix} g^{ij} & 0 \\
    0 & - g^{\bar i\bar j}
  \end{pmatrix}\,.
\end{equation}
This is a severe problem if one tries to compare DFT${}_\mathrm{WZW}$ and DFT. A short calculation shows that the generalized vielbein defined in \eqref{eqn:EhatAI} fixes this problem. It fulfills the relation
\begin{equation}\label{eqn:EhatAIO(D,D)}
  E_A{}^I \eta^{AB} E_B{}^J = \eta^{IJ} = \begin{pmatrix} 0 & \delta_i^j \\
      \delta_j^i & 0
    \end{pmatrix}
\end{equation}
and hence is an O($D,D$) matrix.

This new generalized vielbein should give rise to the constant structure coefficients
\begin{equation}
  F_{ABC} = 2 \Omega_{[AB]C} \quad \text{with} \quad
  \Omega_{ABC} = E_{A}{}^I \partial_I E_{B}{}^J E_{C J}
\end{equation}
from which the derivation of DFT${}_\mathrm{WZW}$ in chapter~\ref{chap:groupdft} starts. Unfortunately, this does not work out because the resulting structure coefficients fail to be constant. A workaround is to consider the covariant fluxes
\begin{equation}\label{eqn:FandOmega}
  \mathcal{F}_{ABC} = 3 \Omega_{[ABC]}
\end{equation}
instead. Backgrounds of DFT${}_\mathrm{WZW}$ correspond to a generalized Scherk-Schwarz ansatz which has by definition constant covariant fluxes. Taking the results from section~\ref{sec:dftnongeofluxes} and remembering that the vielbein $e_a{}^i$ and the $B$-field $B_{ij}$ depend on the momentum coordinates $x^i$ only, we obtain
\begin{align}
  {\mathcal F}_{abc} &= - 3 e_a{}^i e_b{}^j e_c{}^k \partial_{[i} B_{jk]} = - H_{abc} = - F_{abc}
  \quad \text{and} \\
  {\mathcal F}^a{}_{bc} &= 2 e_{[b}{}^i \partial_i e_{c]}{}^j e^a{}_j = 2\Omega_{[bc]}{}^a = F^a{}_{bc} \,.
\end{align}
The remaining independent components $\mathcal{F}^{ab}{}_c$ and $\mathcal{F}^{abc}$ vanish. Next, we switch from $\mathcal{F}_{ABC}$ to $\mathcal{F}_{\bar A\bar B\bar C}$ by applying the transformation $M_{\bar A}{}^B$ defined in \eqref{eqn:MbarAB}. Doing so gives rise to the covariant fluxes
\begin{equation}\label{eqn:FABChat}
  \mathcal{F}_{\bar A\bar B\bar C} = \begin{cases}
    \mathcal{F}_{abc} + \eta_{ad} \mathcal{F}^d{}_{bc} +  \eta_{bd} \mathcal{F}_a{}^d{}_c +  \eta_{cd} \mathcal{F}_{ab}{}^d = 2 F_{abc} \\
    \mathcal{F}_{\bar a b c} - \eta_{\bar a\bar d} \mathcal{F}^{\bar d}{}_{b c} +  \eta_{b d} \mathcal{F}_{\bar a}{}^{d}{}_c +  \eta_{cd} \mathcal{F}_{\bar a b}{}^d = 0 \\
    \mathcal{F}_{\bar a\bar b c} - \eta_{\bar a\bar d} \mathcal{F}^{\bar d}{}_{\bar b c} -  \eta_{\bar b\bar d} \mathcal{F}_{\bar a}{}^{\bar d}{}_c +  \eta_{cd} \mathcal{F}_{\bar a\bar b}{}^d =  -2 F_{\bar a\bar b c} \\
    \mathcal{F}_{\bar a\bar b\bar c} - \eta_{\bar a\bar d} \mathcal{F}^{\bar d}{}_{\bar b\bar c} -  \eta_{\bar b\bar d} \mathcal{F}_{\bar a}{}^{\bar d}{}_{\bar c} -  \eta_{\bar c\bar d} \mathcal{F}_{\bar a\bar b}{}^{\bar d} =  -4 F_{\bar a\bar b\bar c}\,,
  \end{cases}
\end{equation}
which are constant but still do not match the strict left/right separation in the structure coefficients required to formulate DFT${}_\mathrm{WZW}$. However, there is still a way to cure this problem without spoiling the O($D,D$) property \eqref{eqn:EhatAIO(D,D)}. To this end, we apply a coordinate dependent O($D$)$\times$O($D$) transformation which acts on
\begin{equation}
  E_{\bar A}{}^I = M_{\bar A}{}^B E_B{}^I = \begin{pmatrix}
    e_{a i} - e_a{}^j B_{ji} & e_a{}^i \\
    -e_{a i} - e_a{}^j B_{ji} & e_a{}^i
  \end{pmatrix}
  \quad \text{as} \quad
  {\tilde E}_{\bar A}{}^I = T_{\bar A}{}^{\bar B}(x^i) E_{\bar B}{}^I\,.
\end{equation}
In the second row of $E_{\bar A}{}^I$, we drop the bar over the index $a$ of $e_{a i}$ and $e_a{}^i$ respectively to emphasize that in contrast to \eqref{eqn:EAI} we use  the left mover vielbein only. It is connected to the one for the right movers by the $O(D)$ transformation
\begin{equation}
  e_{\bar a}{}^i = t_{\bar a}{}^b e_b{}^i
    \quad \text{with} \quad
  t_{\bar a}{}^b = \mathcal{K}(t_{\bar a}, g t^b g^{-1})\,,
\end{equation}
where $\mathcal{K}$ denotes the Killing form \eqref{eqn:killingform} introduced in section~\ref{sec:wzwmodel}. This transformation is embedded into
\begin{equation}\label{eqn:backgroundvielbein}
  T_{\bar A}{}^{\bar B} = \begin{pmatrix}
    \delta_a^b & 0\\
    0 & t_{\bar a}{}^b
  \end{pmatrix}
  \quad \text{producing} \quad
  \tilde E_{\bar A}{}^I = \begin{pmatrix}
    e_{a i} - e_a{}^j B_{ji} & e_a{}^i \\
    -e_{\bar a i} - e_{\bar a}{}^j B_{ji} & e_{\bar a}{}^i
  \end{pmatrix}
\end{equation}
which recovers the correct index structure. Due to the coordinate dependence of this transformation, it modifies the coefficients of anholonomy according to
\begin{equation}\label{eqn:Omegatilde}
  {\tilde \Omega}_{\bar A\bar B\bar C} = T_{\bar A}{}^{\bar D} T_{\bar B}{}^{\bar E} T_{\bar C}{}^{\bar F} ( \Omega_{\bar D\bar E\bar F} - E_{\bar D}{}^I \partial_I T_{\bar H\bar E} T^{\bar H}{}_{\bar F} )\,.
\end{equation}
After some algebra and keeping the definition $t_a = - t_{\bar a}$ in mind, we obtain
\begin{equation}
  \partial_i t_{\bar db} t^{\bar d}{}_c = \mathcal{K}([t_b, t_c], t_a) e^a{}_i = e^a{}_i F_{abc}
\end{equation}
and finally
\begin{equation}
  E_{\bar A}{}^I \partial_I T_{\bar D\bar B} T^{\bar D}{}_{\bar C} = 2 E_{\bar A}{}^I \begin{pmatrix}
    0 & 0 \\ 0 & - \partial_I t_{\bar d b} t^{\bar d}{}_c
  \end{pmatrix} = - 2 \begin{cases}
    F_{a \bar b\bar c} \\
    F_{\bar a\bar b\bar c} \\
    0 & \text{otherwise.}
  \end{cases}
\end{equation}
This result is nice, because it allows us to fix the problem we encountered with the covariant fluxes $\mathcal{F}_{\bar A\bar B\bar C}$ in \eqref{eqn:FABChat}. After proper antisymmetrization of $\tilde \Omega_{\bar A\bar B\bar C}$, the covariant fluxes for the O($D$)$\times$O($D$) rotated generalized vielbein ${\tilde E}_{\bar A}{}^I$ read
\begin{equation}\label{eqn:covfluxesbackground}
  {\tilde{\mathcal F}}_{\bar A\bar B\bar C} = 2 \begin{cases}
    F_{abc} \\
    F_{\bar a\bar b\bar c} \\
    0 & \text{otherwise}
  \end{cases}
  \quad \text{or in the standard form} \quad
  {\tilde{\mathcal F}}_{\bar A\bar B}{}^{\bar C} = \begin{cases}
    F_{ab}{}^c \\
    - F_{\bar a\bar b}{}^{\bar c} \\
    0 & \text{otherwise.}
  \end{cases}
\end{equation}
They are now compatible with the required left/right separation of the structure coefficients \eqref{eqn:conventFABC}. Thus, via \eqref{eqn:backgroundvielbein} we have succeeded to properly embed the WZW background into the flux formulation of ordinary DFT.

\subsection{Extended strong constraint}\label{sec:extendedstrongconst}
There is still a small but peculiar difference in the two definitions of the structure coefficients
\begin{equation}
  F_{ABC} = 2 \Omega_{[AB]C}
    \quad \text{and the covariant fluxes} \quad
  \mathcal{F}_{ABC} = 3 \Omega_{[ABC]}
\end{equation}
defined in \eqref{eqn:covfluxes}. In order to identify them even so, first note that $\Omega_{ABC}$ is antisymmetric with respect to its last two indices due to O($D,D$) property \eqref{eqn:EhatAIO(D,D)}. Thus, we are able to write
\begin{equation}
  \mathcal{F}_{ABC} = \Omega_{ABC} + \Omega_{CAB} + \Omega_{BCA} = F_{ABC} + \Omega_{CAB}\,.
\end{equation}
Moreover, the purpose of $F_{ABC}$ in DFT${}_\mathrm{WZW}$ is to define the commutator relation
\begin{equation}
  [D_A, D_B] = F_{AB}{}^C D_C
\end{equation}
between flat derivatives. Thus, it is sufficient to study
\begin{equation}\label{eqn:mathcalFABCDC}
  \mathcal{F}_{AB}{}^C D_C \,\cdot = F_{AB}{}^C D_C\, \cdot + \left(D^C E_A{}^I\right) E_B{}_I D_C\, \cdot
\end{equation}
where $\cdot$ denotes arbitrary products of fluctuations $\epsilon^{AB}$, $\tilde d$ and the gauge parameter $\xi^A$ which we also consider as a fluctuation. In DFT${}_\mathrm{WZW}$, the strong constraint only acts on these fluctuations, whereas it does not apply for the background or the relation between background and fluctuations. However, we can of course introduce an additional constraint, the so called extended strong constraint
\begin{equation}\label{eqn:extstrongconst}
  D_A b \, D^A f = 0\,,
\end{equation}
linking background fields $b$ with fluctuations $f$. It restricts all valid field configurations in DFT${}_\mathrm{WZW}$ to a particular subset which allows to cancel the last term in \eqref{eqn:mathcalFABCDC} and therefore to identify $\mathcal{F}_{ABC} = F_{ABC}$. Furthermore, it allows to cancel the last term in the strong constraint in curved indices giving rise to
\begin{equation}\label{eqn:samesc}
  ( \partial_I \partial^I - 2\, \partial_I \bar d\, \partial^I ) \cdot = \partial_I \partial^I \cdot = 0\,,
\end{equation}
which is apparently equivalent to the strong constraint in the traditional DFT formulation.

\subsection{Gauge transformations}\label{sec:=gauge}
Using the covariant fluxes $\mathcal{F}_{ABC}$ instead of the structure coefficients $F_{ABC}$, we have to recalculate the Christoffel symbols of the covariant derivative. To this end, we solve the frame compatibility condition
\begin{equation}
  \nabla_A E_B{}^I = D_A E_B{}^I + \frac{1}{3} \mathcal{F}_{BA}{}^C E_C{}^I + E_A{}^K \Gamma_{KJ}{}^I E_B{}^J = 0
\end{equation}
which gives rise to
\begin{equation}\label{eqn:connectioncovfluxes}
  \Gamma_{IJ}{}^K = -\Omega_{IJ}{}^K + \Omega_{[IJL]}\eta^{LK} = \frac{1}{3} ( -2 \Omega_{IJ}{}^K + \Omega^K{}_{IJ} + \Omega_J{}^K{}_I )\,.
\end{equation}
For this connection, the generalized torsion
\begin{equation}
  \mathcal{T}^I{}_{JK} = 2  \Gamma_{[JK]}{}^I + \Gamma^I{}_{[JK]} = 0
\end{equation}
vanishes. The latter links the C-bracket
\begin{equation}
  [\xi_1, \xi_2]_\mathrm{C}^I = [\xi_1, \xi_2]_\mathrm{DFT,C}^J + \mathcal{T}^I{}_{JK} \xi_1^J \xi_2^K
\end{equation}
of DFT${}_\mathrm{WZW}$ and DFT. Thus, both theories share besides the strong constraint \eqref{eqn:samesc} the same gauge algebra, too. This also holds for the generalized Lie derivative, which can be derived from the C-bracket as
\begin{equation}
  \mathcal{L}_\xi V^I = [\xi, V]^I_\mathrm{C} + \frac{1}{2} \nabla^I (\xi_J V^J) = [\xi, V]^I_\mathrm{DFT,C} + \frac{1}{2} \partial^I (\xi_J V^J) = \mathcal{L}_{\mathrm{DFT,}\xi} V^I \,.
\end{equation}
Even if the Christoffel symbols $\Gamma_{IJ}{}^K$ get modified, they still keep their transformation behavior
\eqref{eqn:DeltaxiGammaIJK} under 2D-diffeomorphisms. In this sense, 2D-diffeomorphisms are still a manifest symmetry of the action and its gauge transformations. However, this symmetry gets partially broken due to the constraint
\begin{equation}
  L_\xi \eta^{IJ} = 0 = \partial^J \xi^I + \partial^I \xi^J
\end{equation}
which preserves the O($D,D$) property \eqref{eqn:EhatAIO(D,D)} of the background generalized vielbein $E_A{}^I$. Furthermore, the strong constraint for $E_A{}^I$ and the extended strong constraint have to transform covariantly which gives rise to the additional restrictions
\begin{align}
  \Delta_\xi ( \partial_I E_A{}^J \partial^I f ) &= -E_A{}^K \partial_K \partial_I \xi^J \partial^I f = 0 \,, \\
  \Delta_\xi ( \partial_I E_A{}^J \partial^I E_B{}^K ) &= - E_A{}^L \partial_L \partial_I \xi^J \partial^I E_B{}^K  - \partial_I E_A{}^J E_B{}^L \partial_L \partial^I \xi^K = 0
\end{align}
requiring
\begin{equation}
  \partial_I \xi^J \partial^I f = 0 \quad \text{and} \quad
  \partial_I \xi^J \partial^I E_A{}^K = 0 \quad \text{or} \quad
  \partial_I \xi^K = \text{const.} \,.
\end{equation}
The latter allows for global O($D,D$) rotations. Besides them, only transformations of the form
\begin{equation}
  L_\xi E_A{}^I = \xi^J \partial_J E_A{}^I + E_A^J \partial_J \xi^I = E_A{}^J \begin{pmatrix}
    0 & 0 \\
    \partial_{[j} \tilde \xi_{i]} & 0 \\
  \end{pmatrix} 
\end{equation}
are possible. They correspond to $B$-field gauge transformations with
\begin{equation}
  B_{ij} \to B_{ij} + \partial_{[i} \xi_{j]}
\end{equation}
and, as well as the global O($D,D$) rotations, can be expressed in terms of generalized diffeomorphisms. Hence, the additional 2D-diffeomorphism invariance of DFT${}_\mathrm{WZW}$ is completely broken by the extended strong constraint \eqref{eqn:extstrongconst} and the O($D,D$) valued background generalized vielbein.

\subsection{Action}\label{sec:=action}
The new connection \eqref{eqn:connectioncovfluxes} has a non-trivial effect on the background dilaton $\bar d$ defined in \eqref{eqn:splitdilaton}, too. To be compatible with integration by parts, $\bar d$ has to fulfill \eqref{eqn:condibp} or equally
\begin{equation}
  \label{eqn:dilatonomega}
  \Omega^J{}_{JI} + 2 \partial_I \bar d = 0
\end{equation}
after using \eqref{eqn:omegaABB=omegaBAB} and the antisymmetry of $\Omega_{IJK}$ in its last two indices.

Subsequently, we show that the action $S$ of DFT${}_\mathrm{WZW}$ in curved indices is equivalent to the traditional DFT action
\begin{align}
  S_\mathrm{DFT} = \int d^{2D} X e^{-2d} \Big( & \frac{1}{8} \mathcal{H}^{KL} \partial_K \mathcal{H}_{IJ} \partial_L \mathcal{H}^{IJ} -\frac{1}{2} \mathcal{H}^{IJ} \partial_{J} \mathcal{H}^{KL} \partial_L \mathcal{H}_{IK} \nonumber \\
  \label{eqn:actionDFT}
  & - 2 \partial_I d \partial_J \mathcal{H}^{IJ} + 4 \mathcal{H}^{IJ} \partial_I d \partial_J d \Big)\,.
\end{align}
Of course,
\begin{equation}
  S = S_\mathrm{DFT}
\end{equation}
only holds under the extended strong constraint \eqref{eqn:extstrongconst}. To prove this identity, we show that 
\begin{equation}
  S - S_\mathrm{DFT} = \int d^{2D} X e^{-2d} \Delta
\end{equation}
vanishes. Expressing all covariant derivatives in terms of partial derivatives and the connection \eqref{eqn:connectioncovfluxes}, $\Delta$ can be simplified to
\begin{align}
  \Delta = & \mathcal{H}^{IJ} \Big( \Omega_{IKL} \Omega^{KL}{}_J - \Omega^K{}_{KI} \Omega^L{}_{LJ} +
    \frac{1}{2} \Omega_{KLI}\Omega^{KL}{}_J \Big) \nonumber \\
    & \quad - \Omega_{IJ}{}^K \partial_K \mathcal{H}^{IJ} + 2 \Omega^K{}_{KI} \mathcal{H}^{IJ} \partial_J \tilde d  - \Omega^K{}_{KI} \partial_J \mathcal{H}^{IJ} + 2 \mathcal{H}^{IJ} \Omega_{IJ}{}^K \partial_K \tilde d\,.
\end{align}
The last term in the first line vanishes under the strong constraint of the background fields. After integration by parts analogous to \eqref{eqn:ibp} and splitting the generalized dilaton according to \eqref{eqn:splitdilaton}, one obtains
\begin{align}
  - \Omega_{IJ}{}^K \partial_K \mathcal{H}^{IJ} &= - 2 \mathcal{H}^{IJ} \Omega_{IJ}{}^K \partial_K \tilde d + \mathcal{H}^{IJ} \Omega_{IJ}{}^K \Omega^L{}_{LK} + \partial_K \Omega_{IJ}{}^K \mathcal{H}^{IJ} \quad \text{and} \\
  - \Omega^{K}_{KI} \partial_J \mathcal{H}^{IJ} &= - 2 \Omega^K{}_{KI} \mathcal{H}^{IJ} \partial_J \tilde d + \mathcal{H}^{IJ} \Omega^K{}_{KI} \Omega^L{}_{LJ} + \mathcal{H}^{IJ} \partial_I \Omega^K{}_{KJ}\,.
\end{align}
Here, we also have applied \eqref{eqn:dilatonomega} to get rid of derivatives acting on $\bar d$. After these substitutions, $\Delta$ reads
\begin{equation}
  \Delta = \mathcal{H}^{IJ} \big( \Omega_{IKL} \Omega^{KL}{}_J + \Omega_{IJ}{}^K \Omega^L{}_{LK} + \partial_K \Omega_{IJ}{}^K + \partial_I \Omega^K{}_{KJ} \big)\,.
\end{equation}
Finally, by taking the definition of $\Omega_{IJK}$ \eqref{eqn:OmegaIJK} into account, it is straightforward to show that
\begin{equation}
  \partial_K \Omega_{IJ}{}^K + \partial_I \Omega^K{}_{KJ} = - \Omega_{IJ}{}^K \Omega^L{}_{LK} -  \Omega_{IKL} \Omega^{KL}{}_J
\end{equation}
holds and thus one obtains the desired result
\begin{equation}
  \Delta = 0\,.
\end{equation}

The calculations shown in this subsection generalize in some sense the endeavor of \cite{Hohm:2010jy} to find a background independent version of the cubic DFT action derived in \cite{Hull:2009mi}. The main idea behind the technically challenging calculations in that paper is: `\ldots one can absorb a constant part of the fluctuation field $e_{ij}$ into a change of the background field $E_{ij}$. The dilaton plays no role in the background dependence; \ldots' (\cite{Hohm:2010jy} page six, first paragraph). In our context, we have a similar situation by splitting the generalized metric into
\begin{equation}
  \mathcal{H}^{IJ} = H^{IJ} + h^{IJ}, \quad \text{where} \quad
  h^{IJ} = \epsilon^{IJ} + \frac{1}{2} \epsilon^{IK}H_{KL} \epsilon^{LJ} + \dots,
\end{equation}
i.e. the background field $H^{IJ}$ and the fluctuation field $h^{IJ}$. As opposed to \cite{Hohm:2010jy}, we consider the generalized dilaton \eqref{eqn:splitdilaton}, too. Furthermore, we are not limited to constant background fields, because $H^{IJ}$ is not constant on an arbitrary group manifold. It is only constant for the special case of a torus. For being a consistent background, it only has to fulfill the field equations of the theory. Still, we were able to reproduce the background independence of ordinary DFT proposed by \cite{Hohm:2010jy}.

As we have seen, for this background independence we have to impose the extended strong constraint, which rules out any solutions beyond SUGRA. To this extend, DFT${}_\mathrm{WZW}$ possesses the same background independence as DFT but still allows to have a glimpse at physics not covered by SUGRA. Moreover, the derivation in this subsection shows that DFT breaks the 2D-diffeomorphism invariance of DFT${}_\mathrm{WZW}$. Especially in the context of the doubled sigma model with a manifest 2D-diffeomorphism invariance like e.g. in \cite{Nibbelink:2013zda}, this could be interesting.

\chapter{Flux formulation}\label{chap:fluxform}
Starting from a worldsheet theory describing a closed string propagating on a group manifold, DFT${}_\mathrm{WZW}$ \cite{Blumenhagen:2014gva,Blumenhagen:2015zma} was derived in the last two chapters. It generalizes some concepts of traditional DFT in an intriguing way and gives rise to an action which is invariant under generalized and $2D$-diffeomorphisms at the same time. At the beginning, this new theory was motivated by studying generalized Scherk-Schwarz compactifications in chapter~\ref{chap:genScherkSchwarz}. Hence, it is natural to ask how they integrate into DFT${}_\mathrm{WZW}$. Central objects in a generalized Scherk-Schwarz compactification are the covariant fluxes $\mathcal{F}_{ABC}$ and $\mathcal{F}_A$. We identify these fluxes in our new framework in section~\ref{sec:covfluxesWZW}. Afterwards, we rewrite the generalized metric action \eqref{eqn:actiongenmetric} through them, yielding a flux formulation \cite{Bosque:2015jda}. Combining it with a slightly adapted generalized Scherk-Schwarz ansatz, section~\ref{sec:genSSWZW} derives the low energy effective theory which arises after the compactification. As expected, this theory describes the bosonic sector of a half-maximal, electrically gauged supergravity. However, we are now able to evade the problems of constructing appropriate twists which were encountered in section~\ref{sec:questiongenSS}. Further, as a top down approach, DFT${}_\mathrm{WZW}$ allows to study the uplift of genuinely non-geometric compactifications to full string theory. Section~\ref{sec:uplift}, discusses this procedure for the duality orbit 1 in table~\ref{tab:solembedding}.

\section{Covariant fluxes}\label{sec:covfluxesWZW}
Before deriving the DFT${}_\mathrm{WZW}$ action in the flux formulation, we first have to fix its constituents, the covariant fluxes. To this end, we introduce the generalized vielbein
\begin{equation}
  \label{eqn:fluxepsilon}
  \mathcal{E}_{\hat{A}}{}^I = \tilde{E}_{\hat{A}}{}^B {E_B}^I
\end{equation}
which combines the background vielbein $E_A{}^I$ with a new vielbein $\tilde{E}_{\hat A}{}^B$ capturing fluctuations around the background. While the former is not O($D,D$) valued, the latter is and thus fulfills
\begin{equation}
  \eta_{AB} = \tilde{E}^{\hat{C}}{}_A \, \eta_{\hat{C}\hat{D}} \, \tilde{E}^{\hat{D}}{}_B
\end{equation}
where $\eta_{AB}$ and $\eta_{\hat A\hat B}$ have exactly the same entries. Further, it allows to express the generalized metric as
\begin{equation}
  \label{eqn:fluxgenmetric}
  \mathcal{H}_{AB} = \tilde{E}^{\hat C}{}_A \, S_{\hat C\hat D} \, \tilde{E}^{\hat D}{}_B\,.
\end{equation}
It is important to distinguish between the different indices of the generalized vielbeins. We already know the curved indices $I$, $J$, $K$, $\dots$ and their flat counter parts. Now, we also use hatted indices like $\hat{A}$, $\hat{B}$, $\hat{C}$, \ldots\,. As we are going to see shortly, these indices are connected to the double Lorentz symmetry we discussed in section~\ref{sec:doublelorentz}.

To get familiar with the new, composite generalized vielbein $\mathcal{E}_{\hat A}{}^I$, we calculate the C-bracket
\begin{align}
  \label{eqn:fluxcbracketepsilon1}
  \big[ \mathcal{E}{}_{\hat{A}}, \mathcal{E}{}_{\hat{B}} \big]_C^J \, \mathcal{E}_{\hat{C}J} &=  2 {\mathcal{E}_{[\hat{A}}}{}^I \partial_I {\mathcal{E}_{\hat{B}]}}{}^J \mathcal{E}_{\hat{C}J} - {\mathcal{E}_{[\hat{A}}}{}^I \partial^J \mathcal{E}_{\hat{B}]I} \mathcal{E}_{\hat{C}J} + {\mathcal{T}^J}_{IK} {\mathcal{E}_{\hat{A}}}{}^I \mathcal{E}_{\hat{B}}{}^K \mathcal{E}_{\hat{C}J} \nonumber\\ 
  &= F_{\hat{A}\hat{B}\hat{C}} +2 D_{[\hat{A}} \, \tilde{E}_{\hat{B]}}{}^D \tilde{E}_{\hat{C}D} - D_{\hat{C}} \, \tilde{E}_{[\hat{B}}{}^D \tilde{E}_{\hat{A}]D}\,.
\end{align}
Note that this result essentially depends on the generalized torsion \eqref{eqn:gentorsion}. Like we use $E_A{}^I$ to switch between flat and curved indices, we apply $\tilde{E}_{\hat A}{}^B$ to obtain the structure coefficients
\begin{equation}
  F_{\hat{A}\hat{B}\hat{C}} = \tilde{E}_{\hat{A}}{}^D \tilde{E}_{\hat{B}}{}^E \tilde{E}_{\hat{C}}{}^F F_{DEF}
\end{equation}
in hatted indices. To simplify \eqref{eqn:fluxcbracketepsilon1}, one defines the coefficients of anholonomy
\begin{equation}
\label{eqn:fluxanholonomy}
  \tilde{\Omega}_{\hat{A}\hat{B}\hat{C}} = \tilde{E}_{\hat{A}}{}^D D_{D} \tilde{E}_{\hat{B}}{}^E \tilde{E}_{\hat{C}E} = D_{\hat A} \tilde{E}_{\hat{B}}{}^E \tilde{E}_{\hat{C}E}
\end{equation}
with
\begin{equation}
  D_{\hat A} = \tilde{E}_{\hat A}{}^B D_B
\end{equation}
for the fluctuations. Because the metric $\eta_{AB}$ is constant and thus can be pulled through flat derivatives, they are antisymmetric with respect to their last two indices:
\begin{equation}
  \tilde{\Omega}_{\hat{A}\hat{B}\hat{C}} = - \tilde{\Omega}_{\hat{A}\hat{C}\hat{B}}\,.
\end{equation}
Finally, we introduce the fluxes
\begin{equation}
  \label{eqn:fluxfluxes}
    \tilde{F}_{\hat{A}\hat{B}\hat{C}} = 3 \tilde{\Omega}_{[\hat{A}\hat{B}\hat{C}]} = \tilde{\Omega}_{\hat{A}\hat{B}\hat{C}} + \tilde{\Omega}_{\hat{B}\hat{C}\hat{A}} + \tilde{\Omega}_{\hat{C}\hat{A}\hat{B}}
\end{equation}
in exactly the same way as they are defined in the flux formulation of traditional DFT. With these definitions,
\eqref{eqn:fluxcbracketepsilon1} simplifies to
\begin{equation}
  \label{eqn:fluxcbracketepsilon2}
  \big[ \mathcal{E}{}_{\hat{A}}, \mathcal{E}{}_{\hat{B}} \big]_C^M \, \mathcal{E}_{\hat{C}M} = F_{\hat{A}\hat{B}\hat{C}} + 2\tilde{\Omega}_{[\hat{A}\hat{B}]\hat{C}} - \tilde{\Omega}_{\hat{C}[\hat{B}\hat{A}]} = F_{\hat{A}\hat{B}\hat{C}} + \tilde{F}_{\hat{A}\hat{B}\hat{C}} = \mathcal{F}_{\hat{A}\hat{B}\hat{C}}
\end{equation}
which allows us to introduce the covariant fluxes $\mathcal{F}_{\hat{A}\hat{B}\hat{C}}$. They decompose into a background part $F_{\hat{A}\hat{B}\hat{C}}$ and a fluctuation part $\tilde{F}_{\hat{A}\hat{B}\hat{C}}$. An alternative way to construct the covariant fluxes makes use of the generalized Lie derivative
\begin{equation}
\mathcal{E}_{\hat{C}M} \, \mathcal{L}_{\mathcal{E}_{\hat{A}}} \mathcal{E}_{\hat{B}}{}^{M} = \big[ \mathcal{E}_{\hat{A}}, \mathcal{E}_{\hat{B}} \big]_C^M \mathcal{E}_{\hat{C}M} + \frac{1}{2} \nabla^M \big( \mathcal{E}_{\hat{A}N} \mathcal{E}_{\hat{B}}{}^N \big) =  \big[ \mathcal{E}_{\hat{A}}, \mathcal{E}_{\hat{B}} \big]_C^M \mathcal{E}_{\hat{C}M} = \mathcal{F}_{\hat{A}\hat{B}\hat{C}}\,.
\end{equation}
By construction, these fluxes are covariant under generalized diffeomorphisms and $2D$-diffeo\-morphisms. Under both, they transform like scalars.

Besides $\mathcal{F}_{ABC}$, the traditional flux formulation contains $\mathcal{F}_A$. Its embedding in the DFT${}_\mathrm{WZW}$ framework follows from the definition
\begin{align}
  \mathcal{F}_{\hat{A}} &= - e^{2d} \mathcal{L}_{\mathcal{E}_{\hat{A}}} e^{-2d} = - e^{2d} \nabla_B \big( \mathcal{E}_{\hat{A}}{}^B e^{-2d} \big) = \tilde{\Omega}^{\hat{B}}{}_{\hat{B}\hat{A}} + 2 D_{\hat{A}} \, \tilde{d} - \mathcal{E}_{\hat A}{}^B e^{2 \bar d} \nabla_B e^{-2\bar d} \nonumber\\  &= 2 D_{\hat{A}} \, \tilde{d} + \tilde{\Omega}^{\hat{B}}{}_{\hat{B}\hat{A}} = \tilde F_{\hat A}\,. \label{eqn:defFA}
\end{align}
Going from the first to the second line, we make use of $\nabla_B e^{-2\bar d} = 0$ which is a direct consequence of the covariant derivative's compatibility with integration by parts. Like the covariant fluxes derived in the last paragraph, $\mathcal{F}_{\hat{A}}$ transforms under generalized and $2D$-diffeomorphisms like a scalar.

\section{Action}
Now, we are ready to derive the action of the DFT${}_\mathrm{WZW}$ flux formulation. Following the original idea \cite{Geissbuhler:2011mx}, we start from the generalized curvature scalar \eqref{eqn:gencurvature} and plug in the generalized metric \eqref{eqn:fluxgenmetric} expressed in terms of the generalized vielbein $\mathcal{E}_{\hat A}{}^I$.

Let us first calculate the term
\begin{align}
\label{eqn:fluxmetrictrafo}
\nabla_{\hat{A}} \mathcal{H}^{\hat{B}\hat{C}} &= \tilde{E}_{\hat{A}}{}^A \tilde{E}^{\hat{B}}{}_B \tilde{E}^{\hat{C}}{}_C \nabla_{A} \mathcal{H}^{BC} \\ &= \tilde{\Omega}_{\hat{A}\hat{D}}{}^{\hat{B}} S^{\hat{D}\hat{C}} + \tilde{\Omega}_{\hat{A}\hat{D}}{}^{\hat{C}} S^{\hat{B}\hat{D}} + \frac{1}{3} F^{\hat{B}}{}_{\hat{A}\hat{D}} S^{\hat{D}\hat{C}} + \frac{1}{3} F^{\hat{C}}{}_{\hat{A}\hat{D}} S^{\hat{B}\hat{D}}
\end{align}
which we are going to need several times in following calculations. Equipped with this result, we obtain for the first two terms in the second line of \eqref{eqn:gencurvature}
\begin{align}
  \frac{1}{8} \mathcal{H}^{CD} \nabla_C \mathcal{H}_{AB} \nabla_D \mathcal{H}^{AB} &= \frac{1}{36} F_{\hat{A}\hat{C}\hat{D}}\, F_{\hat{B}}{}^{\hat{C}\hat{D}} S^{\hat{A}\hat{B}} - \frac{1}{36} F_{\hat{A}\hat{C}\hat{E}}\, F_{\hat{B}\hat{D}\hat{F}}\, S^{\hat{A}\hat{B}} S^{\hat{C}\hat{D}} S^{\hat{E}\hat{F}} \nonumber\\
  &+\frac{1}{4} \tilde{\Omega}_{\hat{A}\hat{C}\hat{D}}\, \tilde{\Omega}_{\hat{B}}{}^{\hat{C}\hat{D}} S^{\hat{A}\hat{B}} - \frac{1}{4} \tilde{\Omega}_{\hat{A}\hat{C}\hat{E}}\, \tilde{\Omega}_{\hat{B}\hat{D}\hat{F}}\, S^{\hat{A}\hat{B}} S^{\hat{C}\hat{D}} S^{\hat{E}\hat{F}} \nonumber \\
  &+ \frac{1}{6} F_{\hat{A}\hat{C}\hat{D}}\, \tilde{\Omega}_{\hat{B}}{}^{\hat{C}\hat{D}} S^{\hat{A}\hat{B}} - \frac{1}{6} F_{\hat{A}\hat{C}\hat{E}}  \,\tilde{\Omega}_{\hat{B}\hat{D}\hat{F}}\, S^{\hat{A}\hat{B}} S^{\hat{C}\hat{D}} S^{\hat{E}\hat{F}}
\end{align}
and
\begin{align}
  - \frac{1}{2} \mathcal{H}^{AB} &\nabla_B \mathcal{H}^{CD} \nabla_D \mathcal{H}_{AC} = \frac{1}{18} F_{\hat{A}\hat{C}\hat{D}}\, F_{\hat{B}}{}^{\hat{C}\hat{D}} S^{\hat{A}\hat{B}}  - \frac{1}{18} F_{\hat{A}\hat{C}\hat{E}}\, F_{\hat{B}\hat{D}\hat{F}}\, S^{\hat{A}\hat{B}} S^{\hat{C}\hat{D}} S^{\hat{E}\hat{F}} \nonumber \\
  &+ \frac{1}{2} \tilde{\Omega}_{\hat{A}\hat{C}\hat{E}}\, \tilde{\Omega}_{\hat{D}\hat{B}\hat{F}}\, S^{\hat{A}\hat{B}} S^{\hat{C}\hat{D}} S^{\hat{E}\hat{F}} - \frac{1}{2} \tilde{\Omega}_{\hat{C}\hat{A}\hat{D}}\, \tilde{\Omega}_{\hat{B}}{}^{\hat{C}\hat{D}} S^{\hat{A}\hat{B}} - \frac{1}{2} \tilde{\Omega}_{\hat{A}\hat{C}\hat{D}}\, \tilde{\Omega}^{\hat{C}}{}_{\hat{B}}{}^{\hat{D}} S^{\hat{A}\hat{B}} \nonumber \\ 
  &- \frac{1}{2} \tilde{\Omega}_{\hat{C}\hat{D}\hat{A}}\, \tilde{\Omega}^{\hat{D}}{}_{\hat{B}}{}^{\hat{C}} S^{\hat{A}\hat{B}} + \frac{1}{3} F_{\hat{A}\hat{C}\hat{D}}\, \tilde{\Omega}_{\hat{B}}{}^{\hat{C}\hat{D}} S^{\hat{A}\hat{B}} - \frac{1}{3} F_{\hat{A}\hat{C}\hat{E}}  \,\tilde{\Omega}_{\hat{B}\hat{D}\hat{F}}\, S^{\hat{A}\hat{B}} S^{\hat{C}\hat{D}} S^{\hat{E}\hat{F}}\,.
\end{align}
The remaining third term in this line yields
\begin{equation}
\label{eqn:fluxterm3}
\frac{1}{6} F_{ACE} F_{BDF} \mathcal{H}^{AB} \eta^{CD} \eta^{EF} = \frac{1}{6} F_{\hat{A}\hat{C}\hat{D}}\, F_{\hat{B}}{}^{\hat{C}\hat{D}} S^{\hat{A}\hat{B}}\,.
\end{equation}
Summing up these three terms and combining appropriate terms into the covariant fluxes $\mathcal{F}_{\hat{A}\hat{B}\hat{C}}$, we find
\begin{gather}
  \frac{1}{8} \mathcal{H}^{CD} \nabla_C \mathcal{H}_{AB} \nabla_D \mathcal{H}^{AB} - \frac{1}{2} \mathcal{H}^{AB} \nabla_B \mathcal{H}^{CD} \nabla_D \mathcal{H}_{AC} +\frac{1}{6} F_{ACE} F_{BDF} \mathcal{H}^{AB} \eta^{CD} \eta^{EF} = \nonumber \\ 
  \frac{1}{4} \mathcal{F}_{\hat{A}\hat{C}\hat{E}} \mathcal{F}_{\hat{B}\hat{D}\hat{F}} S^{\hat{A}\hat{B}} \eta^{\hat{C}\hat{D}} \eta^{\hat{E}\hat{F}} - \frac{1}{12} \mathcal{F}_{\hat{A}\hat{C}\hat{E}}\mathcal{F}_{\hat{B}\hat{D}\hat{F}} S^{\hat{A}\hat{B}} S^{\hat{C}\hat{D}} S^{\hat{E}\hat{F}} \nonumber \\ - \frac{1}{2} \tilde{\Omega}_{\hat{C}\hat{D}\hat{A}}\, \tilde{\Omega}^{\hat{C}\hat{D}}{}_{\hat{B}}\, S^{\hat{A}\hat{B}} - \tilde{\Omega}_{\hat{C}\hat{D}\hat{A}}\, \tilde{\Omega}^{\hat{D}}{}_{\hat{B}}{}^{\hat{C}} S^{\hat{A}\hat{B}} - F_{\hat{A}\hat{C}\hat{D}}\, \tilde{\Omega}^{\hat{C}\hat{D}}{}_{\hat{B}}\, S^{\hat{A}\hat{B}}\,. \label{eqn:fluxresults1}
\end{gather}
Except for the last line, this result looks already quite promising. Subsequently, we calculate the terms in the first line of \eqref{eqn:gencurvature}. They give rise to
\begin{align}
  \label{eqn:fluxterm4}
  4\mathcal{H}^{AB} \nabla_A \nabla_B d &= 4 S^{\hat{A}\hat{B}} D_{\hat{A}} D_{\hat{B}} \, \tilde{d} - 4 S^{\hat{A}\hat{B}} \, \tilde{\Omega}_{\hat{A}\hat{B}}{}^{\hat{C}} D_{\hat{C}} \, \tilde{d}\,, \\
  \label{eqn:fluxterm5}
  -4 \mathcal{H}^{AB} \nabla_A d \, \nabla_B d &= -4 S^{\hat{A}\hat{B}} D_{\hat{A}} \tilde{d} \, D_{\hat{B}} \tilde{d}\,, \\
  \label{eqn:fluxterm6}
  4 \nabla_A d \, \nabla_B \mathcal{H}^{AB} &= - 4 D_{\hat{A}} \tilde{d} \, \tilde{\Omega}^{\hat{C}}{}_{\hat{C}\hat{B}} \, S^{\hat{A}\hat{B}} + 4 S^{\hat{A}\hat{B}} \, \tilde{\Omega}_{\hat{A}\hat{B}}{}^{\hat{C}} D_{\hat{C}} \, \tilde{d} \\
\intertext{and}
  - \nabla_A \nabla_B \mathcal{H}^{AB} &= - S^{\hat{A}\hat{B}}\, \tilde{\Omega}^{\hat{C}}{}_{\hat{C}\hat{A}} \, \tilde{\Omega}^{\hat{D}}{}_{\hat{D}\hat{B}} + S^{\hat{A}\hat{B}} D_{\hat{A}} \, \tilde{\Omega}^{\hat{C}}{}_{\hat{C}\hat{B}} \nonumber\\ 
  &+ \tilde{\Omega}_{\hat{B}\hat{C}}{}^{\hat{A}} S^{\hat{B}\hat{C}} \, \tilde{\Omega}^{\hat{D}}{}_{\hat{D}\hat{A}} - D_{\hat{A}}\, \tilde{\Omega}_{\hat{B}\hat{C}}{}^{\hat{A}} S^{\hat{B}\hat{C}}\,. \label{eqn:fluxterm7}
\end{align}
We rewrite the last two terms of \eqref{eqn:fluxterm7} as
\begin{equation}
  - \tilde{E}_{\hat{A}}{}^A \tilde{E}_{\hat{B}}{}^B \big( D_A D_B \tilde{E}_{\hat{C}}{}^M \big) \tilde{E}^{\hat{A}}{}_M \, S^{\hat{B}\hat{C}} + \tilde{\Omega}_{\hat{C}\hat{D}\hat{A}}\, \tilde{\Omega}^{\hat{D}}{}_{\hat{B}}{}^{\hat{C}} S^{\hat{A}\hat{B}}\,,
\end{equation}
while the last term in the first line of this equation yields
\begin{equation}
  - \tilde{E}_{\hat{A}}{}^A \tilde{E}_{\hat{B}}{}^B \big( D_A D_B \tilde{E}_{\hat{C}}{}^M \big) \tilde{E}^{\hat{A}}{}_M \, S^{\hat{B}\hat{C}} - F_{\hat{A}\hat{C}\hat{D}}\, \tilde{\Omega}^{\hat{C}\hat{D}}{}_{\hat{B}}\, S^{\hat{A}\hat{B}}\,.
\end{equation}
Combining these two results, we find
\begin{align}
  - \nabla_A \nabla_B \mathcal{H}^{AB} &= - S^{\hat{A}\hat{B}}\, \tilde{\Omega}^{\hat{C}}{}_{\hat{C}\hat{A}} \, \tilde{\Omega}^{\hat{D}}{}_{\hat{D}\hat{B}} + 2 S^{\hat{A}\hat{B}} D_{\hat{A}} \, \tilde{\Omega}^{\hat{C}}{}_{\hat{C}\hat{B}} \\
  &+ \tilde{\Omega}_{\hat{C}\hat{D}\hat{A}}\, \tilde{\Omega}^{\hat{D}}{}_{\hat{B}}{}^{\hat{C}} S^{\hat{A}\hat{B}} + F_{\hat{A}\hat{C}\hat{D}}\, \tilde{\Omega}^{\hat{C}\hat{D}}{}_{\hat{B}}\, S^{\hat{A}\hat{B}}\,. \label{eqn:fluxterm7*}
\end{align}
In total, the terms in the first line of \eqref{eqn:gencurvature} give rise to
\begin{gather}
  4 \mathcal{H}^{AB} \nabla_A \nabla_B d - \nabla_A \nabla_B \mathcal{H}^{AB} - 4 \mathcal{H}^{AB} \nabla_A d\, \nabla_B d + 4 \nabla_A d \,\nabla_B \mathcal{H}^{AB} = \nonumber \\
  2 S^{\hat{A}\hat{B}} D_{\hat{A}}\,\mathcal{F}_{\hat{B}} - S^{\hat{A}\hat{B}} \mathcal{F}_{\hat{A}} \mathcal{F}_{\hat{B}} + \tilde{\Omega}_{\hat{C}\hat{D}\hat{A}}\, \tilde{\Omega}^{\hat{D}}{}_{\hat{B}}{}^{\hat{C}} S^{\hat{A}\hat{B}} + F_{\hat{A}\hat{C}\hat{D}}\, \tilde{\Omega}^{\hat{C}\hat{D}}{}_{\hat{B}}\, S^{\hat{A}\hat{B}}\,. \label{eqn:fluxresults2}
\end{gather}
Finally, with \eqref{eqn:fluxresults1} and \eqref{eqn:fluxresults2} we arrive at
\begin{align}
\label{eqn:fluxresults3}
\mathcal{R} =& \frac{1}{4} \mathcal{F}_{\hat{A}\hat{C}\hat{D}} \mathcal{F}_{\hat{B}}{}^{\hat{C}\hat{D}} S^{\hat{A}\hat{B}} - \frac{1}{12} \mathcal{F}_{\hat{A}\hat{C}\hat{E}} \mathcal{F}_{\hat{B}\hat{D}\hat{F}}  S^{\hat A\hat B} S^{\hat{C}\hat{D}} S^{\hat{E}\hat{F}} \nonumber \\ 
  & - \frac{1}{2} \tilde{\Omega}_{\hat{C}\hat{D}\hat{A}}\, \tilde{\Omega}^{\hat{C}\hat{D}}{}_{\hat{B}}\, S^{\hat{A}\hat{B}} + 2 S^{\hat{A}\hat{B}} D_{\hat{A}}\,\mathcal{F}_{\hat{B}} - S^{\hat{A}\hat{B}} \mathcal{F}_{\hat{A}} \mathcal{F}_{\hat{B}}\,.
\end{align}
Applying the strong constraint
\begin{equation}
  D_{\hat C} \tilde{E}_{\hat D}{}^A  D^{\hat C} \tilde{E}^{\hat D}{}_B = 0
\end{equation}
for the fluctuations, the first term in the second line vanishes. Analogous to the generalized metric formulation of DFT${}_\mathrm{WZW}$ discussed in the last chapter, the strong constraint is only required for fluctuation. For the background captured by $F_{\hat{A}\hat{B}\hat{C}}$, the Jacobi identity has to hold.
Performing integration by parts
\begin{equation}\label{eqn:ibpDhatA}
  \int d^{2 D}X \, e^{-2d} D_{\hat A} v \, w = \int d^{2 D}X\, ( \mathcal{F}_{\hat A} v \, w - v\, D_{\hat A} w )\,,
\end{equation}
we obtain the action
\begin{align}\label{eqn:actionfluxform}
  S = \int d^{2 D}X \, e^{-2d} \big( S^{\hat{A}\hat{B}} \mathcal{F}_{\hat{A}} \mathcal{F}_{\hat{B}} + \frac{1}{4} \mathcal{F}_{\hat{A}\hat{C}\hat{D}} \, \mathcal{F}_{\hat{B}}{}^{\hat{C}\hat{D}} \, S^{\hat{A}\hat{B}} - \frac{1}{12} \mathcal{F}_{\hat{A}\hat{C}\hat{E}} \, \mathcal{F}_{\hat{B}\hat{D}\hat{F}} \, S^{\hat{A}\hat{B}} S^{\hat{C}\hat{D}} S^{\hat{E}\hat{F}} \big)\,. 
\end{align}
Because it contains only covariant fluxes and no additional flat derivatives, it is manifestly invariant under
generalized diffeomorphisms and $2D$-diffeomorphisms.

\subsection{Strong constraint violating terms}\label{sec:missingFABCFABCterm}
This action reproduces all terms in the traditional flux formulation \eqref{eqn:Sdftfluxform}, except for the strong constraint violating ones. All fluctuations have to fulfill the strong constraint. Hence, they can not contribute to these missing terms. Still, one would expect to find at least background contributions of the form
\begin{equation}\label{eqn:candiateterms}
  F_A F^A \quad \text{or} \quad \frac{1}{6} F_{ABC} F^{ABC}\,.
\end{equation}
To see why these terms are also missing, we go back to the CSFT origins of DFT${}_\mathrm{WZW}$. First, we have considered CFTs with a constant dilaton. Thus, $F_A = 0$ holds and the first term in \eqref{eqn:candiateterms} drops out. Further, remember the expression for the central charge \eqref{eqn:centralcharge} of the closed string's left moving part. It gives rise to the total central charge
\begin{equation}\label{eqn:ctotleft}
  c_\mathrm{tot} = c + c_\mathrm{gh} = D - \frac{D h^\vee}{k} + c_\mathrm{gh} + \mathcal{O}(k^{-1})\,,
\end{equation}
after adding the ghost contribution $c_\mathrm{gh}$. Terms of the order $k^{-2}$ or higher were excluded in the derivation of the action and its gauge transformation. Hence, we also neglect them while computing the central charge. Using \eqref{eqn:defetaab}, we express the second term in \eqref{eqn:ctotleft},
\begin{equation}
  - \frac{D h^\vee}{k} = \frac{\alpha'}{4} F_{ad}{}^c F_{bc}{}^d \eta^{ab}\,,
\end{equation}
in terms of the structure coefficients defined in \eqref{eqn:liealgebra}. Keeping in mind that the same relations hold for the central charge of the anti-chiral, right moving part, we obtain
\begin{equation}
  c_\mathrm{tot} - \bar c_\mathrm{tot} = \frac{\alpha'}{4} \big( F_{ad}{}^c F_{bc}{}^d \eta^{ab} - F_{ad}{}^c F_{bc}{}^d \eta^{ab} \big) = \frac{\alpha'}{2} F_{ACB} F^{ABC}
\end{equation}
by remembering the definitions \eqref{eqn:etaAB} and \eqref{eqn:conventFABC}. This result is proportional to the second term in \eqref{eqn:candiateterms}. Because the CSFT derivations require that both total central charges $c_\mathrm{tot}$ and $\bar c_\mathrm{tot}$ vanish independently, it has to vanish, too. Another interesting effect of this observation is that the scalar curvature
\begin{equation}
  R = \frac{2}{9} F_{ABC} F^{ABC} = R_{ABC}{}^B \eta^{AC} = 0\,,
\end{equation}
which arises from the Riemann curvature tensor \eqref{eqn:RABCD} induced by the covariant derivative $\nabla_A$, has to vanish.

\subsection{Double Lorentz symmetry}
Whereas in the generalized metric formulation local double Lorentz symmetry is manifest, in the flux formulation it is not and we have to check it explicitly. To this end, we consider the infinitesimal version of \eqref{eqn:localodxodsym}. We denote such transformations by
\begin{equation}
  \delta_\Lambda \mathcal{E}_{\hat{A}}{}^I = \Lambda_{\hat{A}}{}^{\hat{B}} \mathcal{E}_{\hat{B}}{}^I
\end{equation}
and they act on hatted, flat indices only. As a generator of a double Lorentz transformation, $\Lambda_{\hat{A}\hat{B}}$ fulfills the identities
\begin{equation}
  \label{eqn:odxodlambda}
  \Lambda_{\hat{A}\hat{B}} = - \Lambda_{\hat{B}\hat{A}} \quad \text{and} \quad
  \Lambda_{\hat{A}\hat{B}} = S_{\hat{A}\hat{C}} \Lambda^{\hat{C}\hat{D}} S_{\hat{D}\hat{B}}\,.
\end{equation}

A small calculation gives rise to the transformation behavior
\begin{align}
  \delta_\Lambda \mathcal{F}_{\hat{A}\hat{B}\hat{C}} &= 3 \big( D_{\hat{[A}} \Lambda_{\hat{B}\hat{C}]} + \Lambda_{[\hat{A}}{}^{\hat{D}} \mathcal{F}_{\hat{B}\hat{C}]\hat{D}} \big) \\
  \delta_\Lambda \mathcal{F}_{\hat{A}} &= D^{\hat{B}} \Lambda_{\hat{B}\hat{A}} + \Lambda_{\hat{A}}{}^{\hat{B}} \mathcal{F}_{\hat{B}}
\end{align}
of the covariant fluxes. The last term in both equations spoils covariance under double Lorentz transformations. Using these results, it is straightforward to calculate
\begin{equation}
  \label{eqn:odxodaction2}
  \delta_\Lambda S = - \int d^{2n} X \, e^{-2d} \Lambda_{\hat{A}}{}^{\hat{C}} \delta^{\hat{A}\hat{B}} \mathcal{Z}_{\hat{B}\hat{C}}
\end{equation}
with
\begin{equation}
  \label{eqn:odxodz1}
  \mathcal{Z}_{\hat{A}\hat{B}} = D^{\hat{C}} \mathcal{F}_{\hat{C}\hat{A}\hat{B}} + 2 D_{[\hat{A}} \mathcal{F}_{\hat{B}]} - \mathcal{F}^{\hat{C}} \mathcal{F}_{\hat{C}\hat{A}\hat{B}}\,.
\end{equation}
We do not present the intermediate steps of this calculation, because it is analogous to the derivation for the flux formulation of traditional DFT \cite{Geissbuhler:2013uka}. To evaluate $\mathcal{Z}_{\hat A\hat B}$, we split the covariant fluxes $\mathcal{F}_{\hat A\hat B\hat C}$ into their fluctuation and background parts according to \eqref{eqn:fluxcbracketepsilon2}. Further, we have to calculate the terms
\begin{align}
  D^{\hat{C}} \tilde{F}_{\hat{C}\hat{A}\hat{B}} &= D^C \big( D_C \tilde{E}_{[\hat{A}}{}^D \tilde{E}_{\hat{B}]D} \big) + \tilde{\Omega}^{\hat{C}}{}_{\hat{C}\hat{D}} \, \tilde{\Omega}^{\hat{D}}{}_{\hat{A}\hat{B}} + \underline{2 D^{\hat{C}} \tilde{\Omega}_{[\hat{A}\hat{B}]\hat{C}}} \nonumber \\
  D^{\hat{C}} F_{\hat{C}\hat{A}\hat{B}} &= \tilde{E}_{\hat{A}}{}^A \tilde{E}_{\hat{B}}{}^B D^C F_{CAB} + \tilde{\Omega}^{\hat{D}}{}_{\hat{D}}{}^{\hat{C}} F_{\hat{C}\hat{A}\hat{B}} + \underline{2 F_{[\hat{A}\hat{C}\hat{D}} \tilde{\Omega}^{\hat{C}\hat{D}}{}_{\hat{B}]}} \nonumber \\
  2 D_{[\hat{A}} \tilde F_{\hat{B}]} &= 2 F_{\hat{A}\hat{B}}{}^{\hat{C}} D_{\hat{C}} \, \tilde{d} + 4 \tilde{\Omega}_{[\hat{A}\hat{B}]}{}^{\hat{C}} D_{\hat{C}} \, \tilde{d} + \underline{2 D_{[\hat{A}} \tilde{\Omega}^{\hat{C}}{}_{\hat{C}\hat{B}]}} \nonumber \\
  - \tilde F^{\hat{C}} F_{\hat{C}\hat{A}\hat{B}} &= - 2 F_{\hat{A}\hat{B}}{}^{\hat{C}} D_{\hat{C}} \, \tilde{d} - \tilde{\Omega}^{\hat{D}}{}_{\hat{D}}{}^{\hat{C}} F_{\hat{C}\hat{A}\hat{B}} \nonumber \\
  - \tilde F^{\hat{C}} \tilde{F}_{\hat{C}\hat{A}\hat{B}} &= - 2 \tilde{\Omega}^{\hat{C}}{}_{\hat{A}\hat{B}} D_{\hat{C}} \, \tilde{d} - 4 \tilde{\Omega}_{[\hat{A}\hat{B}]}{}^{\hat{C}} D_{\hat{C}} \, \tilde{d} - \tilde{\Omega}^{\hat{D}}{}_{\hat{D}}{}^{\hat{C}} \tilde{\Omega}_{\hat{C}\hat{A}\hat{B}} - \underline{2 \tilde{\Omega}^{\hat{D}}{}_{\hat{D}}{}^{\hat{C}} \tilde{\Omega}_{[\hat{A}\hat{B}]\hat{C}}} \nonumber\,.
\end{align}
The underlined terms cancel due to the identity
\begin{equation}
  2 D^{\hat{C}} \tilde{\Omega}_{[\hat{A}\hat{B}]\hat{C}} - 2 \tilde{\Omega}^{\hat{D}}{}_{\hat{D}}{}^{\hat{C}} \, \tilde{\Omega}_{[\hat{A}\hat{B}]\hat{C}} = - 2 F_{[\hat{A}\hat{C}\hat{D}} \tilde{\Omega}^{\hat{C}\hat{D}}{}_{\hat{B}]} - 2 D_{[\hat{A}} \tilde{\Omega}^{\hat{C}}{}_{\hat{C}\hat{B}]}
\end{equation}
which arises after swapping two flat derivatives. Thus, equation \eqref{eqn:odxodz1} yields
\begin{equation}
  \mathcal{Z}_{\hat{A}\hat{B}} = D^C \big( D_C \tilde{E}_{[\hat{A}}{}^D \tilde{E}_{\hat{B}]D} \big) -2 \tilde{\Omega}^{\hat{C}}{}_{\hat{A}\hat{B}} D_{\hat{C}} \, \tilde{d} + \tilde{E}_{\hat{A}}{}^A \tilde{E}_{\hat{B}}{}^B D^C F_{CAB} \label{eqn:odxodz2}\,,
\end{equation}
where the first two terms vanish under the strong constraint. The remaining term gives rise to
\begin{equation}
  \label{eqn:odxodxz3}
  \mathcal{Z}_{\hat{A}\hat{B}} = \tilde{E}_{\hat{A}}{}^A \tilde{E}_{\hat{B}}{}^B D^C F_{CAB}\,.
\end{equation}
Because the structure coefficients $F_{ABC}$ are constant, we finally find $\mathcal{Z}_{\hat A\hat B}=0$ and prove the invariance of the action \eqref{eqn:actionfluxform} under double Lorentz transformations.

\section{Generalized Scherk-Schwarz compactification}\label{sec:genSSWZW}
Equipped with the flux formulation of DFT${}_\mathrm{WZW}$, we now follow the steps presented in the sections~\ref{sec:KKansatz}--\ref{sec:gaugedgravity} to derive a low energy effective action which arises after a generalized Scherk-Schwarz compactification.

As outlined in section~\ref{sec:KKansatz}, we have to distinguish between compact, internal and the extended, external directions. To make this situation manifest, we split the three different types of indices
\begin{align}\label{eqn:genSSindexconv}
  V^{\hat{\tilde{A}}} &= \begin{pmatrix} W_{\hat{a}} & W^{\hat{a}} & W_{\hat{A}} \end{pmatrix} &
  W^{\tilde{B}} &= \begin{pmatrix} V_b & V^b & V^B \end{pmatrix} &
  X^{\tilde{M}} &= \begin{pmatrix} X_\mu & X^\mu & X^M \end{pmatrix} \\
  \intertext{which are relevant for the flux formulation derived in the last section. The external indices $\hat{a}$, $a$ and $\mu$ run from $0$ to $D-n-1$ and their internal counterparts $\hat A$, $A$ and $M$ parameterize a $2n$-dimensional, doubled space. This index convention gives rise to the three different variants of the $\eta$ metric}
  \eta_{\hat{\tilde{A}}\hat{\tilde{B}}} &= \begin{pmatrix} 0 & \delta^{\hat{a}}_{\hat{b}} & 0 \\  
    \delta_{\hat{a}}^{\hat{b}} & 0 & 0 \\ 0 & 0 & \eta_{\hat{A}\hat{B}} \end{pmatrix} &
  \eta_{\tilde{A}\tilde{B}} &= \begin{pmatrix} 0 & \delta^a_b & 0 \\ 
    \delta_a^b & 0 & 0 \\ 0 & 0 & \eta_{AB} \end{pmatrix} &
  \eta_{\tilde{M}\tilde{N}} &= \begin{pmatrix} 0 & \delta^\mu_\nu & 0 \\
    \delta_\mu^\nu & 0 & 0 \\ 0 & 0 & \eta_{MN} \end{pmatrix}
\end{align}
that are used to lower the indices defined in \eqref{eqn:genSSindexconv}. Further, we use the flat, background generalized metric
\begin{equation}
  S_{\hat{\tilde{A}}\hat{\tilde{B}}}=\begin{pmatrix}
    \eta^{\hat a\hat b} & 0 & 0 \\
    0 & \eta_{\hat a\hat b} & 0 \\
    0 & 0 & S_{\hat A\hat B}
  \end{pmatrix} \quad \text{and its inverse} \quad
  S^{\hat{\tilde{A}}\hat{\tilde{B}}}=\begin{pmatrix}
    \eta_{\hat a\hat b} & 0 & 0 \\
    0 & \eta^{\hat a\hat b} & 0 \\
    0 & 0 & S^{\hat A\hat B}
  \end{pmatrix}\,.
\end{equation}

In the next step, we specify the generalized Scherk-Schwarz ansatz for the composite generalized vielbein
\begin{equation}
  \mathcal{E}_{\hat{\tilde{A}}}{}^{\tilde{M}} = \tilde{E}_{\hat{\tilde{A}}}{}^{\tilde{B}}(\mathds{X}) \, {E_{\tilde{B}}}^{\tilde{M}}(\mathds{Y})\,. \label{eqn:WZWSSansatz}
\end{equation}
Its fluctuation part only depends on the external coordinates $\mathds{X}$, while the background part only depends on the internal ones $\mathds{Y}$. In comparison with the ansatz \eqref{eqn:twistscherkschw}, the background generalized vielbein $E_{\tilde B}{}^{\tilde M}$ takes the part of the twist $U^{\hat N}{}_{\hat M}$. In contrast to the twist, it is not restricted to be O($D,D$) valued. This observation solves the problem of constructing an appropriate twist, which was addressed in section~\ref{sec:questiongenSS}: There is always a straightforward way to construct $E_{\tilde B}{}^{\tilde M}$ as the left-invariant Maurer Cartan form on a group manifold. We went through this process for the example of $S^3$ with H-flux in section~\ref{sec:examplesu(2)}.

For the fluctuation vielbein $\tilde{E}_{\hat{\tilde{A}}}{}^{\tilde{B}}$, the Kaluza-Klein ansatz \eqref{eqn:vielbeinKKansatz} is adapted to the index structure introduced above and gives rise to
\begin{equation}\label{eqn:KKansatzWZW}
  \tilde{E}_{\hat{\tilde A}}{}^{\tilde B}(\mathds{X}) = \begin{pmatrix}
    {e_b}{}^{\hat{a}} & 0 & 0 \\ - {e_{\hat{a}}}^c C_{bc} & {e_{\hat{a}}}^b & -{e_{\hat{a}}}^c \widehat{A}^B{}_c \\ \widehat{E}_{\hat{A}}^C \widehat{A}_{C b} & 0 & \widehat{E}_{\hat A}{}^B
\end{pmatrix}
    \quad \text{with} \quad
    C_{ab} = B_{ab} + \frac{1}{2} \widehat{A}^D{}_a \widehat{A}_{D b}\,.
\end{equation}
In this ansatz, $B_{ab}$ denotes the two-form field appearing in the effective theory and
\begin{equation}
  \widehat{\mathcal H}^{CD} = \widehat{E}_{\hat{A}}{}^C S^{\hat{A}\hat{B}} \widehat{E}_{\hat{B}}{}^D
\end{equation}
represents $n^2$ independent scalar fields which form the moduli of the internal space. In analogy with the twist, the background vielbein has only non-vanishing components in the internal directions and reads
\begin{equation}
\label{eqn:scherckbackgroundvielbein}
  E_{\tilde B}{}^{\tilde M}(\mathds{Y}) = \begin{pmatrix}
\delta^b_\mu & 0 & 0 \\ 0 & \delta_b^\mu & 0 \\ 0 & 0 & E_B{}^M
\end{pmatrix}\,.
\end{equation}

With the Kaluza-Klein ansatz \eqref{eqn:KKansatzWZW} and the partial derivative
\begin{equation}
  \partial^{\tilde M} = \begin{pmatrix} \partial_\mu & \partial^\mu & \partial^M \end{pmatrix}\,,
\end{equation}
it is straightforward to calculate the fluxes $\tilde{F}_{\hat{\tilde{A}}\hat{\tilde{B}}\hat{\tilde{C}}}$ and $\tilde{F}_{\hat{\tilde A}}$ defined in \eqref{eqn:fluxfluxes} and \eqref{eqn:defFA}. After some algebra, we obtain the non-vanishing components
\begin{align}
  \tilde{F}_{\hat a\hat b\hat c} &= e_{\hat{a}}{}^{d} e_{\hat{b}}{}^{e} e_{\hat{c}}{}^{f} \, 3 \big( D_{[d} B_{e f]} + \widehat{A}^D{}_{[d} D_e \widehat{A}_{D f]} \big) & 
  \tilde{F}_{\hat a\hat b}{}^{\hat c} &= 2 e_{[\hat a}{}^d D_{d} e_{\hat b]}{}^e e_{e}{}^{\hat c} = \tilde{f}_{\hat a\hat b}^{\hat c} \nonumber \\ 
  \tilde{F}_{\hat{a}\hat{b}\hat{C}} &= - e_{\hat{a}}{}^d e_{\hat{b}}{}^e \widehat{E}_{\hat{C}D} \, 2 D_{[d} \widehat{A}^D{}_{e]} &
  \tilde{F}_{\hat{a}\hat{B}\hat{C}} &=  e_{\hat{a}}{}^d D_d \widehat{E}_{\hat{B}}{}^D \widehat{E}_{\hat{C}D} \nonumber \\
  \tilde{F}_{\hat{a}} &= \tilde f_{\hat a\hat c}^{\hat c} + 2 e_{\hat{a}}{}^b D_b \phi\,.
\intertext{Furthermore, we use the background generalized vielbein \eqref{eqn:scherckbackgroundvielbein} to switch from flat to curved indices, namely}
  \tilde{F}_{\hat{a}\hat{b}\hat{c}} &= e_{\hat{a}}{}^{\mu} e_{\hat{b}}{}^{\nu} e_{\hat{c}}{}^{\rho} \, 3 \big( \partial_{[\mu} B_{\nu \rho]} + \widehat{A}_{[\mu}{}^M \partial_\nu \widehat{A}_{\rho] M} \big) &
  \tilde{F}_{\hat{a}\hat{b}}{}^{\hat{c}} &= 2 e_{[\hat{a}}{}^\mu \partial_{\mu} e_{\hat{b}]}{}^\nu e_{\nu}{}^{\hat{c}} = \tilde{f}_{\hat a\hat b}^{\hat c} \nonumber \\   
  \tilde{F}_{\hat{a}\hat{b}\hat{C}} &= - e_{\hat{a}}{}^\mu e_{\hat{b}}{}^\nu  \widehat{E}_{\hat{C}M} \, 2 \partial_{[\mu} \widehat{A}^{M}{}_{\nu]} &
  \tilde{F}_{\hat{a}\hat{B}\hat{C}} &= e_{\hat{a}}{}^\mu \partial_\mu \widehat{E}_{\hat{B}}{}^M \widehat{E}_{\hat{C}M} \nonumber \\ 
  \tilde{F}_{\hat{a}} &= \tilde f_{\hat a\hat c}^{\hat c} + 2 e_{\hat{a}}{}^\mu \partial_\mu \phi\,.
  \label{eqn:genSSfluxfluxes}
\end{align}
In order to determine the covariant fluxes $\mathcal{F}_{\hat{\tilde{A}}\hat{\tilde{B}}\hat{\tilde{C}}}$, we also need to evaluate the background contribution $F_{\hat{\tilde{A}}\hat{\tilde{B}}\hat{\tilde{C}}}$. Because the background vielbein \eqref{eqn:scherckbackgroundvielbein} only depends on internal coordinates, the only non-vanishing components of $F_{\tilde A\tilde B\tilde C}$ are
\begin{equation}
 F_{DEF} = 2 \Omega_{[AB]C}\,.
\end{equation}
They give rise to
\begin{equation}
F_{\hat{\tilde{A}}\hat{\tilde{B}}\hat{\tilde{C}}} =  \tilde{E}_{\hat{\tilde{A}}}{}^D \tilde{E}_{\hat{\tilde{B}}}{}^E \tilde{E}_{\hat{\tilde{C}}}{}^F F_{DEF}
\end{equation}
with the non-vanishing components
\begin{align}
  F_{\hat{a}\hat{b}\hat{c}} &= -e_{\hat{a}}{}^{\mu} e_{\hat{b}}{}^{\nu} e_{\hat{c}}{}^{\rho} A_\mu{}^M A_\nu{}^N A_\rho{}^P F_{MNP} &
  F_{\hat{a}\hat{b}\hat{C}}  &= e_{\hat{a}}{}^{\mu} e_{\hat{b}}{}^{\nu} A_\mu{}^M A_\nu{}^N E_{\hat{C}}{}^P F_{MNP} \nonumber \\ 
  F_{\hat{a}\hat{B}\hat{C}}  &= - e_{\hat{a}}{}^{\mu} A_\mu{}^M E_{\hat{B}}{}^N E_{\hat{C}}{}^P F_{MNP} &
  F_{\hat{A}\hat{B}\hat{C}}  &= E_{\hat{A}}{}^M E_{\hat{B}}{}^N E_{\hat{C}}{}^P F_{MNP} \,.
\end{align}
Combining these results with \eqref{eqn:genSSfluxfluxes} and remembering the gauge covariant quantities
\begin{align}
  \widehat{D}_\mu \widehat{E}_{\hat{A}}{}^{M} &= \partial_\mu \widehat{E}_{\hat{A}}{}^{M} - {F^{M}}_{NP} \widehat{A}_\mu{}^N \widehat{E}_{\hat{A}}{}^{P} \nonumber \\ 
  \widehat{F}_{\mu\nu}{}^M &= 2 \partial_{[\mu} \widehat{A}_{\nu]}{}^M - {F^M}_{NP} \widehat{A}_\mu{}^N \widehat{A}_\nu{}^P \nonumber \\ 
  \widehat{G}_{\mu\nu\rho} &= 3 \partial_{[\mu} B_{\nu\rho]} + \widehat{A}_{[\mu}{}^M \partial_\nu \widehat{A}_{\rho]M} - F_{MNP} \widehat{A}_\mu{}^M \widehat{A}_\nu{}^N \widehat{A}_\rho{}^P\,,
\end{align}
introduced in section~\ref{sec:gaugedgravity}, we finally obtain
\begin{align}
  \mathcal{F}_{\hat{a}\hat{b}\hat{c}} &= e_{\hat{a}}{}^{\mu} e_{\hat{b}}{}^{\nu} e_{\hat{c}}{}^{\rho} \widehat{G}_{\mu\nu\rho} &
  \mathcal{F}_{\hat{a}\hat{b}}{}^{\hat{c}} &= 2 e_{[\hat{a}}{}^\mu \partial_{\mu} e_{\hat{b}]}{}^\nu e_{\nu}{}^{\hat{c}} \nonumber \\ 
  \mathcal{F}_{\hat{a}\hat{b}\hat{C}} &= -e_{\hat{a}}{}^\mu e_{\hat{b}}{}^\nu \widehat{E}_{\hat{C}M} \widehat{F}^M{}_{\mu\nu} &
  \mathcal{F}_{\hat{a}\hat{B}\hat{C}} &= e_{\hat{a}}{}^\mu \widehat{D}_\mu \widehat{E}_{\hat{B}}{}^M \widehat{E}_{\hat{C}M} \nonumber \\
  \mathcal{F}_{\hat{A}\hat{B}\hat{C}} &= \widehat{E}_{\hat{A}}{}^D \widehat{E}_{\hat{B}}{}^E \widehat{E}_{\hat{C}}{}^F \, 2 \Omega_{[DE]F} &
  \mathcal{F}_{\hat{a}} &= f_{ab}^b + 2 e_{\hat{a}}{}^\mu \partial_\mu \phi \label{eqn:fluxcompWZW}
\end{align}
and exactly reproduce the results \eqref{eqn:fluxcomponents} stated in section~\ref{sec:gaugedgravity}. Thus, it is hardly surprising that the effective action \eqref{eqn:ddimeffaction} arises after plugging the covariant fluxes \eqref{eqn:fluxcompWZW} into \eqref{eqn:actionfluxform}.

From this point on, all further calculations proceed as explained in chapter~\ref{chap:genScherkSchwarz}. However, our new approach solves two fundamental ambiguities of generalized Scherk-Schwarz compactifications:
\begin{itemize}
  \item In the DFT${}_\mathrm{WZW}$ framework, the twist is equivalent to the background generalized vielbein $E_A{}^I$. It is constructed in the same way as for traditional Scherk-Schwarz compactifications. This is possible, because the theory possesses standard $2D$-diffeomorphisms. Thus, all mathematical tools available for group manifolds are applicable. We immediately lose these tools, when going to the traditional formulation, because the extended strong constraint which is necessary for this transition, breaks $2D$-diffeomorphism invariance. Hence, it is natural that generalized Scherk-Schwarz compactifications in traditional DFT suffer from the problems outlined in section~\ref{sec:questiongenSS}.
  \item All derivations in the last two chapters and in this one are top down. We started from full bosonic closed string theory and reduced it step by step until we finally arrived at the low energy effective action \eqref{eqn:ddimeffaction}. Thus, we are able to uplift solutions along the dashed path in figure~\ref{fig:bigpicture} to string theory at tree-level. Based on the duality orbit 1 in table~\ref{tab:solembedding}, such an uplift is discussed in the next section.
\end{itemize}

\section{Uplift}\label{sec:uplift}
In the introduction, we discussed the relevance of compactifications on genuinely non-geomet\-ric backgrounds. They are assumed to improve moduli stabilization and broaden our perspective on the string theory landscape. A major problem is that most of these backgrounds are motivated from a lower dimensional perspective. Their uplift is not clear. As a top down approach, DFT${}_\mathrm{WZW}$ helps to shed some light on this uplift process. In the following, we underpin this clam with an explicit example.

To this end, remember the duality orbit 1 in table~\ref{tab:solembedding}. Because the generalized Scherk-Schwarz ansatz \eqref{eqn:WZWSSansatz} prohibits fluctuations in the internal directions, the covariant fluxes in these directions
\begin{equation}
  \mathcal{F}_{ABC} = F_{ABC}
\end{equation}
are equivalent to the background fluxes. Using this identity and switching to bared indices, we obtain
\begin{equation}\label{eqn:fluxsu(2)embedding}
  F_{abc} = \sqrt{2} \epsilon_{abc} (\cos\alpha + \sin\alpha) \quad \text{and} \quad
  F_{\bar a\bar b\bar c} = \sqrt{2} \epsilon_{abc} ( \cos\alpha - \sin\alpha )\,,
\end{equation}
where $\epsilon_{abc}$ denotes the totally antisymmetric tensor in three dimensions with $\epsilon_{123}=1$. All other components vanish. For $\alpha=0$, these structure coefficients reproduce the $S^3$ with $H$-flux presented in section~\ref{sec:examplesu(2)}. A T-duality transformation along all internal directions flips the sign of the right movers structure coefficients $F_{\bar a\bar b\bar c}$. It is equivalent to a $-\pi/2$ shift of $\alpha$ and acts as
\begin{equation}
  M_{mn} \leftrightarrow - \tilde M^{mn}
\end{equation}
on the parameters $M_{mn}$ and $\tilde M^{mn}$ of the embedding tensor discussed in section~\ref{sec:solutionsconstr}. Except for $\alpha=0$, all other backgrounds in the orbit are not T-dual to a geometric counterpart. Hence, these background are genuinely non-geometric. They violate the strong constraint of toroidal DFT. To see this, one computes
\begin{equation}
  F_{ABC} F^{ABC} = \frac{1}{2} ( F_{abc} F^{abc} - F_{\bar a\bar b\bar c} F^{\bar a\bar b\bar c} ) = 6 \sin ( 2\alpha )\,.
\end{equation}
As show in section~\ref{sec:twistedgendiff}, this contraction has to vanish if the strong constraint holds. But, this is only the case for
\begin{equation}
  \alpha = \frac{\pi}{2} n \quad \text{with} \quad n \in \mathds{Z}\,,
\end{equation}
corresponding to the $S^3$ with $H$-flux and its T-dual version. All other backgrounds violate the strong constraint. This result is reflected by the fluxes \eqref{eqn:mapMMtildefluxes}, too. For $\alpha\ne0$, there is $H$- and $R$-flux at the same time.

We conclude that left/right asymmetric WZW models are candidates for the uplift of genuinely non-geometric backgrounds. Until now, this uplift was only studied for locally flat backgrounds in terms of asymmetric orbifolds \cite{Condeescu:2012sp,Condeescu:2013yma}. Here, a possible generalization for curved backgrounds emerges. However, such uplifted theories have to be handled with much care. At tree-level, one has to carefully check whether it is possible to cancel the central charges of the chiral and anti-chiral Virasoro algebra. Further, to guarantee consistency also at loop-level, the modular invariance of the CFT's torus partition function has to be checked.

\chapter{Conclusions and Outlook}\label{chap:conclusion}
In the course of this thesis, we developed DFT${}_\mathrm{WZW}$ \cite{Blumenhagen:2014gva,Blumenhagen:2015zma,Bosque:2015jda}. It originates from a group manifold instead of a torus and generalizes the structures of traditional DFT in an intriguing way. Its action is invariant under both standard $2D$-diffeomorphisms and generalized diffeomorphisms. The former are absent in the traditional framework. Breaking them by imposing the extended strong constraint, our new theory automatically reduces to standard DFT. Because this additional constraint is not required for the consistency of the theory, we have indeed found a generalization and not just a rewriting. Using it, we were able to address two major problems arising in generalized Scherk-Schwarz compactifications. The twist, which characterizes the compactification can now be constructed in the same way as in ordinary Scherk-Schwarz compactifications because it is no longer restricted to be an O($n,n$) element. Further, we are able to study the uplift of genuinely non-geometric backgrounds to closed string theory at tree-level.

These results suggest to rethink our current perspective on DFT. Perhaps, the ambitious approach to combine background and fluctuations into a single object which is governed by a doubled geometry is slightly misleading. Of course, it is very appealing because it gives rise to a background independent formulation \cite{Hohm:2010jy}, but it fails to reproduce the results discussed in this thesis. Further, the proposed doubled geometry suffers from several issues like undetermined components of the connection and the Riemann tensor \cite{Jeon:2010rw,Jeon:2011cn,Hohm:2011si}. To avoid these problems, we offer an alternative picture. It is based on two ingredients: a $2D$-dimensional, geometric background space $N$ and a $D$-dimensional subspace  $M$ which supports the physical fields. While the former is characterized by the non-constant $\eta$ metric of DFT${}_\mathrm{WZW}$, the latter arises as an explicit solution of the strong constraint. In general this solution is not unique and there exist several different subspaces $M_i$ which are embedded into $N$. From a worldsheet perspective, they only characterize different representations of the CFT's primaries. But in target space, totally different spaces arise. This effect is the distinguishing property of T-duality. Thus, the doubled space $N$ unifies all different T-dual theories on a certain background. In this sense, it seizes the fundamental idea of unification this thesis started with. While traditional DFT specifies $N$ for tori, we have generalized this result to the family of group manifolds. It includes the torus as the abelian special case.

This picture is still sketchy and there is a lot more work to do. In the following we collect ideas for future projects which further develop the DFT${}_\mathrm{WZW}$ framework and use it to address recent questions in non-commutative/non-associative geometry and non-geometric branes.
\begin{itemize}
  \item As already outlined above, DFT${}_\mathrm{WZW}$ should implement T-duality on group manifolds. On a torus, an abelian group manifold, T-duality is governed by the Buscher rules \eqref{eqn:buscherrules}. Gauging a U($1$) isometry, they implement so called abelian T-duality. The situation for non-abelian group manifolds, which are also covered by our theory, is more challenging. Here, non-abelian T-duality \cite{Giveon:1993ai,Alvarez:1994np} that is less well understood has to be applied. Nonetheless, it should arise natural as different solutions of the strong constraint.
  \item From the worldsheet perspective, non-abelian T-duality is implemented by gauged Wess-Zumino-Witten models \cite{Sfetsos:1994vz,Polychronakos:2010fg,Polychronakos:2010hd}. Their current algebras arise from a GKO coset construction \cite{Goddard:1984vk}. Hence, it would be instructive to generalize the construction we gave for group manifolds to coset spaces. This seems to be feasible, because the sector of coset CFTs needed to derive a tree-level, cubic, low energy effective action can still be handled analytically.
  \item In a similar spirit, one could apply CSFT to orbifolds of CFTs with K\v{a}c-Moody current algebras to study the global, topological properties of the doubled space $N$. These properties are essential for soliton solutions like non-geometric $Q$- or $R$-branes. They arise after applying the T-duality chain \eqref{eqn:t-dualitychain} to a NS5-brane \cite{deBoer:2012ma,Hassler:2013wsa}. In contrast to a torus with $H$-flux, a single NS5-brane lacks U($1$) isometries in its normal directions. In order to still apply the Buscher rules, it has to be smeared. Hence, only smeared $Q$-brane solutions were recently studied. Interestingly, the near horizon geometry of a single NS5-brane corresponds to a $S^3$ with $H$-fluxes. Thus, DFT${}_\mathrm{WZW}$ seems to be  an appropriate tool to study the T-duality chain of localized NS5-branes. Perhaps it is even possible to find branes which are not T-dual to geometric ones.
  \item We were able to identify some components of the embedding tensor in half-maximal, electrically gauged  supergravity with the structure coefficients of a K\v{a}c-Moody algebra on the corresponding worldsheet CFT. Using this link allows to uplift Scherk-Schwarz compactifications on genuinely non-geometric backgrounds. However, these results only hold at tree-level. It would be instructive to investigate the loop-level, e.g. calculate the torus partition function and check its modular invariance. This could prove or disprove the as of now only conjectured uplift from a large class of genuinely non-geometric backgrounds to full closed string theory.
  \item Finally, one should calculate $\alpha'$ corrections for $\mathrm{DFT}_\mathrm{WZW}$. The results promise to be very interesting. On a group manifold, $\alpha'$ corrections are much richer than on the torus where such corrections have been discussed recently \cite{Hohm:2013jaa,Hohm:2014xsa,Hohm:2014eba}. At non-vanishing $\alpha'$, the closed string back-reacts with the curvature of the target space it probes. Further one obtains  a finite level $k$ on a compact group manifold with a fixed volume. It represents a natural cutoff for the primary fields. Take our standard example $S^3$ with $H$-flux. There, the primaries correspond to hyperspherical harmonics \cite{WenAvery85,LachiezeRey:2005hs} which form the Hilbert space of square integrable function on the $S^3$ for $k\to \infty$. For a finite $k$, one is left with a finite number of these functions. They are not sufficient to capture the dynamics of arbitrary fast changing functions on $S^3$ and so naturally give rise to an uncertainty in position space. This observation is essential for the definition of fuzzy spheres, the generalization of ordinary spheres to a non-commutative space. While the fuzzy two-sphere is only non-commutative, higher dimensional fuzzy spheres like $S^3$ are non-commutative and even non-associative \cite{Ramgoolam:2001zx}. Thus, $\alpha'$ corrections in DFT${}_\mathrm{WZW}$ should automatically include non-commutative and non-associative target space geometries.
\end{itemize}
Generally, $\alpha'$ corrections and possible uplifts are very interesting for string inflation and string phenomenology.

In addition to T-duality, there is also S-duality which connects weakly and strongly coupled theories. It plays an important role in the formulation of EFT \cite{Hohm:2013vpa,Hohm:2013uia,Hohm:2014fxa,Godazgar:2014nqa}, recently proposed. Hence, it is natural to ask whether the framework of DFT${}_\mathrm{WZW}$ can be extended to the U-duality group, combining S- and T-duality.

  \backmatter
  \bibliographystyle{JHEP}
  \bibliography{literatur}
  \cleardoublepage

  \markboth{}{}
\end{document}